\documentclass[12pt,a4paper,openright]{book}
\usepackage[italian,english]{babel}
\addto\extrasenglish{}
\usepackage{frontespizio}
\usepackage[toc,page]{appendix}

\usepackage{ulem}

\usepackage{wrapfig}

\newcommand\Chapter[2]{
	\chapter[#1: {#2}]{#1\\[2ex]\Large#2}
}

\usepackage[utf8]{inputenc}

\usepackage{footnote}
\makesavenoteenv{tabular}

\usepackage[a4paper,top=2.2cm,bottom=2.2cm,left=1.9cm,right=1.3cm]{geometry}
\usepackage[tableposition=top,figureposition=bottom,font=small,format=hang]{caption}

\newcommand{\paginavuota}{\newpage\null\thispagestyle{empty}\newpage}

\usepackage{emptypage}
\usepackage{indentfirst}
\usepackage{multicol}
\usepackage{graphicx}
\usepackage{multirow}
\usepackage{verbatim} 
\usepackage{url} 
\usepackage[colorlinks]{hyperref}
\hypersetup{
	colorlinks = true,
	linkcolor = red,
	citecolor = {green},
	urlcolor = {cyan}
}
\usepackage{amsmath}
\usepackage{siunitx}
\usepackage{textcomp}
\usepackage{amssymb}
\usepackage[autoplay,loop]{animate} 
\usepackage{listings} 
\usepackage{color} 
\usepackage{enumitem}
\usepackage{pdfpages}

\newenvironment{listamia}%
{\begin{itemize}\setlength{\itemsep}{0cm}}
	{\end{itemize}
}

\lstnewenvironment{codice_c}[1][]
{\lstset{basicstyle=\small, columns=fullflexible,
		keywordstyle=\color{red}\bfseries, commentstyle=\color{blue}, language=C, numbers=left, numberstyle=\tiny, stepnumber=2, tabsize=5, numbersep=5pt, showstringspaces=true, frame=tbrl, captionpos=t, float=htbp, #1}}{}
\lstnewenvironment{codice_c_pic}[1][]
{\lstset{basicstyle=\footnotesize, columns=fullflexible,
		keywordstyle=\color{red}\bfseries, commentstyle=\color{blue}, language=C, numbers=left, numberstyle=\tiny, stepnumber=2, tabsize=5, numbersep=5pt, showstringspaces=true, frame=tbrl, captionpos=t, float=htbp, #1}}{}
\lstnewenvironment{codice_c_piccolo}[1][]
{\lstset{basicstyle=\footnotesize, columns=fullflexible,
		keywordstyle=\color{red}\bfseries, commentstyle=\color{blue}, language=C, numbers=left, numberstyle=\tiny, stepnumber=2, tabsize=3, numbersep=5pt, showstringspaces=true, frame=tbrl, captionpos=t, float=htbp, #1}}{}
\lstnewenvironment{codice_m_piccolo}[1][]
{\lstset{language=Matlab,%
		basicstyle=\footnotesize,
		keywordstyle=\color{red}\bfseries, commentstyle=\color{blue},
		numbers=left,
		numberstyle=\tiny,
		stepnumber=2,
		frame=lines,
		captionpos=t, #1
}}{}

\usepackage{setspace}
\usepackage{amsthm}
\theoremstyle{plain}

\theoremstyle{definition}

\theoremstyle{remark}

\newcommand{\dpar}{\mathop{}\!\partial}

\usepackage{fancyhdr}
\makeatletter
\renewcommand*\l@chapter[2]{%
	\ifnum \c@tocdepth >\m@ne
	\addpenalty{-\@highpenalty}%
	\vskip 1.0em \@plus\p@
	\setlength\@tempdima{1.5em}%
	\begingroup
	\parindent \z@ \rightskip \@pnumwidth
	\parfillskip -\@pnumwidth
	\leavevmode \bfseries
	\advance\leftskip\@tempdima
	\hskip -\leftskip
	{#1}\nobreak
	\bgroup\mdseries
	\leaders\hbox{$\m@th
		\mkern \@dotsep mu\hbox{.}\mkern \@dotsep
		mu$}\hfill
	\egroup
	\nobreak\hb@xt@\@pnumwidth{\hss #2}\par
	\penalty\@highpenalty
	\endgroup
	\fi}
\makeatother

\begin{document}

\frontmatter	

\includepdf[pages={1}]{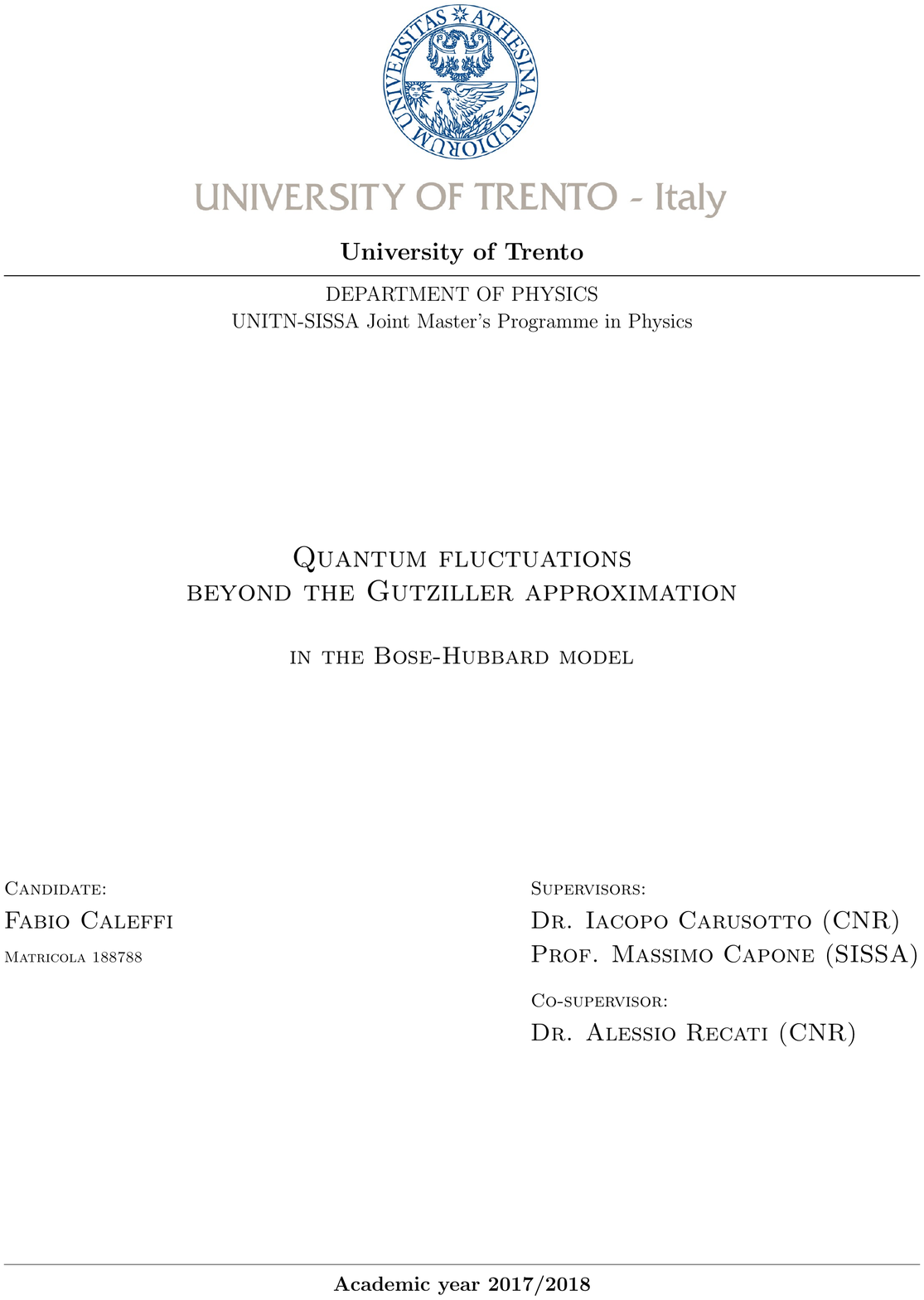}

\selectlanguage{english}

\fontsize{4.5mm}{4.5mm}\selectfont

\paginavuota

\begin{flushright}
	\null\vspace{\stretch{1}}
	\textit{\large{Dedicated to my parents \\ and Luana}}
	\null\vspace{\stretch{3}}
\end{flushright}

\chapter*{Acknowledgements}
The thesis research included in this work could not have been performed if not for the assistance, patience and support of many individuals. \\
I would like to extend my gratitude first and foremost to my thesis supervisors Dr. Iacopo Carusotto and Prof. Massimo Capone for having proposed me to undertake this interdisciplinary project inspired by different but communicating perspectives on low-energy physics. They have helped me over the course of the development and writing of this document by providing me with precious suggestions and always supporting my personal contribution to this research; for these reasons, I sincerely thank them for their confidence in me. \\
I would additionally like to thank Dr. Alessio Recati for his enthusiastic support in both the thesis research and especially the understanding process that has lead to the completion of this work. His knowledge of the physical problems addressed by this work has allowed me to fully express the concepts behind the results of this research. \\
I would also like to extend my appreciation to all the people, from fellow students to professors, which have actively enriched my study experience through useful discussions, comments and remarks. Their important contribution has played a fundamental role. \\
Finally, I would like to extend my deepest gratitude to my parents Davide and Cristina and to my girlfriend Luana, without whose love, support and understanding I could never have completed this master degree.

\tableofcontents

\chapter*{Abstract}
\addcontentsline{toc}{chapter}{Abstract}
Since its first formulation the Bose-Hubbard model has been representing an inexhaustible source of scientific interest because of its peculiar physical properties. In spite of the simplicity of its Hamiltonian, the discover of the complex phase diagram of this model, introduced in 1963 by Gersch and Knollman for studying granular superconductors, allowed to investigate a large class of physical phenomena and, in particular, the physics of ultracold atoms in optical lattices. As a milestone in the study of their dynamical behaviour, the celebrated quantum phase transition from the superfluid into the Mott insulator phase \cite{fisher} is accompanied by the opening of a gap in the excitation spectrum. Exact analytical results can be obtained only  in the weakly interacting limit or for hard-core boson in one dimensions, but relatively recent experimental efforts \cite{nature_1}-\cite{lattice_depth_modulation_2} have confirmed several theoretical predictions performed in different physical regimes using a variety of diverse approaches. \\
Taking inspiration from the state-of-the art knowledge of the model and recent developments proposed in the fermionic case \cite{fabrizio}, this work deals with the study of the collective dynamics of a lattice Bose gas beyond the mean-field picture through a quantum description of its elementary excitations. The Hamiltonian quantization, performed via a Bogoliubov quadratization of the Bose-Hubbard action within the Gutzwiller approach, allows to expand the effective action of the theory up to second order in the fluctuations around the mean-field solution, as well as to prefigure the possibility of identifying the main decay vertices and other effects that are not evident at the second-order level. \\
This quantum description extends the standard Bogoliubov approach to the study of superfluid Bose systems for comprising higher excitation branches, including the Higgs mode in the superfluid phase, and identify their physical meaning together with appropriate observables which could be taken into consideration for their experimental characterization. The ultimate aim of the quantization procedure is the determination of fundamental quantities as the depletion of the condensate and the effective superfluid fraction, which are not accessible by a mean-field description or not completely characterized in the regime of strong interactions.
\begin{flushright}
	\vspace{1.5cm}
	Trento, 24/10/2018
\end{flushright}

\mainmatter

\chapter{Introduction}\label{Introduction}
\markboth{\MakeUppercase{1. Introduction}}{\MakeUppercase{1. Introduction}}
The Hubbard model was originally introduced as a description of the motion of electrons in transition metals, with the motivation of understanding their magnetic properties. In 1981 the outstanding paper by Fisher \text{et al.} \cite{fisher} marked the beginning of a flourishing period of scientific research focusing on strongly-correlated boson systems, of which the Bose-Hubbard model is the most important representative. \\
Apart from its direct physical applications, the importance of the bosonic Hubbard model lies in providing one of the simplest realizations of a quantum phase transition that does not map onto a previously studied classical phase transition, as well as a fundamental picture of the excitation dynamics of a quantum strongly-correlated system.

\medskip

This chapter consists in a brief introduction to the physical model under study based on a reworked version of the picture offered by Sachdev \cite{sachdev}, in order to outline its general properties on the basis of simple considerations. After presenting a short summary of the theoretical and experimental techniques historically used to study the Bose-Hubbard model, we explain the motivations and purposes of this thesis work in connection with crucial questions left almost unresolved by the traditional approaches.

\section{The Bose-Hubbard model}
Let us define the degrees of freedom of the model of interest. We introduce the particle operator $\hat{b}_i$, which annihilates a boson on the site $i$ of a regular lattice in $d$ dimensions. The Bose operators and their hermitian conjugate obey the standard commutation relations:
\begin{equation}\label{1.1}
\left[ \hat{b}_i, \hat{b}^{\dagger}_j \right] = \delta_{i j} \quad , \quad \left[ \hat{b}_i, \hat{b}_j \right] = 0
\end{equation}
We allow an arbitrary number of bosons on each site, so that the Hilbert space is composed by eigenstates of the number operator $\hat{n}_i = \hat{b}^{\dagger}_i \, \hat{b}_i$. \\
The Hamiltonian of the Bose-Hubbard model is:
\begin{equation}\label{1.2}
\hat{H} = -J \sum_{\langle i, j \rangle} \left( \hat{b}^{\dagger}_i \, \hat{b}_j + \text{h.c.} \right) + \frac{U}{2} \sum_{i} \hat{n}_i \left( \hat{n}_i - 1 \right) - \mu \sum_{i} \hat{n}_i
\end{equation}
The first term of (\ref{1.2}) represents the hopping of bosons between nearest neighbour sites, in analogy with the Josephson tunnelling process; the second term is associated to the simplest on-site repulsive contact interaction between bosons; the last term is the chemical potential and its value fixes the mean number of particles in the system. \\
It is important to observe that the Hamiltonian (\ref{1.2}) is invariant under a $U{\left( 1 \right)} \, \equiv O{\left( 2 \right)}$ transformation under which:
\begin{equation}\label{1.3}
\hat{b}_i \to \hat{b}_i \, e^{i \varphi}
\end{equation}
We observe also that the hopping term couples the neighbouring sites in a manner that favours a state that breaks the global symmetry. This term competes with the local interaction, whose eigenstates are invariant under the symmetry transformations. This is the reason why we expect a quantum phase transition to occur as a function of $J/U$ between an invariant state and a phase that breaks spontaneously the $U{\left( 1 \right)}$ symmetry.

\section{Mean-field theory}\label{simple_MF}
\begin{wrapfigure}{l}{0.5\textwidth}
	\centering
	\includegraphics[width=0.5\textwidth]{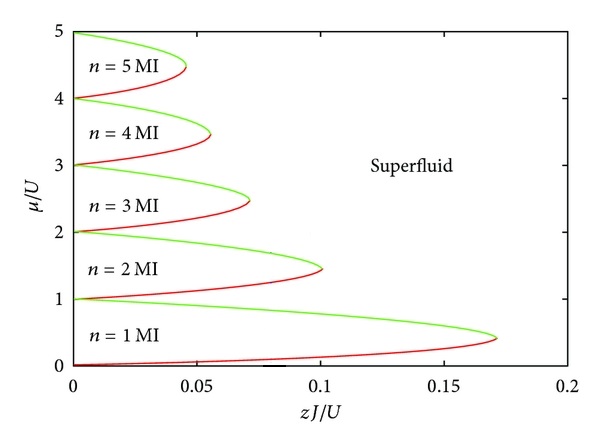}
	\caption{Mean-field phase diagram of the Bose-Hubbard model.}
	\label{Figure}
\end{wrapfigure}

The main strategy of a mean-field theory consists in studying the properties of a physical model by approximating its Hamiltonian as a sum of single-site contributions. In the present case:
\begin{equation}\label{1.4}
\hat{H}_{MF} =\sum_{i} \left[ -\psi^{*} \, \hat{b}_i - \psi \, \hat{b}^{\dagger}_i + \frac{U}{2} \, \hat{n}_i \left( \hat{n}_i - 1 \right) - \mu \hat{n}_i \right]
\end{equation}
where the field $\psi$ represents the mean influence of the neighbouring sites with respect to a given position and has to be determined self-consistently. It is crucial to notice that the presence of this term breaks the $U{\left( 1 \right)}$ symmetry and does not conserve the number of particles; on the other hand symmetric phases appear for $\psi = 0$. We will see that the state that breaks the global symmetry of the system will have a non-zero stiffness to rotations of the order parameter, that is the superfluid density characterizing the ground state condensate of bosons. \\
The optimum value of the mean-field order parameter can be calculated by a standard procedure. The ground state wave function of $\hat{H}_{MF}$ is simply a product of single-site wave functions. One can compute the expectation value of $\hat{H}$ on this mean-field wave function and, by adding and subtracting the analogous quantity of $\hat{H}_{MF}$, obtain the following expression:
\begin{equation}\label{1.5}
\frac{E_0}{M} = \frac{\langle \Psi_{MF} | \hat{H} | \Psi_{MF} \rangle}{M} = \frac{\langle \Psi_{MF} | \hat{H}_{MF} | \Psi_{MF} \rangle}{M} - J \, z \, \langle \hat{b}^{\dagger} \rangle \langle \hat{b} \rangle + \psi^{*} \, \langle \hat{b} \rangle + \psi \, \langle \hat{b}^{\dagger} \rangle
\end{equation}
where $M$ is the number of sites of the lattice, $z = 2 d$ is the coordination number and the average values are calculated with respect to the ground state of $\hat{H}_{MF}$. The final step is to minimize numerically the mean-field energy (\ref{1.5}) with respect to variations of $\psi$. The derivative of the mean-field energy (\ref{1.5}) with respect to $\psi$ returns the optimum value of the order parameter:
\begin{equation}\label{1.6}
\psi = J \, z \, \langle \hat{b} \rangle
\end{equation}
according to our conventions. \\
Every single site can exhibit an infinite number of occupation states. Although the numerical estimation of the mean-field phase diagram requires a truncation of the maximum number of available Fock states, the errors turn out to be not difficult to control beyond a certain number threshold. Nevertheless, we will show now that all the essential properties of the phase diagram can be studied analytically (see Figure \ref{Figure}).

\medskip

First, let us consider the limit $J = 0$. In this case the sites are decoupled, so that the mean-field theory is exact. Since $\psi = 0$ the minimization is performed only on the on-site terms of the Hamiltonian, which depend only on the number operator $\hat{n}_i$. Then the ground state wave function is simply given by:
\begin{equation}\label{1.7}
\Psi{\left( J = 0 \right)} = \prod_i | n_i = n_0{\left( \mu/U \right)} \rangle
\end{equation}
where the integer-value function $n_0{\left( \mu/U \right)}$ has the form:
\begin{equation}\label{1.8}
\begin{aligned}
&n_0{\left( \mu/U \right)} = 0 \quad \text{for} \ \mu/U < 0 \\
&n_0{\left( \mu/U \right)} = n \quad \text{for} \ n - 1 < \mu/U < n \in \mathbb{N}
\end{aligned}
\end{equation}
Thus each site has exactly the same integer number of bosons, which jump discontinuously whenever $\mu/U$ goes through a positive integer. When $\mu/U$ is exactly equal to a positive integer, there are two degenerate state on each site; this large degeneracy implies a macroscopic entropy which is lifted when the hopping parameter is non-zero. \\
We consider now the effect of turning on a non-zero hopping amplitude $J$. The phase diagram shows that the symmetric phase survives in lobes delimiting states with a definite mean local occupation, with the exception of the high-degeneracy points described above. The mean-field wave function associated to the lobe states are again of the form (\ref{1.7}). More exactly and beyond the mean-field point of view, the expectation value of the number of bosons on each site is always:
\begin{equation}\label{1.9}
\langle \hat{b}^{\dagger}_i \, \hat{b}_i \rangle = n_0{\left( \mu/U \right)}
\end{equation}
The previous result is justified in terms of two important facts: the existence of an energy gap in the spectrum of the symmetric phase and the commutation between $\hat{H}$ and $\hat{N}$, that is the particle number operator. The former is identified with the energy separation between the unique ground state in the symmetric phase and all the other occupation eigenstates.  As a result, when we turn on a small non-zero $J$, the ground state shifts adiabatically without undergoing any level crossing with any other state. Now, since the $J = 0$ state is an exact eigenstate of $\hat{N}$ and the perturbation arising from a non-zero $J$ commutes with $\hat{N}$, the ground state will remain an eigenstate of the same operator with the same eigenvalue. This reasoning explains the non-trivial statement (\ref{1.9}). \\
The lobe regions with a quantized value of the density and an energy gap for all the excitations are known as \textsl{Mott insulators}. Their exact ground states involve also terms with bosons undergoing virtual fluctuations between pairs of sites, creating particles and holes. The Mott insulators are also known to be \textsl{incompressible} because their density does not change under a variation of the chemical potential or other parameters in the Hamiltonian:
\begin{equation}\label{1.10}
\frac{\dpar \langle \hat{N} \rangle}{\dpar \mu} = 0
\end{equation}
From the perspective of classical critical phenomena, it is most unusual that the expectation value of any observable is pinned at a quantized value over a finite region of the phase diagram. Quantum field theories of a specific structure allows this kind of phenomenon and, in this sense, the existence of observable that commute with $\hat{H}$ seems to be a crucial requirement. \\
As far as finite temperatures are concerned, strictly speaking there is no true Mott insulating state due to thermal fluctuations. However, there exist quasi-MI regions which have finite but very small compressibility, while the density remains
close to an integer.

\medskip

Analytical results and numerical analyses reveal that the boundary of the Mott lobes is connected to a second-order quantum phase transition, where the order parameter $\psi$ turns on continuously. This information can be used for determining the position of the phase boundaries at a mean field level. \\
Resorting to the Landau-Ginzburg theory of phase transitions, the ground-state energy can be expanded in powers of $\psi$:
\begin{equation}\label{1.11}
E_0 = E^{\left( 0 \right)}_0 + r \left| \psi \right|^2 + \text{O}{\left( \left| \psi \right|^4 \right)}
\end{equation}
and the phase boundary appears when $r$ changes sign. The functional value of $r$ can be computed through second-order perturbation theory applied to the calculation of (\ref{1.5}) (see Appendix \ref{A}), hence:
\begin{equation}\label{1.12}
r = \chi_0{\left( \mu/U \right)} \left[ 1 - J \, z \, \chi_0{\left( \mu/U \right)} \right]
\end{equation}
where:
\begin{equation}\label{1.13}
\chi_0{\left( \mu/U \right)} \equiv \frac{n_0{\left( \mu/U \right)} + 1}{U \, n_0{\left( \mu/U \right)} - \mu} + \frac{n_0{\left( \mu/U \right)}}{\mu - U \, \left[ n_0{\left( \mu/U \right)} - 1 \right]}
\end{equation}
Standard arguments allow to find the phase boundaries by solving the simple equation $r = 0$. \\
The phase with $\psi \neq 0$ presents a continuous variation of the order parameter as the parameters of the theory are varied. As a result all the thermodynamic variables change and the average density does not take a quantized value. It follows that this state is compressible:
\begin{equation}\label{1.14}
\frac{\dpar \langle \hat{N} \rangle}{\dpar \mu} \neq 0
\end{equation}
As anticipated earlier, the presence of a non-zero $\psi$ corresponds the a $U{\left( 1 \right)}$ symmetry breaking, as well as to a non-zero stiffness to twists in the orientation of the order parameter. Appropriate analytical methods show that the broken-symmetry state is a superfluid system whose density is exactly the stiffness to rotations of the systems. \\
The mean-field physical picture discussed so far is quite inaccurate in describing phenomena related to strong correlations, especially in the vicinity of the critical line. On the other hand, Monte Carlo simulations have been carried out systematically and confirm the qualitative value of this first approach.

\section{Quantum field theory: the MI-SF transition}
The emergence of a quantum phase transition in the Bose-Hubbard phase diagram opens new questions about the low energy properties of the system in the vicinity of criticality. As shown by the first literature on the Bose-Hubbard transition \cite{fisher}, it is crucial to distinguish between two different critical behaviours, each characterized by its own universality class. The fingerprint of this distinction can be found in the density trend across the critical line. Passing from the Mott insulator to the superfluid state two possible behaviours occur:
\begin{itemize}
	\item the density locally conserves its integer value when crossing the critical line at the Mott lobe tips; this kind of transition is driven by the onset of boson hopping due to the increase of the ratio $J/U$, i.e. by quantum fluctuations;
	\item the transition is accompanied by a continuous change in the density at any other point of the critical line, which acquires a mean-field character.
\end{itemize}
The first case is described by the $O{\left( 2 \right)}$ universality class, while the second one is associated to a Gaussian critical field theory which recalls the Gross-Pitaevskii action. \\
Let us introduce the partition function of the Hamiltonian (\ref{1.2}) in the well-known coherent state path integral representation continued over imaginary time:
\begin{equation}\label{1.15}
Z = \text{Tr}{e^{-\hat{H}/T}} = \int \mathcal{D} b_i{\left( \tau \right)} \mathcal{D} b^{\dagger}_i{\left( \tau \right)} \exp{\left[ -\int_{0}^{1/T} d\tau \, \mathcal{L} \right]}
\end{equation}
where the Lagrangian density $\mathcal{L}$ is:
\begin{equation}\label{1.16}
\mathcal{L}  = \sum_i \left( b^{\dagger}_i \frac{\dpar}{\dpar \tau} b_i + \frac{U}{2} \, b^{\dagger}_i \, b^{\dagger}_i \, b_i \, b_i - \mu \, b^{\dagger}_i \, b_i \right) - J \sum_{\langle i, j \rangle} \left( b^{\dagger}_i \, b_j + b^{\dagger}_j \, b_i \right)
\end{equation}
where the notation $b_i$ for the bosonic field must not be confused with its operator counterpart in the Hamiltonian language. \\
The critical field theory of the superfluid-insulator transition should be expressed in terms of space-time dependent field $\Psi{\left( x, \tau \right)}$, which is introduced by a Hubbard-Stratanovich transformation on the coherent state path integral. Namely, we decouple the hopping term proportional to $J$ by introducing an auxiliary field $\Psi_i{\left( \tau \right)}$ in the action, so that:
\begin{equation}\label{1.17}
Z = \text{Tr}{e^{-\hat{H}/T}} = \int \mathcal{D} b_i{\left( \tau \right)} \, \mathcal{D} b^{\dagger}_i{\left( \tau \right)} \, \mathcal{D} \Psi_i{\left( \tau \right)} \, \mathcal{D} \Psi^{\dagger}_i{\left( \tau \right)} \, \exp{\left[ -\int_{0}^{1/T} d\tau \, \mathcal{L}' \right]}
\end{equation}
\begin{equation}\label{1.18}
\mathcal{L}' = \sum_i \left( b^{\dagger}_i \frac{\dpar}{\dpar \tau} b_i + \frac{U}{2} \, b^{\dagger}_i \, b^{\dagger}_i \, b_i \, b_i - \mu \, b^{\dagger}_i \, b_i - \Psi_i{\left( \tau \right)} \, b^{\dagger}_i - \Psi^{\dagger}_i{\left( \tau \right)} \, b_i \right) + \sum_{i, j} \Psi^{\dagger}_i{\left( \tau \right)} \, J^{-1}_{i j} \, \Psi_j{\left( \tau \right)}
\end{equation}
where we have introduced the symmetric matrix $J_{i j}$, whose elements are equal to $J$ if $i$ and $j$ are nearest neighbours and vanish otherwise. The path integrals (\ref{1.15}) and (\ref{1.17}) are connected by a straightforward Gaussian integration over $\Psi_i{\left( \tau \right)}$, while irrelevant normalization factors are included in the respective integration measure. Strictly speaking, the transformation connecting (\ref{1.15}) and (\ref{1.17}) requires that all the eigenvalues of $J_{i j}$ are positive. For example, this is not the case for the hypercubic lattice. This inconvenience can be repaired by adding a positive constant to all the diagonal elements of $J_{i j}$ and subtracting the same constant from the on-site of the Hamiltonian.

\vspace{0.2cm}

Let us analyse the relevant symmetries of the path integral action. We notice that the functional integrand (\ref{1.18}) is invariant under the following time-dependent $U{\left( 1 \right)}$ transformation:
\begin{equation}\label{1.19}
\begin{aligned}
&b_i \to b_i \, e^{i \phi{\left( \tau \right)}} \\
&\Psi_i{\left( \tau \right)} \to \Psi_i{\left( \tau \right)} \, e^{i \phi{\left( \tau \right)}} \\
&\mu \to \mu + i \frac{\dpar \phi}{\dpar \tau}
\end{aligned}
\end{equation}
Although a time-dependent chemical potential takes one out of the physical parameter regime, the above class of symmetries places important restrictions on the subsequent manipulations of $Z$. \\
The next step consists in integrating out the bosonic degrees of freedom $b_i$ and $b^{\dagger}_i$. This can be done by expanding the Lagrangian density (\ref{1.18}) in terms of powers of the field $\Psi_i{\left( \tau \right)}$ whose coefficients are the Green's functions of the original bosonic fields. The latter can be determined in closed form because the $\Psi_i{\left( \tau \right)}$-independent part of (\ref{1.18}) is simply a sum of single-site Hamiltonians. These have been diagonalized in the previous section and all the single-site Green's functions can be easily calculated. Finally, we re-exponentiate the resulting series in powers of $\Psi_i{\left( \tau \right)}$ and expand the terms in spatial and temporal gradients. The partition function turns into:
\begin{equation}\label{1.20}
Z = \text{Tr}{e^{-\hat{H}/T}} = \int \mathcal{D} \Psi_i{\left( \tau \right)} \mathcal{D} \, \Psi^{\dagger}_i{\left( \tau \right)} \, \exp{\left[ -\frac{V \mathcal{F}_0}{T}  -\int_{0}^{1/T} d\tau \int d^d x \, \mathcal{L}_{\Psi} \right]}
\end{equation}
\begin{equation}\label{1.21}
\mathcal{L}_{\Psi} = K_1 \Psi^{\dagger}{\left( x, \tau \right)} \frac{\dpar \Psi{\left( x, \tau \right)}}{\dpar \tau} + K_2 \left| \frac{\dpar \Psi{\left( x, \tau \right)}}{\dpar \tau} \right|^2 + K_3 \left| \nabla \Psi{\left( x, \tau \right)} \right|^2 + \tilde{r} \left| \Psi{\left( x, \tau \right)} \right|^2 + \frac{u}{2} \left| \Psi{\left( x, \tau \right)} \right|^4 + \dots
\end{equation}
where $V = M a^d$ is the total volume of the lattice and $a^d$ is the single-site volume. $\mathcal{F}_0$ is the free energy density of a system of decoupled sites; its derivative with respect to the chemical potential gives the density of the Mott insulating state:
\begin{equation}\label{1.22}
n_0{\left( \mu/U \right)} = -a^d \, \frac{\dpar \mathcal{F}_0}{\dpar \mu}
\end{equation}
All the parameters in (\ref{1.21}) are functions of $J$, $U$ and $\mu$. Most important is the parameter $\tilde{r}$, which can be shown to be:
\begin{equation}\label{1.23}
\tilde{r} a^d = \frac{1}{J z} - \chi_0{\left( \mu/U \right)}
\end{equation}
It is important to notice that $\tilde{r}$ is proportional to $r$ defined in (\ref{1.12}). In particular, they have the same sign and vanish in the same points. This is not surprising, because we are analysing the critical points of two communicating mean-field theories. \\
Among the other couplings in (\ref{1.21}), $K_1$ also plays a crucial role. It can be simply fixed by requiring that (\ref{1.21}) is invariant under the gauge transformation (\ref{1.19}) for a small phase $\phi$, so that:
\begin{equation}\label{1.24}
K_1 = -\frac{\dpar \tilde{r}}{\dpar \mu}
\end{equation}
First, we notice that $K_1$ vanishes when $\tilde{r}$ is $\mu$-independent. This is precisely the condition that the Mott insulator-superfluid phase boundary has a vertical tangent, that is at the tips of the Mott lobes. This significant fact leads to the identification of these critical points with the $O{\left( 2 \right)}$ universality class. \\
On the other hand, the case $K_1 \neq 0$ corresponds to a different critical field theory. We can drop the term proportional to $K_2$ as it involves two time derivatives and turns to be an irrelevant field with respect to the single time derivative one. \\
As a final discussion, it is interesting to look at the relationship between this distinction of universality classes at the Mott insulator-superfluid critical line and the behaviour of the boson density across the transition. The latter can be calculated by taking the derivative of the full free energy with respect to the chemical potential:
\begin{equation}\label{1.25}
\langle \hat{b}^{\dagger}_i \, \hat{b}_i \rangle = -a^d \left( \frac{\dpar \mathcal{F}_0}{\dpar \mu} + \frac{\dpar \mathcal{F}}{\dpar \mu} \right) = n_0{\left( \mu/U \right)} - a^d \, \frac{\dpar \mathcal{F}}{\dpar \mu}
\end{equation}
where $\mathcal{F}$ is the free energy given by the functional integral over the order parameter field $\Psi{\left( x, \tau \right)}$ in (\ref{1.20}). From the mean-field theory point of view $\Psi{\left( x, \tau \right)} = 0$ for $\tilde{r} > 0$, hence $\mathcal{F} = 0$ and:
\begin{equation}\label{1.26}
\langle \hat{b}^{\dagger}_i \, \hat{b}_i \rangle = n_0{\left( \mu/U \right)} \quad \text{for} \ \tilde{r} > 0
\end{equation}
This is clearly the equation qualifying the Mott insulator and, as argued in the previous section, it is an exact result. \\
For $\tilde{r} < 0$ one has:
\begin{equation}\label{1.27}
\Psi{\left( x, \tau \right)} = \sqrt{-\frac{\tilde{r}}{u}}
\end{equation}
and, computing the resulting free energy:
\begin{equation}\label{1.28}
\langle \hat{b}^{\dagger}_i \, \hat{b}_i \rangle = n_0{\left( \mu/U \right)} + a^d \, \frac{\dpar}{\dpar \mu} \left( \frac{\tilde{r}^2}{2 u} \right) \simeq n_0{\left( \mu/U \right)} + \frac{a^d \, \tilde{r}}{u} \frac{\dpar \tilde{r}}{\dpar \mu} \quad \text{for} \ \tilde{r} < 0
\end{equation}
where in the second expression we have ignored the derivative of $u$ as it is less singular than $\tilde{r}$ when the critical point is approached. At the tips of the Mott lobes, where $K_1 = 0$, the leading correction to (\ref{1.28}) vanishes, so that locally the density remains equal to $n_0{\left( \mu/U \right)}$ when crossing the tip along a path parallel to the $J/U$ axis of the phase diagram, in accordance with the condition $K_1 = 0$. Conversely, for the case $K_1 \neq 0$, the transition is always accompanied by a small density change, meaning that its first derivative with respect to the Hamiltonian parameters is discontinuous at the transition. This evidence suggests that the density variation itself could be considered as an order parameter of the transition.

\newpage

\markboth{\MakeUppercase{1. Introduction}}{\MakeUppercase{1.4 Review of modern studies}}
\section{Modern theoretical studies and exploration methods}
\markboth{\MakeUppercase{1. Introduction}}{\MakeUppercase{1.4 Review of modern studies}}
While in Section \ref{simple_MF} we have shown how the a simple mean-field theory is able to capture qualitatively the fundamental properties of the model, its reformulation in terms of a quantum field theory provides exact information about the physics hidden around the criticality separating the Mott and superfluid phases. \\
Since the Bose-Hubbard model is not integrable, more exact analytical results regarding the its ground state properties and excitation dynamics can be obtained only in a few special cases, such as in the limit of weakly-interacting gas \cite{moseley_fialko_ziegler, rey} or hard-core bosons in one dimension \cite{cazalilla}. However, in general, exact results can be derived only with the aid of numerical methods. \\
A large amount of scientific effort has been dedicated to study the spectrum of low-energy collective excitations. Firstly, this has been achieved by means of exact numerical diagonalizations, which, although they can be applied only to small lattices that are far from the thermodynamic limit, allow to capture all the main characteristic features of realistic systems. Larger systems have been analysed through Monte Carlo simulations \cite{qmc_1}-\cite{qmc_3}, but real-time dynamics within quantum Monte Carlo has not yet been performed for experimentally relevant cases \cite{qmc_4, qmc_5}. Nevertheless, ground state properties as well as the real-time dynamics of one-dimensional systems subject to external perturbations were studied by the powerful DMRG method. \\
As far as approximate theoretical and analytical approaches are concerned, the equilibrium features and excitation structure of lattice bosons have been studied using the random phase approximation (RPA) \cite{sheshandri_krishnamurthy_pandit_ramakrishnan}-\cite{hazzard_mueller}, the Schwinger boson approach \cite{huber_altman_buchler_blatter}, the time-dependent variational principle with subsequent quantization within the quantum phase model \cite{huber_theiler_altman_blatter}, the slave boson representation of the Bose-Hubbard model \cite{slave_boson}, the standard-basis operator method \cite{konabe_nikuni_nakamura, ohashi_kitaura_matsumoto} and the Ginzburg-Landau theory \cite{grass_santos_pelster}. All these methods have revealed that the Mott phase is characterized by a gapped spectrum populated by particle-hole modes, while the superfluid phase presents a gapless Goldstone mode and gapped higher branches, among which the so-called Higgs mode, strictly related to the occurrence of Lorentz invariance and particle-hole symmetry \cite{pekker_varma}, plays a special role. However, different approximations adopted by a variety approaches do not always lead to the same results.

\medskip

A number of these theoretical studies have been inspirations for important experimental investigations. In particular, the ground state properties of the Bose-Hubbard model have been thoroughly investigated through the celebrated time-of-flight imaging technique \cite{nature_1}, measure of noise correlations \cite{nature_2}, and single-site microscopy \cite{nature_3}. The observation of collective excitations and dynamical properties, with particular attention to the detection of the Goldstone and Higgs modes in the density response channel, have been also addressed through tilting of the lattice \cite{nature_1}, Bragg spectroscopy \cite{bragg}, and lattice depth modulation \cite{lattice_depth_modulation_1, lattice_depth_modulation_2}.
\begin{figure}[!htp]
	\centering
	\includegraphics[width=0.5\textwidth]{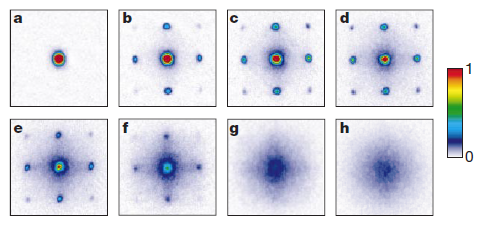}
	\caption{Absorption images of multiple matter wave interference patterns obtained by releasing the atoms of a 3D optical lattice potential. These were obtained after suddenly releasing the atoms from an optical lattice potential after a time of flight of 15 ms. The images are ordered according to the increase of the on-site interaction. We can notice the loss of phase coherence when passing from the superfluid regime to the Mott phase. Figure taken from \cite{nature_1}.}
\end{figure}

\section{Motivations and purposes}
Depending on their domain of applicability, the theoretical approaches briefly presented above have provided a large amount of information about the physics of bosonic lattice systems; however, since some of these methods involve the implementation of different types of approximations and are sometimes affected by analytical or numerical limitations, a comprehensive picture of the Bose-Hubbard dynamical phenomena due to quantum fluctuations remains unattained. For instance, exact approaches based on continuum quantum field theories have the advantage of capturing the long wavelength physics near the phase transition point, but the relationship between what is being observed and what is being computed is not straightforward; on the other hand, quantum Monte Carlo simulations are successfully applicable both outside and inside the quantum critical regime; unfortunately, a direct and systematic comparison of Monte Carlo calculations with the experiments has yet to be made, especially at the level of quantum dynamical effects and response functions, which are usually computed through complicated Hilbert transforms of imaginary-time correlation functions in a few cases. \\
It follows that there are no exhaustive answers to all the open questions regarding the most peculiar feature of the Bose-Hubbard model, that is the existence of a strongly-interacting superfluid phase and the underlying physics of dynamical excitations. In particular, there is still debate about the exact connection between strong correlations and the physical role of high-energy collective modes, especially at the boundary between the critical region and the non-trivial regime where the kinetic and interaction terms have comparable amplitudes. \\
In this sense, from the practical and experimental point of view, recent findings regarding the observation of the collective dynamics and the indirect detection of the Higgs mode near the MI-SF transition \cite{nature_3, lattice_depth_modulation_1, lattice_depth_modulation_2} call for first-principles theoretical treatments of the Bose-Hubbard response dynamics in a wider range of physical parameters where quantum fluctuations cannot be neglected. Moreover, a rigorous theoretical description of the quantum dynamical effects arising in non-uniform systems or due to a confinement potential is still lacking and strongly required for the interpretation of new experimental data and the exploration of possible interesting aspects to be examined in depth. \\
The increasing demand of innovative theoretical instruments and predictions does not originate only in the field of ultracold atoms, but corresponds also to the exceptionally fast advancements in quantum simulation technology, which nowadays allow to recreate complex experimental situations in order to gain new insights into the physical phenomena appearing in strongly-correlated systems. It follows that the large amount of data that is expected to be produced by highly-controllable simulations of real systems needs to be compared with the theoretical predictions provided by simple and systematic speculative methods.

\medskip

Given these premises, the necessity of a more coherent description of the quantum fluctuations occurring in strongly-correlated lattice bosons is strongly required by the progressive emergence of new experimental data open to original interpretations and justifies the search of general and flexible methods which could bridge the gap between the opposite limits of the Bose-Hubbard phase diagram. Since its first formulation for Fermi systems the Gutzwiller approximation has produced important predictions about the behaviour of both strongly-correlated and weakly-interacting systems; these successful results obtained in the context of real materials, together with the observation that the Gutzwiller ansatz matches exactly with the wave function of the pure Mott insulator and the non-interacting Bose gas, has led to the tentative application of the same formalism to Bose lattice systems \cite{rokhsar_kotliar}. Proceeding along this line of research, quite recent results \cite{krutitsky_navez} have revealed the considerable predictive power of the Gutzwiller mean-field treatment, both in describing ground state properties and also extending the spectrum of many-body collective excitations on an equal footing, as well as in pointing out the strict correlation between the particle-hole character of such modes and the oscillations of observable fluctuations \cite{BEC}. \\
Inspired by the methodological path traced these recent developments, the present work aims at improving the Gutzwiller mean-field approach and the promising outcomes of its application to Bose-Hubbard systems by a formal study of the quantum fluctuations living above the ground state level of the approximation. Adopting the Bogoliubov theory of weakly-interacting condensates as a reference, we will draw on a perturbative scheme which could take advantage of the parametric adaptability of the Gutzwiller ansatz and, at least for dimensions larger that $1$, provide reliable results on the quantum and thermal effects due to strong interactions, their relationship with different excitation modes in distinct physical regimes and their impact on the non-linear dynamical properties of the system. The ultimate purpose of this work is to provide a simple and versatile method for studying quantum fluctuations which could provide new evidences in the regime of strong interactions and a more comprehensive picture of the Bose-Hubbard excitation dynamics with a reduced computational complexity, in contrast with more sophisticated theoretical and numerical approaches with a limited range of application or effectiveness.

\section*{Thesis outline}
\hspace{-0.8cm} The outline of this thesis is organized as follows. \\
In Chapter \ref{Bogoliubov} we review the Bogoliubov treatment of weakly-interacting atomic gases in a lattice, with particular focus on the quantum depletion and the corresponding effect on the superfluid density. Chapter \ref{Bogoliubov} ends with a brief discussion on the limitations of the standard Bogoliubov theory in estimating the behaviour of quantum corrections when strong interactions occur. \\
The bosonic formulation of the Gutzwiller approximation is introduced in Chapter \ref{MF}. After a general analysis of the Gutzwiller equations and their connection with the discrete Gross-Pitaevskii equation, we present the mean-field ground state solutions for the Mott and superfluid phase. The mean-field study of the Bose-Hubbard excitation dynamics is illustrated through the theoretical work of Krutitsky and Navez \cite{krutitsky_navez}; in particular, we discuss the spectral features of collective excitations provided by the Gutzwiller approach within linear response theory and how these contribute to the linear fluctuations of local observables. In conclusion, the renormalization of the order parameter and sound velocity due to strong interactions is highlighted. \\
Taking inspiration from the studies performed at the mean-field level demonstrating the versatility of the Gutzwiller method, in Chapter \ref{quantum} we follow a conservative and perturbative scheme for expanding the action of the Gutzwiller degrees of freedom around the saddle-point solution and successively promoting the would-be quantum fluctuations to operators via a Bogoliubov quadratization. The second part of Chapter \ref{quantum} is dedicated to the analysis of quantum corrections to local operators and the physical content of the particle-hole amplitudes defining the Bose field operator derived by the quantization procedure. \\
In Chapter \ref{quantum_corrections} the present Gutzwiller quantum theory is applied to the determination of quantities as the quantum depletion of the condensate, extending the Bogoliubov results in the deep superfluid regime, and the superfluid fraction, which are not accessible by the mean-field treatment of the excitations introduced in Chapter \ref{MF}. After discussing new results about the contribution of different excitation modes to these quantum corrections, we point out the predictive weaknesses of the Gutzwiller approach near the critical transition and in the Mott phase. On the basis of simple considerations we argue that a possible explanation could rely in the lack of self-consistency affecting our calculation of quantum corrections with respect to the mean-field solution. Finally, we investigate the possibility of giving non-trivial quantitative predictions for the one-body and pair correlation functions which account for quantum fluctuations due to interaction effects. \\
The study of quadratic corrections to the mean-field theory is concluded in Chapter \ref{response_functions}, which focuses on a rigorous treatment of density susceptibilities to different probes within the formalism of linear response theory. A particular attention is dedicated to the study of the physical nature of high-energy modes living above the Goldstone and Higgs excitations. After recognizing their strong particle-hole character in localized regions of the phase diagram, we search for a suitable physical probe of these modes in the hopping modulation channel. The consequent results are discussed in relation to possible experimental implementations, such as the long-debated indirect detection of the Higgs mode. \\
Conceptually concluding the discussion about the predictive power of our Gutzwiller theory, Chapter \ref{third_order} is finally devoted to test the capability of our quantum description of describing non-linear quantum effects in the domain of strong interactions. To this purpose, we expand the Gutzwiller reformulation of the Bose-Hubbard Hamiltonian up to the third order in the fluctuation operators and consider only the terms corresponding to the decay of the Higgs excitation. After determining the damping rate of the amplitude mode for different dimensions near the quantum critical point corresponding to the Mott tips, we compare our results with exact predictions deriving from field theory \cite{altman_auerbach} and discuss the possible experimental implications of our findings. \\
The final chapter \textit{Conclusions and perspectives} contains a review of the main results of this thesis work and discusses future directions for further research in this field and possible experimental implementations. \\
The original results derived in this thesis are entirely presented in Chapters \ref{quantum}-\ref{third_order}.

\Chapter{Quantum theory of weak interactions}{Bogoliubov approach to superfluidity in an optical lattice}\label{Bogoliubov}
\markboth{\MakeUppercase{2. Weak interactions}}{\MakeUppercase{2. Weak interactions}}
The present chapter focuses on the analysis the weakly-interacting limit of the Bose-Hubbard model by the use of the Bogoliubov approximation on optical lattices, which turns to provide a direct way to determine key physical quantities such as the superfluid fraction and its relation with the quantum depletion of the condensate. The results derived for dilute gases will constitute part of the background knowledge for the extended approach to quantum fluctuations that we propose in this thesis work. \\
As references for the Bogoliubov description of weakly-interacting condensates, we adopt the benchmark works by Moseley, Rey et al. \cite{moseley_fialko_ziegler, rey} as references.

\section{The Bose-Hubbard model and superfluidity}
The concept of superfluidity is closely related to the existence of a condensate in a interacting many-body system. Formally, the one-body density matrix $\rho^{\left( 1 \right)}{\left( \mathbf{r}, \mathbf{r}' \right)}$ has to have exactly one macroscopic eigenvalue, which defines the number of particles in the condensate; the corresponding eigenvector describes the condensate wave function $\phi_0{\left( \mathbf{r} \right)} = \left| \phi_0{\left( \mathbf{r} \right)} \right| \exp{\left[ i \, \Theta{\left( \mathbf{r} \right)} \right]}$. The spatially-varying condensate phase $\Theta{\left( \mathbf{r} \right)}$ is associated to a condensate velocity field given by:
\begin{equation}\label{2.1}
\mathbf{v}_s{\left( \mathbf{r} \right)} = \frac{\hbar}{m} \nabla \Theta{\left( \mathbf{r} \right)}
\end{equation}
where $m$ is the mass of a particle belonging to the system. This irrotational velocity field is identified with the velocity of the \textsl{superfluid flow} and enables a first direct determination of the superfluid fraction $f_s$. \\
Let us consider a Bose-Hubbard optical lattice with a linear dimension $L$ along the $x$-direction and a ground-state energy $E_0$ calculated with periodic boundary conditions. Now we impose a linear phase variation $\Theta{\left( \mathbf{r} \right)} = \theta \, x/L$ with a total twist angle $\theta$ over the linear size of the system. The resulting ground-state energy $E_{\theta}$ will depend on the phase twist. For small twist angles, that is $\theta \ll \pi$, the energy difference $E_{\theta} - E_0$ can be attributed to the kinetic energy term $T_s$ generated by the superflow due to the phase gradient:
\begin{equation}\label{2.2}
E_{\theta} - E_0 = T_s = \frac{1}{2} N m f_s \mathbf{v}^2_s
\end{equation}
where $N$ is the total number of particles in the lattice, so that $N m f_s$ is the total mass of the superfluid component. Replacing the superfluid velocity $\mathbf{v}_s$ with the phase gradient according to equation (\ref{2.1}) leads to a fundamental relation for the superfluid fraction:
\begin{equation}\label{2.3}
f_s = \frac{2 \, m}{\hbar^2} \frac{L^2}{N} \frac{E_{\theta} - E_0}{\theta^2} = \frac{1}{N} \frac{E_{\theta} - E_0}{J \, \Delta \theta^2}
\end{equation}
where the second equality applies to a lattice system where a linear phase variation has been imposed. Here the intersite distance is $a$, the phase variation over this distance is $\Delta \theta$ and the number of sites is $M$, hence $J = \hbar^2/\left( 2 \, m \, a^2 \right)$. \\
In the context of the discrete Bose-Hubbard model it is convenient to map the phase variations by means of a unitary transformation onto the Hamiltonian, whose twisted form is:
\begin{equation}\label{2.4}
\hat{H}_{\theta} = -J \sum_{\langle \mathbf{r}, \mathbf{s} \rangle} \left( e^{-i \Delta \theta} \hat{a}^{\dagger}_{\mathbf{s}} \, \hat{a}_{\mathbf{r}} + e^{i \Delta \theta} \hat{a}^{\dagger}_{\mathbf{r}} \, \hat{a}_{\mathbf{s}} \right) + \frac{U}{2} \sum_{\mathbf{r}} \hat{a}^{\dagger}_{\mathbf{r}} \, \hat{a}^{\dagger}_{\mathbf{r}} \, \hat{a}_{\mathbf{r}} \, \hat{a}_{\mathbf{r}} - \mu \sum_{\mathbf{r}} \hat{a}^{\dagger}_{\mathbf{r}} \, \hat{a}_{\mathbf{r}}
\end{equation}
where the hopping term exhibits the so-called Peierls phase factors $e^{\pm i \Delta \theta}$. Interestingly, these demonstrate that the phase twist is equivalent to the imposition of an acceleration on the lattice for a finite time interval or, gauge equivalently, the application of a constant vector potential $\mathcal{A} = \Delta \theta$ along the flux direction. \\
We can calculate the energy change $E_{\theta} - E_0$ under the assumption that $\Delta \theta \ll \pi$, so that we can write:
\begin{equation}\label{2.5}
e^{-i \Delta \theta} \approx 1 - i \Delta \theta - \frac{\Delta \theta^2}{2}
\end{equation}
It follows that the transformed Hamiltonian (\ref{2.4}) takes the form:
\begin{equation}\label{2.6}
\hat{H}_{\theta} = \hat{H} + \Delta \theta \hat{J} - \frac{\Delta \theta^2}{2} \hat{T} = \hat{H} + \hat{H}_{pert}
\end{equation}
where $\hat{H}$ is the original Bose-Hubbard Hamiltonian, $\hat{J}$ is the current operator and $\hat{H}_{pert} = \Delta \theta \hat{J} - \frac{\Delta \theta^2}{2} \hat{T}$ can be treated as a perturbation. In particular:
\begin{equation}\label{2.7}
\hat{J} \equiv i \, J \sum_{\langle \mathbf{r}, \mathbf{s} \rangle} \left( \hat{a}^{\dagger}_{\mathbf{s}} \, \hat{a}_{\mathbf{r}} - \hat{a}^{\dagger}_{\mathbf{r}} \, \hat{a}_{\mathbf{s}} \right)
\end{equation}
\begin{equation}\label{2.8}
\hat{T} = -J \sum_{\langle \mathbf{r}, \mathbf{s} \rangle} \left( \hat{a}^{\dagger}_{\mathbf{s}} \, \hat{a}_{\mathbf{r}} + \hat{a}^{\dagger}_{\mathbf{r}} \, \hat{a}_{\mathbf{s}} \right)
\end{equation}
Retaining only terms up to the second-order in $\Delta \theta$, the energy shift $E_{\theta} - E_0$ due to the imposed phase twist can be now calculated within second-order perturbation theory as the sum of two contributions:
\begin{equation}\label{2.9}
E_{\theta} - E_0 \approx \Delta E_1 + \Delta E_2
\end{equation}
The first-order contribution $\Delta E_1$ is proportional to the expectation value of the hopping operator over the original ground-state solution:
\begin{equation}\label{2.10}
\Delta E_1 = \langle \Psi_0 | \hat{H}_{pert} | \Psi_0 \rangle = -\frac{\Delta \theta^2}{2} \langle \Psi_0 | \hat{T} | \Psi_0 \rangle
\end{equation}
The second-order term $\Delta E_2$ is related to the matrix elements of the current operator involving the excited states and the ground state of the original Bose-Hubbard model:
\begin{equation}\label{2.11}
\Delta E_2 = -\sum_{\alpha} \frac{\left| \langle \Psi_{\alpha} | \hat{H}_{pert} | \Psi_0 \rangle \right|^2}{E_{\alpha} - E_0} = -\Delta \theta^2 \sum_{\alpha} \frac{\left| \langle \Psi_{\alpha} | \hat{J} | \Psi_0 \rangle \right|^2}{E_{\alpha} - E_0}
\end{equation}
Thus we obtain the energy change $E_{\theta} - E_0$ up to second order in the phase twist $\Delta \theta$:
\begin{equation}\label{2.12}
E_{\theta} - E_0 = -\Delta \theta^2 \left( \frac{1}{2} \langle \Psi_0 | \hat{T} | \Psi_0 \rangle + \sum_{\alpha} \frac{\left| \langle \Psi_{\alpha} | \hat{J} | \Psi_0 \rangle \right|^2}{E_{\alpha} - E_0} \right) \equiv M \, \Delta \theta^2 \, D
\end{equation}
where $D$ is defined to be:
\begin{equation}\label{2.13}
D \equiv -\frac{1}{M} \left( \frac{1}{2} \langle \Psi_0 | \hat{T} | \Psi_0 \rangle + \sum_{\alpha} \frac{\left| \langle \Psi_{\alpha} | \hat{J} | \Psi_0 \rangle \right|^2}{E_{\alpha} - E_0} \right)
\end{equation}
and is formally equivalent to the Drude weight used to specify the DC conductivity of charged fermionic systems. Then, the superfluid fraction reads:
\begin{equation}\label{2.14}
f_s = f^{\left( 1 \right)}_s + f^{\left( 2 \right)}_s
\end{equation}
where:
\begin{equation}\label{2.15}
f^{\left( 1 \right)}_s \equiv -\frac{1}{2 \, N \, J} \langle \Psi_0 | \hat{T} | \Psi_0 \rangle
\end{equation}
\begin{equation}\label{2.16}
f^{\left( 2 \right)}_s \equiv \frac{1}{N \, J} \sum_{\alpha} \frac{\left| \langle \Psi_{\alpha} | \hat{J} | \Psi_0 \rangle \right|^2}{E_{\alpha} - E_0}
\end{equation}
In general both the terms included in (\ref{2.14}) contribute significantly; however, for a translationally invariant lattice the second term $f^{\left( 2 \right)}_s$ vanishes in the Bogoliubov limit (as it is going to be shown later). \\
One can further understand and check the consistency of this approach for studying the superfluid density by calculating the flow that is produced by the application of a phase twist. To this purpose we work out the expectation value of the current operator expressed in terms of the twisted degrees of freedom:
\begin{equation}\label{2.17}
\hat{J}_{\theta} \equiv i \, J \sum_{\langle \mathbf{r}, \mathbf{s} \rangle} \left( e^{-i \Delta \theta} \hat{a}^{\dagger}_{\mathbf{s}} \, \hat{a}_{\mathbf{r}} - e^{i \Delta \theta} \hat{a}^{\dagger}_{\mathbf{r}} \, \hat{a}_{\mathbf{s}} \right)
\end{equation}
Expanding (\ref{2.17}) to find the lowest-order contributions:
\begin{equation}\label{2.18}
\hat{J}_{\theta} \approx \hat{J} - \Delta \theta \, \hat{T}
\end{equation}
we can use first-order perturbation theory on the twisted wave function for obtaining the following expression:
\begin{equation}\label{2.19}
\langle \Psi{\left( \Delta \theta \right)} | \hat{J}_{\theta} | \Psi{\left( \Delta \theta \right)} \rangle \approx -2 \, \Delta \theta \left( \frac{1}{2} \langle \Psi_0 | \hat{T} | \Psi_0 \rangle + \sum_{\alpha} \frac{\left| \langle \Psi_{\alpha} | \hat{J} | \Psi_0 \rangle \right|^2}{E_{\alpha} - E_0} \right) = 2 \, N \, J \, f_s \, \Delta \theta
\end{equation}
Remembering that the kinetic energy has the quadratic dispersion $J \, a^2 \, \left| \mathbf{q} \right|^2$ in the phononic regime, we can define the effective lattice mass as:
\begin{equation}\label{2.20}
m \equiv \frac{\hbar^2}{2 \, J \, a^2}
\end{equation}
Therefore, the physical current (\ref{2.19}) can be expressed in the form:
\begin{equation}\label{2.21}
\langle \Psi{\left( \Delta \theta \right)} | \hat{J}_{\theta} | \Psi{\left( \Delta \theta \right)} \rangle = \frac{N \, \hbar^2 \, \Delta \theta}{m \, a^2} f_s
\end{equation}
In conclusion, the flux density is provided by:
\begin{equation}\label{2.22}
\frac{1}{M} \langle \Psi{\left( \Delta \theta \right)} | \hat{J}_{\theta} | \Psi{\left( \Delta \theta \right)} \rangle = \frac{\hbar^2 \, \Delta \theta}{m \, a} \frac{N \, f_s}{M \, a} = n_s \, \mathbf{v}_s
\end{equation}
Equality (\ref{2.22}) demonstrates that the Drude formulation of the superfluid fraction is consistent with an intuitively satisfying expression for the amount of flowing superfluid.

\section{The Bogoliubov approximation to the model}
As it occurs in the case of continuous Bose gases, we can resort to the Bogoliubov approximation for the Bose-Hubbard model in the limit where quantum fluctuations, or equivalently the depletion of the condensate, are small compared with the fraction of condensed particles. \\
In the non-interacting case, where quantum fluctuations can be completely ignored, we can replace the creation and annihilation operators $\hat{a}^{\dagger}_{\mathbf{r}}$ and $\hat{a}_{\mathbf{r}}$ with a \textit{c}-number $\psi_0{\left( \mathbf{r} \right)}$ on each site. In the weakly-interacting regime his prescription leads to a set of coupled discrete non-linear Schr\"odinger or Gross-Pitaevskii equations for such amplitudes:
\begin{equation}\label{2.23}
i \hbar \dpar_t \psi_0{\left( \mathbf{r} \right)} = -J \sum_{a = 1}^{d} \left[ \psi_0{\left( \mathbf{r} + \mathbf{e}_a \right)} + \psi_0{\left( \mathbf{r} - \mathbf{e}_a \right)} \right] + U \left| \psi_0{\left( \mathbf{r} \right)} \right|^2 \psi_0{\left( \mathbf{r} \right)} - \mu \, \psi_0{\left( \mathbf{r} \right)}
\end{equation}
where $\mathbf{e}_a$ is the unit versor along the $a$-direction. Equation (\ref{2.23}) can be used to study the properties of the condensate loaded into the lattice when the kinetic energy is large enough when compared to the interaction one. Quantum fluctuations are included in this description by rewriting the full annihilation operator as a combination of the condensate amplitude part and a fluctuation operator:
\begin{equation}\label{2.24}
\hat{a}_{\mathbf{r}} \approx \left[ \psi{\left( \mathbf{r} \right)} + \hat{\beta}{\left( \mathbf{r}, t \right)} \right] e^{-i \, \mu \, t/\hbar}
\end{equation}
where the time-dependence derives from the solution of the stationary Gross-Pitaevskii equation (\ref{2.23}). \\
The effectiveness of the Bogoliubov method consists in providing expectation values of observables involving second-order combinations of the fluctuation operator and determining whether the assumption of small fluctuations is verified on the basis of the results.

\subsection{Bogoliubov theory for the translationally invariant lattice}
Looking at (\ref{2.23}), it is straightforward to see that the ground state solution for the translationally invariant lattice is associated to the eigenvalue:
\begin{equation}\label{2.25}
\mu = \left| \psi_0 \right|^2 U - J \, z
\end{equation}
where $z$ is the lattice coordination number and $\left| \psi_0 \right|^2 = N/M$ is the mean local density. The Bogoliubov equations for the quantum fluctuation degrees of freedom present the form:
\begin{equation}\label{2.26}
i \hbar \dpar_t \hat{\beta}{\left( \mathbf{r}, t \right)} = \left( 2 \, \left| \psi_0 \right|^2 U - \mu \right) \hat{\beta}{\left( \mathbf{r}, t \right)} - J \sum_{a = 1}^{d} \left[ \hat{\beta}{\left( \mathbf{r} + \mathbf{e}_a, t \right)} + \hat{\beta}{\left( \mathbf{r} - \mathbf{e}_a, t \right)} \right] + \left| \psi_0 \right|^2 U \, \hat{\beta}^{\dagger}{\left( \mathbf{r}, t \right)}
\end{equation}
These equations are simply solved by constructing a suitable Bogoliubov rotation which transforms (\ref{2.24}) to quasi-particle operators:
\begin{equation}\label{2.27}
\hat{\beta}{\left( \mathbf{r}, t \right)} = \frac{1}{\sqrt{M}} \sum_{\mathbf{q}} \left\{ u{\left( \mathbf{q} \right)} \, \hat{\alpha}_{\mathbf{q}} \, e^{i \left[ \mathbf{q} \cdot \mathbf{r} - \omega{\left( \mathbf{q} \right)} t \right]} - v^{*}{\left( \mathbf{q} \right)} \, \hat{\alpha}^{\dagger}_{\mathbf{q}} \, e^{-i \left[ \mathbf{q} \cdot \mathbf{r} - \omega{\left( \mathbf{q} \right)} t \right]} \right\}
\end{equation}
\begin{equation}\label{2.28}
\hat{\beta}^{\dagger}{\left( \mathbf{r}, t \right)} = \frac{1}{\sqrt{M}} \sum_{\mathbf{q}} \left\{ u^{*}{\left( \mathbf{q} \right)} \, \hat{\alpha}^{\dagger}_{\mathbf{q}} \, e^{-i \left[ \mathbf{q} \cdot \mathbf{r} - \omega{\left( \mathbf{q} \right)} t \right]} - v{\left( \mathbf{q} \right)} \, \hat{\alpha}_{\mathbf{q}} \, e^{i \left[ \mathbf{q} \cdot \mathbf{r} - \omega{\left( \mathbf{q} \right)} t \right]} \right\}
\end{equation}
in order to diagonalize the expanded Bose-Hubbard Hamiltonian where only quadratic terms in the fluctuation operators are retained. It follows that the quasi-particle operators satisfy bosonic commutation relations:
\begin{equation}\label{2.29}
\left[ \hat{\alpha}_{\mathbf{k}}, \hat{\alpha}^{\dagger}_{\mathbf{p}} \right] = \delta_{\mathbf{k}, \mathbf{p}}
\end{equation}
and their thermodynamic state is determined by Bose statistics:
\begin{equation}\label{2.30}
\langle \hat{\alpha}^{\dagger}_{\mathbf{k}} \,\hat{\alpha}_{\mathbf{p}} \rangle = \delta_{\mathbf{k}, \mathbf{p}} \frac{1}{\exp{\left( \frac{\hbar \, \omega{\left( \mathbf{q} \right)}}{k_B \, T} \right)} - 1}
\end{equation}
Inserting (\ref{2.27}) and (\ref{2.28}) into the Bogoliubov equations (\ref{2.26}) we find two coupled relations for the excitation amplitudes and frequencies:
\begin{equation}\label{2.31}
\hbar \, \omega{\left( \mathbf{q} \right)} \, u{\left( \mathbf{q} \right)} = \left[ \left| \psi_0 \right|^2 U + \varepsilon{\left( \mathbf{q} \right)} \right] u{\left( \mathbf{q} \right)} - \left| \psi_0 \right|^2 U \, v{\left( \mathbf{q} \right)}
\end{equation}
\begin{equation}\label{2.32}
\hbar \, \omega{\left( \mathbf{q} \right)} \, v{\left( \mathbf{q} \right)} = -\left[ \left| \psi_0 \right|^2 U + \varepsilon{\left( \mathbf{q} \right)} \right] v{\left( \mathbf{q} \right)} + \left| \psi_0 \right|^2 U \, u{\left( \mathbf{q} \right)}
\end{equation}
where:
\begin{equation}\label{2.33}
\varepsilon{\left( \mathbf{q} \right)} \equiv 4 \, J \sum_{i = 1}^{d} \sin^2{\left( \frac{q_i \, a}{2} \right)}
\end{equation}
is the free-particle dispersion relation on the lattice. The explicit expressions for the quasi-particle square amplitudes read:
\begin{equation}\label{2.34}
\left| u{\left( \mathbf{q} \right)} \right|^2 = \frac{\varepsilon{\left( \mathbf{q} \right)} + \left| \psi_0 \right|^2 U + \hbar \, \omega{\left( \mathbf{q} \right)}}{2 \, \hbar \, \omega{\left( \mathbf{q} \right)}}
\end{equation}
\begin{equation}\label{2.35}
\left| v{\left( \mathbf{q} \right)} \right|^2 = \frac{\varepsilon{\left( \mathbf{q} \right)} + \left| \psi_0 \right|^2 U - \hbar \, \omega{\left( \mathbf{q} \right)}}{2 \, \hbar \, \omega{\left( \mathbf{q} \right)}}
\end{equation}
where the phonon excitation eigenfrequencies are given by:
\begin{equation}\label{2.36}
\hbar \, \omega{\left( \mathbf{q} \right)} = \sqrt{\varepsilon{\left( \mathbf{q} \right)} \left[ \varepsilon{\left( \mathbf{q} \right)} + 2 \left| \psi_0 \right|^2 U \right]}
\end{equation}

\subsection{Superfluid fraction in the translationally invariant lattice}
Having obtained the expressions for the excitation amplitudes, we can now determine crucial observables such as the superfluid fraction. \\
A straightforward calculation reveals that the second-order contribution (\ref{2.16}) to the superfluid fraction:
\begin{equation}\label{2.37}
\begin{aligned}
f^{\left( 2 \right)}_s &= \frac{1}{N \, J} \sum_{\alpha} \frac{\left| \langle \Psi_{\alpha} | \hat{J} | \Psi_0 \rangle \right|^2}{E_{\alpha} - E_0} = \\
&= \frac{J}{N} \sum_{\alpha} \sum_{\mathbf{p}, \mathbf{q}} \Bigg\{ \frac{\left| \sum_{\mathbf{r}} \sum_a \left[ u{\left( \mathbf{r} + \mathbf{e}_a, \mathbf{p} \right)} \, v{\left( \mathbf{r}, \mathbf{q} \right)} - u{\left( \mathbf{r}, \mathbf{p} \right)} \, v{\left( \mathbf{r} + \mathbf{e}_a, \mathbf{q} \right)} \right] \right|^2}{\hbar \left[ \omega{\left( \mathbf{p}\right)} + \omega{\left( \mathbf{p}\right)} \right]} + \\
&+ \delta_{\mathbf{p}, \mathbf{q}} \, \frac{\left| \sum_{\mathbf{r}} \sum_a \left[ u{\left( \mathbf{r} + \mathbf{e}_a, \mathbf{p} \right)} \, v{\left( \mathbf{r}, \mathbf{p} \right)} - u{\left( \mathbf{r}, \mathbf{p} \right)} \, v{\left( \mathbf{r} + \mathbf{e}_a, \mathbf{p} \right)} \right] \right|^2}{2 \, \hbar \, \omega{\left( \mathbf{p}\right)}} \Bigg\}
\end{aligned}
\end{equation}
where:
\begin{equation}\label{2.38}
u{\left( \mathbf{r}, \mathbf{k} \right)} \equiv u{\left( \mathbf{q} \right)} \, e^{i \mathbf{k} \cdot \mathbf{r}} \quad , \quad v{\left( \mathbf{r}, \mathbf{k} \right)} \equiv v{\left( \mathbf{q} \right)} \, e^{i \mathbf{k} \cdot \mathbf{r}}
\end{equation}
vanishes for a translationally invariant system, so that we only need to calculate only the following expression:
\begin{equation}\label{2.39}
f_s = -\frac{1}{2 \, N \, J} \langle \Psi_0 | \hat{T} | \Psi_0 \rangle
\end{equation}
As already observed by Rey et al. in (\cite{rey}), \uline{it is crucial to stress that the current contribution (\ref{2.37}) is identically zero in the standard Bogoliubov approximation, whereas exact calculations and the self-consistent Hartree-Fock-Bogoliubov-Popov (HFB-Popov) theory produces non-vanishing results}. Since this second-order contribution $f^{\left( 2 \right)}_s$ is extremely small in the deep superfluid regime, \uline{the Bogoliubov approximation provides a good description of $f_s$ only over a limited region of the physical parameters}. \\
On the other hand, it is also important to specify that the definition of superfluid density given by equation (\ref{2.12}) does not distinguish between the longitudinal and transversal current responses to a global phase twist applied to the system; in fact, only the latter is expected to give a finite contribution in superfluid systems, while the former should vanish identically when summing the terms $f^{\left( 2 \right)}_s$ and $f^{\left( 1 \right)}_s$: this is exactly the feature that is missing in the Bogoliubov description. More details about the distinction between longitudinal and transversal current responses in superfluid systems are provided in Appendix \ref{E} in the context of the Gutzwiller theory developed in this thesis.

\bigskip

Returning to the Bogoliubov description, the explicit form of (\ref{2.39}) is given by:
\begin{equation}\label{2.40}
\begin{aligned}
f_s &= \frac{1}{2 \, N} \sum_{\langle \mathbf{r}, \mathbf{s} \rangle} \big\langle \Psi_0 \big| \left[ \psi^{*}_0{\left( \mathbf{s} \right)} + \hat{\beta}^{\dagger}{\left( \mathbf{s} \right)} \right] \left[ \psi_0{\left( \mathbf{r} \right)} + \hat{\beta}{\left( \mathbf{r} \right)} \right] + \left[ \psi_0^{*}{\left( \mathbf{r} \right)} + \hat{\beta}^{\dagger}{\left( \mathbf{r} \right)} \right] \left[ \psi_0{\left( \mathbf{s} \right)} + \hat{\beta}{\left( \mathbf{s} \right)} \right] \big| \Psi_0 \big\rangle \approx \\
&\approx \frac{1}{2 \, N} \sum_{\langle \mathbf{r}, \mathbf{s} \rangle} \big\langle \Psi_0 \big| 2 \, \left| \psi_0 \right|^2 + \hat{\beta}^{\dagger}{\left( \mathbf{s} \right)} \, \hat{\beta}{\left( \mathbf{r} \right)} + \hat{\beta}^{\dagger}{\left( \mathbf{r} \right)} \, \hat{\beta}{\left( \mathbf{s} \right)} \big| \Psi_0 \big\rangle
\end{aligned}
\end{equation}
Replacing the fluctuation operators with their quasi-particle rotations (\ref{2.27}) and (\ref{2.28}) that diagonalize the weakly-interacting quadratic approximation to the Bose-Hubbard Hamiltonian, we obtain:
\begin{equation}\label{2.41}
\begin{aligned}
f_s &= \frac{M}{N} \left( \left| \psi_0 \right|^2 + \frac{1}{M} \sum_{\mathbf{q}} \left\{ \left[ \left| u{\left( \mathbf{q} \right)} \right|^2 + \left| v{\left( \mathbf{q} \right)} \right|^2 \right] \langle \hat{\alpha}^{\dagger}_{\mathbf{q}} \,\hat{\alpha}_{\mathbf{q}} \rangle + \left| v{\left( \mathbf{q} \right)} \right|^2 \right\} j{\left( \mathbf{q} \right)} \right) \longrightarrow \\
&\xrightarrow[T = 0]{} \frac{M}{N} \left[ \left| \psi_0 \right|^2 + \frac{1}{M} \sum_{\mathbf{q}} \left| v{\left( \mathbf{q} \right)} \right|^2 j{\left( \mathbf{q} \right)} \right]
\end{aligned}
\end{equation}
where:
\begin{equation}\label{2.42}
j{\left( \mathbf{q} \right)} \equiv \sum_{i = 1}^{d} \cos{\left( q_i \, a \right)}
\end{equation}
This final result shows that at $T = 0$ and in the limit of zero lattice spacing the superfluid fraction tends to unity, as the following normalization condition holds:
\begin{equation}\label{2.43}
N \equiv M \langle \hat{a}^{\dagger}_{\mathbf{r}} \,\hat{a}_{\mathbf{r}} \rangle \approx M \left| \psi_0 \right|^2 + \sum_{\mathbf{q}} \left| v{\left( \mathbf{q} \right)} \right|^2
\end{equation}
The quantity:
\begin{equation}\label{2.44}
\mathcal{D} \equiv \frac{1}{M} \sum_{\mathbf{q}} \left| v{\left( \mathbf{q} \right)} \right|^2
\end{equation}
is the so-called quantum depletion of the condensate at zero temperature.

\bigskip

The expressions calculated above give a direct insight into the change of the superfluid fraction as the atoms are pushed out of the condensate due to interactions. In equation (\ref{2.41}) the sum involving the Bogoliubov amplitudes $v{\left( \mathbf{q} \right)}$ characterizes the difference between the condensate fraction, which is given by the first term, and the superfluid fraction. For weak interactions and a small depletion, which fills only the lower half of the energy band (\ref{2.33}), where the factor $j{\left( \mathbf{q} \right)}$ has a positive sign, the superfluid fraction is larger than the condensate one. Thus the depletion of the condensate has initially little effect on superfluidity. When the depleted population spreads into the upper part of the energy band, where $j{\left( \mathbf{q} \right)}$ is negative, the superfluid fraction is reduced and might even become smaller than the condensate fraction. \\
In a sense, the interactions are playing a role akin to the so-called \textsl{Fermi exclusion pressure} in the case of the electronic flow in the corresponding band spectrum. This, however, can lead to perfect filling and a cancellation of the flow. In the case of our Bogoliubov description, we can only see a reduction of the superfluid flow, not a perfect switching off of its particle fraction. A perfect cancellation of the flow due to interactions is expected to happen in the Mott insulator state, which cannot be described by the bare Bogoliubov approximation.

\Chapter{The time-dependent \texorpdfstring{\\}{} Gutzwiller mean-field approach}{A short review}\label{MF}
\markboth{\MakeUppercase{3. A benchmark theory}}{\MakeUppercase{3. A benchmark theory}}
\textbf{Note:} The contents of this chapter are loosely based on the work by Krutitsky and Navez \cite{krutitsky_navez} focusing on the excitation dynamics of the Bose-Hubbard model from the point of view of time-dependent Gutzwiller mean-field ansatz. All the figures proposed along the dissertation have been extracted from \cite{krutitsky_navez} only for informational purposes.

\vspace{1cm}

Mean-field theories in dimensions higher than one allow a self-consistent study of the fundamental excitations and the system dynamics in the lowest order with respect to quantum fluctuations. In the weakly-interacting regime the atoms are fully condensed and the system is satisfactorily described by the time-dependent discrete Gross-Pitaevskii equation (DGPE) that we have introduced in Chapter \ref{Bogoliubov}, while the excitation spectrum is provided by the corresponding Bogoliubov-de Gennes equations (BdGE). This kind of theory takes into account the Goldstone mode alone, whose long-wavelength behaviour is governed by the sound velocity given by:
\begin{equation}\label{3.1}
c = \sqrt{\frac{\left| \psi \right|^2}{\kappa \, m}}
\end{equation}
where $\left| \psi \right|^2$ is the condensate density, $\kappa$ is the compressibility, given by the density derivative with respect to the chemical potential:
\begin{equation}\label{3.2}
\kappa = \frac{\dpar n}{\dpar \mu}
\end{equation}
and $m$ is the lattice effective mass defined in (\ref{2.20}). \\
In the strongly-interacting regime the DGPE is no more valid and on the mean-field level they must be replaced by general Gutzwiller equations (GE) which turn to be exact in the limit of infinite dimensions. Obtaining similar results, excitations above the ground state described by the GE were studied using also the random phase approximation (RPA) \cite{sheshandri_krishnamurthy_pandit_ramakrishnan}-\cite{hazzard_mueller}, the Schwinger boson approach \cite{huber_altman_buchler_blatter}, the time-dependent variational principle with subsequent quantization \cite{huber_theiler_altman_blatter}, the slave boson representation of the Bose-Hubbard model \cite{slave_boson}, the standard-basis operator method \cite{konabe_nikuni_nakamura, ohashi_kitaura_matsumoto} and the Ginzburg-Landau theory \cite{grass_santos_pelster}. All these methods have revealed that the Mott phase is characterized by a gapped spectrum populated by particle-hole modes, while the superfluid phase presents a gapless Goldstone mode and gapped higher branches. However, different approximations adopted by a variety approaches do not always lead to the same final results. \\
Excitations above the Gutzwiller ground state can also be investigated by resorting to a generalization of the BdGE derived from the GE within the linear responds theory framework. This last approach, which has been widely used only in the fermionic Hubbard model, allows to obtain results that are consistent with the other mean-field predictions mentioned above and study the ground state properties, the stationary excitation modes and the real-time dynamics on an equal footing.

\bigskip

The purpose of this third chapter is to provide an insight into the description of the Bose-Hubbard collective excitations as they emerge within the bosonic time-dependent Gutzwiller ansatz, which is \textsl{gapless} and satisfies the basic conservations laws, for instance the \textit{f}-sum rule as shown by Krutitsky and Navez in $\cite{krutitsky_navez}$. Solutions of the GE based on the Gutzwiller ansatz allows us not only to calculate the dispersion relations of the excitations, but also to determine the transition amplitudes of Bragg scattering processes that are currently used for probing the properties of Bose lattice systems, as well as the response to external perturbations. \\
It is important to emphasise that the implementation of the Gutzwiller ansatz is the only approximation used in the following sections, so that the results are valid over the whole range of physical parameters entering the model Hamiltonian. This is one of the reasons that have suggested to take inspiration from these findings for developing the Gutzwiller quantum theory at the centre of the present work.

\vspace{0.7cm}
\section{The time-dependent Gutzwiller ansatz}
Let us again consider a system of ultracold interacting bosons in a $d$-dimensional isotropic lattice described by the Bose-Hubbard Hamiltonian:
\begin{equation}\label{3.3}
\hat{H} = -J \sum_{\langle \mathbf{r}, \mathbf{s} \rangle} \left( \hat{a}^{\dagger}_{\mathbf{r}} \, \hat{a}_\mathbf{s} + \text{h.c.} \right) + \frac{U}{2} \sum_{\mathbf{r}} \hat{n}_{\mathbf{r}} \left( \hat{n}_{\mathbf{r}} - 1 \right) - \mu \sum_{i} \hat{n}_{\mathbf{r}}
\end{equation}
where $\mathbf{r}$ is a $d$-dimensional spatial vector, $J$ is the hopping rate, $U$ is the on-site interaction strength and $\mu$ is the chemical potential. The creation and annihilation operators in (\ref{3.3}) are always required to satisfy bosonic commutation relations. \\
Our analysis employs the time-dependent Gutzwiller ansatz, that is the system wave functions is imposed to be a tensor product of local states:
\begin{equation}\label{3.4}
| \Psi \rangle = \otimes_{\mathbf{r}} | \phi{\left( \mathbf{r} \right)} \rangle \quad , \quad | \phi{\left( \mathbf{r} \right)} \rangle = \sum_{n = 0}^{\infty} c_n{\left( \mathbf{r} \right)} | n \rangle_{\mathbf{r}}
\end{equation}
where $| n \rangle_{\mathbf{r}}$ is the Fock state with $n$ particles at site $\mathbf{r}$ and $c_n{\left( \mathbf{r} \right)}$ is the corresponding statistical weight. Normalization of the states $| \phi{\left( \mathbf{r} \right)}$ leads to the identity:
\newpage
\begin{equation}\label{3.5}
\sum_{n = 0}^{\infty} \left| c_n{\left( \mathbf{r} \right)} \right|^2 = 1
\end{equation}
In this model the mean number of condensed atoms at site $\mathbf{r}$ is given by the square modulus of the average of the annihilation operator $\hat{a}_{\mathbf{r}}$ over the wave function ansatz (\ref{3.4}):
\begin{equation}\label{3.6}
\psi{\left( \mathbf{r} \right)} \equiv \langle \Psi | \hat{a}_{\mathbf{r}} | \Psi \rangle = \sum_{n = 0}^{\infty} \sqrt{n + 1} \, c^{*}_n{\left( \mathbf{r} \right)} \, c_{n + 1}{\left( \mathbf{r} \right)}
\end{equation}
where $\psi{\left( \mathbf{r} \right)}$ is the condensate order parameter. On the other hand, the mean occupation number for site $\mathbf{r}$ reads:
\begin{equation}\label{3.7}
n{\left( \mathbf{r} \right)} \equiv \langle \Psi | \hat{n}_{\mathbf{r}} | \Psi \rangle = \sum_{n = 0}^{\infty} n \left| c_n{\left( \mathbf{r} \right)} \right|^2
\end{equation}
hence one can easily demonstrate that:
\begin{equation}\label{3.8}
n{\left( \mathbf{r} \right)} \ge \left| \psi{\left( \mathbf{r} \right)} \right|^2
\end{equation}
by means of the Schwarz inequality. \\
The following step consists in calculating the action of the Bose-Hubbard model over the Gutzwiller ansatz:
\begin{equation}\label{3.9}
\begin{aligned}
S{\left[ c^{*}_n{\left( \mathbf{r}, t \right)}, c_n{\left( \mathbf{r}, t \right)} \right]} &= \int dt \, \langle \Psi | i \hbar \dpar_t - \hat{H} | \Psi \rangle = \\
&= \frac{i \hbar}{2} \sum_{\mathbf{r}} \sum_{n = 0}^{\infty} \left[  c^{*}_n{\left( \mathbf{r}, t \right)} \dpar_t c_n{\left( \mathbf{r}, t \right)} - c_n{\left( \mathbf{r}, t \right)} \dpar_t c^{*}_n{\left( \mathbf{r}, t \right)} \right] - \langle \Psi | \hat{H} | \Psi \rangle
\end{aligned}
\end{equation}
The saddle-point minimization of the functional in (\ref{3.9}) leads to the following time-dependent system of GE:
\begin{equation}\label{3.10}
i \hbar \frac{\dpar c_n{\left( \mathbf{r}, t \right)}}{\dpar t} = \sum_m H_{n m} \, c_m{\left( \mathbf{r}, t \right)}
\end{equation}
where:
\begin{equation}\label{3.11}
\begin{aligned}
H_{n m} &= \left[ \frac{U}{2} n \left( n - 1 \right) - \mu \, n \right] \delta_{n, m} - J \, \sqrt{m} \, \delta_{n + 1, m} \sum_{\alpha = 1}^{d} \left[ \psi^{*}{\left( \mathbf{r} + \mathbf{e}_{\alpha}, t \right)} + \psi^{*}{\left( \mathbf{r} - \mathbf{e}_{\alpha}, t \right)} \right] + \\
&- J \, \sqrt{n} \, \delta_{n, m + 1} \sum_{\alpha = 1}^{d} \left[ \psi{\left( \mathbf{r} + \mathbf{e}_{\alpha}, t \right)} + \psi{\left( \mathbf{r} - \mathbf{e}_{\alpha}, t \right)} \right]
\end{aligned}
\end{equation}
In the matrix element (\ref{3.11}) $\mathbf{e}_{\alpha}$ is the unit vector in the lattice direction $\alpha$. \\
These equations are called to be \textsl{conserving} because they do not violate any conservation law of the original Bose-Hubbard model. Actually, the fact that the coefficients $c_n{\left( \mathbf{r}, t \right)}$ for different sites are coupled to each other hints at the possibility of studying the real-time dynamics of the system excitations. \\
As it follows form the definition of the state (\ref{3.4}), at the mean-field level the Gutzwiller approximation neglects quantum correlations between different lattice sites, but takes into account local quantum fluctuations. This is supposed to be sufficient for describing the main features of the MI-SF quantum phase transition. \\
Equation (\ref{3.10}) allows to deduce also the following relation for the order parameter:
\begin{equation}\label{3.12}
\small
i \hbar \frac{\dpar \psi{\left( \mathbf{r}, t \right)}}{\dpar t} = -J \sum_{\alpha = 1}^{d} \left[ \psi{\left( \mathbf{r} + \mathbf{e}_{\alpha}, t \right)} + \psi{\left( \mathbf{r} - \mathbf{e}_{\alpha}, t \right)} \right] - \mu \, \psi{\left( \mathbf{r}, t \right)} + U \sum_{n = 0}^{\infty} n \, \sqrt{n + 1} \ c^{*}_n{\left( \mathbf{r}, t \right)} \, c_{n + 1}{\left( \mathbf{r}, t \right)}
\end{equation}
This equation assumes a closed form if the Gutzwiller coefficients are given by a coherent state distribution:
\begin{equation}\label{3.13}
c_n{\left( \mathbf{r} \right)} \equiv c^{\text{coh}}_n{\left( \mathbf{r} \right)} = e^{-\left|\psi{\left( \mathbf{r}, t \right)} \right|^2/2} \frac{\psi^n{\left( \mathbf{r}, t \right)}}{\sqrt{n!}}
\end{equation}
which is precisely an exact solution of (\ref{3.12}) for $U = 0$. The substitution of (\ref{3.13}) in (\ref{3.12}) leads to:
\begin{equation}\label{3.14}
i \hbar \frac{\dpar \psi{\left( \mathbf{r}, t \right)}}{\dpar t} = - J \sum_{\alpha = 1}^{d} \left[ \psi{\left( \mathbf{r} + \mathbf{e}_{\alpha}, t \right)} + \psi{\left( \mathbf{r} - \mathbf{e}_{\alpha}, t \right)} \right] - \mu \psi{\left( \mathbf{r}, t \right)} + U \left| \psi{\left( \mathbf{r}, t \right)} \right|^2 \psi{\left( \mathbf{r}, t \right)}
\end{equation}
which is exactly the DGPE for small values of the relative on-site interaction $U/J$. It follows that the bosonic Gutzwiller ansatz (3.3) is able to reproduce exactly the non-interacting limit as it coincides with a Glauber coherent state, while the GE match with the equations holding in the weakly-interacting limit.

\section{Ground state}
As long as the system is homogeneous, the ground state coefficients $c_n{\left( \mathbf{r} \right)}$ do not depend on the site position, so that the stationary solution of (\ref{3.10}) has the form:
\begin{equation}\label{3.15}
c^0_n{\left( \mathbf{r}, t \right)} = c^0_n \, e^{-i \omega_0 t}
\end{equation}
The coefficients $c^0_n$ are calculated by exact diagonalization of equation (\ref{3.10}). Example results of this calculation are shown in Figure \ref{FigureA}. In particular, $c^0_n$ has a broad distribution in the SF phase, where the order parameter $\psi{\left( \mathbf{r}, t \right)}$ does not vanish, whereas in the MI phase:
\begin{equation}\label{3.16}
c^0_n = \delta_{n, \overline{n}}
\end{equation}
where $\overline{n}$ is the particle filling of a given Mott lobe. The energy eigenvalue of the stationary solution (\ref{3.15}) reads:
\begin{equation}\label{3.17}
\small
\hbar \omega_0 = -4 \, d \, J \left| \psi_0 \right|^2 + \sum_{n = 0}^{\infty} \left[ \frac{U}{2} \, n \left( n - 1 \right) - \mu \, n \right] \left| c^0_n \right|^2
\end{equation}
In Figure \ref{FigureA} we compare the numerical solution of (\ref{3.10}) for the coefficients $c^0_n$ with the corresponding outcomes for the coherent state distribution calculated as in (\ref{3.13}). Figure \ref{FigureA} shows that the $c^0_n$ and $c^{0, \text{coh}}_n$ coincide in the limit $J/U \to \infty$ at constant $\mu/U$, so that the use of the DGPE in this regime is completely justified by the match with the Gutzwiller equations.

\subsection{Considerations on the diagonalization of the saddle-point equations}
The stationary version of equation (\ref{3.10}) has the form of a self-consistent eigenvalue problem which should be solved by iteration. Nevertheless, we can perform a much simpler calculation by solving a different eigenvalue problem which involves the cancellation of the hopping energy in (\ref{3.17}) from the main diagonal of the saddle-point matrix in (\ref{3.11}):
\begin{equation}\label{3.18}
\hbar \, \tilde{\omega}_0 \, c^0_n = \sum_m \tilde{H}_{n m} \, c^0_m
\end{equation}
\begin{equation}\label{3.19}
\tilde{H}_{n m} = \left[ \frac{U}{2} n \left( n - 1 \right) - \mu \, n + 2 \, d \,  J \left| \psi_0 \right|^2 \right] \delta_{n, m} - 2 \, d \, J \, \sqrt{m} \, \psi^{*}_0 \, \delta_{n + 1, m} - 2 \, d \, J \, \sqrt{n} \, \psi_0 \, \delta_{n, m + 1}
\end{equation}
\begin{wrapfigure}{r}{0.5\textwidth}
	\centering
	\includegraphics[width=0.5\textwidth]{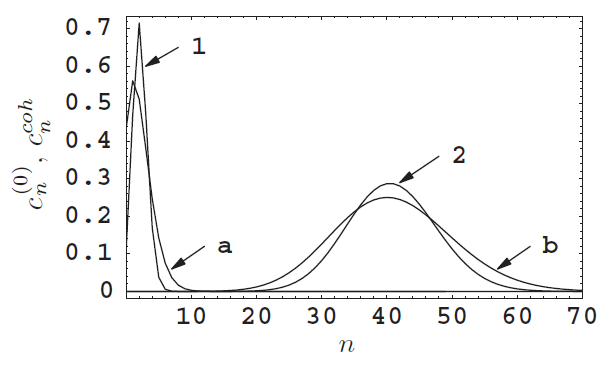}
	\caption{Comparison of $c^0_n$ (1, 2) with $c^{0, \text{coh}}$ (a, b) for the scaled chemical potential $\mu/U = 1.2$ and the tunnelling rates $2 d J/U = 0.5$ (1, a), $50$ (2, b). Figure taken from \cite{krutitsky_navez}.}
	\label{FigureA}
\end{wrapfigure}
We can observe that the hopping term subtracted from the main diagonal of (\ref{3.19}) in square brackets and the hopping contribution to (\ref{3.17}) differs by a factor $2$. This is due to the fact that the eigenvalue of the Gutzwiller equations $\hbar \omega_0$ is not the exact ground state energy associated to the original Hamiltonian density. The same feature occurs in the case of the Gross-Pitaevskii equation, whose eigenvalue is the chemical potential, or DFT calculations. It follows that the correct minimization has to take into account the true ground state energy, which contains only half of the hopping term contributing to $\hbar \, \omega_0$. \\
Secondly, one has to minimise the lowest eigenvalue $\tilde{\omega}_0$ with respect to $\psi_0$ and finally use the latter result for solving the original stationary equations.

\medskip

\begin{figure}[!htp]
	\centering
	\includegraphics[width=0.5\textwidth]{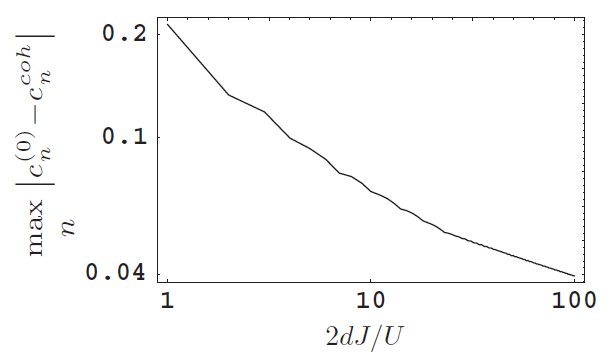}
	\caption{Deviation of the exact ground state distribution from the coherent state one for the scaled chemical potential $\mu/U = 1.2$. Figure taken from \cite{krutitsky_navez}.}
	\label{FigureB}
\end{figure}

In the numerical calculations presented in this chapter and future applications, $n$ was restricted by some finite cut-off $n_{max}$, that is $c^0_n \equiv 0$ for $n > n_{max}$. The cut-off number has been chosen so that its influence on the solutions to the eigenvalue problems is negligible. For example, for the plots shown in Figures \ref{FigureA} and \ref{FigureB} it has been enough to use $n_{max} = 500$.

\section{Excitations}
Let us consider a small perturbation of the ground state stationary solution in the form:
\begin{equation}\label{3.20}
c_n{\left( \mathbf{r}, t \right)} = \left[ c^0_n + c^{\left( 1 \right)}{\left( \mathbf{r}, t \right)} + \dots \right] e^{-i \omega_0 t}
\end{equation}
where:
\begin{equation}\label{3.21}
c^{\left( 1 \right)}{\left( \mathbf{r}, t \right)} \equiv u_{\mathbf{k}, n} e^{i \left( \mathbf{k} \cdot \mathbf{r} - \omega_{\mathbf{k}} t \right)} + v^{*}_{\mathbf{k}, n} e^{-i \left( \mathbf{k} \cdot \mathbf{r} - \omega_{\mathbf{k}} t \right)}
\end{equation}
where $u_{\mathbf{k}, n}$ and $u_{\mathbf{k}, n}$ are complex amplitudes in general. Substituting the expression (\ref{3.21}) in the GE equations (\ref{3.10}) and retaining only linear terms with respect to $u_{\mathbf{k}, n}$ and $u_{\mathbf{k}, n}$, we obtain the following system of linear equations:
\begin{equation}\label{3.22}
\hbar \omega_{\mathbf{k}}
\begin{pmatrix}
\vec{u}_{\mathbf{k}} \\
\vec{v}_{\mathbf{k}}
\end{pmatrix}
=
\begin{pmatrix}
A_{\mathbf{k}} & B_{\mathbf{k}} \\
-B_{\mathbf{k}} & -A_{\mathbf{k}}
\end{pmatrix}
\begin{pmatrix}
\vec{u}_{\mathbf{k}} \\
\vec{v}_{\mathbf{k}}
\end{pmatrix}
\end{equation}
where the vector notation refers to the Fock occupation states, while:
\begin{equation}\label{3.23}
\begin{aligned}
A^{n m}_{\mathbf{k}} &= -J{\left( \mathbf{0} \right)} \, \psi_0 \left( \sqrt{m} \, \delta_{n + 1, m} + \sqrt{n} \, \delta_{n, m + 1} \right) + \left[ \frac{U}{2} \, n \left( n - 1 \right) - \mu \, n - \hbar \, \omega_0 \right] \delta_{n, m} + \\
&- J{\left( \mathbf{k} \right)} \left( \sqrt{n + 1} \sqrt{m + 1} \, c^0_{n + 1} \, c^0_{m + 1} + \sqrt{n} \sqrt{m} \, c^0_{n - 1} \, c^0_{m - 1} \right)
\end{aligned}
\end{equation}
\begin{equation}\label{3.24}
B^{n m}_{\mathbf{k}} = -J{\left( \mathbf{k} \right)} \left( \sqrt{n + 1} \sqrt{m} \, c^0_{n + 1} \, c^0_{m - 1} + \sqrt{n} \sqrt{m + 1} \, c^0_{n - 1} \, c^0_{m + 1} \right)
\end{equation}
\begin{equation}\label{3.25}
J{\left( \mathbf{k} \right)} \equiv 2 \, d \, J - \varepsilon{\left( \mathbf{k} \right)} \quad , \quad \varepsilon{\left( \mathbf{k} \right)} = 4 \, J \sum_{\alpha = 1}^{d} \sin^2{\left( \frac{k_{\alpha}}{2} \right)}
\end{equation}
where $\varepsilon{\left( \mathbf{k} \right)}$ is the energy of a free particle. The linear system (\ref{3.22}) is valid for both phases and generalizes the BdGE (\ref{3.11}) and (\ref{3.14}) previously derived for coherent states. \\
The energy increase due to the perturbation (\ref{3.21}) has the form:
\begin{equation}\label{3.26}
\Delta E = \left( \left| \vec{u}_{\mathbf{k}} \right|^2 - \left| \vec{v}_{\mathbf{k}} \right|^2 \right) \hbar \omega_{\mathbf{k}}
\end{equation}
as the eigenvalue problem in (\ref{3.22}) is characterized by a pseudo-hermitian quadratic form. Formally, equations (\ref{3.22}) have solutions with positive and negative energies that are equivalent under the symmetry transformations $\omega_{\mathbf{k}} \to -\omega_{\mathbf{k}}$, $\mathbf{k} \to -\mathbf{k}$, $\vec{u}_{\mathbf{k}} \to \vec{v}^{*}_{\mathbf{k}}$, $\vec{v}^{*}_{\mathbf{k}} \to \vec{u}_{\mathbf{k}}$, so that only solutions with positive energies will be considered in the following. The corresponding eigenvectors are chosen to satisfy the orthonormality relations:
\begin{equation}\label{3.27}
\vec{u}^{*}_{\mathbf{k}, \alpha} \cdot \vec{u}_{\mathbf{k}, \beta} - \vec{v}^{*}_{\mathbf{k}, \alpha} \cdot \vec{v}_{\mathbf{k}, \beta} = \delta_{\alpha, \beta}
\end{equation}
where $\alpha$ and $\beta$ label different eigenmodes. Perturbation (\ref{3.21}) gives rise to plane waves for the order parameter:
\begin{equation}\label{3.28}
\psi{\left( \mathbf{r}, t \right)} = \psi_0 + \psi^{\left( 1 \right)}{\left( \mathbf{r}, t \right)} + \dots
\end{equation}
where:
\begin{equation}\label{3.29}
\psi^{\left( 1 \right)}{\left( \mathbf{r}, t \right)} = U_{\mathbf{k}} e^{i \left( \mathbf{k} \cdot \mathbf{r} - \omega_{\mathbf{k}} t \right)} + V^{*}_{\mathbf{k}} e^{-i \left( \mathbf{k} \cdot \mathbf{r} - \omega_{\mathbf{k}} t \right)}
\end{equation}
\begin{equation}\label{3.30}
U_{\mathbf{k}} = \sum_{n = 0}^{\infty} \sqrt{n} \left( c^0_n \, u_{\mathbf{k}, n + 1} + c^0_{n + 1} \, v_{\mathbf{k}, n} \right) \quad , \quad V_{\mathbf{k}} = \sum_{n = 0}^{\infty} \sqrt{n} \left( c^0_{n + 1} \, u_{\mathbf{k}, n} + c^0_n \, v_{\mathbf{k}, n + 1} \right)
\end{equation}
The perturbations of the total density and the condensate density are given by the following expression:
\begin{equation}\label{3.31}
n{\left( \mathbf{r}, t \right)} = n_0 + \left[ \mathcal{A}_{\mathbf{k}} \, e^{i \left( \mathbf{k} \cdot \mathbf{r} - \omega_{\mathbf{k}} t \right)} + \text{c.c.} \right]
\end{equation}
\begin{equation}\label{3.32}
\mathcal{A}_{\mathbf{k}} = \sum_{n = 0}^{\infty} n \, c^0_n \left( u_{\mathbf{k}, n} + v_{\mathbf{k}, n} \right)
\end{equation}
and:
\begin{equation}\label{3.33}
\rho_c{\left( \mathbf{r}, t \right)} = \psi^2_0 + \left[ \mathcal{B}_{\mathbf{k}} \, e^{i \left( \mathbf{k} \cdot \mathbf{r} - \omega_{\mathbf{k}} t \right)} + \text{c.c.} \right]
\end{equation}
\begin{equation}\label{3.34}
\mathcal{B}_{\mathbf{k}} = \psi_0 \sum_{n = 0}^{\infty}\sqrt{n} \left[ c^0_n \left( u_{\mathbf{k}, n + 1} + v_{\mathbf{k}, n + 1} \right) + c^0_{n + 1} \left( u_{\mathbf{k}, n} + v_{\mathbf{k}, n} \right) \right]
\end{equation}
The following sections will be dedicated to the analysis of the properties of the excitations emerging from (\ref{3.21}) and (\ref{3.22}) in relation with their perturbative effect on the main observables of the model.

\subsection{Mott insulator}
For the MI phase the coefficients $c^0_n$ have the simple analytical form (\ref{3.16}), so that the diagonalization problem (\ref{3.22}) reduces to the solution of a $2 \times 2$ linear system that couples only $u_{\mathbf{k}, \overline{n} - 1}$ to $v_{\mathbf{k}, \overline{n} + 1}$ and $u_{\mathbf{k}, \overline{n} + 1}$ to $v_{\mathbf{k}, \overline{n} - 1}$. The lowest-energy excitation spectrum consists of two branches with energies:
\begin{equation}\label{3.35}
\hbar \omega^{\pm}_{\mathbf{k}} = \frac{1}{2} \sqrt{U^2 - 4 \, U \, J{\left( \mathbf{k} \right)} \left( \overline{n} + \frac{1}{2} \right) + J^2{\left( \mathbf{k} \right)}} \pm \left[ U \left( \overline{n} - \frac{1}{2} \right) - \mu - \frac{ J{\left( \mathbf{k} \right)}}{2} \right]
\end{equation}
The same result has been obtained by using the Hubbard-Stratonovich transformation and within the Schwinger boson approach. \\
These two branches are presented in the following figure and display a gap. Equations (\ref{3.29}) and (\ref{3.31}) show that no density waves are created in the two modes, although the perturbation of the order parameter does not vanish. A suitable rewriting:
\begin{equation}\label{3.36}
\begin{aligned}
&\hbar \omega^+_{\mathbf{k}} = \varepsilon^p_{\mathbf{k}} - \mu \\
&\hbar \omega^-_{\mathbf{k}} = -\varepsilon^h_{\mathbf{k}} + \mu
\end{aligned}
\end{equation}
\begin{figure}
	\centering
	\includegraphics[width=0.5\textwidth]{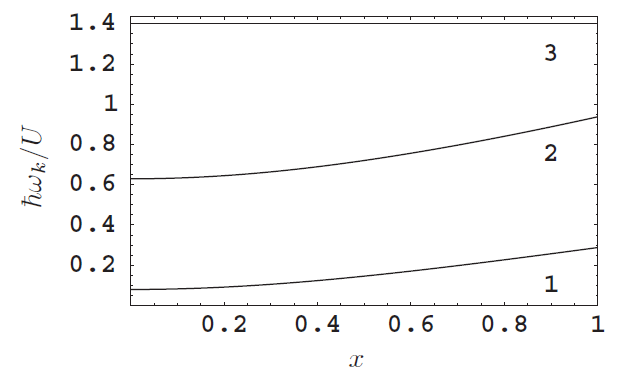}
	\caption{First three branches $\hbar \omega^-_{\mathbf{k}}$ (1), $\hbar \omega^+_{\mathbf{k}}$ (2), $\hbar \omega_{\lambda = 0}$ (3) of the excitation spectrum of the Mott phase for $\mu/U = 1.2$ and $2 d J/U = 0.05$, which correspond to $\overline{n} = 2$. $x$ is the reduced momentum variable $x = \left[ 1/d \, \sum_{a = 1}^{d} \sin^2{\left( k_a/2 \right)} \right]^{1/2}$. Figure taken from \cite{krutitsky_navez}.}
	\label{FigureC}
\end{figure}
allows to identify the lowest modes in the Mott phase with pure particle and hole excitations. At low momentum the additional particles or holes exhibit a free-particle dispersion, since the energy penalty for double occupation is the same at every site. \\
Other solutions of equations (\ref{3.22}) are independent of the momentum $\mathbf{k}$ and have the following narrow-banded energies:
\begin{equation}\label{3.37}
\hbar \, \omega_{\alpha} = \frac{U}{2} \left[ \alpha \left( \alpha - 1 \right) - \overline{n} \left( \overline{n} - 1 \right) \right] - \mu \left( \alpha - \overline{n} \right)
\end{equation}
where $\alpha$ is a non-negative integer such that $\alpha \neq \overline{n}, \overline{n} \pm 1$. If $\overline{n}$ is the smallest integer larger than $\mu/U$, these excitation energies are always positive. The corresponding eigenvectors have the simple form $u_{\mathbf{k}, n, \alpha} = \delta_{n, \alpha}$ and $u_{\mathbf{k}, n, \alpha} = 0$, so that all the amplitudes of the perturbation waves defined in (\ref{3.29}), (\ref{3.31}) and (\ref{3.33}) vanish. \\
The mean-field boundary between the MI and the SF phases is determined by the disappearance of the gap of the first excitation mode in the MI phase, i.e. when $\hbar \omega^{\pm}_{\mathbf{0}} = 0$ depending on the type of the lowest excitation, hence one recovers the critical hopping-interaction ratio:
\begin{equation}\label{3.38}
2 d \left( J/U \right)_c = \frac{\left( \overline{n} - \mu/U \right) \left( \mu/U - \overline{n} + 1 \right)}{\left( 1 + \mu/U \right)}
\end{equation}
whose maximal value identifies the tip of the Mott lobe and reads:
\begin{equation}\label{3.39}
2 d \left( J/U \right)^{tip}_c = \left( \sqrt{\overline{n} + 1} - \sqrt{\overline{n}} \right)^2
\end{equation}
for a chemical potential given by:
\begin{equation}\label{3.40}
\left( \mu/U \right)^{tip}_c = \sqrt{\overline{n} \left( \overline{n} + 1 \right)} - 1
\end{equation}
For $J/U > \left( J/U \right)_c$ at a given chemical potential the lowest eigenfrequency $\omega^{\pm}_{\mathbf{0}}$ becomes negative, so that the Mott state does not correspond to the stable ground state anymore. \\
It is worth mentioning the interesting features exhibited by the excitation spectrum on the critical boundary. On the tip of the Mott lobe, i.e. for $J/U = \left( J/U \right)^{tip}_c$ and $\mu/U = \left( \mu/U \right)^{tip}_c$, the excitation energies can be rewritten as:
\begin{equation}\label{3.41}
\hbar \omega^{\pm}_{\mathbf{k}} = \left[ \overline{n} \left( \overline{n} + 1 \right) U \, \varepsilon{\left( \mathbf{k} \right)} + \frac{\varepsilon^2{\left( \mathbf{k} \right)}}{4} \right]^{1/2} \pm \frac{\varepsilon{\left( \mathbf{k} \right)}}{2}
\end{equation}
In the low energy limit the two branches are degenerate and have a linear dispersion $\hbar \omega^{\pm}_{\mathbf{k}} = c^{tip}_s \left| \mathbf{k} \right|$, where the tip sound velocity has the expression:
\begin{equation}\label{3.42}
c^{tip}_s = \frac{U}{\hbar} \sqrt{\left( J/U \right)^{tip}_c} \left[ \overline{n} \left( \overline{n} + 1 \right) \right]^{1/4}
\end{equation}
expressed in units of number of sites per second. For other points laying on the boundary no low-energy degeneracy appears, while the sound velocity vanishes as the excitation energies assume a Galilean dispersion $\hbar \omega^{\pm}_{\mathbf{k}} \sim \mathbf{k}^2$.

\subsection{Superfluid}
In the SF phase the diagonalization procedure related to (\ref{3.22}) can be solved only numerically. The subsequent results reveal that the excitation spectrum has the form of a band structure whose lowest branch has no gap at $\mathbf{k} = \mathbf{0}$. This can be identified with a Goldstone mode associated with the spontaneous breaking of the phase symmetry. Remarkably, in the phase mode the amplitude of the total-density wave (\ref{3.32}) is larger than the amplitude of the condensate-density wave (\ref{3.34}). In particular, the condition $\mathcal{A}_{\mathbf{k}}/\mathcal{B}_{\mathbf{k}} > 1$ means that the condensed part and the normal part oscillate in phase. \\
Higher modes emerging from the Gutzwiller approach are not comprised in the Gross-Pitaevskii theory. They have gaps $\Delta_{\alpha} = \hbar \omega_{\alpha, \mathbf{k}}$ which grow monotonically with the increase of the hopping strength. As shown in Appendix \ref{B}, for large $J$ one can derive the following asymptotic identity:
\begin{equation}\label{3.43}
\Delta_{\alpha} = 2 \, d \, J \, \alpha + \frac{U}{2} \, \alpha \left( n_0 + \alpha - 1 \right)
\end{equation}
We note that only the first two excitation modes have a strong dependence on the momentum, while higher branches inherit their narrow-banded structure from the corresponding modes in the MI phase.

\medskip

For the second mode $\left( \alpha = 2 \right)$ the amplitude of the total-density wave is much less than that of the condensate density wave, which means that the oscillations of the condensate and normal components are out-of-phase. This does not necessarily mean that there is an exchange of particles between the two components. \\
Due to the reasons explained above, the second mode is called \textit{amplitude} or \textit{Higgs mode}, as this excitation emerges at the $\text{O}{\left( 2 \right)}$ quantum critical point and within the realm of strong interactions as a consequence of exact or approximate particle-hole symmetry (Lorentz invariance) in the Bose-Hubbard dynamics and his associated to amplitude oscillations of the order parameter of the spontaneously-symmetry-broken phase \cite{lattice_depth_modulation_2, sachdev_higgs, zwerger_higgs, podolsky_higgs}. In general, Higgs modes appear as collective excitations in quantum many-body systems and are coupled to Goldstone modes depending on the parameters of the model under study.

\medskip

Let us return to the properties of the Goldstone mode. As derived in Appendix \ref{B}, this gapless excitation presents a linear dispersion relation $\hbar \omega_{1, \mathbf{k}} = c^0_s \left| \mathbf{k} \right|$ for $\left| \mathbf{k} \right| \to 0$, where:
\begin{equation}\label{3.44}
c^0_s = \frac{1}{\hbar} \sqrt{\frac{2\, J}{\kappa}} \psi_0
\end{equation}
is again the sound velocity and $\kappa = \frac{\dpar n_0}{\dpar \mu}$ is the mean-field compressibility. All these results demonstrate that the Gutzwiller approximation in gapless and provides predictions compatible with the BdGE within the Gross-Pitaevskii theory. \\
The fact that the sound velocity is identically zero on the phase boundary except for the tips of the Mott lobes can be understood by considering the properties of $\psi_0$ and $\kappa$. As the critical line is approached from the SF part of the phase diagram, the order parameter $\psi_0$ tends always continuously to zero. On the other hand, the compressibility reaches a finite value at every point of the boundary except the tip of the MI lobes, where it tends continuously to zero such that the ratio $\psi_0/\sqrt{\kappa}$ is finite. \\
\begin{wrapfigure}{r}{0.5\textwidth}
	\centering
	\includegraphics[width=0.5\textwidth]{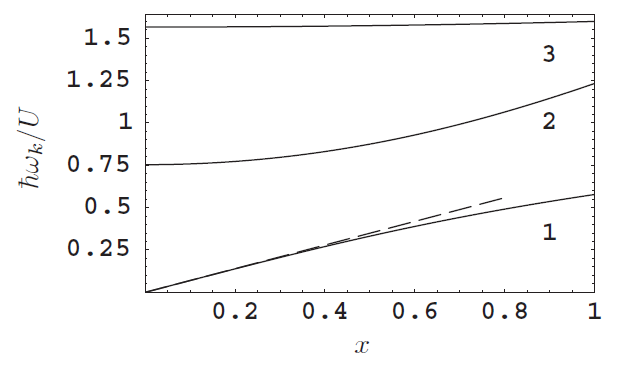}
	\caption{First three branches $\hbar \omega_{\mathbf{k}, \lambda}$ ($\lambda = 1, 2, 3$) of the excitation spectrum of the superfluid phase for $\mu/U = 1.2$ and $2 d J/U = 0.15$. The straight dashed line represents the linear approximation with the sound velocity (\ref{3.44}). $x$ is the reduced momentum variable $x = \left[ 1/d \, \sum_{a = 1}^{d} \sin^2{\left( k_a/2 \right)} \right]^{1/2}$. Figure taken from \cite{krutitsky_navez}.}
	\label{FigureD}
\end{wrapfigure}
For a weakly-interacting gas $\left( J \gg U \right)$ we have $\left| \psi_0 \right|^2 \simeq n_0$ and $\kappa \sim 1/U$, so that we recover the Bogoliubov dispersion relation (see again Appendix \ref{B}) and the expected identity for the sound velocity:
\begin{equation}\label{3.45}
c^B_s = \frac{1}{\hbar} \sqrt{2\, J \, U \, n_0}
\end{equation}
In the opposite limit $\left( J \ll U \right)$ the superfluid regions with atomic densities $\overline{n} < n_0 < \overline{n} + 1$ are confined in the regions defined by $\mu_- < \mu < \mu_+$, where $\mu_-$ is the upper boundary of the Mott lobe with density $\overline{n}$ and $\mu_+$ is the lower boundary of the Mott lobe with density $\overline{n} + 1$. For $J \to 0$ the chemical potential turns out to be a linear function of the density:
\begin{equation}\label{3.46}
\mu = \mu_- + \left( \mu_+ - \mu_- \right) \left( n_0 - \overline{n} \right)
\end{equation}
Using equation (\ref{3.38}) up to the first order in $J$ we obtain:
\begin{equation}\label{3.47}
\mu_{\pm} = \overline{n} \, U \pm 2 \, d \, J \left( \overline{n} + 1 \right)
\end{equation}
hence:
\begin{equation}\label{3.48}
\kappa = \left[ 4 \, d \, J \left( \overline{n} + 1 \right) \right]^{-1}
\end{equation}
Using the thermodynamic relation $\mu = \frac{\dpar E}{\dpar n_0} = $ and the condition $E_{\pm} = \overline{n} \left( \overline{n} \pm 1 \right) U/2$ on the boundaries of the Mott lobe, the total energy in the strong-interacting limit becomes:
\begin{equation}\label{3.49}
E = \overline{n} \, U \left( n_0 - \frac{\overline{n} + 1}{2} \right) - 2 \, d \, J \left( \overline{n} + 1 \right) \left( n_0 - \overline{n} \right) \left( \overline{n} - n_0 + 1 \right)
\end{equation}
The relation $\frac{\dpar E}{\dpar J} = -2 \, J \left| \psi_0 \right|^2$ allows to deduce the order parameter expression:
\begin{equation}\label{3.50}
\left| \psi_0 \right|^2 = \left( \overline{n} + 1 \right) \left( n_0 - \overline{n} \right) \left( \overline{n} - n_0 + 1 \right)
\end{equation}
Equations (\ref{3.49}) and (\ref{3.50}) coincide the ones obtained through a perturbation theory approach around the critical region. Substituting (\ref{3.48}) and (\ref{3.50}) into (\ref{3.44}) we obtain finally:
\begin{equation}\label{3.51}
c^0_s = \frac{2 \, J \left( \overline{n} + 1 \right)}{\hbar} \sqrt{2 d \left( n_0 - \overline{n} \right) \left( \overline{n} - n_0 + 1 \right)}
\end{equation}
As expected, in the strong-interacting limit the sound velocity vanishes at $n_0 = \overline{n}, \overline{n} + 1$ and takes maximal values at $n_0 = \overline{n} + \frac{1}{2}$. This qualitative behaviour is the same as in the case of hard-core bosons in 1D, where the sound velocity has the form:
\begin{equation}\label{3.52}
c^{HC}_s = \frac{2 \, J}{\hbar} \sin{\left( \pi \langle \hat{n} \rangle \right)}
\end{equation}
with $0 < \langle \hat{n} \rangle < 1$. It is interesting to notice that (\ref{3.52}) recalls the expression of the Fermi velocity of the corresponding non-interacting Fermi-Hubbard model below half filling. In fact, since the bare energy band is given by:
\begin{equation}\label{3.53}
\omega{\left( \mathbf{k} \right)} = -\frac{2 \, J}{\hbar} \sum_{a = 1}^{d} \cos{\left( k_a \right)}
\end{equation}
one can derive the local velocity of the higher energy state from:
\begin{equation}\label{3.54}
v{\left( \mathbf{k} \right)} = \frac{\dpar \omega{\left( \mathbf{k} \right)}}{\dpar \mathbf{k}} = \frac{2 \, J}{\hbar} \sum_{a = 1}^{d} \sin{\left( k_a \right)} \, \mathbf{e}_a
\end{equation}
Finally, parametrizing the momentum value through the filling number $0 < \langle n \rangle < 1$ as $k = \pi \, \langle n \rangle$, one obtains the result (\ref{3.52}). This fact comes as no surprise, as hard core bosons in 1D can be exactly studied through a so-called fermionization procedure.

\chapter{Quantum theory of the excitations}\label{quantum}
\markboth{\MakeUppercase{4. Quantum theory}}{\MakeUppercase{4. Quantum theory}}
In Chapter \ref{MF} we have shown how a simple manipulation of the Gutzwiller ansatz in terms of momentum-dependent linear perturbations of the occupation weights $c_n{\left( \mathbf{r} \right)}$ leads to a complete characterization of the many-body excitation spectrum of the Bose-Hubbard model in both the phases of the phase diagram, as well as to yield significant results on the parameters of the low-energy collective modes, in particular the scaling of the Higgs gap and the behaviour of the sound velocity at the transition. \\
On the other hand, the mean-field formalism alone does not allow a formal and clear study of genuine quantum effects due to the occurence of strong interactions, for example the quantum depletion and the effective superfluid fraction of the systems previously determined at the level of the weak interactions in Chapter \ref{Bogoliubov}. \\
As first original content of the present thesis research, in the following Chapter we illustrate a formal procedure for quantizing the fluctuations of the Gutzwiller weights $c_n{\left( \mathbf{r} \right)}$, which to this purpose are treated as complex conjugate variables at the Lagrangian level and canonically promoted to operators. After passing to the quasi-particle description through a Bogoliubov diagonalization of the Gutzwiller Hamiltonian problem, the following step consists in analysing the fluctuation part of the density and Bose field operators.

\section{Quantization of the Gutzwiller theory}
Summarising the most important conceptual nodes addressed in the previous chapter, the Gutzwiller approach corresponds formally to a mean-field solution of a quantum problem based on a spatially factorized ground state ansatz for the wave function:
\begin{equation}\label{4.1}
| \Psi \rangle = \prod_{\mathbf{r}} \sum_n c_n{\left( \mathbf{r} \right)} | n, \mathbf{r} \rangle
\end{equation}
The optimization of the energy expectation value corresponds to the minimization of the corresponding single-site Hamiltonian with respect to the Gutzwiller parameters $c_n{\left( \mathbf{r} \right)}$. \\
The central aim of this chapter is to study the quantum fluctuations living above the Gutzwiller variational ground state (\ref{4.1}) and their impact on the expectation value of physical observables to the lowest order in the fluctuations. \\
As a first step, we can define small fluctuations above the mean-field ground state by perturbing its parameters $c^0_n$ in terms of small variations $\delta c_n{\left( \mathbf{r} \right)}$:
\begin{equation}\label{4.2}
c_n{\left( \mathbf{r} \right)} = A \, c^0_n + \delta c_n{\left( \mathbf{r} \right)}
\end{equation}
in a similar way as done in equation (\ref{3.19}) for studying the mean-field excitation dynamics and in analogy with the approach followed by Fabrizio in \cite{fabrizio}. As it will be demonstrated later, the small perturbations $\delta c_n{\left( \mathbf{r} \right)}$ can be defined to be orthogonal to the ground state solution in the occupation number subspace:
\begin{equation}\label{4.3}
\sum_n \left( c^0_n \right)^{*} \delta c_n{\left( \mathbf{r} \right)} \equiv 0
\end{equation}
while the coefficient $A$ is intended as a renormalizing factor due to the perturbation of the ground state distribution $c^0_n$, in accordance with the conservation of probability. The value of $A$ is fixed by the usual normalization condition (\ref{3.4}) for the coefficients $c_n{\left( \mathbf{r} \right)}$, which in our theory turns out to work also as a gauge-fixing condition:
\begin{equation}\label{4.4}
\sum_n \left| c_n{\left( \mathbf{r} \right)} \right|^2 = \left| A \right|^2 \sum_n \left| c^0_n \right|^2 + \sum_n \left| \delta c_n{\left( \mathbf{r} \right)} \right|^2 = \left| A \right|^2 + \sum_n \left| \delta c_n{\left( \mathbf{r} \right)} \right|^2 = 1
\end{equation}
hence we can choose:
\begin{equation}\label{4.5}
A = \sqrt{1 - \sum_n \left| \delta c_n{\left( \mathbf{r} \right)} \right|^2}
\end{equation}
Formulating a theory of the quantum fluctuations above the ground state requires the quantization of the Lagrangian degrees of freedom. Actually, it is easy to realize that the Gutzwiller ansatz allows to rewrite the Bose-Hubbard Hamiltonian in the form of a functional of the parameters $c_n{\left( \mathbf{r} \right)}$ only:
\begin{equation}\label{4.6}
E{\left[ c_n{\left( \mathbf{r} \right)}, c^{*}_n{\left( \mathbf{r} \right)} \right]} = -J \sum_{\langle \mathbf{r}, \mathbf{s} \rangle} \left[ \psi^{*}{\left( \mathbf{r} \right)} \, \psi{\left( \mathbf{s} \right)} + \text{c.c.} \right] + \sum_{\mathbf{r}} \sum_n H_n \left| c_n{\left( \mathbf{r} \right)} \right|^2
\end{equation}
where:
\begin{equation}\label{4.7}
\begin{aligned}
&\psi{\left( \mathbf{r} \right)} = \sum_n \sqrt{n} \, c^{*}_{n - 1}{\left( \mathbf{r} \right)} \, c_n{\left( \mathbf{r} \right)} \\
&H_n = \frac{U}{2} \, n \left( n - 1 \right) - \mu \, n
\end{aligned}
\end{equation}
In particular, the Lagrangian functional of the Bose-Hubbard model calculated over the Gutzwiller wave function reads explicitly:
\begin{equation}\label{4.8}
L{\left[ c_n{\left( \mathbf{r}, t \right)}, c^{*}_n{\left( \mathbf{r}, t \right)} \right]} = i \hbar \sum_{\mathbf{r}} \sum_n c^{*}_n{\left( \mathbf{r} \right)} \dpar_t c_n{\left( \mathbf{r} \right)} - E{\left[ c_n{\left( \mathbf{r} \right)}, c^{*}_n{\left( \mathbf{r} \right)} \right]}
\end{equation}
As we have already observed while discussing the mean-field solution to the many-body problem, the variational ground state parameters $c^0_n$ correspond to the lowest-energy solution of the saddle-point equations extracted from the action of the Gutzwiller parameters, namely:
\begin{equation}\label{4.9}
\begin{aligned}
&\frac{\delta S{\left[ c_n{\left( \mathbf{r}, t \right)}, c^{*}_n{\left( \mathbf{r}, t \right)} \right]}}{\delta c^{*}_n{\left( \mathbf{r}, t \right)}} = 0 \quad \longrightarrow \quad i \, \hbar \, c_n{\left( \mathbf{r} \right)} = \frac{\delta E{\left[ c_n{\left( \mathbf{r}, t \right)}, c^{*}_n{\left( \mathbf{r}, t \right)} \right]}}{\delta c_n{\left( \mathbf{r}, t \right)}} \\
&\frac{\delta S{\left[ c_n{\left( \mathbf{r}, t \right)}, c^{*}_n{\left( \mathbf{r}, t \right)} \right]}}{\delta c_n{\left( \mathbf{r}, t \right)}} = 0 \quad \longrightarrow \quad i \, \hbar \, c^{*}_n{\left( \mathbf{r} \right)} = -\frac{\delta E{\left[ c_n{\left( \mathbf{r}, t \right)}, c^{*}_n{\left( \mathbf{r}, t \right)} \right]}}{\delta c_n{\left( \mathbf{r}, t \right)}}
\end{aligned}
\end{equation}
where $S{\left[ c_n{\left( \mathbf{r}, t \right)}, c^{*}_n{\left( \mathbf{r}, t \right)} \right]} = \int dt \, L{\left[ c_n{\left( \mathbf{r}, t \right)}, c^{*}_n{\left( \mathbf{r}, t \right)} \right]}$. \\
Looking at the RHS of equations (\ref{4.9}), we can recognize that $c_n{\left( \mathbf{r} \right)}$ and $c^{*}_n{\left( \mathbf{r} \right)}$ behave as canonically conjugate variables which satisfy classical Poisson brackets with respect to the Hamiltonian reformulation given by (\ref{4.6}):
\begin{equation}\label{4.10}
\left\{ c_n{\left( \mathbf{r} \right)}, c^{*}_m{\left( \mathbf{s} \right)} \right\}_P = \delta_{n, m} \delta{\left( \mathbf{r} - \mathbf{s} \right)} \quad , \quad i \, \hbar \, \dot{c}_n{\left( \mathbf{r} \right)} = \left\{ c_n{\left( \mathbf{r} \right)}, E \right\}_P  \quad , \quad i \, \hbar \, \dot{c}^{*}_n{\left( \mathbf{r} \right)} = -\left\{ c^{*}_n{\left( \mathbf{r} \right)}, E \right\}_P
\end{equation}
Having identified the conjugate variables of the Lagrangian functional (\ref{4.8}) by  and their classical Hamiltonian with the energy density (\ref{4.6}), \uline{we can perform a canonical quantization procedure by treating $c_n{\left( \mathbf{r} \right)}$ and $c^{*}_n{\left( \mathbf{r} \right)}$ as conjugate complex scalar fields and requiring that these operators satisfy canonical commutation relations at equal time}, given by:
\begin{equation}\label{4.11}
\left[ c_n{\left( \mathbf{r} \right)}, c_m{\left( \mathbf{s} \right)} \right] = 0 \quad , \quad \left[ c^{\dagger}_n{\left( \mathbf{r} \right)}, c^{\dagger}_m{\left( \mathbf{s} \right)} \right] = 0 \quad , \quad \left[ c_n{\left( \mathbf{r} \right)}, c^{\dagger}_m{\left( \mathbf{s} \right)} \right] = i \, \hbar \, \delta_{n, m} \, \delta{\left( \mathbf{r} - \mathbf{s} \right)}
\end{equation}
which substitute the first Poisson bracket in equation (\ref{4.10}). Therefore, introducing the operator notation, the perturbative expression of the Gutzwiller variables (\ref{4.2}) turns into the sum of a $c$-number (multiplied by a renormalization operator) and a fluctuation operator:
\begin{equation}\label{4.12}
\hat{c}_n{\left( \mathbf{r} \right)} = \hat{A} \, c^0_n + \delta \hat{c}_n{\left( \mathbf{r} \right)}
\end{equation}
where $\hat{A} = \left[ 1 - \sum_n \delta \hat{c}^{\dagger}_n{\left( \mathbf{r} \right)} \, \delta \hat{c}_n{\left( \mathbf{r} \right)} \right]^{1/2}$, so that from equations (\ref{4.11}):
\begin{equation}\label{4.13}
\left[ \delta \hat{c}_n{\left( \mathbf{r} \right)}, \delta \hat{c}_m{\left( \mathbf{s} \right)} \right] = 0 \quad , \quad \left[ \delta \hat{c}^{\dagger}_n{\left( \mathbf{r} \right)}, \delta \hat{c}^{\dagger}_m{\left( \mathbf{s} \right)} \right] = 0 \quad , \quad \left[ \delta \hat{c}_n{\left( \mathbf{r} \right)}, \delta \hat{c}^{\dagger}_m{\left( \mathbf{s} \right)} \right] = i \, \hbar \, \delta_{n, m} \, \delta{\left( \mathbf{r} - \mathbf{s} \right)}
\end{equation}
at the second-order in the fluctuation operators, since $\left[ \hat{A}, \hat{A} \right] = 0$ and we are interested only into Gaussian quantum fluctuations. \\
Being the Gutzwiller Hamiltonian (\ref{4.6}) not diagonal in real space, the quantization procedure is simplified if we move to momentum space. Let us consider a regular $d$-dimensional cubic lattice composed by $M$ sites with unit volume. Exploiting the translational invariance of the system, the fluctuation operators $\delta \hat{c}_n{\left( \mathbf{r} \right)}$ can expanded in terms of a plane-wave basis defined over the first Brillouin zone:
\begin{equation}\label{4.14}
\delta \hat{c}_n{\left( \mathbf{r} \right)} = \frac{1}{\sqrt{M}} \sum_{\mathbf{k} \in \text{BZ}} e^{i \mathbf{k} \cdot \mathbf{r}} \, \delta \hat{C}_n{\left( \mathbf{k} \right)}
\end{equation}
Inserting the expression (\ref{4.15}) into equations (\ref{4.13}), we find the commutation relations for the Fourier components of the Gutzwiller operators:
\begin{equation}\label{4.15}
\left[ \delta \hat{C}_n{\left( \mathbf{k} \right)}, \delta \hat{C}_m{\left( \mathbf{p} \right)} \right] = 0 \quad , \quad \left[ \delta \hat{C}^{\dagger}_n{\left( \mathbf{k} \right)}, \delta \hat{C}^{\dagger}_m{\left( \mathbf{p} \right)} \right] = 0 \quad , \quad \left[ \delta \hat{C}_n{\left( \mathbf{k} \right)}, \delta \hat{C}^{\dagger}_m{\left( \mathbf{p} \right)} \right] = i \, \hbar \, \delta_{n, m} \, \delta_{\mathbf{k}, \mathbf{p}}
\end{equation}
By inserting the operator expansion into (\ref{4.12}) into the Hamiltonian density (\ref{4.6}) using the Fourier decomposition (\ref{4.14}) and keeping inly terms up to the quadratic order in the fluctuations, we obtain the following Hamiltonian operator:
\newpage
\begin{equation}\label{416}
\hat{H}^{\left( 2 \right)} = E_0 + \frac{1}{2} \sum_{\mathbf{k}}
\begin{bmatrix}
\delta \underline{\hat{C}}^{\dagger}{\left( \mathbf{k} \right)} & -\delta \underline{\hat{C}}{\left( -\mathbf{k} \right)}
\end{bmatrix}
\underbrace{
	\begin{pmatrix}
	\overline{A}_{\mathbf{k}} & \overline{B}_{\mathbf{k}} \\
	-\overline{B}^{*}_{\mathbf{k}} & -\overline{A}^{*}_{\mathbf{k}}
	\end{pmatrix}
}_{\hat{\mathcal{L}}_{\mathbf{k}}}
\begin{bmatrix}
\delta \underline{\hat{C}}{\left( \mathbf{k} \right)} \\
\delta \underline{\hat{C}}^{\dagger}{\left( -\mathbf{k} \right)}
\end{bmatrix}
\end{equation}
where $E_0$ is the ground-state energy given obtained by the saddle point solutions $c^0_n$ and the vector notation $\underline{c}$ refers to the occupation number space and the infinite-dimensional pseudo-hermitian matrix $\hat{\mathcal{L}}_{\mathbf{k}}$ is the same object introduced in equation (\ref{3.22}) and whose solutions have been numerically treated by Krutitsky and Navez in \cite{krutitsky_navez}:
\begin{equation}\label{417}
\begin{aligned}
\overline{A}^{n m}_{\mathbf{k}} &= -J{\left( \mathbf{0} \right)} \, \psi_0 \left( \sqrt{m} \, \delta_{n + 1, m} + \sqrt{n} \, \delta_{n, m + 1} \right) + \left[ \frac{U}{2} \, n \left( n - 1 \right) - \mu \, n - \hbar \, \omega_0 \right] \delta_{n, m} + \\
&- J{\left( \mathbf{k} \right)} \left( \sqrt{n + 1} \sqrt{m + 1} \, c^0_{n + 1} \, c^0_{m + 1} + \sqrt{n} \sqrt{m} \, c^0_{n - 1} \, c^0_{m - 1} \right)
\end{aligned}
\end{equation}
\begin{equation}\label{418}
\overline{B}^{n m}_{\mathbf{k}} = -J{\left( \mathbf{k} \right)} \left( \sqrt{n + 1} \sqrt{m} \, c^0_{n + 1} \, c^0_{m - 1} + \sqrt{n} \sqrt{m + 1} \, c^0_{n - 1} \, c^0_{m + 1} \right)
\end{equation}
It is clear that the first-order contributions to (\ref{416}) vanish identically because of the definition of the parameters $c^0_n$, which are the saddle-point solutions to the Gutzwiller dynamical equations given by the right hand side of (\ref{4.9}).

\vspace{0.3cm}

\fbox{\begin{minipage}{16.5cm}
\textbf{\textcolor{blue}{Remark} - } \textsl{The diagonal part of $\overline{A}_{\mathbf{k}}$ presents again a constant shift of the chemical potential $\hbar \omega_0$, which is equal to the variational ground state energy.} In the linear response formalism adopted in \cite{krutitsky_navez} such term derives from the temporal dependence (\ref{3.19}) attributed to the perturbed parameters $c_n{\left( \mathbf{r}, t \right)}$, while \textsl{it emerges naturally in the actual context as a consequence of the renormalizing factor $A$ and, remarkably, does not violate the Goldstone theorem, as it leads to a gapless theory}. \\
Nevertheless, this means that the equations of motion of the parametric perturbations $\delta c_n{\left( \mathbf{r} \right)}$ and the mean-field distribution $c^0_n$ are coupled with each other, so that formally \textsl{a rigorous quantization procedure would corresponds to the solution of a self-consistency problem}. Therefore, \uline{the quantization of the Gutzwiller equations has strong similarities with the Hartree-Fock-Bogoliubov-Popov (HFBP) theory} \cite{popov}, which improves the standard Bogoliubov approximation for weakly-interacting ultracold Bose systems by calculating the condensate fraction $\psi^2_0$ allowing for depletion effects (see also the series of works \cite{proukakis_morgan_choi_burnett}-\cite{morgan}).\footnote{The standard HFB-Popov of weakly-interacting condensates on translationally invariant lattices consists searching the solution to the self-consistency equation $\mu = \left( \left| \psi_0 \right|^2 + 2 \, D \right) U - 2 \, d \, J$, where $D$ is the quantum depletion, together with the Bogoliubov diagonalization problem. We expect that a generalization of this set of equations can be applied to the case of the Gutzwiller equations.} \\
\uline{In the following we will neglect the role of self-consistency solutions on the quantization scheme, as only the effects due to linear-order terms are considered.} Section \ref{second_order_corrections} of this chapter will be dedicated to show that the correction due to the operators $\delta \hat{c}^{\dagger}_n{\left( \mathbf{r} \right)} \, \delta \hat{c}_n{\left( \mathbf{r} \right)}$ in $\hat{A}$ is not always small compared to the saddle-point solutions and how the lack of a more refined quantum theory of the fundamental excitations comes to light in higher-order corrections to common physical observables.
\end{minipage}}

\vspace{0.5cm}

The matrix $\hat{\mathcal{L}}_{\mathbf{k}}$ is pseudo-hermitian because of the minus sign characterizing the matrix elements of its second line. It is important to notice that this sign feature is a direct consequence of Bose statistics, as the second-order expansion (\ref{416}) derives the application of the commutations relations (\ref{4.15}) to the operator form of the Hamiltonian density (\ref{4.6}). \\
The spectral properties of the class of operators represented by $\hat{\mathcal{L}}_{\mathbf{k}}$ are well known from the mathematical theory of weakly-interacting Bose-Einstein condensates (\cite{castin}), as the corresponding right and left eigenvectors are strictly related to the algebra of Bogoliubov rotations. A more detailed discussion about the spectral decomposition of $\hat{\mathcal{L}}_{\mathbf{k}}$ is presented in Appendix \ref{C}. \\
Knowing the eigenspace of the matrix $\hat{\mathcal{L}}_{\mathbf{k}}$, the quadratic form in (\ref{416}) can be diagonalized by applying a suitable Bogoliubov rotation of the Gutzwiller operators:
\begin{equation}\label{419}
\begin{aligned}
&\delta \hat{C}_n{\left( \mathbf{k} \right)} = \sum_{\alpha} u_{\alpha, \mathbf{k}, n} \, \hat{b}_{\alpha, \mathbf{k}} + \sum_{\alpha} v^{*}_{\alpha, -\mathbf{k}, n} \, \hat{b}^{\dagger}_{\alpha, -\mathbf{k}} \\
&\delta \hat{C}^{\dagger}_n{\left( \mathbf{k} \right)} = \sum_{\alpha} u^{*}_{\alpha, \mathbf{k}, n} \, \hat{b}^{\dagger}_{\alpha, \mathbf{k}} + \sum_{\alpha} v_{\alpha, -\mathbf{k}, n} \, \hat{b}_{\alpha, -\mathbf{k}}
\end{aligned}
\end{equation}
where the vectors $\left( u_{\alpha, \mathbf{k}, n}, v_{\alpha, -\mathbf{k}, n} \right)$ and $\left( u^{*}_{\alpha, \mathbf{k}, n}, -v^{*}_{\alpha, -\mathbf{k}, n} \right)$ are chosen to be respectively the right and left eigenvectors of $\hat{\mathcal{L}}_{\mathbf{k}}$ belonging to so-called $+$\textsl{-family} of the $\hat{\mathcal{L}}_{\mathbf{k}}$-eigenspace. It follows that the mode eigenfrequencies $\omega_{\alpha, \mathbf{k}}$ are the positive eigenvalues of $\hat{\mathcal{L}}_{\mathbf{k}}$ and coincide with the well-known excitation spectrum that we have analysed at the mean-field level. \\
The index $\alpha$ refers to the excitation modes associated to such eigenvectors and $\hat{b}_{\alpha, \mathbf{k}}$ $\left( \hat{b}^{\dagger}_{\alpha, \mathbf{k}} \right)$ are defined to be the corresponding quasi-particle annihilation (creation) operators. Specifically, the operator $\hat{b}_{\alpha, \mathbf{k}}$ annihilates a $\alpha$-mode quasi-particle with momentum $\mathbf{k}$, but it can be also seen as a coherent superposition of $\delta \hat{c}_n{\left( \mathbf{k} \right)}$ and $\delta \hat{c}^{\dagger}_n{\left( \mathbf{k} \right)}$. \\
The Bogoliubov transformation (\ref{419}) is required to preserve the bosonic commutation relations of the Gutzwiller fluctuation operators (\ref{4.15}) through the quasi-particle operators:
\begin{equation}\label{420}
\begin{aligned}
\left[ \hat{b}_{\alpha, \mathbf{k}}, \hat{b}_{\beta, \mathbf{p}} \right] = 0 \quad , \quad \left[ \hat{b}^{\dagger}_{\alpha, \mathbf{k}}, \hat{b}^{\dagger}_{\beta, \mathbf{p}} \right] = 0 \quad , \quad \left[ \hat{b}_{\alpha, \mathbf{k}}, \hat{b}^{\dagger}_{\beta, \mathbf{p}} \right] = \delta_{\alpha, \beta} \, \delta_{\mathbf{k}, \mathbf{p}}
\end{aligned}
\end{equation}
Inserting equations (\ref{419}) into the commutation relations (\ref{4.15}) and taking into account the conditions (\ref{420}), we obtain the following normalization condition for the Bogoliubov coefficients:
\begin{equation}\label{421}
\underline{u}^{*}_{\alpha, \mathbf{k}} \underline{u}_{\beta, \mathbf{k}} - \underline{v}^{*}_{\alpha, -{\mathbf{k}}} \underline{v}_{\beta, -\mathbf{k}} = \delta_{\alpha \beta}
\end{equation}
compatibly with the structure of the $+$-family eigenspace of $\hat{\mathcal{L}}_{\mathbf{k}}$ (see again Appendix \ref{C}). \\
\uline{It is interesting to notice that the orthogonality between the ground state weights $c^0_n$ and their perturbations $\delta c_n{\left( \mathbf{k} \right)}$ is a direct consequence of the spectral properties of $\hat{\mathcal{L}}_{\mathbf{k}}$}: in fact, the distribution $c^0_n$ corresponds to the only right eigenvector with zero energy for all $\mathbf{k}$. This justifies the choice of considering perturbations that are orthogonal to  the ground-state solution (\ref{4.3}).

\medskip

Since the operator rotation (\ref{419}) has been chosen to correspond to the spectral decomposition of $\hat{\mathcal{L}}_{\mathbf{k}}$ and observing that $v_{\alpha, n, \mathbf{k}} = v_{\alpha, n, -\mathbf{k}}$ by symmetry, as $\hat{\mathcal{L}}_{\mathbf{k}}$ depends only on the moduli of the momentum components, the diagonalized quadratic Hamiltonian is directly obtained:
\begin{equation}\label{422}
\begin{aligned}
\hat{H}^{\left( 2 \right)} &= E_0 + \hbar \sum_{\alpha > 0} \sum_{\mathbf{k}} \left( \left| u_{\alpha, \mathbf{k}} \right|^2 - \left| v_{\alpha, \mathbf{k}} \right|^2 \right) \omega_{\alpha, \mathbf{k}} \, \hat{b}^{\dagger}_{\alpha, \mathbf{k}} \hat{b}_{\alpha, \mathbf{k}} + \frac{\hbar}{2} \sum_{\alpha > 0} \sum_{\mathbf{k}} \omega_{\alpha, \mathbf{k}} = \\
&= E_0 + \hbar \sum_{\alpha > 0} \sum_{\mathbf{k}} \omega_{\alpha, \mathbf{k}} \, \hat{b}^{\dagger}_{\alpha, \mathbf{k}} \hat{b}_{\alpha, \mathbf{k}} + \frac{\hbar}{2} \sum_{\alpha > 0} \sum_{\mathbf{k}} \omega_{\alpha, \mathbf{k}}
\end{aligned}
\end{equation}
as a set of uncoupled quantum harmonic oscillators corresponding to the many-body excitation modes of the system. \\
As usual, the harmonic zero-point energy in (\ref{422}) derives from the commutation of the operators $\hat{b}_{\alpha, \mathbf{k}}$ and $\hat{b}^{\dagger}_{\alpha, \mathbf{k}}$. Moreover, it is important to specify that the momentum summation in (\ref{4.14}) comprises the overall Brillouin zone, except for the Goldstone mode ($\alpha = 1$), which turns to be analytically singular in $\mathbf{k} = \mathbf{0}$. \\
It is important to notice that the introduction of quantum fluctuations through the operators (\ref{419}) do not change the features of the mean-field phase diagram given by the underlying Gutzwiller approximation, as the matrix elements of $\hat{\mathcal{L}}_{\mathbf{k}}$ strongly depend on the mean-field solution $c^0_n$. On the other hand, it has been demonstrated that a generalization of the Gutzwiller method in terms of a cluster treatment gives significant corrections to the mean-field location of the MI-SF transition, in accordance with the exact numerical results provided by quantum Monte Carlo techniques \cite{cluster_gutzwiller}, so that we take into consideration the extension of the present canonical quantization approach to more recent and advanced versions of the Gutzwiller-based approaches.

\newpage
\markboth{\MakeUppercase{4. Quantum theory}}{\MakeUppercase{4.2 Quantum corrections of relevant operators}}
\section{Quantum fluctuations of the Gutzwiller operators}
\markboth{\MakeUppercase{4. Quantum theory}}{\MakeUppercase{4.2 Quantum corrections of relevant operators}}
\textbf{Note:} The numerical solution to the evolution equation of the operators (\ref{419}) requires a truncation of the single-site Hilbert space up to a maximum occupation number $n_{max}$, which has been fixed in order to minimize its influence on the saddle-point parameters $c^0_n$ and the solutions of the numerical diagonalization of $\hat{\mathcal{L}}_{\mathbf{k}}$. Moreover, in the following developments the ground state Gutzwiller parameters $c^0_n$ and the Bogoliubov coefficients $u_{\alpha, \mathbf{k}, n}$, $v_{\alpha, \mathbf{k}, n}$ will be chosen to be purely real for simplicity.

\subsection{Operator fluctuations and quasi-particle amplitudes}
The quadratic quantization of the Gutzwiller ansatz leading to the harmonic Hamiltonian operator (\ref{422}) has allowed to identify the algebraic form of the quantum fluctuations which give rise to the zero-point physics beyond the mean-field solution. \\
As we pointed out at the beginning of Chapter \ref{MF}, the simplest observable that we can calculate within the Gutzwiller approach is the local density. In our quantization scheme the density operator can be expanded up to the first order in the form:
\begin{equation}\label{4.16}
\hat{n}{\left( r \right)} = \sum_n n \, \hat{c}^{\dagger}_n{\left( \mathbf{r} \right)} \, \hat{c}_n{\left( \mathbf{r} \right)} \approx n_0 + \delta_1 \hat{n}{\left( \mathbf{r} \right)}
\end{equation}
where:
\begin{equation}\label{4.17}
\begin{aligned}
&n_0 = \sum_n n \left| c^0_n \right|^2 \\
&\delta_1 \hat{n}{\left( \mathbf{r} \right)} = \sum_n n \, c^0_n \, \delta \hat{c}_n{\left( \mathbf{r} \right)} + \text{h.c.} = \\
&= \frac{1}{\sqrt{M}} \sum_n n \sum_{\alpha > 0} \sum_{\mathbf{k}} \left[ c^0_n \left( u_{\alpha, \mathbf{k}, n} \, \hat{b}_{\alpha, \mathbf{k}} + v_{\alpha, -\mathbf{k}, n} \, \hat{b}^{\dagger}_{\alpha, -\mathbf{k}} \right) e^{i \mathbf{k} \cdot \mathbf{r}} + \text{h.c.} \right] = \\
&= \frac{1}{\sqrt{M}} \sum_{\alpha > 0} \sum_{\mathbf{k}}  N_{\alpha, \mathbf{k}} \left( e^{i \mathbf{k} \cdot \mathbf{r}} \, \hat{b}_{\alpha, \mathbf{k}} + e^{-i \mathbf{k} \cdot \mathbf{r}} \, \hat{b}^{\dagger}_{\alpha, \mathbf{k}} \right)
\end{aligned}
\end{equation}
where $n_0$ is the mean-field density according to the saddle-point equations and the amplitude $N_{\alpha, \mathbf{k}}$ is defined to be:
\begin{equation}\label{4.18}
N_{\alpha, \mathbf{k}} \equiv \sum_n n \, c^0_n \left( u_{\alpha, \mathbf{k}, n} + v_{\alpha, \mathbf{k}, n} \right)
\end{equation}
so that the occupation degree of freedom is integrated out. \\
A similar reasoning can be applied also to the Gutzwiller expression for the order parameter, which after quantization assumes the role of Bose field operator:
\begin{equation}\label{4.19}
\hat{\psi}{\left( \mathbf{r} \right)} = \sum_n \sqrt{n} \, \hat{c}^{\dagger}_{n - 1}{\left( \mathbf{r} \right)} \, \hat{c}_n{\left( \mathbf{r} \right)} \approx \psi_0 + \delta_1 \hat{\psi}{\left( \mathbf{r} \right)}
\end{equation}
where:
\begin{equation}\label{4.20}
\begin{aligned}
&\psi_0 = \sum_n \sqrt{n} \, c^0_{n - 1} \, c^0_n \\
&\delta_1 \hat{\psi}{\left( \mathbf{r} \right)} = \sum_n \sqrt{n} \, c^0_{n - 1} \, \delta \hat{c}_n{\left( \mathbf{r} \right)} + \sum_n \sqrt{n} \, c^0_n \, \delta \hat{c}^{\dagger}_{n - 1}{\left( \mathbf{r} \right)} =
\end{aligned}
\end{equation}
\begin{equation}\label{4.21}
\begin{aligned}
&= \frac{1}{\sqrt{M}} \sum_n \sqrt{n} \sum_{\alpha > 0} \sum_{\mathbf{k}} \bigg[ c^0_{n - 1} \left( u_{\alpha, \mathbf{k}, n} \, \hat{b}_{\alpha, \mathbf{k}} + v_{\alpha, -\mathbf{k}, n} \, \hat{b}^{\dagger}_{\alpha, -\mathbf{k}} \right) e^{i \mathbf{k} \cdot \mathbf{r}} + \\
&+ c^0_n \left( u_{\alpha, \mathbf{k}, n - 1} \, \hat{b}^{\dagger}_{\alpha, \mathbf{k}} + v_{\alpha, -\mathbf{k},  n - 1} \, \hat{b}_{\alpha, -\mathbf{k}} \right) e^{-i \mathbf{k} \cdot \mathbf{r}} \bigg] = \\
&= \frac{1}{\sqrt{M}} \sum_{\alpha > 0} \sum_{\mathbf{k}} \left( U_{\alpha, \mathbf{k}} \, e^{i \mathbf{k} \cdot \mathbf{r}} \, \hat{b}_{\alpha, \mathbf{k}} + V_{\alpha, \mathbf{k}} \, e^{-i \mathbf{k} \cdot \mathbf{r}} \, \hat{b}^{\dagger}_{\alpha, \mathbf{k}} \right)
\end{aligned}
\end{equation}
The quantity $\psi_0$ is again the mean-field prediction for the order parameter, while the particle-hole amplitudes $U_{\alpha, \mathbf{k}}$ and $V_{\alpha, \mathbf{k}}$ are given by:
\begin{equation}\label{4.22}
\begin{aligned}
&U_{\alpha, \mathbf{k}} = \sum_n \sqrt{n} \left( c^0_{n - 1} \, u_{\alpha, \mathbf{k}, n}  + c^0_n \, v_{\alpha, \mathbf{k}, n - 1} \right) \\
&V_{\alpha, \mathbf{k}} = \sum_n \sqrt{n} \left( c^0_n \, u_{\alpha, \mathbf{k}, n - 1}  + c^0_{n - 1} \, v_{\alpha, \mathbf{k}, n} \right)
\end{aligned}
\end{equation}
It is interesting to notice that the operator form of $\delta_1 \hat{\psi}{\left( \mathbf{r} \right)}$ strongly recalls the expression of the Bogoliubov rotation of the Bose annihilation operator within the standard theory of the weakly-interacting limit. \\
A first good check of the quantum theory under study is provided by the calculation of the commutation relation:
\begin{equation}\label{4.23}
\left[ \hat{\psi}{\left( \mathbf{r} \right)}, \hat{\psi}^{\dagger}{\left( \mathbf{s} \right)} \right] \approx \left[ \delta_1 \hat{\psi}{\left( \mathbf{r} \right)}, \delta_1 \hat{\psi}^{\dagger}{\left( \mathbf{s} \right)} \right]
\end{equation}
as long as only quadratic fluctuations are concerned. Taking into consideration the more transparent rewriting of $\delta_1 \hat{\psi}{\left( \mathbf{r} \right)}$ in (\ref{4.21}), a straightforward calculation exploiting the inversion symmetry $\mathbf{k} \to -\mathbf{k}$ leads to:
\begin{equation}\label{4.24}
\left[ \delta_1 \hat{\psi}{\left( \mathbf{r} \right)}, \delta_1 \hat{\psi}^{\dagger}{\left( \mathbf{s} \right)} \right] = \frac{1}{M} \sum_{\mathbf{k}} e^{i \mathbf{k} \cdot \left( \mathbf{r} - \mathbf{s} \right)} \sum_{\alpha > 0} \left( \left| U_{\alpha, \mathbf{k}} \right|^2 - \left| V_{\alpha, \mathbf{k}} \right|^2 \right) = \delta{\left( \mathbf{r} - \mathbf{s} \right)}
\end{equation}
where \uline{the last equality is numerically justified by the fact that the quantity} $\sum_{\alpha > 0} \left( \left| U_{\alpha, \mathbf{k}} \right|^2 - \left| V_{\alpha, \mathbf{k}} \right|^2 \right)$ \uline{turns out to be always identically equal to $1$ over the whole phase diagram, provided that all the excitation modes are considered}, as already noticed by Menotti and Trivedi in \cite{menotti_trivedi}. \\
As a first conclusion, the operator $\hat{\psi}{\left( \mathbf{r} \right)}$ can be interpreted as the quantized Bose field without ambiguity, as it satisfies the correct commutation relations.

\medskip

The consistency of the operator expansion (\ref{4.21}) can be directly tested by comparing the Goldstone mode particle-hole weights $U_{1, \mathbf{k}}$ and $V_{1, \mathbf{k}}$ with their counterparts determined by the conventional Bogoliubov approximation for $J/U \to \infty$. \\
\uline{More interestingly, since the Gutzwiller approximation is able to catch the physical properties deriving from strong interactions, we can deduce how the particle-hole amplitudes of each excitation mode behave in the vicinity of the Mott lobes.} \\
This analysis allows us to make significant predictions about the most important physical observables away from the deep-superfluid region, where the standard Bogoliubov theory has no access.

\subsection{Particle-hole amplitudes \texorpdfstring{$U_{\alpha, \mathbf{k}}$ and $V_{\alpha, \mathbf{k}}$}{}}\label{U_V}
In the following we propose a study of the momentum behaviour of the particle-hole amplitudes $U_{\alpha, \mathbf{k}}$ and $V_{\alpha, \mathbf{k}}$ depending on the region of physical parameters and the excitation mode. Such quantities have been calculated in the 1D case only, which is equivalent to choose an arbitrary momentum direction in the 3D reciprocal lattice, and along horizontal lines at constant $\mu/U$ for which the weakly-interacting limit condition $\mu/U \simeq \left| \psi_0 \right|^2 - 2 \, d \, J/U$ is automatically satisfied when $2 \, d \, J/U \gg 1$ in accordance with the Bogoliubov equations.
\vspace{1cm}
\begin{listamia}
	\item \textbf{Deep superfluid regime + Boundary between weak/strong interactions}
	\begin{figure}[!htp]
		\begin{minipage}[t]{7.7cm}
			\centering
			\includegraphics[width=9.1cm]{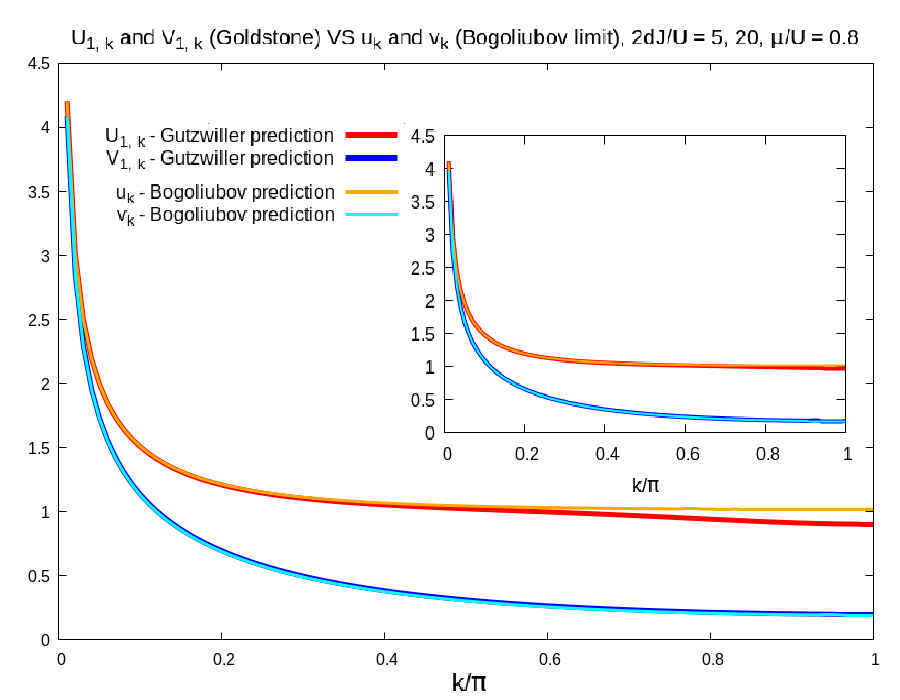}
			\caption{Particle-hole amplitudes $U_{1, \mathbf{k}}$ and $V_{1, \mathbf{k}}$ (Goldstone mode) for $2 d J/U = 5, 20$ (inset) and $\mu/U = 0.8$, compared with the Bogoliubov prediction. The inset shows a zoom of low-momentum behaviour.}
			\label{Figure1}
		\end{minipage}
		\ \hspace{4mm} \hspace{4mm} \
		\begin{minipage}[t]{7.7cm}
			\centering
			\includegraphics[width=9.1cm]{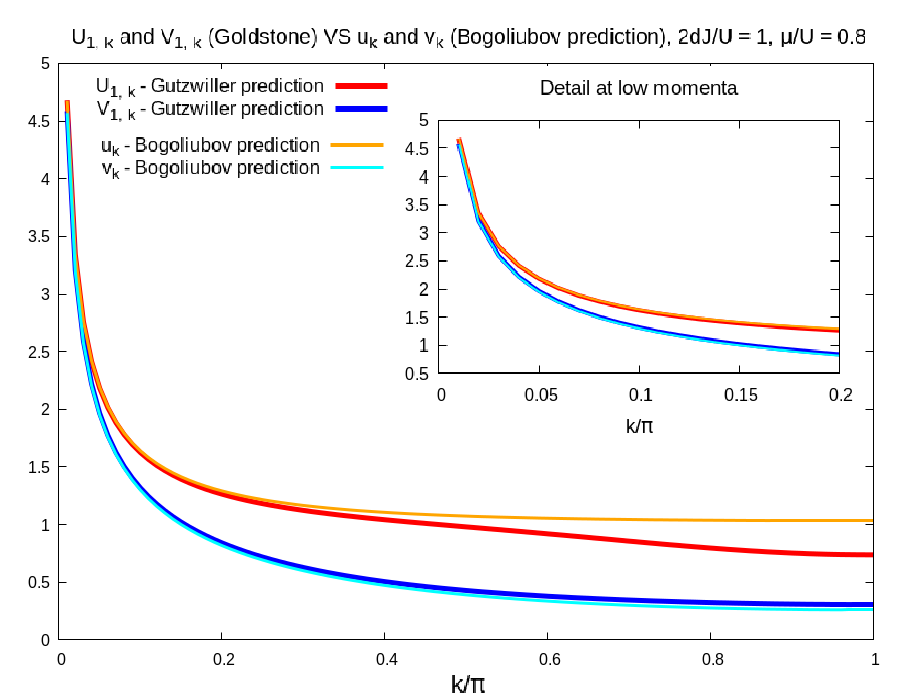}
			\caption{Particle-hole amplitudes $U_{1, \mathbf{k}}$ and $V_{1, \mathbf{k}}$ (Goldstone mode) for $2 d J/U = 1$ and $\mu/U = 0.8$, compared with the Bogoliubov prediction.}
			\label{Figure2}
		\end{minipage}
	\end{figure} \\
	Far from being a trivial result, we can observe that the Gutzwiller and Bogoliubov predictions for the particle-hole amplitudes within the Goldstone mode match exactly in the weakly-interacting limit as expected. Therefore, there is no need to multiply the quadratic Hamiltonian (\ref{422}) by any additional multiplicative constant. \\
	More interestingly, \uline{the discrepancy between the two descriptions starts to increase at large momenta as the hypothesis of weak interactions is softened}, as in Figure \ref{Figure1} for $2 \, d \, J/U \approx 1$, while the phononic regime is still compatible with the Bogoliubov theory. This means that the weight of higher modes is larger at the border of the Brillouin zone, that is the upper part of the bare single-particle energy band, and offers an important clue about their observability in experiments focusing on response functions. \\
	\uline{These features are due to the fact that the hydrodynamic theory of collective excitations, which underlies the Bogoliubov approach as a more general model, is particularly robust in the low-energy limit even in the presence of strong interactions, while the higher-energy domain contains partial information about the spatial discretization of the lattice.} \\
	More generally, \uline{we can claim that the quadratic action of a general hydrodynamic theory:}
	\begin{equation}\label{4.25}
	\mathcal{S}_H = \frac{1}{2} \int dt \int d\mathbf{r} \left\{ 2 \, i \dpar_t \phi{\left( \mathbf{r}, t \right)} + \rho_s \left| \nabla \phi{\left( \mathbf{r}, t \right)} \right|^2 + \gamma_{\pi} \, \Pi^2{\left( \mathbf{r}, t \right)} \right\}
	\end{equation}
	\uline{can be always used as an effective description of the low-energy physics of an interacting superfluid system by an appropriate renormalization of the parameters of the theory} when the Bogoliubov approach produces unreliable results. This reasoning finds strong correspondence with the formal procedure followed by Popov \cite{popov} based on the integration of fast fields in a perturbative scheme. \\
	The renormalization of the parameters of the hydrodynamic theory is equivalent to say that the second term of the action (\ref{4.25}), obtained by substituting the density-phase representation of the Bose field in the interaction term for weak interactions, is dressed by higher-order contributions in presence of stronger interactions. \\
	Here $\phi{\left( \mathbf{r}, t \right)}$ is the velocity potential or phase field and $\Pi{\left( \mathbf{r}, t \right)}$ is the corresponding conjugate field linked to the fluid density. The parameters $\rho_s$ and $\gamma_{\pi} = \kappa^{-1}$, related to the sound velocity by $c^2_s = \rho_s \, \gamma_{\pi}$, are the superfluid fraction and the inverse of the compressibility respectively. More precisely, \uline{the values of these parameters} are given by neither the wrong Bogoliubov theory nor the Gutzwiller mean-field approach described in Chapter \ref{MF}, but \uline{have to be calculated consistently with the inclusion of second-order quantum fluctuations} (see the determination of the effective superfluid fraction in Section \ref{superfluid_fraction}). \\
	As a second observation, from the inset of Figure \ref{Figure2} we can deduce that the Bogoliubov or hydrodynamic prediction and the Gutzwiller result for the amplitudes $U_{1, \mathbf{k}}$ and $V_{1, \mathbf{k}}$ are almost equal in the low-momentum limit, but not exactly coincident. The disagreement between the Gutzwiller and Bogoliubov results is due to the fact that the latter theory is applied in a regime where it is not expected to hold; however this convergence trend seems to suggest that a quantity underlying the very definition of $U_{1, \mathbf{k}}$ and $V_{1, \mathbf{k}}$ is almost conserved by the two theories. \\
	In particular, since at low momenta the hydrodynamic amplitudes associated to the Goldstone mode given by (\ref{4.25}) behave as:
	\begin{equation}\label{4.26}
	U^H_{1, \mathbf{k}} \sim \frac{1}{2} \left( \frac{\gamma_{\pi}}{\rho_s} \right)^{1/4} \left| \mathbf{k} \right|^{-1/2} \quad , \quad V^H_{1, \mathbf{k}} \sim -\frac{1}{2} \left( \frac{\gamma_{\pi}}{\rho_s} \right)^{1/4} \left| \mathbf{k} \right|^{-1/2} \quad \text{for} \ \left| \mathbf{k} \right| \to 0
	\end{equation}
	we can presume that the quantity $\gamma_{\pi}/\rho_s$ is almost the same in the Gutzwiller and Bogoliubov calculations, where the theory parameters are considered again to be dressed by the inclusion of quantum fluctuations.
	
	\medskip

	As a final remark, it is important to highlight that the formulation of an effective hydrodynamic theory of strong interactions would lead to the identification of the relevant low-energy decay vertices of the model in terms of the particle and hole amplitudes only, in analogy with the traditional treatment of the Beliaev decay presented in \cite{pitaevskii_stringari}.
	\begin{figure}[!htp]
		\centering
		\includegraphics[width=10cm]{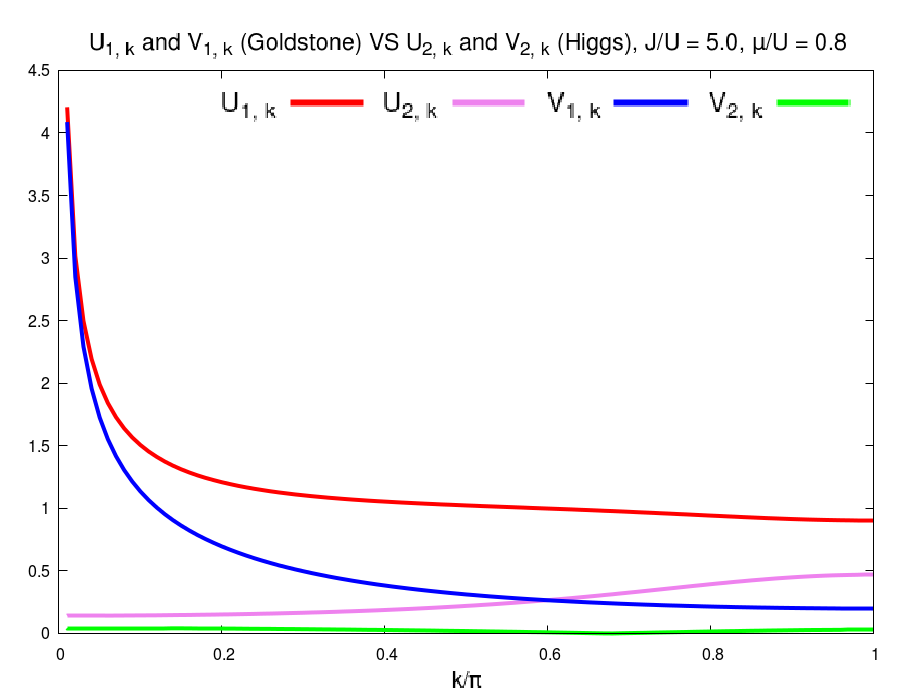}
		\caption{Particle-hole amplitudes $U_{1, \mathbf{k}}$ and $V_{1, \mathbf{k}}$ (Goldstone mode) compared with $U_{2, \mathbf{k}}$ and $V_{2, \mathbf{k}}$ (Higgs mode) for $2 d J/U = 5$ and $\mu/U = 0.8$. Here we can appreciate the emergence of the non-negligible contribution of the Higgs mode slightly away from the weakly-interacting limit in analogy with Figure \ref{Figure2}.}
		\label{Figure3}
	\end{figure}
	\newpage
	\item \textbf{Strong interactions and near criticality}
	\begin{figure}[!htp]
		\begin{minipage}[t]{7.5cm}
			\centering
			\includegraphics[width=9.1cm]{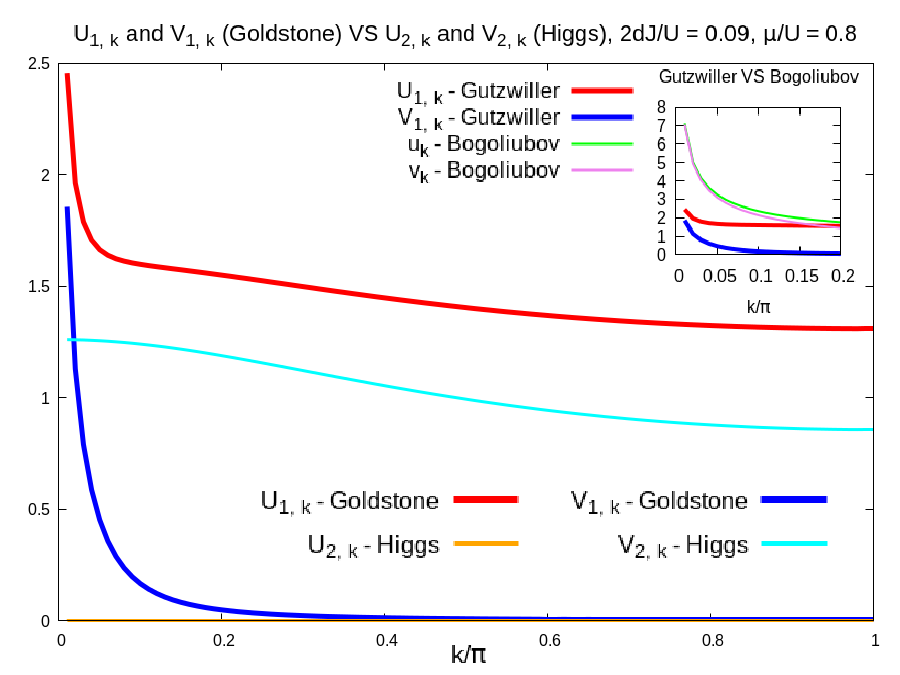}
			\caption{Particle-hole amplitudes $U_{1, \mathbf{k}}$ and $V_{1, \mathbf{k}}$ (Goldstone mode) compared with $U_{2, \mathbf{k}}$ and $V_{2, \mathbf{k}}$ (Higgs mode) for $2 d J/U = 0.09$ and $\mu/U = 0.8$. The inset shows the comparison with the low-momentum Bogoliubov prediction.}
			\label{Figure4}
		\end{minipage}
		\ \hspace{5mm} \hspace{5mm} \
		\begin{minipage}[t]{8.6cm}
			\centering
			\includegraphics[width=9.1cm]{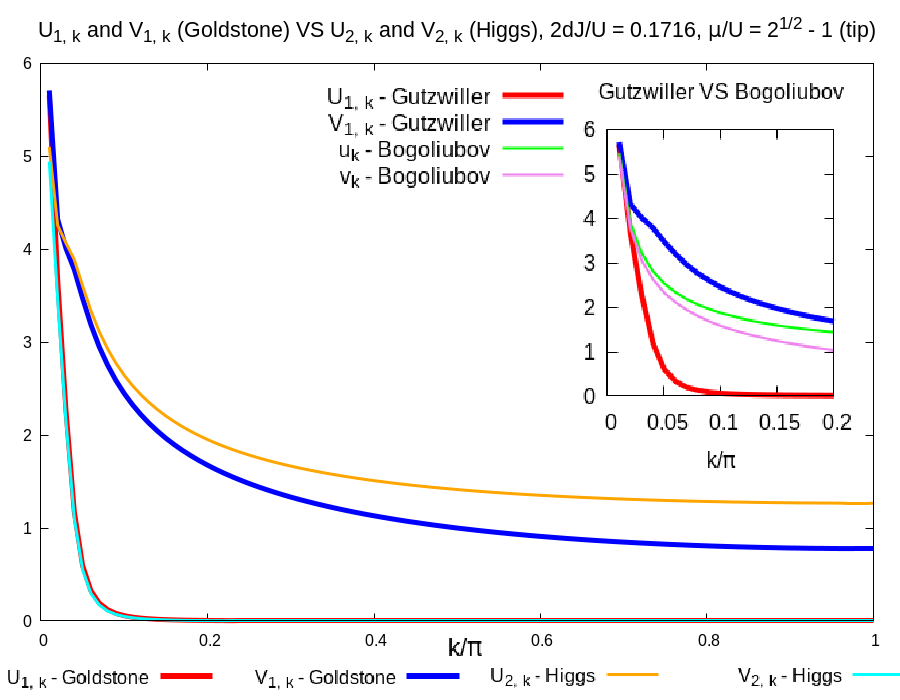}
			\caption{Particle-hole amplitudes $U_{1, \mathbf{k}}$ and $V_{1, \mathbf{k}}$ (Goldstone mode) compared with $U_{2, \mathbf{k}}$ and $V_{2, \mathbf{k}}$ (Higgs mode) for $2 d J/U = 0.1716$ and $\mu/U = \sqrt{2} - 1$ (tip of the first Mott lobe). The inset shows the comparison with the low-momentum Bogoliubov prediction.}
			\label{Figure5}
		\end{minipage}
	\end{figure} \\
	Moving away from the regime where the hopping and interaction terms interplay at the same level, Figures \ref{Figure4} and \ref{Figure5} show how the Gutzwiller predictions for the amplitudes $U_{\alpha, \mathbf{k}}$ and $V_{\alpha, \mathbf{k}}$ behave within the Goldstone and Higgs modes for superfluid systems dominated by strong interactions near criticality. \\
	In accordance with the discussion made above, the Gutzwiller approximation conserves the hydrodynamic description in the low-energy limit even in the critical region, where we can recognize again the phononic behaviour of $U_{1, \mathbf{k}}, V_{1, \mathbf{k}} \sim \left| \mathbf{k} \right|^{-1/2}$ through a comparison with the wrong Bogoliubov result, which we know to follow an hydrodynamic trend at low energies. \\
	In particular, this feature can be immediately noticed when approaching the Lorentz-invariant points of the critical line (see Figure \ref{Figure5}), where the Goldstone and Higgs mode become degenerate and, more importantly, the sound velocity does not vanish. On the contrary, a renormalized hydrodynamic theory seems to be turn out to be less stable in other points of the critical region (see Figure \ref{Figure4}) where the compressibility reaches a finite value and the sound velocity vanishes scaling as the condensate density, see equation (\ref{3.44}). \\
	\textsl{As a final observation, it is also important to highlight that the hole amplitudes $V_{1, \mathbf{k}}$ and $V_{2, \mathbf{k}}$ (determining the physical properties connected to the particle depletion) do not present any particular symmetry with respect to their momentum dependence in the vicinity of the transition.}
	\vspace{0.8cm}
	\item \textbf{Mott phase}
	\begin{figure}[!htp]
		\begin{minipage}[b]{7.8cm}
			\centering
			\includegraphics[width=9.1cm]{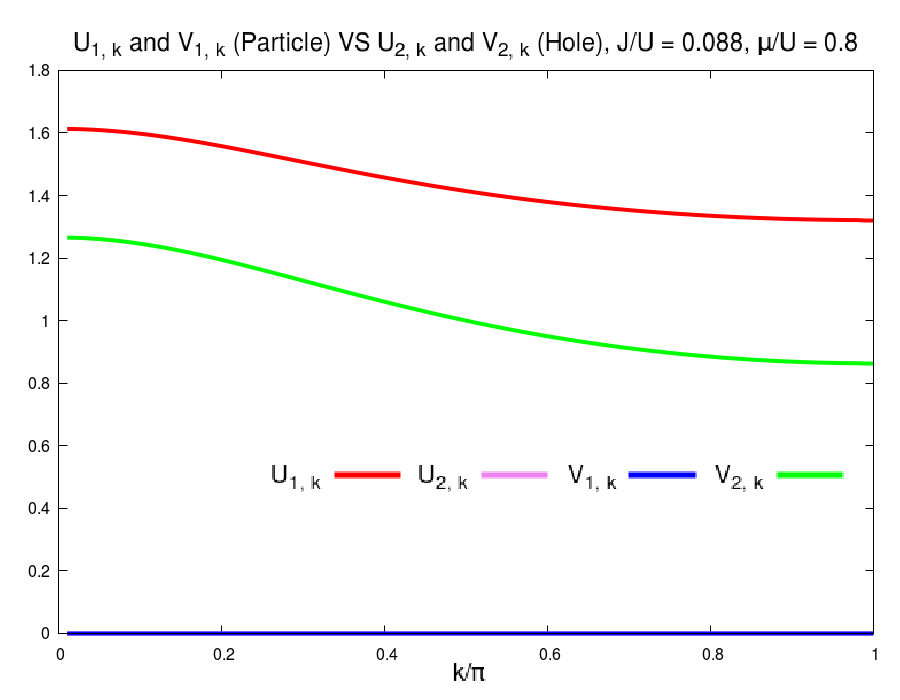}
			\caption{Particle-hole amplitudes $U_{1, \mathbf{k}}$ and $V_{1, \mathbf{k}}$ (particle mode) compared with $U_{2, \mathbf{k}}$ and $V_{2, \mathbf{k}}$ (hole mode) for $2 d J/U = 0.088$ and $\mu/U = 0.8$.}
			\label{Figure6}
		\end{minipage}
		\ \hspace{2mm} \hspace{2mm} \
		\begin{minipage}[b]{7.8cm}
			\centering
			\includegraphics[width=9.1cm]{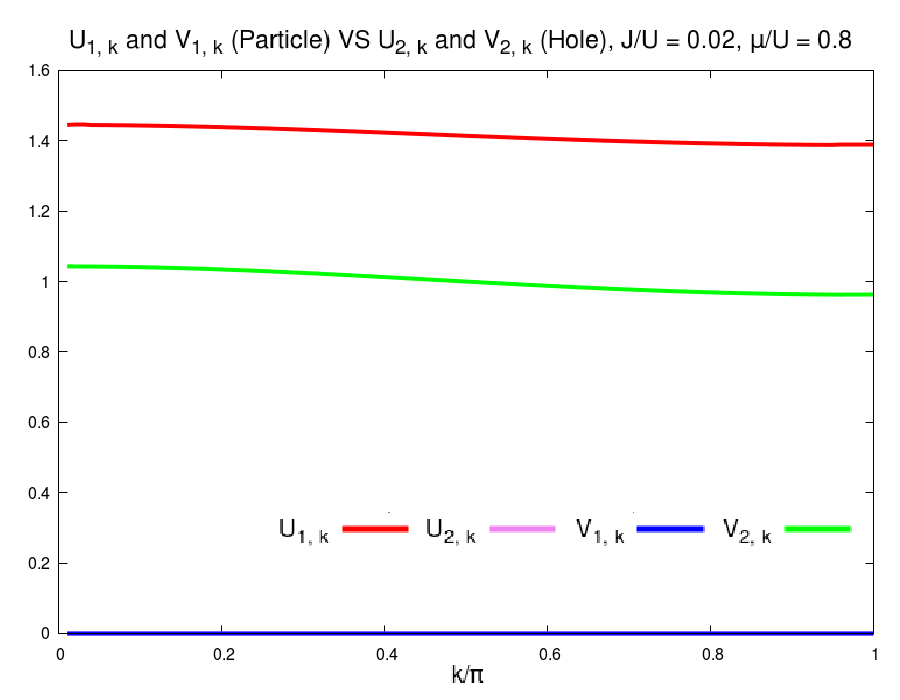}
			\caption{Particle-hole amplitudes $U_{1, \mathbf{k}}$ and $V_{1, \mathbf{k}}$ (particle mode) compared with $U_{2, \mathbf{k}}$ and $V_{2, \mathbf{k}}$ (hole mode) for $2 d J/U = 0.02$ and $\mu/U = 0.8$.}
			\label{Figure7}
		\end{minipage}
	\end{figure} \\
	Figures \ref{Figure6} and \ref{Figure7} reveal that in the Mott phase only the particle-mode amplitude $U_{p, \mathbf{k}}$ and the hole-mode one $V_{h, \mathbf{k}}$ survive in a natural way, as we expect from the pure particle or hole nature of the fundamental excitations of an insulator. This fact is due to the very definitions (\ref{4.21}) and the peculiar form of the Bogoliubov coefficients $u_{\alpha, \mathbf{k}, n}$ and $v_{\alpha, \mathbf{k}, n}$ in the Mott phase. \\
	\textsl{In addition, we notice that the hole-amplitude $V_{h, \mathbf{k}}$ tends to be a constant as the pure Mott state is approached. We will see how this trend affects the behaviour of the so-called Drude weight calculated upon our quantum theory.}
\end{listamia}

\subsection{Particle-hole symmetry}
Let us consider again the first-order correction to the order parameter in terms of the quasi-particle operators:
\begin{equation}\label{4.27}
\delta_1 \hat{\psi}{\left( \mathbf{r} \right)} = \frac{1}{\sqrt{M}} \sum_{\alpha > 0} \sum_{\mathbf{k}} \left( U_{\alpha, \mathbf{k}} \, \hat{b}_{\alpha, \mathbf{k}} \, e^{i \mathbf{k} \cdot \mathbf{r}} + V_{\alpha, \mathbf{k}} \, \hat{b}^{\dagger}_{\alpha, \mathbf{k}} \, e^{-i \mathbf{k} \cdot \mathbf{r}} \right)
\end{equation}
Equation (\ref{4.27}) allows us to identify immediately the following particle-hole symmetry condition:
\begin{equation}\label{4.28}
\left| U_{\alpha, \mathbf{k}} \right| = \left| V_{\alpha, \mathbf{k}} \right|
\end{equation}
whose implications have been widely analysed in the work written by Di Liberto et al. \cite{BEC} that has inspired this thesis work. \\
Along this direction more interesting aspects emerge when we consider the first-order corrections to the density operator:
\begin{equation}\label{4.29}
\delta_1 \hat{n}{\left( \mathbf{r} \right)} = \frac{1}{\sqrt{M}} \sum_{\alpha > 0} \sum_{\mathbf{k}} N_{\alpha, \mathbf{k}} \left( \hat{b}_{\alpha, \mathbf{k}} \, e^{i \mathbf{k} \cdot \mathbf{r}} + \hat{b}^{\dagger}_{\alpha, \mathbf{k}} \, e^{-i \mathbf{k} \cdot \mathbf{r}} \right)
\end{equation}
\begin{figure}[!htp]
	\centering
	\includegraphics[width=0.7\textwidth]{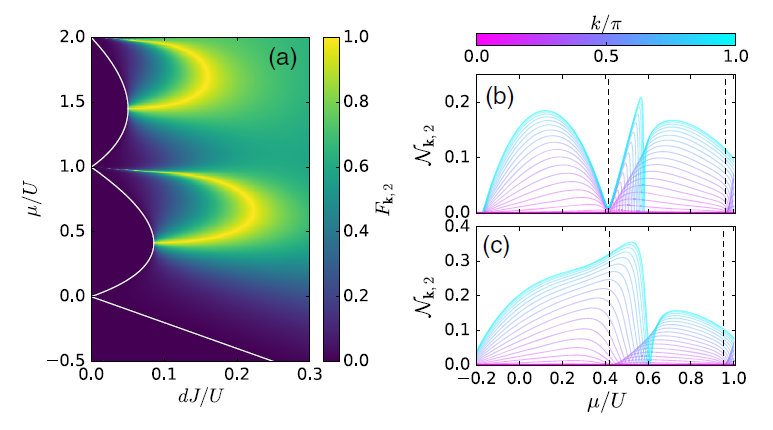}
	\caption{(a) Flatness $F_{2, \mathbf{k}} = V_{2, \mathbf{k}}/U_{2, \mathbf{k}}$ of the first gapped mode or Higgs excitation as a function of $d J/U$ and $\mu/U$ for $k_x = k_y = k_z = \pi/100$; the bright yellow curves are the points where this mode corresponds to pure amplitude oscillations of the order parameter. As a reference, the white lines indicate the Mott to superfluid boundaries. (b) Density oscillation $N_{2, \mathbf{k}}$ as a function of $\mu/U$ for fixed $d J/U = 0.0858$, corresponding to the tip of the first Mott lobe. (c) As in (b) for $d J/U = 0.115$. Figure taken from \cite{BEC}, courtesy of the authors.}
	\label{FigureI}
\end{figure} \\
Drawing on the continuity equation, it is straightforward to verify that the amplitude of the first-order corrections to the density operator (\ref{4.17}) vanish if the equality $U_{\alpha, \mathbf{k}} = V_{\alpha, \mathbf{k}}$ is satisfied. This constraint is compatible with the particle-hole symmetry condition (\ref{4.28}) and corresponds to a vanishing imaginary part of the overall amplitude of (\ref{4.27}). \\
As observed in \cite{BEC}, this property identifies the pure amplitude character of a mode $\alpha$ with an exchange of particles between the condensate and the normal fraction of the system. The suppression of the total density oscillations has been numerically verified for the Higgs mode, whose amplitude character is obtained on slightly different curves depending on the momentum of the excitations. Moreover, the density response is significant only in the short wavelength limit and it is suppressed for $\left| \mathbf{k} \right| \to 0$. These facts should be taken into account when looking for the Higgs mode in possible experiments. \\
A similar reasoning can be followed for the first-order oscillations of the superfluid density:
\begin{equation}\label{4.30}
\delta_1 \hat{\rho}{\left( \mathbf{r} \right)} = 2 \, \psi_0 \, \delta_1 \hat{\psi}{\left( \mathbf{r} \right)} = \frac{2 \, \psi_0}{\sqrt{M}} \sum_{\alpha > 0} \sum_{\mathbf{k}} \left( U_{\alpha, \mathbf{k}} \, \hat{b}_{\alpha, \mathbf{k}} \, e^{i \mathbf{k} \cdot \mathbf{r}} + V_{\alpha, \mathbf{k}} \, \hat{b}^{\dagger}_{\alpha, \mathbf{k}} \, e^{-i \mathbf{k} \cdot \mathbf{r}} \right)
\end{equation}
whose total amplitude with respect to its particle and hole components vanish under the condition $U_{\alpha, \mathbf{k}} + V_{\alpha, \mathbf{k}} = 0$, associated to a pure phase oscillation of the order parameter. Vanishing condensate density oscillations are found again on arc-shaped lines in the phase diagram in the vicinity of and, in particular, below the half-filling level of each Mott lobe.
\begin{figure}[!htp]
	\centering
	\includegraphics[width=0.7\textwidth]{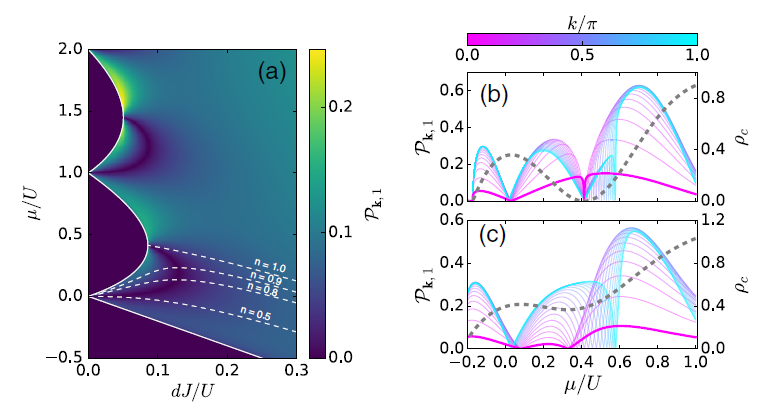}
	\caption{(a) Amplitude of the condensate density fluctuations $P_{1, \mathbf{k}}$ for the Goldstone mode as a function of $d J/U$  and $\mu/U$ for $k_x = k_y = k_z = \pi/100$; constant-density $n_0 = 0.5, 0.8, 0.9, 1$ contours (white lines). (b) $P_{1, \mathbf{k}}$ as a function of $\mu/U$ for fixed $d J/U = 0.0858$, corresponding to the tip of the first Mott lobe. (c) As in (b) for $d J/U = 0.115$. Grey dashed lines show the condensate density $\rho_c = \psi^2_0$ in the ground state. Figure taken from \cite{BEC}, courtesy of the authors.}
	\label{FigureII}
\end{figure}

\chapter{Quantum fluctuations of physical observables}\label{quantum_corrections}
\markboth{\MakeUppercase{5. Quantum fluctuations}}{\MakeUppercase{5. Quantum fluctuations}}
In Chapter \ref{quantum} we have developed a suitable method for treating quantum corrections to the Gutzwiller mean-field solutions as coherent superpositions of creation and annihilations operators of the excitation modes determining the spectral properties of the system. \\
In analogy with the purposes of standard Bogoliubov approach to dilute gases, Chapter \ref{quantum_corrections} is devoted to the presentation of the Gutzwiller predictions for the quantum depletion of the condensate and the superfluid density across the whole range of the parametric ratio $J/U$ determining the superfluid phase, with particular attention to new information deriving from the modal composition of these observables.

\section{Quantum depletion}\label{depletion}
As emphasized at the beginning of Chapter \ref{MF}, the Gutzwiller equations represent a generalization of the Bogoliubov-de Gennes solution to the problem of excitation dynamics in the Bose-Hubbard model which extends to the strongly-interacting regime. In this sense a quantum theory based on the Gutzwiller ansatz could improve the predictions of the usual Bogoliubov approximation, whose validity is limited to the weakly-interacting limit. \\
A simple but illustrative example is represented by the quantum depletion of particles out of the condensate fraction. In our formalism this quantity can be calculated from the second-order expansion of the local density operator defined by the quantized Bose-field:
\begin{equation}\label{4.31}
\begin{aligned}
\langle \hat{\psi}^{\dagger}{\left( \mathbf{r} \right)} \, \hat{\psi}{\left( \mathbf{r} \right)} \rangle &\approx \Bigg\langle \left[ \psi_0 + \frac{1}{\sqrt{M}} \sum_{\alpha > 0} \sum_{\mathbf{k}} \left( U_{\alpha, \mathbf{k}} \, e^{-i \mathbf{k} \cdot \mathbf{r}} \, \hat{b}^{\dagger}_{\alpha, \mathbf{k}} + V_{\alpha, \mathbf{k}} \, e^{i \mathbf{k} \cdot \mathbf{r}} \, \hat{b}_{\alpha, \mathbf{k}} \right) \right] \times \\
&\times \left[ \psi_0 + \frac{1}{\sqrt{M}} \sum_{\beta > 0} \sum_{\mathbf{p}} \left( U_{\beta, \mathbf{p}} \, e^{i \mathbf{p} \cdot \mathbf{r}} \, \hat{b}_{\beta, \mathbf{p}} + V_{\beta, \mathbf{p}} \, e^{-i \mathbf{p} \cdot \mathbf{r}} \, \hat{b}^{\dagger}_{\beta, \mathbf{p}} \right) \right] \Bigg\rangle = \\
&= \psi^2_0 + \frac{1}{M} \sum_{\alpha > 0} \sum_{\mathbf{k}} \left[ \left( \left| U_{\alpha, \mathbf{k}} \right|^2 + \left| V_{\alpha, \mathbf{k}} \right|^2 \right) \langle \hat{b}^{\dagger}_{\alpha, \mathbf{k}} \, \hat{b}_{\alpha, \mathbf{k}} \rangle + \left| V_{\alpha, \mathbf{k}} \right|^2 \right]
\end{aligned}
\end{equation}
where the vacuum state is referred as the \textit{pseudo-vacuum} of the quasi-particle excitations and we have used the fact that the quadratic Hamiltonian (\ref{422}) does not couple different modes and the mean occupation number of the quasi-particle excitations:
\begin{equation}\label{4.32}
\langle \hat{b}^{\dagger}_{\alpha, \mathbf{k}} \hat{b}_{\alpha, \mathbf{k}} \rangle = \frac{1}{\exp{\left( \hbar \, \omega_{\alpha, \mathbf{k}}/T \right)} - 1}
\end{equation}
is determined by Bose statistics. \\
The quantum depletion $D_g$ based on Gutzwiller's theory is defined to be provided by the second term on the RHS of (\ref{4.31}):
\begin{equation}\label{4.33}
D_g \equiv \frac{1}{M} \sum_{\alpha > 0} \sum_{\mathbf{k}} \left[ \left( \left| U_{\alpha, \mathbf{k}} \right|^2 + \left| V_{\alpha, \mathbf{k}} \right|^2 \right) \langle \hat{b}^{\dagger}_{\alpha, \mathbf{k}} \, \hat{b}_{\alpha, \mathbf{k}} \rangle + \left| V_{\alpha, \mathbf{k}} \right|^2 \right] \xrightarrow[T = 0]{} \frac{1}{M} \sum_{\alpha > 0} \sum_{\mathbf{k}} \left| V_{\alpha, \mathbf{k}} \right|^2
\end{equation}
where we have considered the limit of zero temperature as a final step. Expression (\ref{4.33}) is quite similar to the relation that quantifies the standard Bogoliubov depletion $\mathcal{D}$ given by (\ref{2.44}). Nevertheless, the particle and hole amplitudes $U_{\alpha, \mathbf{k}}$ and $V_{\alpha, \mathbf{k}}$ derive from a descriptive approach that is able to catch both the limits of the Bose-Hubbard physics, so that they could give more information about the fraction of depleted particles in the strongly-interacting regime. As far as the occupation of the excited states is concerned, the Goldstone mode is expected to provide the dominant contribution to (\ref{4.33}) for $J/U \to \infty$.

\subsection{Numerical results: quantum depletion \texorpdfstring{$D_g$}{} in 1D and 3D}
This section deals with the presentation of the numerical results concerning the Gutzwiller quantum depletion $D_g$ across the Bose-Hubbard phase diagram. Numerical calculations have been performed both in the 1D and in the 3D case for a relatively large number of particle in order to minimize finite-size effects. \footnote{Once again, the numerical calculations have been performed along horizontal lines at constant $\mu/U$ across the phase diagram for simplicity.} On the other hand, interesting results are expected to emerge even away from the thermodynamic limit.

\begin{listamia}
	\item \textbf{1D case,} $10^2$ \textbf{sites}
	\begin{figure}[!htp]
		\begin{minipage}[t]{7.8cm}
			\centering
			\includegraphics[width=9.2cm]{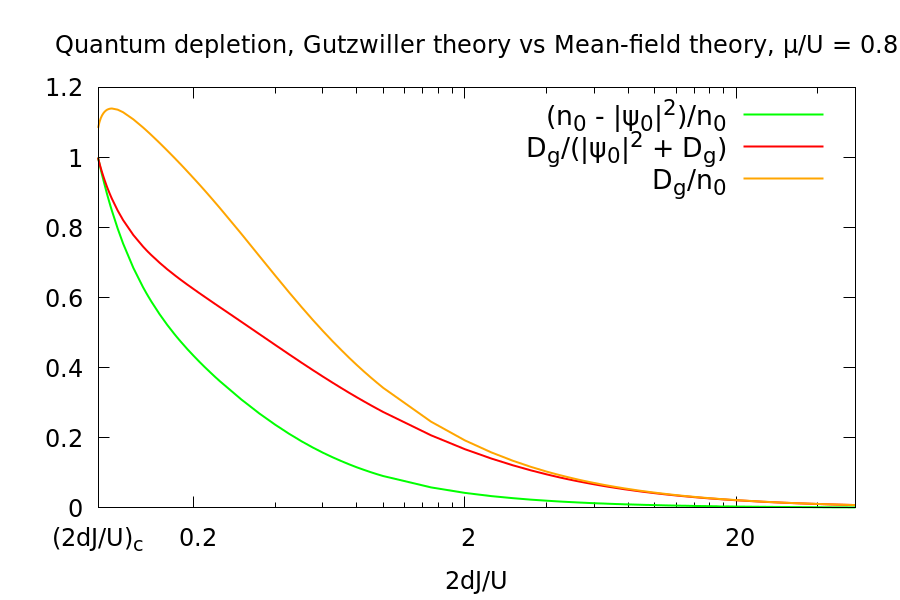}
			\caption{Comparison between the Gutzwiller prediction for the quantum depletion $D_g$ and the mean-field discrepancy between the density $n_0$ and the condensate fraction $\psi^2_0$ for $\mu/U = 0.8$ in 1D.}
			\label{Figure10}
		\end{minipage}
		\ \hspace{3.5mm} \hspace{3.5mm} \
		\begin{minipage}[t]{7.5cm}
			\centering
			\includegraphics[width=9.1cm]{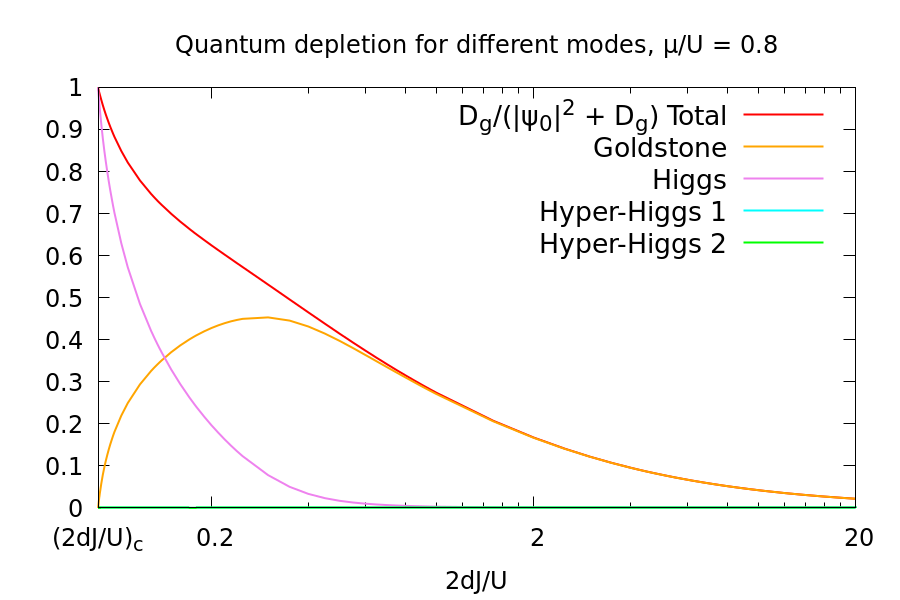}
			\caption{Contributions of the first four excitation modes to the Gutzwiller quantum depletion $D_g$ for $\mu/U = 0.8$ in 1D.}
			\label{Figure11}
		\end{minipage}
	\end{figure} \\
	The Gutzwiller mean-field theory is able to provide a first estimation of the mean local density together with the condensate fraction on an equal footing, so that one could think about their difference as an approximation to the depletion of the condensate. Figure \ref{Figure10} aims at comparing the Gutzwiller prediction for the quantum depletion with the corresponding mean-field picture. \\
	At first sight, \uline{we can notice that the depleted fraction given by $D_g$ matches with the mean-field prediction both in the vicinity of the Mott phase and in the deep superfluid regime, where the saddle-point equations tend to be an exact description of the Bose-Hubbard physics}. On the other hand, their discrepancy becomes important where the superfluid phase is characterized by strong interactions. \\
	It is crucial to observe that only the normalization:
	\begin{equation}\label{4.34}
	D_g \longrightarrow \frac{D_g}{\psi_0^2 + D_g}
	\end{equation}
	assures the physical constraint $D_g \le 1$ for the normalized fraction of depleted particles via the Schwarz inequality. The same principle does not apply to the orange curve in Figure \ref{Figure10}, which has been obtained by adopting the mean-field density $n_0$ as the normalizing factor. \\
	This fact demonstrates that the Gutzwiller depletion goes in the right direction in quantifying the fraction of non-condensed particles near the transition, as $D_g$ saturates to 1 by construction, but the identification between the operators $\hat{\psi}^{\dagger}{\left( \mathbf{r} \right)} \, \hat{\psi}{\left( \mathbf{r} \right)}$ and $\hat{n}{\left( \mathbf{r} \right)}$ is not consistent, as they have different definitions within the Gutzwiller approach, see equations (\ref{4.17}) and (\ref{4.19}) in the previous Chapter. \\
	On the other hand, as already noticed in \cite{menotti_trivedi} in the context of the RPA approximation, the normalization of the local density is expected to be not preserved when adding quantum fluctuations due to interactions on top of the condensate density. \\
	\textsl{Actually, a complete analysis should require to take into account of second-order quantum corrections to $n_0$ expanding the Gutzwiller density operator (\ref{4.17}); unfortunately, as explained later in Section \ref{second_order_corrections}, these corrections present problematic aspects which do not allow their inclusion in $\langle \hat{n} \rangle$. More generally, this problem regards the correct quantization procedure for the density operator within our quantum theory and, in general, the possible formulation of a self-consistent approach to the study of quantum fluctuations with respect to the saddle-point solution under depletion effects.} In Sections \ref{second_order_corrections} and \ref{quantization_discussion} we will address these issues by offering a more detailed discussion about second-order fluctuations of the Gutzwiller operators and how they are related to the quantization problem.
	\begin{figure}[!htp]
		\begin{minipage}[t]{7.5cm}
			\centering
			\includegraphics[width=9.1cm]{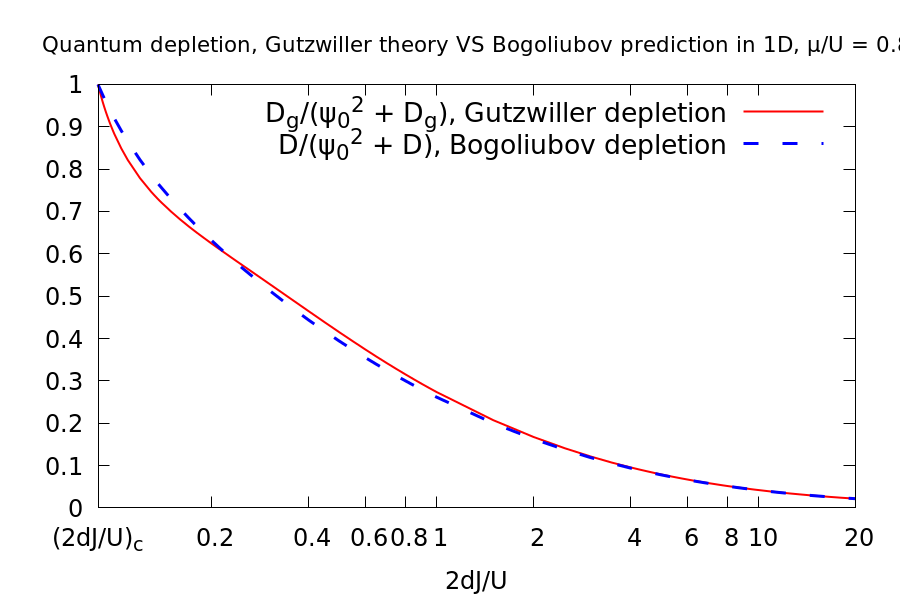}
			\caption{Comparison between $D_g$ and the standard Bogoliubov prediction (dashed blue line) for the quantum depletion for $\mu/U = 0.8$ in 1D.}
			\label{Figure12}
		\end{minipage}
		\ \hspace{4mm} \hspace{4mm} \
		\begin{minipage}[t]{7.4cm}
			\centering
			\includegraphics[width=9.1cm]{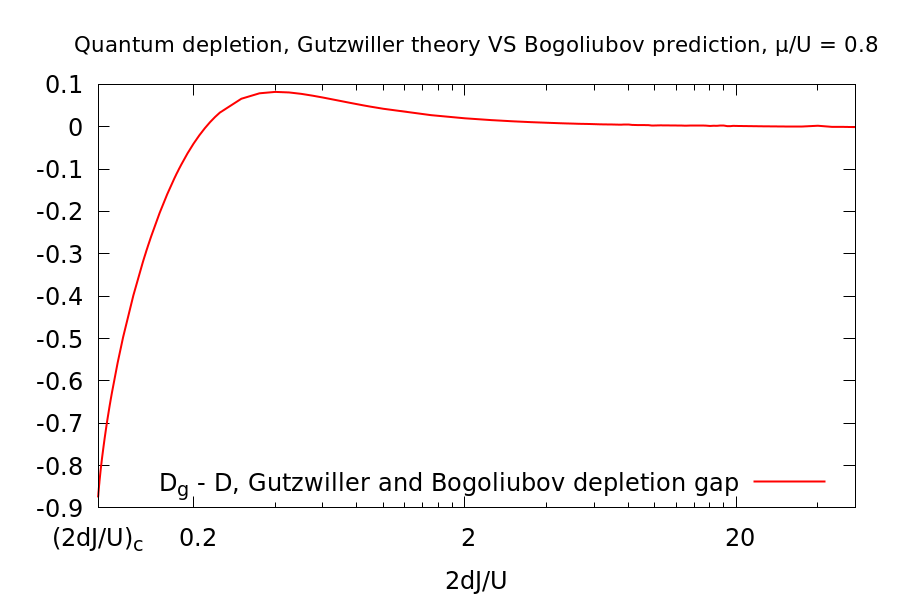}
			\caption{Plot of the gap $D_g - \mathcal{D}$ separating and the quantum depletion within the Gutzwiller theory $D_g$ and the Bogoliubov approximation $\mathcal{D}$ for $\mu/U = 0.8$ in 1D.}
			\label{Figure13}
		\end{minipage}
	\end{figure}

	\medskip

	Important results emerge in Figure \ref{Figure11}, which offers an interesting picture of the modal composition of the quantum depletion. In the deep superfluid regime the Goldstone mode alone saturates the fraction of depleted particles, while the Higgs mode acquires a non-negligible weights exactly at the boundary between weak and strong interactions ($J/U \approx 1$) and becomes the dominant mode when approaching the critical region, where the Goldstone contribution vanishes. On the other hand, higher-energy excitation modes do not contribute significantly to the depletion of the condensate.\footnote{The first two modes occupying the energy spectrum above the Higgs excitation have been called \textit{hyper-Higgs modes} only for classification purposes.} \uline{Remarkably, despite its simplicity and methodological analogy with the Bogoliubov theory, our Gutzwiller approach is able to provide non-trivial and unexpected information about the role of high-energy bands in presence of strong interactions. In this sense, our Gutzwiller theory turns to be a good candidate for investigating quantum fluctuations due to pure interaction effects.}
	
	\medskip
	
	The arguments of the above discussion are further supported by the comparison between the results of the Gutzwiller theory with the weakly-interacting prediction proposed in Figures \ref{Figure12} and \ref{Figure13}. As expected from the very beginning, they match exactly for $J/U \to \infty$, while their gap is maximum when approaching the transition, where the Bogoliubov approximation fails in capturing the number of depleted particles.
	\vspace{1cm}
	\item \textbf{3D case,} $10^3$ \textbf{sites}
	\begin{figure}[!htp]
		\begin{minipage}[t]{7.6cm}
			\centering
			\includegraphics[width=9.2cm]{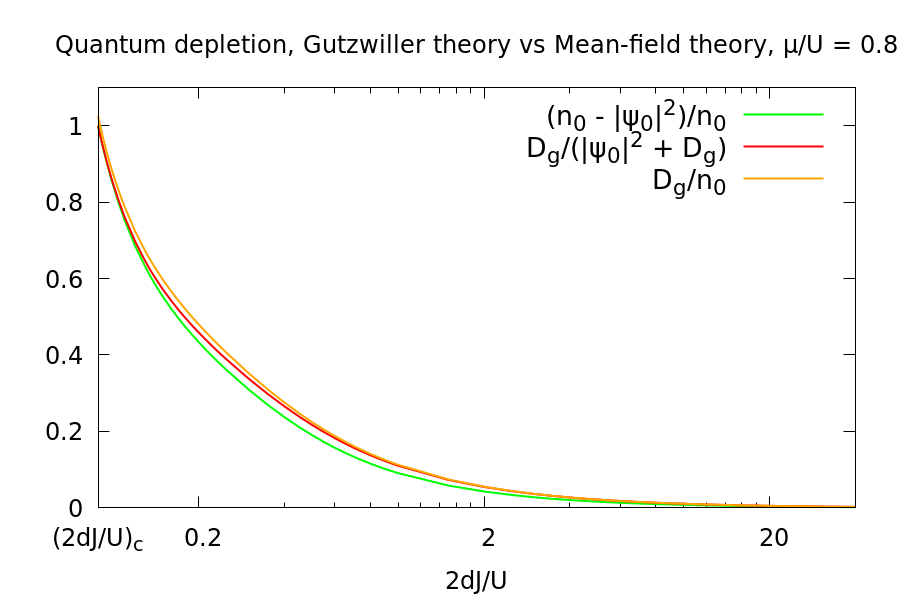}
			\caption{Comparison between the Gutzwiller prediction for the quantum depletion $D_g$ and the mean-field discrepancy between the density $n_0$ and the condensate fraction $\psi^2_0$ for $\mu/U = 0.8$ in 3D.}
			\label{Figure14}
		\end{minipage}
		\ \hspace{4mm} \hspace{5mm} \
		\begin{minipage}[t]{7.5cm}
			\centering
			\includegraphics[width=9.1cm]{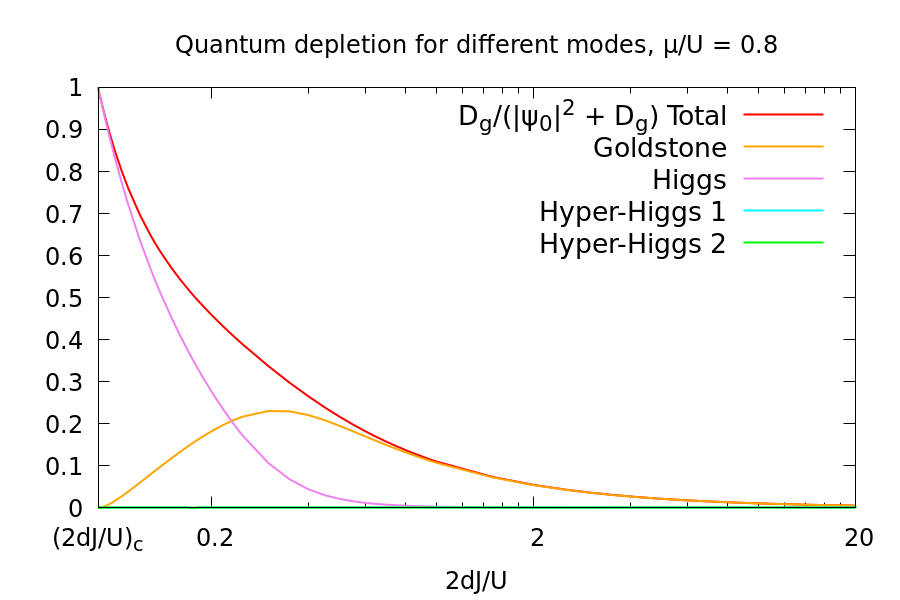}
			\caption{Contributions of the first four excitation modes to the Gutzwiller quantum depletion $D_g$ for $\mu/U = 0.8$ in 3D.}
			\label{Figure15}
		\end{minipage}
	\end{figure}
	\begin{figure}[!htp]
		\begin{minipage}[t]{7.5cm}
			\centering
			\includegraphics[width=9.1cm]{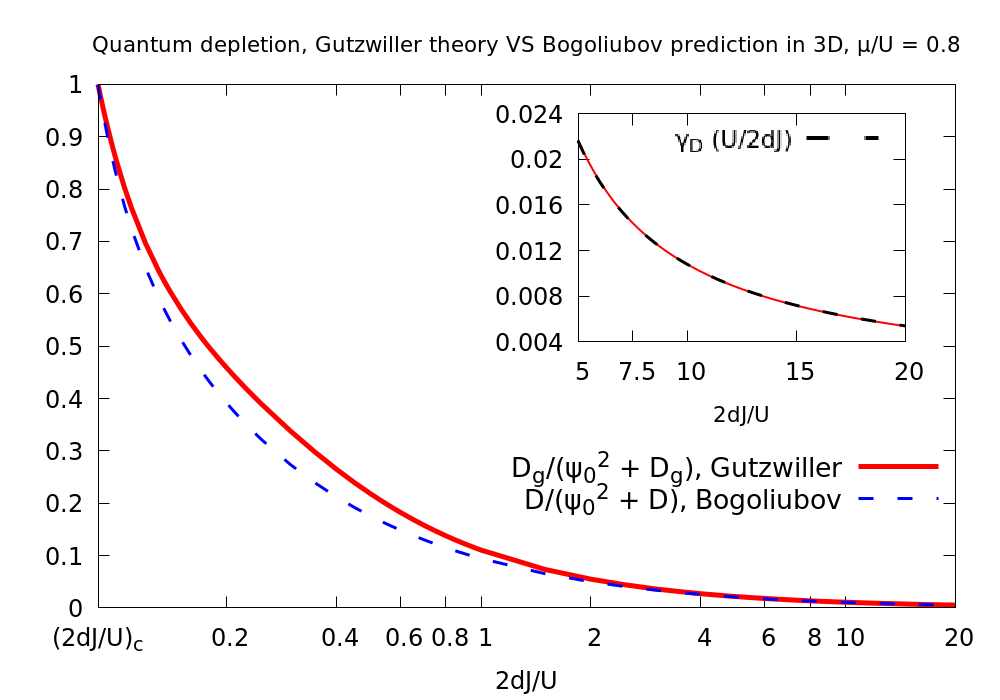}
			\caption{Comparison between $D_g$ and the standard Bogoliubov prediction (dashed blue line) for the quantum depletion for $\mu/U = 0.8$ in 3D. The inset displays the fit of the scaling (\ref{4.37}) to the weak-interaction data.}
			\label{Figure16}
		\end{minipage}
		\ \hspace{5mm} \hspace{5mm} \
		\begin{minipage}[t]{7.5cm}
			\centering
			\includegraphics[width=9cm]{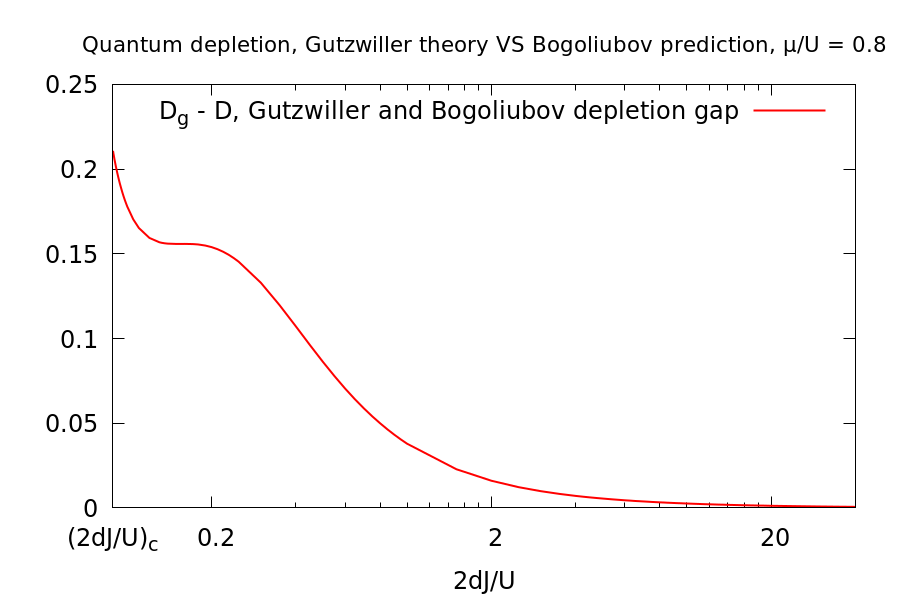}
			\caption{Plot of the gap $D_g - \mathcal{D}$ separating and the quantum depletion within the Gutzwiller theory $D_g$ and the Bogoliubov approximation $\mathcal{D}$ for $\mu/U = 0.8$ in 3D.}
			\label{Figure17}
		\end{minipage}
	\end{figure} \\
	The same observations made on the 1D calculations can be mostly applied to the 3D case, although evident differences come to light in Figures \ref{Figure14} and \ref{Figure16}. Not only the depletion fraction turns to be a small correction to the mean-field prediction, but also $D_g$ tends to reach integer filling almost exactly near the critical transition. \\
	\uline{We suppose that such an improvement is due to the scaling properties of the Gutzwiller approximation behind our quantum theory, as the former is expected to yield exact results in the limit of infinite dimensions.}
	
	\medskip
	
	In 3D the consistency of our numerical results in the weakly-interacting limit can be further tested by loosely comparing the asymptotic behaviour of the fraction of depleted particles for $J/U \to \infty$ with the well-known Lee-Huang-Yang expansion:
	\begin{equation}\label{4.35}
	\frac{D_{LHY}}{n} = \frac{n - \left| \psi_0 \right|^2}{n} = \frac{8}{3 \sqrt{\pi}} \left( n \, a^3 \right)^{1/2}
	\end{equation}
	where $n$ is the mean local density and $a$ is the scattering length. Since $D_{LHY}$ is valid only for continuous condensates, we expect that only the qualitative behaviour of expression (\ref{4.35}) matches with our predictions for weak interactions on the lattice, namely when the Goldstone oscillations have low energy and have Galilean dispersion. Indeed, knowing that:
	\begin{equation}\label{4.36}
	n \sim \left| \psi_0 \right|^2 \approx \mu/U + 2 \, d \, J/U \quad , \quad a = \frac{m \, U}{4 \, \pi \, \hbar^2} \quad \text{for} \quad J/U \gg 1
	\end{equation}
	we obtain:
	\begin{equation}\label{4.37}
	\frac{D_{LHY}}{n} \sim \gamma_D \left( 2 \, d \, J/U \right)^{-1} \quad \text{for} \quad J/U \gg 1
	\end{equation}
	where $\gamma_D$ is an irrelevant multiplicative constant. \\
	Fitting the asymptotic curve (\ref{4.37}) to our data, we finally observe that the scaling of the quantum depletion derived form the Gutzwiller theory perfectly match with the general behaviour of the depleted fraction for weakly-interacting condensates in the continuous case (see the inset of Figure \ref{Figure16}).
\end{listamia}

\newpage

\section{Superfluid fraction}\label{superfluid_fraction}
The quantum depletion of the condensate can be determined on an equal footing with respect to another fundamental observable, that is the superfluid density, which is again defined as the stiffness of the system with respect to a phase twist of the condensate wave function. \\
Since our quantum theory generalizes the Bogoliubov approach beyond the phonon excitation over a wider region of the physical parameters, the Drude expression for the superfluid fraction $f_s$ applied to the weakly-interacting limit in Chapter \ref{Bogoliubov} can be directly reformulated in terms of the actual Gutzwiller particle-hole amplitudes:
\begin{equation}\label{4.38}
\begin{aligned}
f_s &= -\frac{1}{2 \, N \, J} \langle \hat{T} \rangle = \frac{1}{2 \, N} \big\langle \sum_{\langle \mathbf{r}, \mathbf{s} \rangle} \left[ \hat{\psi}^{\dagger}{\left( \mathbf{r} \right)} \, \hat{\psi}{\left( \mathbf{s} \right)} + \text{h.c.} \right] \big\rangle \approx \\
&\approx \frac{M}{N} \left\{ \left| \psi_0 \right|^2 + \frac{1}{M} \sum_{\alpha > 0} \sum_{\mathbf{k}} \left[ \left( \left| U_{\alpha, \mathbf{k}} \right|^2 + \left| V_{\alpha, \mathbf{k}} \right|^2 \right) \langle \hat{b}^{\dagger}_{\alpha, \mathbf{k}} \,\hat{b}_{\alpha, \mathbf{k}} \rangle + \left| V_{\alpha, \mathbf{k}} \right|^2 \right] j{\left( \mathbf{k} \right)} \right\} \longrightarrow \\
&\xrightarrow[T = 0]{} \frac{M}{N} \left[ \left| \psi_0 \right|^2 + \frac{1}{M} \sum_{\alpha > 0} \sum_{\mathbf{k}} \left| V_{\alpha, \mathbf{k}} \right|^2 j{\left( \mathbf{k} \right)} \right]
\end{aligned}
\end{equation}
where:
\begin{equation}\label{4.39}
\hat{T} \equiv -J \sum_{\langle \mathbf{r}, \mathbf{s} \rangle} \left[ \hat{\psi}^{\dagger}{\left( \mathbf{r} \right)} \, \hat{\psi}{\left( \mathbf{s} \right)} + \text{h.c.} \right]
\end{equation}
and:
\begin{equation}\label{4.40}
j{\left( \mathbf{k} \right)} \equiv \sum_{i = 1}^{d} \cos{\left( k_i \, a \right)}
\end{equation}
It is important to notice that second-order current term of the Drude expression for $f_s$ introduced in Chapter \ref{Bogoliubov} in the case of the Bogoliubov approximation:
\begin{equation}\label{4.41}
\begin{aligned}
&f^{\left( 2 \right)}_s \approx \frac{1}{N \, J} \sum_{\nu} \frac{\left| \langle \Psi_{\nu} | \hat{J} | \Psi_0 \rangle \right|^2}{E_{\nu} - E_0} = \\
&= -\frac{J}{N} \sum_{\alpha, \beta} \sum_{\mathbf{p}, \mathbf{q}} \Bigg\{ \frac{\left| \sum_{\mathbf{r}} \sum_a \left[ U_{\alpha}{\left( \mathbf{r} + \mathbf{e}_a, \mathbf{p} \right)} \, V_{\beta}{\left( \mathbf{r}, \mathbf{q} \right)} - U_{\alpha}{\left( \mathbf{r}, \mathbf{p} \right)} \, V_{\beta}{\left( \mathbf{r} + \mathbf{e}_a, \mathbf{q} \right)} \right] \right|^2}{\omega_{\alpha, \mathbf{p}} + \omega_{\beta, \mathbf{q}}} + \\
&+ \delta_{\alpha, \beta} \, \delta_{\mathbf{p}, \mathbf{q}} \, \frac{\left| \sum_{\mathbf{r}} \sum_a \left[ U_{\alpha}{\left( \mathbf{r} + \mathbf{e}_a, \mathbf{p} \right)} \, V_{\alpha}{\left( \mathbf{r}, \mathbf{p} \right)} - U_{\alpha}{\left( \mathbf{r}, \mathbf{p} \right)} \, V_{\alpha}{\left( \mathbf{r} + \mathbf{e}_a, \mathbf{p} \right)} \right] \right|^2}{2 \, \omega_{\alpha, \mathbf{p}}} \Bigg\}
\end{aligned}
\end{equation}
where $\nu$ indicates the many-body excited states and:
\begin{equation}\label{4.42}
U_{\alpha}{\left( \mathbf{r}, \mathbf{k} \right)} \equiv U_{\alpha, \mathbf{k}} \, e^{i \mathbf{k} \cdot \mathbf{r}} \quad , \quad V_{\alpha}{\left( \mathbf{r}, \mathbf{k} \right)} \equiv V_{\alpha, \mathbf{k}} \, e^{i \mathbf{k} \cdot \mathbf{r}}
\end{equation}
vanishes also within the Gutzwiller formalism because of translational invariance. \\ \newpage
The contribution of the condensate depletion to (\ref{4.38}) due to the mode $\alpha$ is quantified by:
\begin{equation}\label{4.43}
f^{\alpha}_s = \frac{1}{M} \sum_{\mathbf{k}} \left| V_{\alpha, \mathbf{k}} \right|^2 j{\left( \mathbf{k} \right)}
\end{equation}
In evaluating expression (\ref{4.38}) non-diagonal averages with respect to the momentum and mode indices have been discarded as always, while the finite result at $T = 0$ derives again from the bosonic commutation relations satisfied by the quasi-particle operators. \\
\textsl{According to the observations concerning the quantum depletion, the normalization factor $N$ has to be fixed in order to satisfy the constraint $f_s \le 1$.} At zero temperature this condition on the local density is again ensured by the Schwarz inequality through the following choice:
\begin{equation}\label{4.44}
N \equiv \psi^2_0 + D_g = M \left( \psi^2_0 + \frac{1}{M} \sum_{\alpha > 0} \sum_{\mathbf{k}} \left| V_{\alpha, \mathbf{k}} \right|^2 \right)
\end{equation}

\bigskip
\vspace{-0.2cm}

In analogy with the conclusions deduced by the results of the standard Bogoliubov approach, we can make some predictions about the expected behaviour of the superfluid fraction $f_s$ with respect to the condensate density $\left| \psi_0 \right|^2$. \\ \textsl{For weak interactions and a small quantum depletion, which fills only the lower half of the band where the factor $j{\left( \mathbf{k} \right)}$ is positive, the superfluid fraction is larger than the condensate density and tends exactly to $1$ as $J/U \to \infty$. Thus, the depletion of the condensate should have little effect on superfluidity in the deep superfluid regime. When the depleted population spreads into the upper part of the energy band, where the function $j{\left( \mathbf{k} \right)}$ has a negative sign, the superfluid fraction could be reduced and might even become smaller than the condensate fraction. This, however, can lead to perfect filling and a possible cancellation of the flow. Whereas in the case of the ordinary Bogoliubov description one cannot observe a perfect switching off of the superfluid, our theory based on the Gutzwiller equations seems to be a good candidate for detecting this effect, which is expected to happen in correspondence of the Mott phase.}

\vspace{1cm}

\subsection{Numerical results: superfluid fraction \texorpdfstring{$f_s$}{} in 1D and 3D}
In the following we report the numerical results concerning the superfluid fraction $f_s$ extracted from our quantized Gutzwiller theory across the Bose-Hubbard phase diagram for different values of the relative chemical potential $\mu/U$. Numerical calculations have been performed again both in the 1D and 3D case for systems composed by a relatively large number of particle without reaching the thermodynamic limit.

\newpage
\subsubsection{3D case, \texorpdfstring{$10^3$}{} sites}
\begin{listamia}
	\item $\mu/U = 0.2$
	\begin{figure}[!htp]
		\begin{minipage}[t]{8.5cm}
			\centering
			\includegraphics[width=9.1cm]{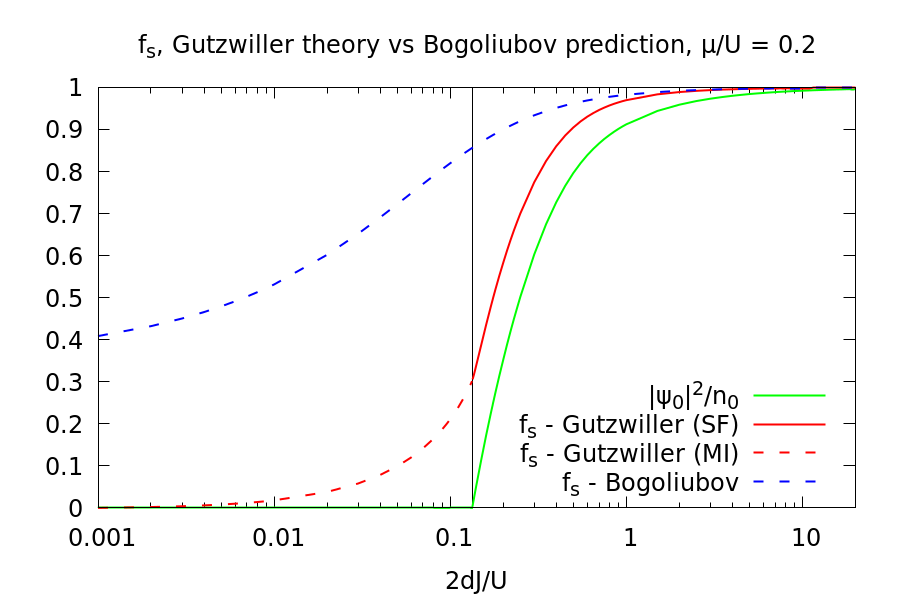}
			\caption{Plot of $f_s$ as a function of $2 d J/U$ for $\mu/U = 0.2$ in 3D compared with the weakly-interacting Bogoliubov prediction (blue dashed line). The vertical black line indicates the transition point $\left( 2 d J/U \right)_c \simeq 0.133$.}
			\label{Figure18}
		\end{minipage}
		\ \hspace{1mm} \hspace{1mm} \
		\begin{minipage}[t]{8cm}
			\centering
			\includegraphics[width=9.1cm]{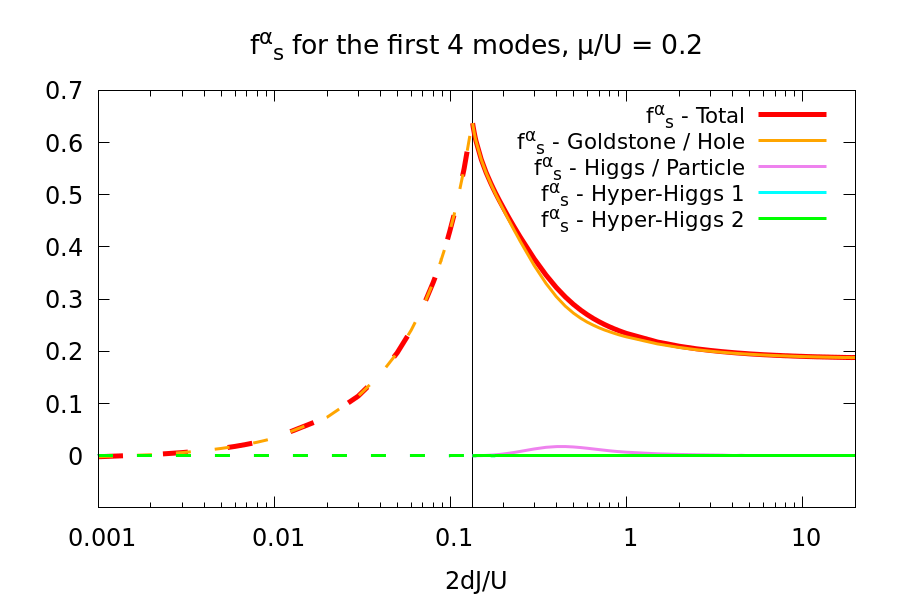}
			\caption{Plot of the mode contribution $f^{\alpha}_s$ as a function of strength $2 d J/U$ for $\mu/U = 0.2$ in 3D. The vertical black line indicates the transition point $\left( 2 d J/U \right)_c \simeq 0.133$.}
			\label{Figure19}
		\end{minipage}
	\end{figure}
	\vspace{1.5cm}
	\item $\mu/U = \sqrt{2} - 1$ (across the tip of the first Mott lobe)
	\begin{figure}[!htp]
		\begin{minipage}[t]{8.5cm}
			\centering
			\includegraphics[width=9.1cm]{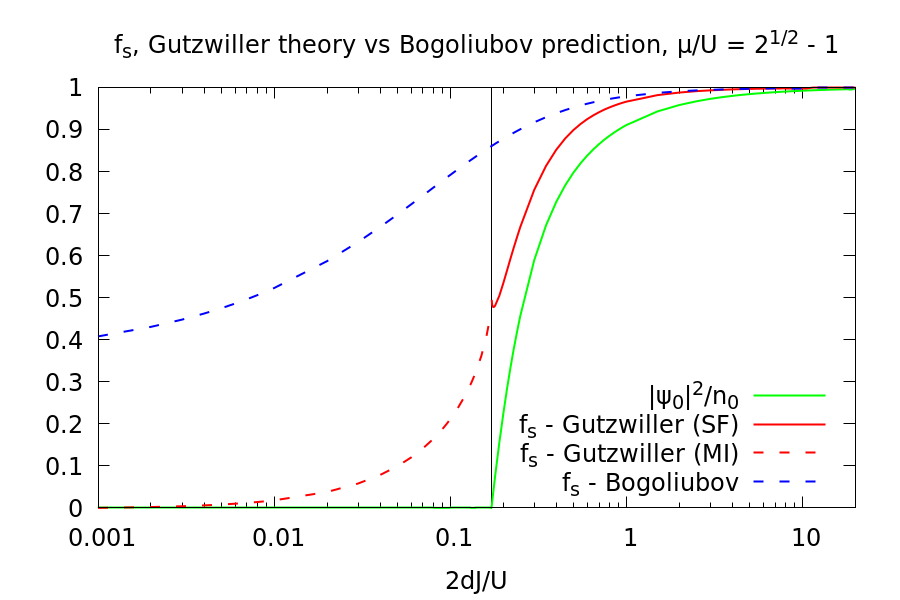}
			\caption{Plot of $f_s$ as a function of $2 d J/U$ for $\mu/U = \sqrt{2} - 1$ in 3D compared with the weakly-interacting Bogoliubov prediction (blue dashed line). The vertical black line indicates the transition point $\left( 2 d J/U \right)_c \simeq 0.172$.}
			\label{Figure20}
		\end{minipage}
		\ \hspace{1mm} \hspace{1mm} \
		\begin{minipage}[t]{8cm}
			\centering
			\includegraphics[width=9.1cm]{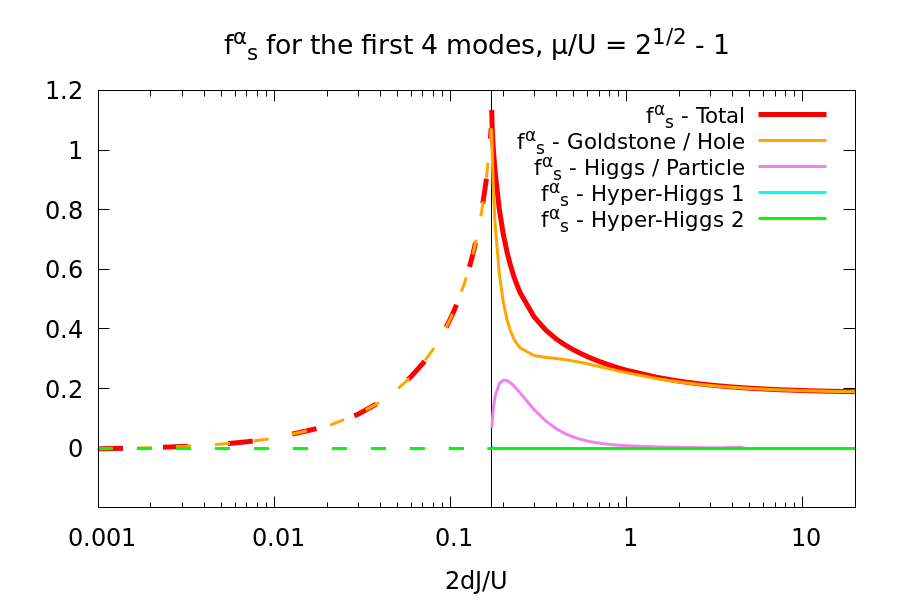}
			\caption{Plot of the mode contribution $f^{\alpha}_s$ as a function of strength $2 d J/U$ for $\mu/U = \sqrt{2} - 1$ in 3D. The vertical black line indicates the transition point $\left( 2 d J/U \right)_c \simeq 0.172$.}
			\label{Figure21}
		\end{minipage}
	\end{figure}
	\newpage
	\item $\mu/U = 0.8$
	\begin{figure}[!htp]
		\begin{minipage}[t]{8.5cm}
			\centering
			\includegraphics[width=9.1cm]{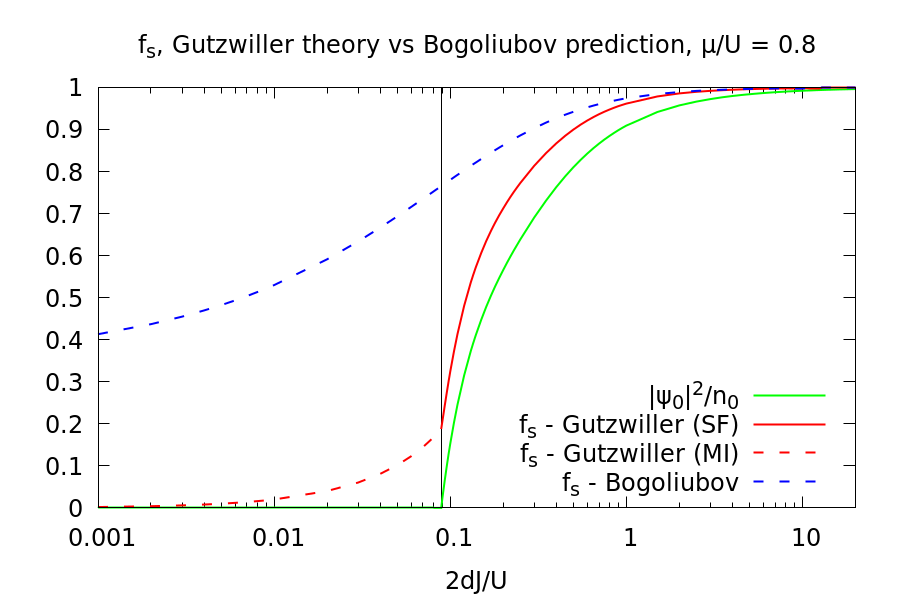}
			\caption{Plot of $f_s$ as a function of $2 d J/U$ for $\mu/U = 0.8$ in 3D compared with the weakly-interacting Bogoliubov prediction (blue dashed line). The vertical black line indicates the transition point $\left( 2 d J/U \right)_c \simeq 0.089$.}
			\label{Figure22}
		\end{minipage}
		\ \hspace{1mm} \hspace{1mm} \
		\begin{minipage}[t]{8cm}
			\centering
			\includegraphics[width=9cm]{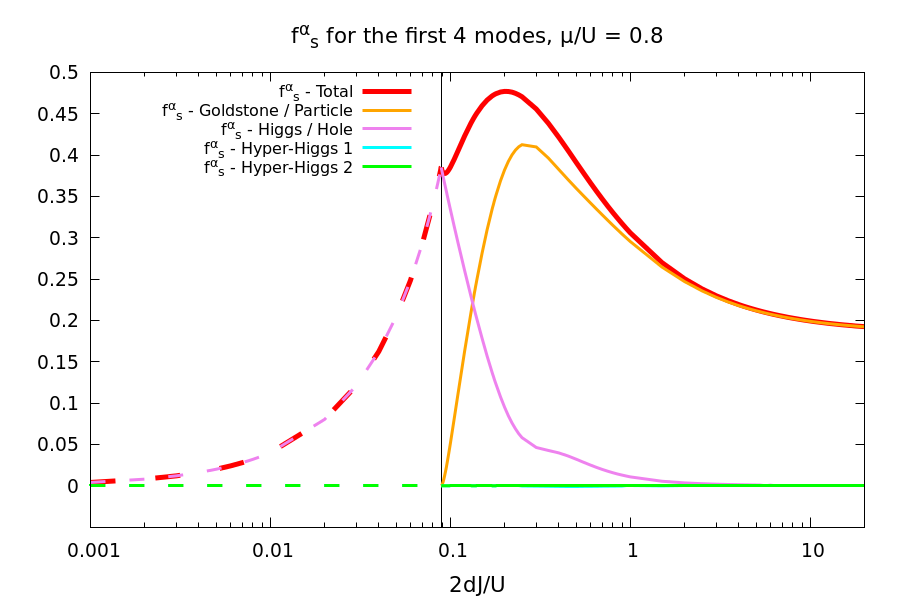}
			\caption{Plot of the mode contribution $f^{\alpha}_s$ as a function of strength $2 d J/U$ for $\mu/U = 0.8$ in 3D. The vertical black line indicates the transition point $\left( 2 d J/U \right)_c \simeq 0.089$.}
			\label{Figure23}
		\end{minipage}
	\end{figure}
\end{listamia}

\subsection{Observations and discussion about the results}\label{superfluid_fraction_discussion}
\paragraph{General features}
In accordance with the analysis of the quantum depletion data, Figures \ref{Figure18}, \ref{Figure20} and \ref{Figure22} show a perfect match between the superfluid fraction given by equation (\ref{4.38}) and the Bogoliubov prediction in the weakly-interacting limit. Moreover, \uline{we can see how the increasing quantum depletion is responsible for a reduction of the superfluid component of the system as the strongly-interacting regime is approached from the superfluid side}.

\smallskip

\uline{Far from being a trivial result, the superfluid fraction $f_s$ turns out to be always larger than the condensate density, as it occurs with the more complex case of $^4\text{He}$. According to the discussion made above, this means that the impact of non-condensed particles on superfluidity is mainly due to the lower part of the single-particle energy band given by (\ref{4.35}).} The role of interactions appears not only in the reduction of $f_s$, but also in the increasing gap between $f_s$ and the mean-field condensate fraction $\psi^2_0/n_0$. \\
This does not mean that the system cannot be found in a state presenting the opposite situation. In fact, it is possible that a superfluid system prepared in an initial state with a significant occupation of the high-energy part of the band would evolve into states with $f_s \le \psi^2_0/n_0$ within suitable evolution processes such as quench dynamics.

\bigskip

\paragraph{Peculiarities of the Gutzwiller approximation}
As an interesting product of the Gutzwiller approach, \uline{we can notice that the weight of the Goldstone or Higgs contribution to $f_s$ in the superfluid phase is inherited by the weight of the hole mode in the Mott phase. This eigenmode saturates the Drude weight alone because of the very definition of the hole amplitudes $V_{\alpha, \mathbf{k}}$ in (\ref{4.22}) and, from a physical point of view, are the only excitations which can live on top of the fully-occupied band of the insulating state.} \\
As expected, the Goldstone mode dominates the superfluid behaviour in the weakly-interacting limit.

\medskip

As a second observation, it is worth highlighting the analytical properties of $f_s$ around the transition point: \textsl{it turns to be a continuous function of the relative hopping strength, but its first derivative displays evident discontinuity that changes its character at the Lorentz-invariant points of the Mott lobe} (see Figure \ref{Figure20}). Since the superfluid fraction is related to the first-order derivative of the free energy with respect to $J/U$, a discontinuity in its second-order derivative seems to be compatible with order of the phase transitions characterizing the Bose-Hubbard model. As shown in \cite{lacki}, \textsl{singularities of simple observables such as $f_s$ can help locate the quantum critical boundary without ambiguities}. \\
\uline{The fact that our Gutzwiller theory is able to capture a further reduction of $f_s$ for $2 \, d \, J \sim 1$ and a signature of the critical point in the strongly-interacting regime through the estimation of the superfluid density confirms that it represents a significant improvement with respect to the Bogoliubov approximation, which does not present similar analytical structures at the transition.}

\bigskip

\paragraph{The interpretation of $f_s$ in the strongly-interacting phases}
Interestingly, we can notice that $f_s$ does not vanish at the Mott-superfluid transition, as one could have deduced also from the structure of the hole amplitudes $V_{\alpha, \mathbf{k}}$ near the transition and in the Mott phase. Actually, the superfluid stiffness decreases monotonically as the system reaches the pure Mott state at $J/U = 0$ where the effective mass $m^* \equiv 1/(2 J)$ is infinite: in fact, \uline{at $J/U \neq 0$ the hopping operator (\ref{4.34}) always presents a non-vanishing expectation value when calculated with respect to the Gutzwiller quantum ground state}. As already observed at the mean-field level \cite{krutitsky_navez}, this effect can be attributed to the fact that at the linear order the first-order fluctuations of the Bose field presents non-vanishing amplitudes within the Mott lobes. \\
Nevertheless, the exact meaning of this result is linked to a more profound explanation.

\bigskip

As a first remark, it is important to remind that \uline{the very definition of $f_s$ given by equation (\ref{4.38}) is not able to distinguish between the effective superfluid fraction of the system and the uniform conductivity or Drude weight}. Formally speaking, in the Mott phase the full quantity $f_s = f^{\left( 1 \right)}_s - f^{\left( 2 \right)}_s$ provides and estimation of the Drude weight, as we know that the system cannot exhibit superfluidity. \\
The problem of distinguishing the effective superfluid fraction $f_s$ from the Drude weight or conductivity $D_c$ has been already addressed by the works of Scalapino et al. \cite{scalapino_white_zhang_1, scalapino_white_zhang_2} and Hetényi \cite{hetenyi}, hence we know that it depends on important mathematical subtleties. \\
Let us suppose to study the Bose-Hubbard model in the formal thermodynamic limit. \uline{Remembering that $f_s$ has been determined as the second-order derivative of the energy variation with respect to the corresponding phase twist applied to the system, we are not guaranteed that this operation commute with the bulk limit. The order of these mathematical steps is exactly what discriminates between the Drude weight and superfluid fraction for a superconductor ($f_s \neq 0$, $D_c \neq 0$), a metal ($f_s = 0$, $D_c \neq 0$) or an insulator ($f_s = D_c = 0$).} In particular, the former requires firstly the twist derivative and then the bulk limit as a second step, while the latter corresponds to the inverse procedure. The precise explanation of this subtle difference relies on the number of adiabatically-evolved many-body states that are accounted by the calculation of the flux response $\dpar \langle \hat{H} \rangle/\dpar \theta^2$.

\bigskip

Having clarified that the physical response provided by our calculations in the Mott phase has to be identified with the Drude weight $D_c$, a non-vanishing conductivity clearly contrasts with the presence of a gap in the Mott excitation spectrum. \\
\uline{In fact, as demonstrated by Scalapino et al. \cite{scalapino_white_zhang_2} in the case of the half-filled fermionic Hubbard model, \textit{a posteriori} we know the Drude weight of a generic insulator must vanish because of the perfect counterbalance between the kinetic term $f^{\left( 1 \right)}_s$ and the current contribution $f^{\left( 2 \right)}_s$. The absence of conductivity in the Mott phase can be also investigated by a simple perturbative calculation of $f_s$ with respect to the hopping strength}, as we show in Appendix \ref{D}. \\
From this perspective, our Gutzwiller approach provides only the response term due to zero-point fluctuations due to the kinetic energy, while remarkably the second-order contribution due to current-current correlations vanishes identically even if calculated according to the formal procedure suggested in \cite{scalapino_white_zhang_1, scalapino_white_zhang_2} (see Appendix \ref{E}).

\bigskip

In conclusion, \uline{it follows that the actual Gutzwiller quantum theory suffers from the same drawbacks of the standard Bogoliubov approximation highlighted in Chapter \ref{Bogoliubov}}. The perturbative calculation of the Drude weight in Mott phase suggests that, through the application global translational invariance, \uline{the current correlations derived from these approaches do not take into account the local dimension of particle-hole pairs responsible for a vanishing conductivity, although the two-point correlation function $g_1{\left( \mathbf{r}, \mathbf{s} \right)}$ presents the correct range in the Mott phase} (see Section \ref{correlations}). \\
As underlined by Rey et al. \cite{rey}, the problem of vanishing current correlations for strong interactions can be fixed by passing to self-consistent approach in the quantization of the fluctuations, which is exactly what is missing in our Gutzwiller theory. \\
Hence \uline{we expect that the formulation of a self-consistent Gutzwiller quantum theory would lead to a suppression of the superfluid fraction at the transition and a vanishing Drude conductivity in the Mott phase}. \\
Nevertheless, \uline{our numerical data can be assumed as a genuine proof of the improvement in the description of strongly-interacting superfluids provided by a Gutzwiller-based quantum theory}.

\vspace{1cm}

\subsection{1D case, \texorpdfstring{$10^2$}{} sites}
As we have observed also in the case of the quantum depletion, the results of the 1D case for the full value of $f_s$ do not present significant variations with respect the 1D results. This outcome is not surprising: our quantum theory has been constructed over a mean-field ansatz which is known to provide exact results only in the limit of infinite dimensions. \\
Indeed, a variation in the predictions of the Gutzwiller theory for larger dimensionalities seems to come to light when we analyse the mode composition of $f_s$ and its peculiar singularities. While the discontinuity character at $\mu/U = \sqrt{2} - 1$ appears unchanged, other points of the critical line present different analytical properties with respect to the 3D case.
\newpage
\begin{figure}[!htp]
	\begin{minipage}[b]{8.5cm}
		\centering
		\includegraphics[width=9cm]{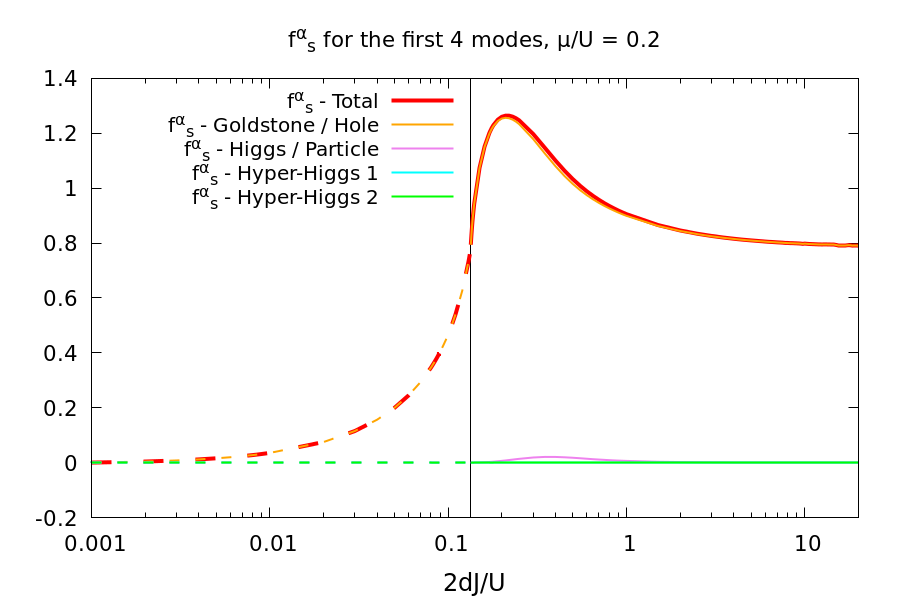}
		\caption{Plot of the mode contribution $f^{\alpha}_s$ as a function of strength $2 d J/U$ for $\mu/U = 0.2$ in 1D. The vertical black line indicates the transition point $\left( 2 d J/U \right)_c \simeq 0.133$.}
		\label{Figure24}
	\end{minipage}
	\ \hspace{1mm} \hspace{1mm} \
	\begin{minipage}[b]{8.5cm}
		\centering
		\includegraphics[width=9cm]{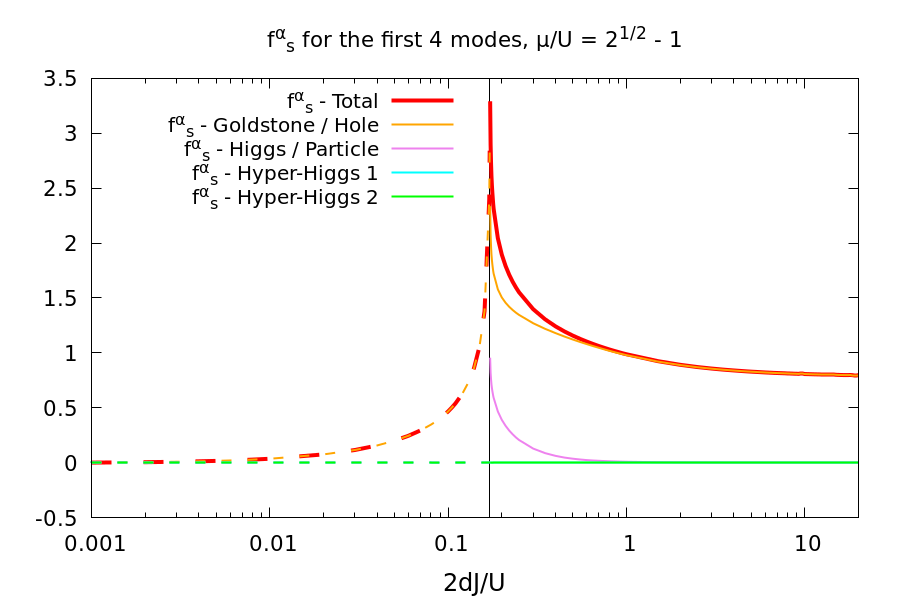}
		\caption{Plot of the mode contribution $f^{\alpha}_s$ as a function of strength $2 d J/U$ for $\mu/U = \sqrt{2} - 1$ in 1D. The vertical black line indicates the transition point $\left( 2 d J/U \right)_c \simeq 0.172$.}
		\label{Figure25}
	\end{minipage}
\end{figure}
\begin{figure}[!htp]
	\centering
	\includegraphics[width=10cm]{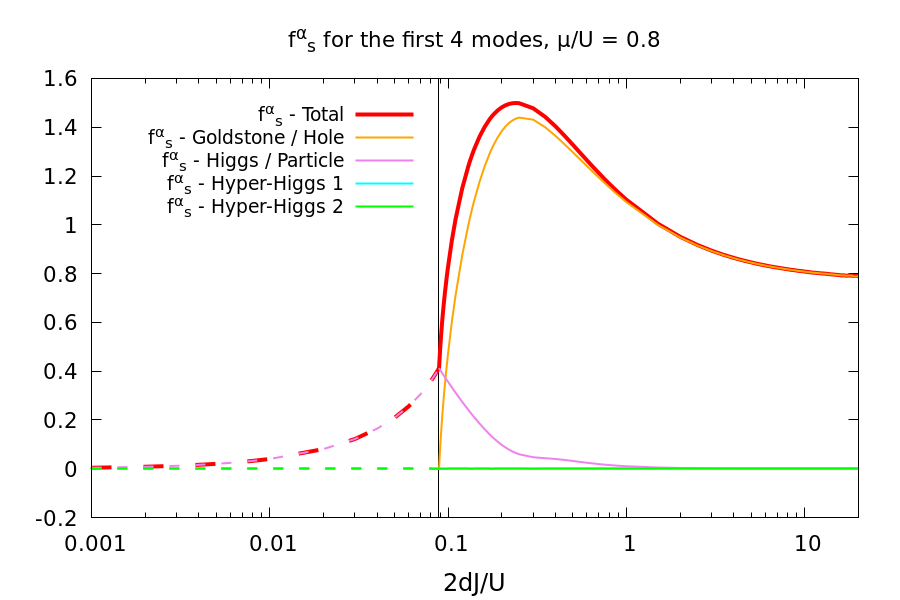}
	\caption{Plot of the mode contribution $f^{\alpha}_s$ as a function of strength $2 d J/U$ for $\mu/U = 0.8$ in 1D. The vertical black line indicates the transition point $\left( 2 d J/U \right)_c \simeq 0.089$.}
	\label{Figure26}
\end{figure}

\newpage

\markboth{\MakeUppercase{5. Quantum fluctuations}}{\MakeUppercase{5.3 Second-order operator corrections}}
\section{Operator fluctuations up to second order}\label{second_order_corrections}
\markboth{\MakeUppercase{5. Quantum fluctuations}}{\MakeUppercase{5.3 Second-order operator corrections}}
The quadratic quantization of the Gutzwiller ansatz leading to the harmonic Hamiltonian operator (\ref{422}) has allowed to identify the algebraic form of the quantum fluctuations which give rise to the zero-point physics beyond the mean-field solution. \\
Recovering the brief discussion on the normalization of the quantum depletion data in Section \ref{depletion}, the following developments focus on the effects produced by the introduction of second-order operatorial terms into the fluctuation part of the main physical observables. As already observed while analysing the role of the renormalizing operator $\hat{A}$ defined in (\ref{4.12}), it is likely that full second-order corrections will show unexpected results due to the fact that our quantization procedure is not self-consistent with respect to the consequent second-second corrections. Nevertheless, numerical calculations could also reveal interesting features which are supposed to go in the right direction in describing higher-order quantum fluctuations above the mean-field solutions.

\subsection{Estimating the weight of second-order corrections}
\begin{figure}[!htp]
	\begin{minipage}[t]{8.2cm}
		\centering
		\includegraphics[width=7.4cm]{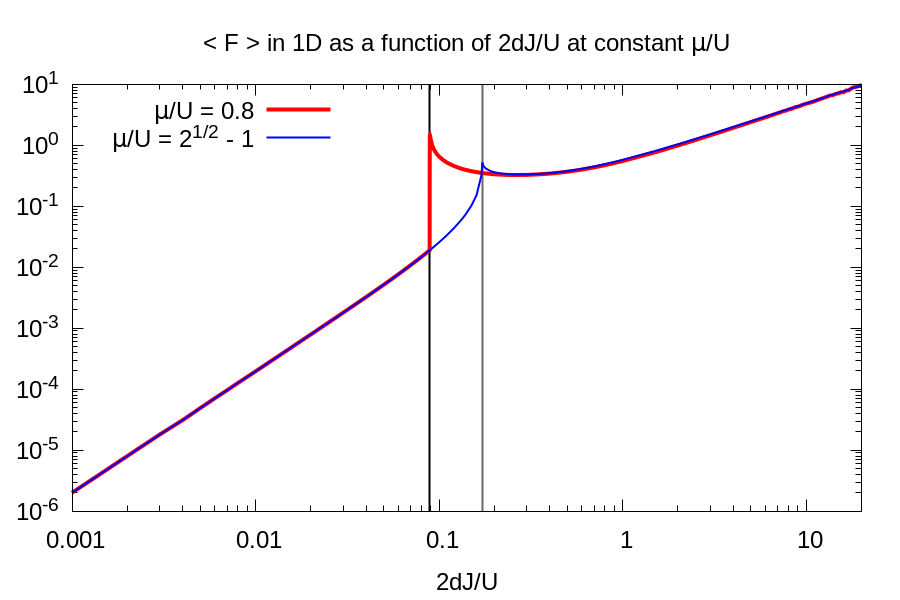}
		\caption{Average value of the fluctuation operator $\langle \hat{F} \rangle$ as a function of $2 \, d \, J/U$ for $\mu/U = \sqrt{2} - 1$ (blue line) and for $\mu/U = 0.8$ (red line) in 1D. The black and grey lines indicate the position of the critical line.}
		\label{Figure56}
	\end{minipage}
	\ \hspace{4mm} \hspace{4mm} \
	\begin{minipage}[t]{8.2cm}
		\centering
		\includegraphics[width=7.4cm]{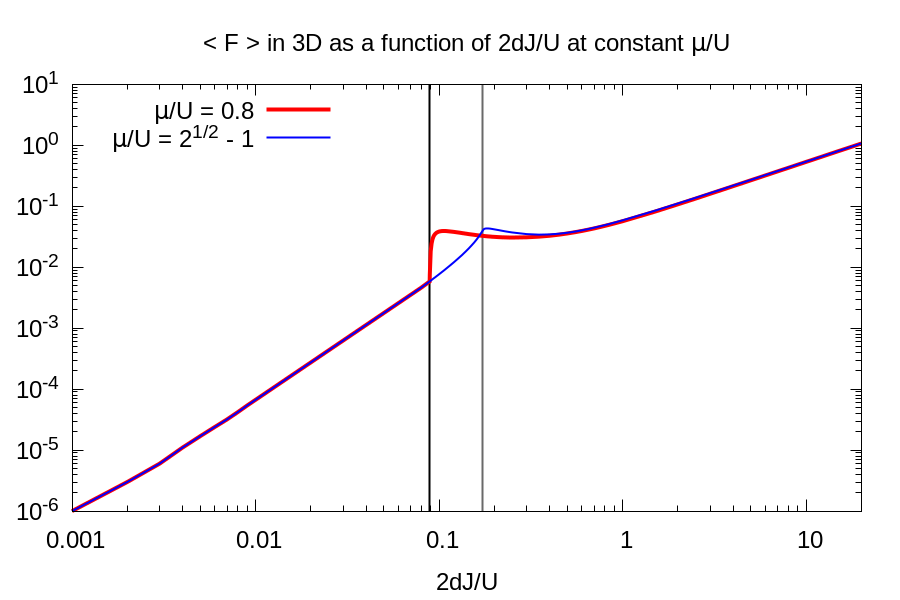}
		\caption{Average value of the fluctuation operator $\langle \hat{F} \rangle$ as a function of $2 \, d \, J/U$ for $\mu/U = \sqrt{2} - 1$ (blue line)  and for $\mu/U = 0.8$ (red line) in 3D. The black and grey lines indicate the position of the critical line.}
		\label{Figure57}
	\end{minipage}
\end{figure}

The necessity of a self-consistent quantum theory can be justified by a more accurate analysis of the weight of higher-order corrections with respect to the mean-field parameters. The simplest way for estimating the contribution of the second-order corrections given by the non-self-consistent theory based on equation (\ref{416}) is to consider the renormalizing operator $\hat{A}$ defined in (\ref{4.12}):
\begin{equation}\label{4.73}
A = \sqrt{1 - \sum_n \delta \hat{c}^{\dagger}_n{\left( \mathbf{r} \right)} \, \delta \hat{c}_n{\left( \mathbf{r} \right)}}
\end{equation}
and calculate the expectation value of the argument of the square root in (\ref{4.73}):
\begin{equation}\label{4.74}
\langle \hat{F} \rangle \equiv \sum_n \langle \delta \hat{c}^{\dagger}_n{\left( \mathbf{r} \right)} \, \delta \hat{c}_n{\left( \mathbf{r} \right)} \rangle = \frac{1}{M} \sum_n \sum_{\alpha > 0} \sum_{\mathbf{k}} \left[ \left( \left| u_{\alpha, {\mathbf{k}}, n} \right|^2 + \left| v_{\alpha, \mathbf{k}, n} \right|^2 \right) \langle \hat{b}^{\dagger}_{\alpha, \mathbf{k}} \hat{b}_{\alpha, \mathbf{k}} \rangle + \left| v_{\alpha, \mathbf{k}, n} \right|^2 \right]
\end{equation}
Figures (\ref{Figure56}) and (\ref{Figure57}) show the behaviour of the quantity (\ref{4.74}) in 1D and 3D at $T = 0$ as a function of $2 \, d \, J/U$ for different values of the relative chemical potential.
\vspace{-0.47cm}
\newpage
As expected, the average value of the second-order term in $\hat{A}$ turns out to be comparable to 1 over a large range of values the hopping strength and in particular in the superfluid phase when approaching the weakly-interacting limit. \\
Both the 1D and 3D cases present pathological results, although the 3D data show a significant improvement with respect to the relative amplitude of the second-order fluctuations. Once again, this result seems to hint at the role of the dimensionality in a quantum theory based on the Gutzwiller approximation.

\subsection{Density and Bose field operators}
Firstly, let us consider again the simplest observable that we can calculate within the Gutzwiller approach, that is the density. Formally its expression can be further expanded up to higher orders in the form:
\begin{equation}\label{4.75}
\hat{n}{\left( r \right)} = \sum_n n \, \hat{c}^{\dagger}_n{\left( \mathbf{r} \right)} \, \hat{c}_n{\left( \mathbf{r} \right)} = n_0 + \delta_1 \hat{n}{\left( \mathbf{r} \right)} + \delta_2 \hat{n}{\left( \mathbf{r} \right)} + \delta_3 \hat{n}{\left( \mathbf{r} \right)} + \dots
\end{equation}
where:
\begin{equation}\label{4.76}
\begin{aligned}
&\delta_2 \hat{n}{\left( r \right)} = \sum_n n \, \delta \hat{c}^{\dagger}_n{\left( \mathbf{r} \right)} \, \delta \hat{c}_n{\left( \mathbf{r} \right)} - n_0 \sum_n \delta \hat{c}^{\dagger}_n{\left( \mathbf{r} \right)} \, \delta \hat{c}_n{\left( \mathbf{r} \right)} \\
&\delta_3 \hat{n}{\left( \mathbf{r} \right)} = -\frac{1}{2} \sum_ n n \, \delta \hat{c}^{\dagger}_n{\left( \mathbf{r} \right)} \sum_m \delta \hat{c}^{\dagger}_m{\left( \mathbf{r} \right)} \, \delta \hat{c}_m{\left( \mathbf{r} \right)} + \text{h.c.}
\end{aligned}
\end{equation}
where $n_0$ is the mean density according to the saddle-point solution as always. \\
The second term of $\delta_2 \hat{n}{\left( \mathbf{r} \right)}$ derive from the overall operator $\left| c^0_n \right|^2 \hat{A}^2$ given by the substitution of (\ref{4.12}) into (\ref{4.75}), while the third order correction $\delta_3 \hat{n}{\left( \mathbf{r} \right)}$ is given by the combination between an operator $\hat{c}_n{\left( \mathbf{r} \right)}$ and the second-order Taylor expansion of the renormalization operator $\hat{A}$:
\begin{equation}\label{4.77}
\hat{A} = \sqrt{1 - \sum_n \delta \hat{c}^{\dagger}_n{\left( \mathbf{r} \right)} \delta \hat{c}_n{\left( \mathbf{r} \right)}} \simeq 1 - \frac{1}{2} \sum_n \delta \hat{c}^{\dagger}_n{\left( \mathbf{r} \right)} \delta \hat{c}_n{\left( \mathbf{r} \right)}
\end{equation}
Actually, as the operator (\ref{4.77}) can be expanded up to an arbitrary even order, the density operator can be expressed up to an arbitrary odd order similarly to $\delta_3 \hat{n}{\left( \mathbf{r} \right)}$. \\
Always taking into account that the harmonic quadratic Hamiltonian (\ref{422}) does not couple different modes, the structure of the average of second-order contributions of the form shown in (\ref{4.76}) are expected to simplify in a form similar to the quantum depletion determined in Section \ref{depletion}, whereas the expectation value of first-order fluctuations turn to zero, as they are linear combinations of single quasi-particle operators.

\medskip

Similar expressions as (\ref{4.76}) hold also for the expansion of the superfluid order parameter, whose complete expression in the quantization framework reads:
\begin{equation}\label{4.78}
\hat{\psi}{\left( \mathbf{r} \right)} = \sum_n \sqrt{n} \, \hat{c}^{\dagger}_{n - 1}{\left( \mathbf{r} \right)} \, \hat{c}_n{\left( \mathbf{r} \right)} = \psi_0 + \delta_1 \hat{\psi}{\left( \mathbf{r} \right)} + \delta_2 \hat{\psi}{\left( \mathbf{r} \right)} + \delta_3 \hat{\psi}{\left( \mathbf{r} \right)} + \dots
\end{equation}
where:
\begin{equation}\label{4.79}
\begin{aligned}
&\delta_2 \hat{\psi}{\left( \mathbf{r} \right)} = \sum_n \sqrt{n} \, \delta \hat{c}^{\dagger}_{n - 1}{\left( \mathbf{r} \right)} \, \delta \hat{c}_n{\left( \mathbf{r} \right)} - \psi_0 \sum_n \delta \hat{c}^{\dagger}_n{\left( \mathbf{r} \right)} \, \delta \hat{c}_n{\left( \mathbf{r} \right)} \\
&\delta_3 \hat{\psi}{\left( \mathbf{r} \right)} = -\frac{1}{2} \sum_ n \sqrt{n} \, c^0_n \, \delta \hat{c}^{\dagger}_{n - 1}{\left( \mathbf{r} \right)} \sum_m \delta \hat{c}^{\dagger}_m{\left( \mathbf{r} \right)} \, \delta \hat{c}_m{\left( \mathbf{r} \right)} + \\
&- \frac{1}{2} \sum_m \delta \hat{c}^{\dagger}_m{\left( \mathbf{r} \right)} \, \delta \hat{c}_m{\left( \mathbf{r} \right)} \sum_ n \sqrt{n} \, c^0_{n - 1} \, \delta \hat{c}_n{\left( \mathbf{r} \right)}
\end{aligned}
\end{equation}
where $\psi_0$ is again the mean-field prediction for the order parameter. \\

\subsection{Density and order parameter averages}
Since the averages of the first-order expansion terms in (\ref{4.75}) and (\ref{4.78}) vanish identically, it is necessary to determine the expectation values of the second-order corrections. It is again useful to express the such terms of the density and the superfluid order parameter expansions with respect to the quasi-particle operators introduced in (\ref{419}):
\begin{equation}\label{4.80}
\begin{aligned}
\delta_2 \hat{n}{\left( \mathbf{r} \right)} &= \sum_n n \, \delta \hat{c}^{\dagger}_n{\left( \mathbf{r} \right)} \, \delta \hat{c}_n{\left( \mathbf{r} \right)} - n_0 \sum_n \delta \hat{c}^{\dagger}_n{\left( \mathbf{r} \right)} \, \delta \hat{c}_n{\left( \mathbf{r} \right)} = \\
&= \frac{1}{M} \sum_n n \sum_{\alpha, \beta > 0} \sum_{\mathbf{k}, \mathbf{p}} e^{i \left( \mathbf{p} - \mathbf{k} \right) \cdot \mathbf{r}} \bigg( u_{\alpha, \mathbf{k}, n} \, u_{\beta, \mathbf{p}, n} \, \hat{b}^{\dagger}_{\alpha, \mathbf{k}} \, \hat{b}_{\beta, \mathbf{p}} + u_{\alpha, \mathbf{k}, n} \, v_{\beta, -\mathbf{p}, n} \, \hat{b}^{\dagger}_{\alpha, \mathbf{k}} \, \hat{b}^{\dagger}_{\beta, -\mathbf{p}} + \\
&+ u_{\beta, \mathbf{p}, n} \, v_{\alpha, -\mathbf{k}, n} \, \hat{b}_{\alpha, -\mathbf{k}} \, \hat{b}_{\beta, \mathbf{p}} + v_{\alpha, -\mathbf{k}, n} \, v_{\beta, -\mathbf{p}, n} \, \hat{b}_{\alpha, -\mathbf{k}} \, \hat{b}^{\dagger}_{\beta, -\mathbf{p}} \bigg) + \\
&- \frac{n_0}{M} \sum_n \sum_{\alpha > 0} \sum_{\mathbf{k}, \mathbf{p}} e^{i \left( \mathbf{p} - \mathbf{k} \right) \cdot \mathbf{r}} \bigg( u_{\alpha, \mathbf{k}, n} \, u_{\beta, \mathbf{p}, n} \, \hat{b}^{\dagger}_{\alpha, \mathbf{k}} \, \hat{b}_{\beta, \mathbf{p}} + u_{\alpha, \mathbf{k}, n} \, v_{\beta, -\mathbf{p}, n} \, \hat{b}^{\dagger}_{\alpha, \mathbf{k}} \, \hat{b}^{\dagger}_{\beta, -\mathbf{p}} + \\
&+ u_{\beta, \mathbf{p}, n} \, v_{\alpha, -\mathbf{k}, n} \, \hat{b}_{\alpha, -\mathbf{k}} \, \hat{b}_{\beta, \mathbf{p}} + v_{\alpha, -\mathbf{k}, n} \, v_{\beta, -\mathbf{p}, n} \, \hat{b}_{\alpha, -\mathbf{k}} \, \hat{b}^{\dagger}_{\beta, -\mathbf{p}} \bigg)
\end{aligned}
\end{equation}
According to (\ref{422}), at the quadratic level the average operation applied to (\ref{4.80}) discards again non-diagonal elements in the momentum and mode labels and terms composed by two annihilation or creation operators, so that:
\begin{equation}\label{4.81}
\begin{aligned}
\langle \delta_2 \hat{n}{\left( \mathbf{r} \right)} \rangle &= \frac{1}{M} \sum_n n \sum_{\alpha > 0} \sum_{\mathbf{k}} \bigg( \left| u_{\alpha, \mathbf{k}, n} \right|^2 \langle \hat{b}^{\dagger}_{\alpha, \mathbf{k}} \hat{b}_{\alpha, \mathbf{k}} \rangle + \left| v_{\alpha, \mathbf{k}, n} \right|^2 \langle \hat{b}_{\alpha, \mathbf{k}} \hat{b}^{\dagger}_{\alpha, \mathbf{k}} \rangle \bigg) + \\
&- \frac{n_0}{M} \sum_n \sum_{\alpha > 0} \sum_\mathbf{k} \bigg( \left| u_{\alpha, \mathbf{k}, n} \right|^2 \langle \hat{b}^{\dagger}_{\alpha, \mathbf{k}} \hat{b}_{\alpha, \mathbf{k}} \rangle + \left| v_{\alpha, \mathbf{k}, n} \right|^2 \langle \hat{b}_{\alpha, \mathbf{k}} \hat{b}^{\dagger}_{\alpha, \mathbf{k}} \rangle \bigg) = \\
&= \frac{1}{M} \sum_n \left( n - n_0 \right) \sum_{\alpha > 0} \sum_{\mathbf{k}} \left[ \left( \left| u_{\alpha, {\mathbf{k}}, n} \right|^2 + \left| v_{\alpha, \mathbf{k}, n} \right|^2 \right) \langle \hat{b}^{\dagger}_{\alpha, \mathbf{k}} \hat{b}_{\alpha, \mathbf{k}} \rangle + \left| v_{\alpha, \mathbf{k}, n} \right|^2 \right]
\end{aligned}
\end{equation}
where we notice the emergence of zero-point terms due to the bosonic commutation relations which in principle are not null even at zero temperature. Although the expression (\ref{4.81}) recalls the form of the quantum depletion within the standard Bogoliubov theory of weakly-interacting Bose gases, it is important to specify that it does not refer to the depletion of particles out of the condensate phase. Actually, the quantity (\ref{4.81}) measures the number of particles which do not fulfill the ground state occupation distribution given by the parameters $c^0_n$ and contribute to the zero-point collective excitation modes. \\
As the quasi-particle degrees of freedom are distributed according to Bose statistics, thermal fluctuations can be always accessed by inserting:
\begin{equation}\label{4.82}
\langle \hat{b}^{\dagger}_{\alpha, \mathbf{k}} \hat{b}_{\alpha, \mathbf{k}} \rangle = \frac{1}{\exp{\left( \hbar \, \omega_{\alpha, \mathbf{k}}/T \right)} - 1}
\end{equation}
into equation (\ref{4.81}).

\medskip

The calculation of the quantum corrections to the average superfluid order parameter leads to similar results:
\begin{equation}\label{4.83}
\small
\begin{aligned}
\delta_2 \hat{\psi}{\left( \mathbf{r} \right)} &= \sum_n \sqrt{n} \, \delta \hat{c}^{\dagger}_{n - 1}{\left( \mathbf{r} \right)} \, \delta \hat{c}_n{\left( \mathbf{r} \right)} - \psi_0 \sum_n \delta \hat{c}^{\dagger}_n{\left( \mathbf{r} \right)} \, \delta \hat{c}_n{\left( \mathbf{r} \right)} = \\
&= \frac{1}{M} \sum_n \sqrt{n} \sum_{\alpha, \beta > 0} \sum_{\mathbf{k}, \mathbf{p}} e^{i \left( \mathbf{p} - \mathbf{k} \right) \cdot \mathbf{r}} \bigg( u_{\alpha, \mathbf{k}, n - 1} \, u_{\beta, \mathbf{p}, n} \, \hat{b}^{\dagger}_{\alpha, \mathbf{k}} \, \hat{b}_{\beta, \mathbf{p}} + u_{\alpha, \mathbf{k}, n - 1} \, v_{\beta, -\mathbf{p}, n} \, \hat{b}^{\dagger}_{\alpha, \mathbf{k}} \, \hat{b}^{\dagger}_{\beta, -\mathbf{p}} + \\
&+ u_{\beta, \mathbf{p}, n} \, v_{\alpha, -\mathbf{k}, n - 1} \, \hat{b}_{\alpha, -\mathbf{k}} \, \hat{b}_{\beta, \mathbf{p}} + v_{\alpha, -\mathbf{k}, n - 1} \, v_{\beta, -\mathbf{p}, n} \, \hat{b}_{\alpha, -\mathbf{k}} \,  \hat{b}^{\dagger}_{\beta, -\mathbf{p}} \bigg) + \\
&- \frac{\psi_0}{M} \sum_n \sum_{\alpha > 0} \sum_{\mathbf{k}, \mathbf{p}} e^{i \left( \mathbf{p} - \mathbf{k} \right) \cdot \mathbf{r}} \bigg( u_{\alpha, \mathbf{k}, n} \, u_{\beta, \mathbf{p}, n} \, \hat{b}^{\dagger}_{\alpha, \mathbf{k}} \, \hat{b}_{\beta, \mathbf{p}} + u_{\alpha, \mathbf{k}, n} \, v_{\beta, -\mathbf{p}, n} \, \hat{b}^{\dagger}_{\alpha, \mathbf{k}} \, \hat{b}^{\dagger}_{\beta, -\mathbf{p}} + \\
&+ u_{\beta, \mathbf{p}, n} \, v_{\alpha, -\mathbf{k}, n} \, \hat{b}_{\alpha, -\mathbf{k}} \, \hat{b}_{\beta, \mathbf{p}} + v_{\alpha, -\mathbf{k}, n} \, v_{\beta, -\mathbf{p}, n} \, \hat{b}_{\alpha, -\mathbf{k}} \, \hat{b}^{\dagger}_{\beta, -\mathbf{p}} \bigg)
\end{aligned}
\end{equation}
\begin{equation}\label{4.84}
\small
\begin{aligned}
\langle \delta_2 \hat{\psi}{\left( r \right)} \rangle &= \frac{1}{M} \sum_n \sqrt{n} \sum_{\alpha > 0} \sum_{\mathbf{k}} \bigg( u_{\alpha, \mathbf{k}, n - 1} \, u_{\alpha, \mathbf{k}, n} \langle \hat{b}^{\dagger}_{\alpha, \mathbf{k}} \hat{b}_{\alpha, \mathbf{k}} \rangle + v_{\alpha, \mathbf{k}, n - 1} \, v_{\alpha, \mathbf{k}, n} \langle \hat{b}_{\alpha, \mathbf{k}} \hat{b}^{\dagger}_{\alpha, \mathbf{k}} \rangle \bigg) + \\
&- \frac{\psi_0}{M} \sum_n \sum_{\alpha > 0} \sum_\mathbf{k} \bigg( \left| u_{\alpha, \mathbf{k}, n} \right|^2 \langle \hat{b}^{\dagger}_{\alpha, \mathbf{k}} \hat{b}_{\alpha, \mathbf{k}} \rangle + \left| v_{\alpha, \mathbf{k}, n} \right|^2 \langle \hat{b}_{\alpha, \mathbf{k}} \hat{b}^{\dagger}_{\alpha, \mathbf{k}} \rangle \bigg) = \\
&= \frac{1}{M} \sum_n \sqrt{n} \sum_{\alpha > 0} \sum_{\mathbf{k}} \left[ \left( u_{\alpha, \mathbf{k}, n - 1} \, u_{\alpha, \mathbf{k}, n} + v_{\alpha, \mathbf{k}, n - 1} \, v_{\alpha, \mathbf{k}, n} \right) \langle \hat{b}^{\dagger}_{\alpha, \mathbf{k}} \hat{b}_{\alpha, \mathbf{k}} \rangle + v_{\alpha, \mathbf{k}, n - 1} \, v_{\alpha, \mathbf{k}, n} \right] + \\
&- \frac{\psi_0}{M} \sum_n \sum_{\alpha > 0} \sum_{\mathbf{k}} \left[ \left( \left| u_{\alpha, {\mathbf{k}}, n} \right|^2 + \left| v_{\alpha, \mathbf{k}, n} \right|^2 \right) \langle \hat{b}^{\dagger}_{\alpha, \mathbf{k}} \hat{b}_{\alpha, \mathbf{k}} \rangle + \left| v_{\alpha, \mathbf{k}, n} \right|^2 \right]
\end{aligned}
\end{equation}
The last quantity to be calculated is the condensate density, which is simply derived as the modulus square of the order parameter. Its quadratic expansion simply reads:
\begin{equation}\label{4.85}
\rho{\left( \mathbf{r} \right)} = \langle \hat{\psi}^{\dagger}{\left( \mathbf{r} \right)} \rangle \langle \hat{\psi}{\left( \mathbf{r} \right)} \rangle \approx \psi^2_0 + 2 \psi_0 \langle \delta_2 \hat{\psi}{\left( \mathbf{r} \right)} \rangle
\end{equation}

\medskip

For sake of concreteness, the following subsections will be dedicated to present the numerical results concerning the second-order density and order parameter corrections in 1D and 3D at zero temperature.\footnote{Here the mean-field density is always assumed as the reference point with the respect to second-order fluctuations, since $\psi_0$ vanishes at the MI-SF transition.}

\newpage
\subsection{Numerical results: 1D case, T = 0, \texorpdfstring{$10^2$}{} sites}
\begin{listamia}
	\item $\mu/U = 0.2$
	\begin{figure}[!htp]
		\begin{minipage}[t]{7.7cm}
			\centering
			\includegraphics[width=9cm]{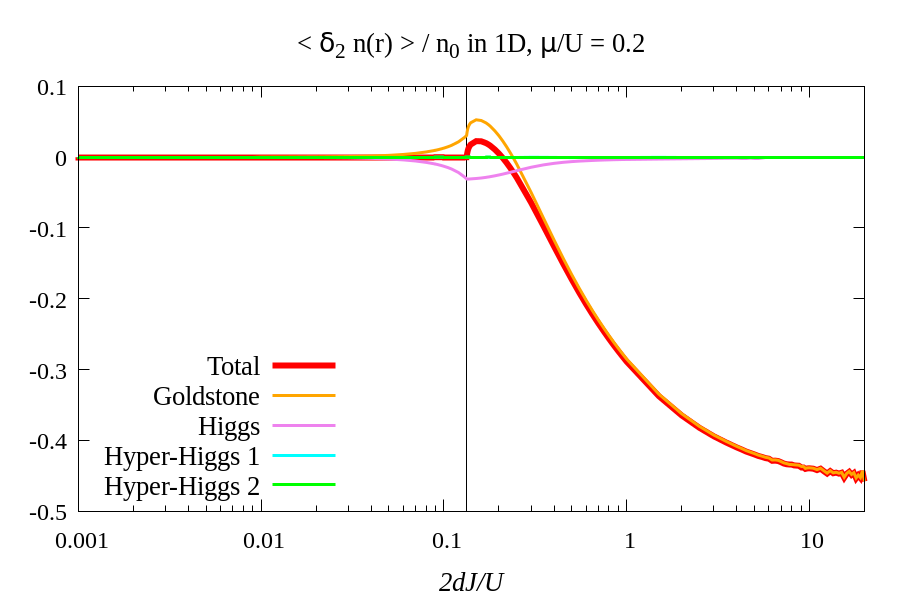}
			\caption{Second-order correction to the relative average local density $\langle \delta_2 \hat{n}{\left( \mathbf{r} \right)} \rangle/n_0$ for $\mu/U = 0.2$ in 1D. The vertical black line indicates the transition point between the Mott and the superfluid phase.}
			\label{Figure58}
		\end{minipage}
		\ \hspace{3mm} \hspace{3mm} \
		\begin{minipage}[t]{7.7cm}
			\centering
			\includegraphics[width=9cm]{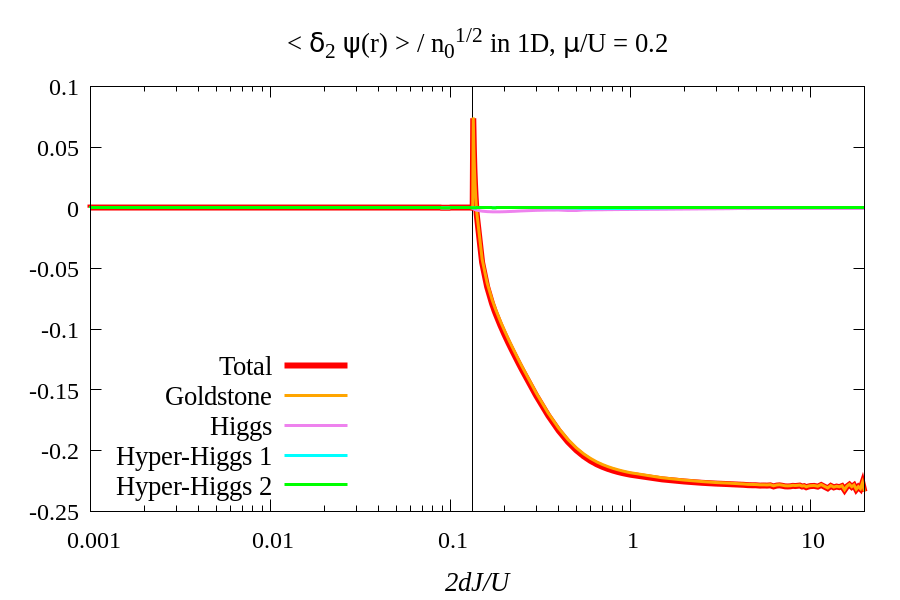}
			\caption{Second-order correction to the relative order parameter $\langle \delta_2 \hat{\psi}{\left( \mathbf{r} \right)} \rangle/\sqrt{n_0}$ for $\mu/U = 0.2$ in 1D.  The vertical black line indicates the transition point between the Mott and the superfluid phase.}
			\label{Figure59}
		\end{minipage}
	\end{figure}
	\item $\mu/U = \sqrt{2} - 1$ (across the tip of the first Mott lobe)
	\begin{figure}[!htp]
		\begin{minipage}[t]{7.7cm}
			\centering
			\includegraphics[width=9cm]{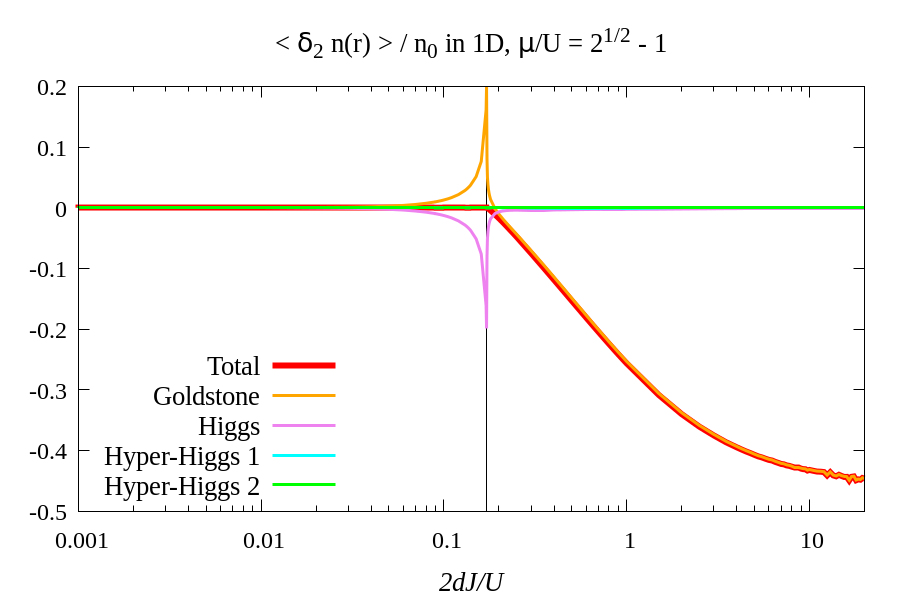}
			\caption{Second-order correction to the relative average local density $\langle \delta_2 \hat{n}{\left( \mathbf{r} \right)} \rangle/n_0$ for $\mu/U = \sqrt{2} - 1$ in 1D. The vertical black line indicates the transition point between the Mott and the superfluid phase.}
			\label{Figure60}
		\end{minipage}
		\ \hspace{3mm} \hspace{3mm} \
		\begin{minipage}[t]{7.7cm}
			\centering
			\includegraphics[width=9cm]{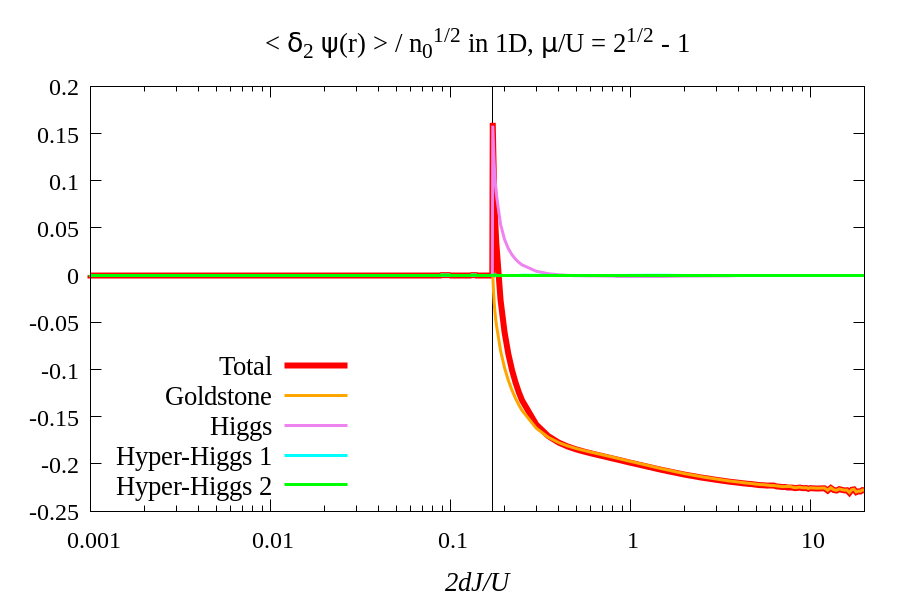}
			\caption{Second-order correction to the relative order parameter $\langle \delta_2 \hat{\psi}{\left( \mathbf{r} \right)} \rangle/\sqrt{n_0}$ for $\mu/U = \sqrt{2} - 1$ in 1D.  The vertical black line indicates the transition point between the Mott and the superfluid phase.}
			\label{Figure61}
		\end{minipage}
	\end{figure}
	\newpage
	\item $\mu/U = 0.8$
	\begin{figure}[!htp]
		\begin{minipage}[t]{7.7cm}
			\centering
			\includegraphics[width=9cm]{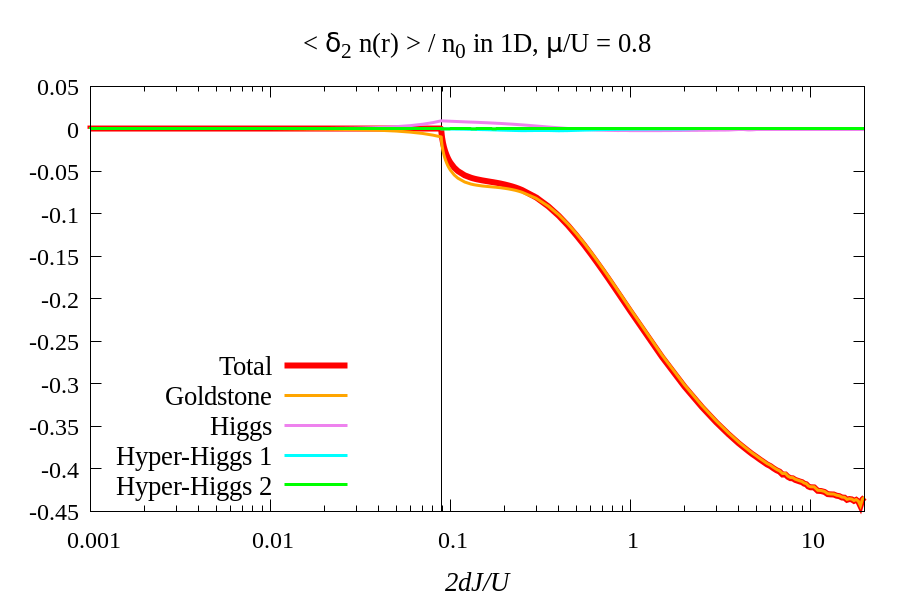}
			\caption{Second-order correction to the relative average local density $\langle \delta_2 \hat{n}{\left( \mathbf{r} \right)} \rangle/n_0$ for $\mu/U = 0.8$ in 1D. The vertical black line indicates the transition point between the Mott and the superfluid phase.}
			\label{Figure62}
		\end{minipage}
		\ \hspace{3mm} \hspace{3mm} \
		\begin{minipage}[t]{7.7cm}
			\centering
			\includegraphics[width=9cm]{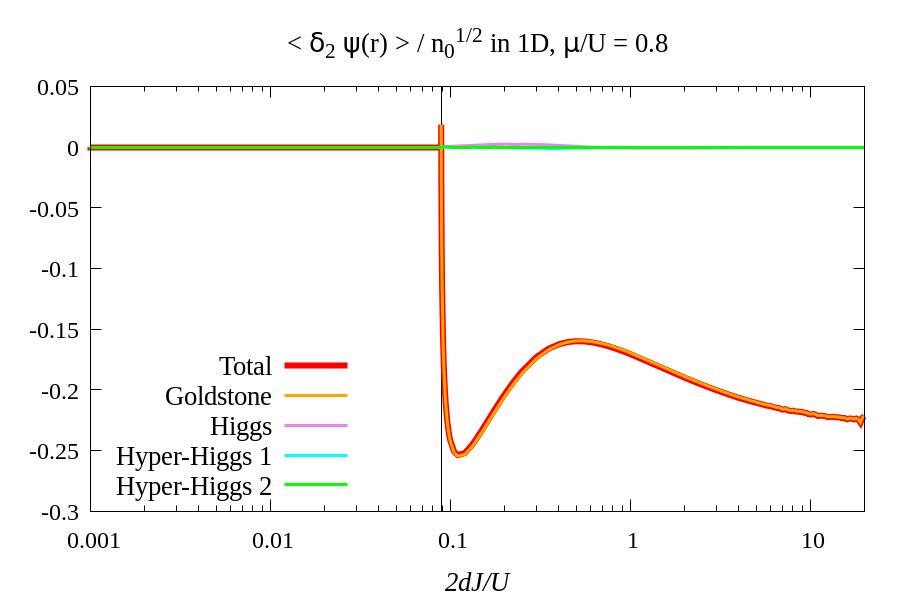}
			\caption{Second-order correction to the relative order parameter $\langle \delta_2 \hat{\psi}{\left( \mathbf{r} \right)} \rangle/\sqrt{n_0}$ for $\mu/U = 0.8$ in 1D.  The vertical black line indicates the transition point between the Mott and the superfluid phase.}
			\label{Figure63}
		\end{minipage}
	\end{figure}
\end{listamia}
\subsubsection{Mott phase}
As one would have expected from the very beginning, the density and order parameter corrections vanish identically in the Mott phase, confirming that the \textsl{Mottness} (Mott physics) is robust against quantum fluctuations. This result can be well synthesised conceptually when considering the equation for the corrections of the density channel at $T = 0$:
\begin{equation}\label{4.86}
\langle \delta_2 \hat{n}{\left( \mathbf{r} \right)} \rangle = \frac{1}{M} \sum_n \left( n - n_0 \right) \sum_{\alpha > 0} \sum_{\mathbf{k}} \left| v_{\alpha, \mathbf{k}, n} \right|^2 = 0
\end{equation}
which can be reorganised in the form:
\begin{equation}\label{4.87}
n_0 = \frac{\sum_n n \sum_{\alpha > 0} \sum_{\mathbf{k}} \left| v_{\alpha, \mathbf{k}, n} \right|^2}{\sum_{\alpha > 0} \sum_{\mathbf{k}} \left| v_{\alpha, \mathbf{k}, n} \right|^2}
\end{equation}
\textsl{Equation (\ref{4.87}), which strongly recalls the well-known formula providing the centre of mass of a many-body system, suggests that the Mott phase tends to conserve its mean-field density even if quantum fluctuations due to particle-hole excitations are included in the theory.} It is interesting to notice that this result depends on the interplay between $\left| v_{h, \mathbf{k}, n} \right|^2$ (hole mode) and $\left| v_{p, \mathbf{k}, n} \right|^2$ (particle mode), as their contributions to (\ref{4.86}) have opposite signs. On the other hand, the other modes ($\alpha > 2$) have quasi-hole weight $v_{\alpha, \mathbf{k}, n} = 0$, so that they do not contribute to (\ref{4.86}).

\subsubsection{Superfluid phase}
Although the Mott physics is well described by the results shown in Figures \ref{Figure58}-\ref{Figure63}, second-order corrections exhibit pathological aspects as the deep superfluid regime is approached. This result comes as no surprise, since from Figures \ref{Figure56}-\ref{Figure57} we know that second-order corrections are always small in the Mott phase. \\
Both the relative average fluctuations $\langle \delta_2 \hat{n}{\left( \mathbf{r} \right)} \rangle/n_0$ and $\langle \delta_2 \hat{\psi}{\left( \mathbf{r} \right)} \rangle/\sqrt{n_0}$ present non-trivial behaviours in the strong-interacting region and increase monotonically in the weakly-interacting limit, finally tending to a constant value. \\
\textsl{Despite the quantum corrections near criticality show interesting features, such as the infrared divergence of the order parameter near criticality due to the reduced dimensionality, we cannot completely rely on our numerical results, since a self-consistent calculation of quantum fluctuations would require to take into account the shift of the chemical potential due to the depletion effects implicitly contained in the renormalizing operator $\hat{A}$.} Moreover, the inconsistency of the results obtained for $2 d J/U \gg 1$ is proved also by the fact that in the deep superfluid regime the Gutzwiller saddle-point equations correspond to the DGPE, so that quantum corrections should be only small perturbations in this limit.

\subsection{Numerical results: 3D case, T= 0, \texorpdfstring{$10^3$}{} sites}
\begin{figure}[!htp]
	\begin{minipage}[t]{7.7cm}
		\centering
		\includegraphics[width=9cm]{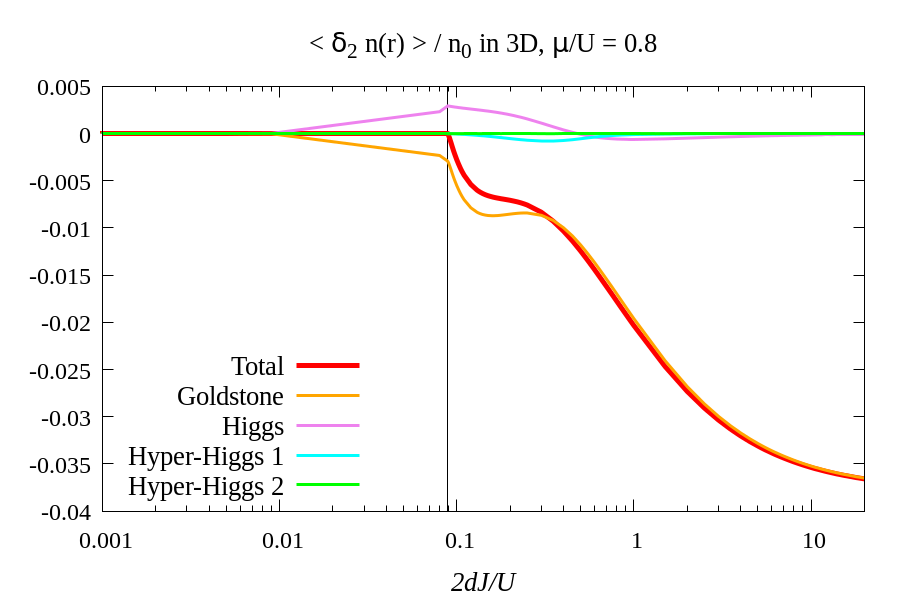}
		\caption{Second-order correction to the relative average local density $\langle \delta_2 \hat{n}{\left( \mathbf{r} \right)} \rangle/n_0$ for $\mu/U = 0.8$ in 3D. The vertical black line indicates the transition point between the Mott and the superfluid phase.}
		\label{Figure64}
	\end{minipage}
	\ \hspace{3mm} \hspace{3mm} \
	\begin{minipage}[t]{7.7cm}
		\centering
		\includegraphics[width=9cm]{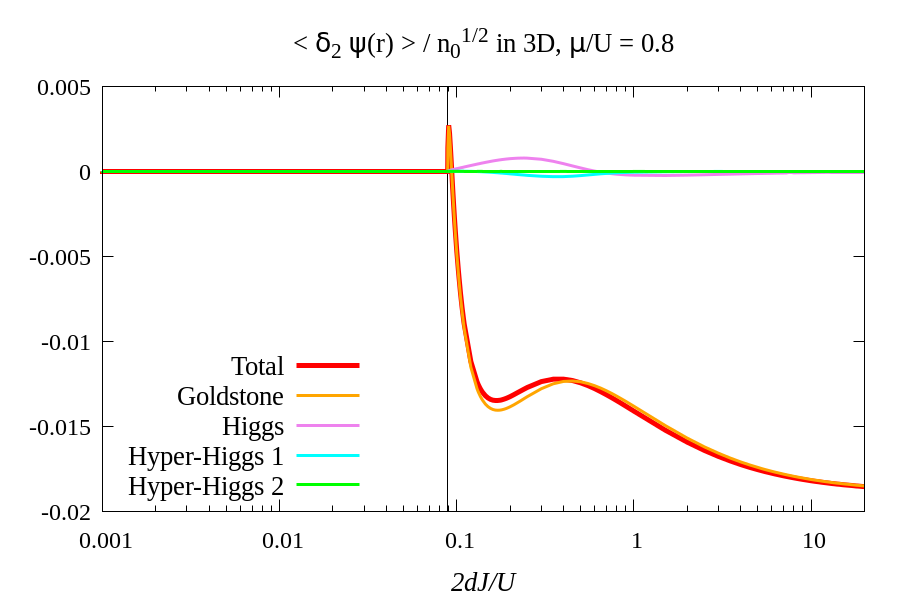}
		\caption{Second-order correction to the relative order parameter $\langle \delta_2 \hat{\psi}{\left( \mathbf{r} \right)} \rangle/\sqrt{n_0}$ for $\mu/U = 0.8$ in 3D.  The vertical black line indicates the transition point between the Mott and the superfluid phase.}
		\label{Figure65}
	\end{minipage}
\end{figure}
\begin{figure}[!htp]
	\centering
	\includegraphics[width=10cm]{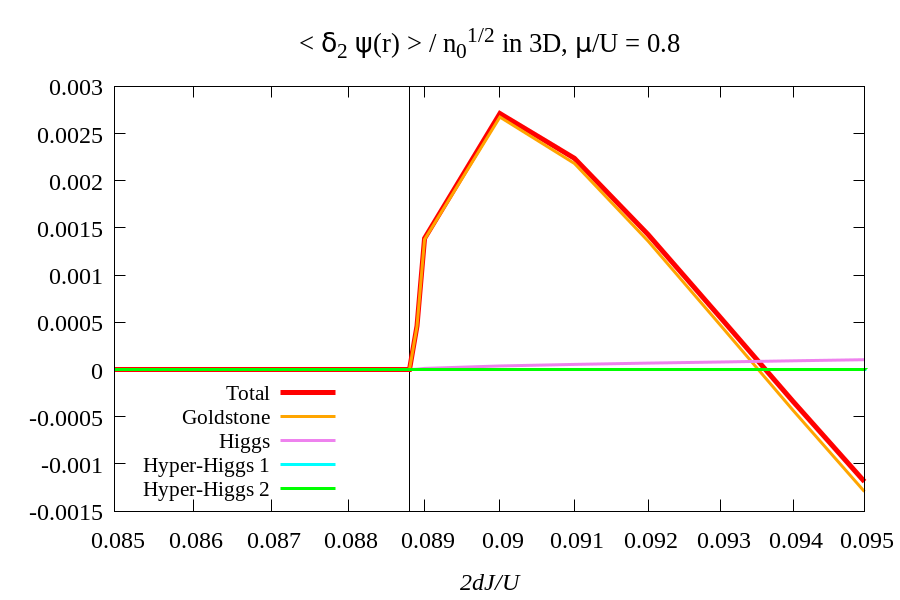}
	\caption{Detail of second-order correction to the relative order parameter $\langle \delta_2 \hat{\psi}{\left( \mathbf{r} \right)} \rangle/\sqrt{n_0}$ at criticality for $\mu/U = 0.8$ in 3D. The vertical black line indicates the transition point between the Mott and the superfluid phase.}
	\label{Figure66}
\end{figure}

Comparing Figures \ref{Figure64} and \ref{Figure65} with Figures \ref{Figure60} and \ref{Figure61}, one can verify that there are no qualitative differences between the 1D and 3D corrections. On the contrary, the order of magnitude of $\langle \delta_2 \hat{n}{\left( \mathbf{r} \right)} \rangle$ and $\langle \delta_2 \hat{\psi}{\left( \mathbf{r} \right)} \rangle$ decreases significantly, in accordance with the expectation that the Gutzwiller approximation improves as the dimensionality of the system is increased. \\
Interestingly, as shown in Figure \ref{Figure66}, the infrared divergence of the order parameter correction at criticality is effectively cured by the change of dimensionality.
\newpage

\newpage

\markboth{\MakeUppercase{5. Quantum fluctuations}}{\MakeUppercase{5.4 Discussion: quantizing Gutzwiller's ansatz}}
\section{Discussion: Gutzwiller observables and normalization}\label{quantization_discussion}
\markboth{\MakeUppercase{5. Quantum fluctuations}}{\MakeUppercase{5.4 Discussion: quantizing Gutzwiller's ansatz}}
The numerical calculations based on the quantized Gutzwiller theory have shown that the quantum depletion of the condensate can be calculated starting from the Bose field operator $\hat{\psi}{\left( \mathbf{r} \right)}$ in analogy with the Bogoliubov approximation in the weakly-interacting limit. \\
It has been also demonstrated that the Gutzwiller prediction for the quantum depletion is able to catch qualitatively the expected behaviour of the local density when approaching the Mott-superfluid transition. \textsl{On the other hand, we cannot rely completely on the Gutzwiller results near the critical region, as higher-order contributions have been neglected and there is a significant relative discrepancy between $\psi^2_0 + D_g$ and $n_0$, that is the saddle-point density directly provided by the Gutzwiller ansatz as additional information.} \uline{In this sense, we can only conclude that our Gutzwiller model generalizes the Bogoliubov approximation in describing quantum fluctuations over the whole phase diagram and gives new insights into the physics of strong interactions.} \\
We can guess that a good check of the depletion prediction (\ref{4.31}) would be to calculate the average value of the Gutzwiller density operator $\langle \hat{n}{\left( \mathbf{r} \right)} \rangle$ up to the second order in the fluctuation operators. Actually, as underlined in Section \ref{second_order_corrections}, a correct calculation of $\langle \delta_2 \hat{n}{\left( \mathbf{r} \right)} \rangle$ should formally involve a self-consistent treatment of the quantization problem, that is not in the ultimate purpose of this work. \\
Nevertheless, similar problems concerning the normalization of the local density when adding quantum fluctuations has been already noticed by Menotti and Trivedi \cite{menotti_trivedi} in the context of the RPA approximation, which turns to be quite compatible with the Gutzwiller mean-field predictions.

\medskip

Moreover, \uline{it is important to notice that in principle the identification of $\hat{n}{\left( \mathbf{r} \right)}$ with $\hat{\psi}^{\dagger}{\left( \mathbf{r} \right)} \, \hat{\psi}{\left( \mathbf{r} \right)}$ is not valid for every point of the phase diagram within our quantum theory, as demonstrated \textit{a posteriori} by calculating the quantum depletion $D_g$}. \\
\textsl{A number of evidences, such as the calculation of the response of the density to a modulation of the on-site interaction (see Section \ref{response_density_interaction} in the following Chapter), seem to suggest that in some contexts the correct recipe for quantizing observables on top of the Gutzwiller mean-field theory is to recover their expressions in the original second quantization formalism as a whole, calculate the corresponding expectation values over the Gutzwiller wave function and promote the parameters $c_n{\left( \mathbf{r} \right)}$ to operators according to the quantization procedure. This prescription reveals to be important when in second quantization a given operator is composed by a combination of two or more sub-operators acting on the same lattice position.} \\
This reasoning obviously does not apply if one calculates the expectation value of two or more operators defined over different lattice sites. As the Gutzwiller wave function ansatz is factorized with respect to the particle positions in real space, the operators $\delta \hat{c}_n{\left( \mathbf{r} \right)}$ commute for different lattice sites, so that the Bose field and density operators are well-defined with respect to their commutation relations. \\
It follows that one has to be careful assuming the value of the expectation values of multiple operators for coincident points as good results, for example in determining the relevant correlation functions.
However, \uline{as far as the quantum depletion data are concerned, the gap between $\hat{n}{\left( \mathbf{r} \right)}$ and $\hat{\psi}^{\dagger}{\left( \mathbf{r} \right)} \, \hat{\psi}{\left( \mathbf{r} \right)}$ appears to decrease as the dimensionality of the system is raised. Therefore, we cannot also exclude that it is a residual of the use of the Gutzwiller approximation, whose mean-field predictions are known to be exact in the limit of infinite dimensions.}

\newpage
\section{Quantum correlation functions}\label{correlations}
\subsection{Two-point one-body correlation functions}
As we are always interested in Gaussian fluctuations due to the quadratic quantization performed in (\ref{422}), we aim at estimating the simplest relevant quantum correlation functions up to the second order in the quasi-particle operators. \uline{Unexpectedly, a quantum theory derived from the Gutzwiller ansatz, which neglects spatial correlations between different sites, leads to non-vanishing correlation functions at non-zero separation distances.} \\
Firstly, let us consider the normalized one-body correlation function given by:
\begin{equation}\label{4.45}
g_1{\left( \mathbf{r}, \mathbf{s} \right)} = \frac{\langle \hat{\psi}^{\dagger}{\left( \mathbf{r} \right)} \, \hat{\psi}{\left( \mathbf{s} \right)} \rangle}{\langle \hat{n}{\left( \mathbf{r} \right)} \rangle}
\end{equation}
The quadratic expansion of (\ref{4.45}) involves the evaluation of second-order average values of the fluctuation operators in analogy with the calculation of the quantum depletion and the superfluid fraction. In particular, the numerator of (\ref{4.45}) reads:
\begin{equation}\label{4.46}
\langle \hat{\psi}^{\dagger}{\left( \mathbf{r} \right)} \, \hat{\psi}{\left( \mathbf{s} \right)} \rangle =  \frac{1}{M} \sum_{\alpha} \sum_{\mathbf{k}} \left[ \left( \left| U_{\alpha, \mathbf{k}} \right|^2 + \left| V_{\alpha, \mathbf{k}} \right|^2 \right) \langle \hat{b}^{\dagger}_{\alpha, \mathbf{k}} \,\hat{b}_{\alpha, \mathbf{k}} \rangle + \left| V_{\alpha, \mathbf{k}} \right|^2 \right] \cos{\left[ \mathbf{k} \cdot \left( \mathbf{r} - \mathbf{s} \right) \right]}
\end{equation}
On the other hand, the expectation value of the density operator $\langle \hat{n}{\left( \mathbf{r} \right)} \rangle$, which fixes the normalization of (\ref{4.45}), can be chosen to be given by $\langle \hat{\psi}^{\dagger}{\left( \mathbf{r} \right)} \, \hat{\psi}{\left( \mathbf{r} \right)} \rangle$. A second option would consist in making the substitution $\langle \hat{n}{\left( \mathbf{r} \right)} \rangle \approx n_0$ compatibly with the orders treated by our theory, but this choice is not guaranteed to satisfy the inequality $g_1{\left( \mathbf{r}, \mathbf{s} \right)} \le g_1{\left( \mathbf{r}, \mathbf{r} \right)} = 1$, according to the digression opened after discussing the numerical results of the quantum depletion (Section \ref{quantization_discussion}) and the discussion of Section \ref{quantization_discussion}. Moreover, the normalization choice $\langle \hat{\psi}^{\dagger}{\left( \mathbf{r} \right)} \, \hat{\psi}{\left( \mathbf{r} \right)} \rangle = \psi^2_0 + D_g$ is justified by the fact that second-order quantum corrections to the density average value $\langle \delta_2 \hat{n}{\left( \mathbf{r} \right)} \rangle$ have been shown to present problematic behaviours, as previously shown in Section \ref{second_order_corrections}.

\subsubsection{1D case, \texorpdfstring{$10^2$}{} sites, T = 0}
\begin{figure}[!htp]
	\begin{minipage}[t]{8.0cm}
		\centering
		\includegraphics[width=9cm]{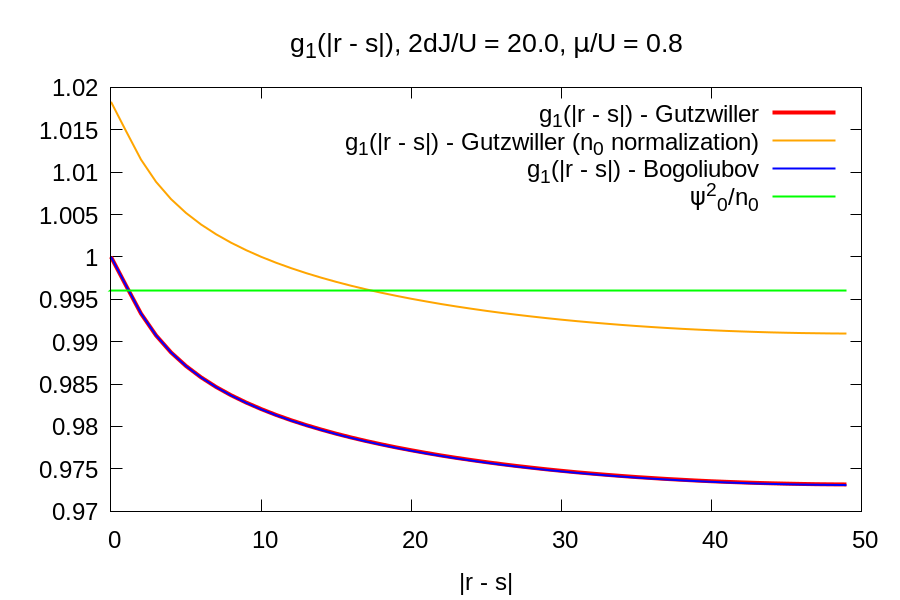}
		\caption{Two-point correlation function $g_1{\left( \mathbf{r}, \mathbf{s} \right)}$ compared with the weakly-interacting Bogoliubov prediction in the deep superfluid regime for $2 d J/U = 20$ in 1D. The red line indicates the quantity (\ref{4.45}) naturally normalized by $N = \psi^2_0 + D_g$, while the orange one refers to the correlation function normalized by the mean-field density $n_0$.}
		\label{Figure27}
	\end{minipage}
	\ \hspace{2mm} \hspace{2mm} \
	\begin{minipage}[t]{8.0cm}
		\centering
		\includegraphics[width=9cm]{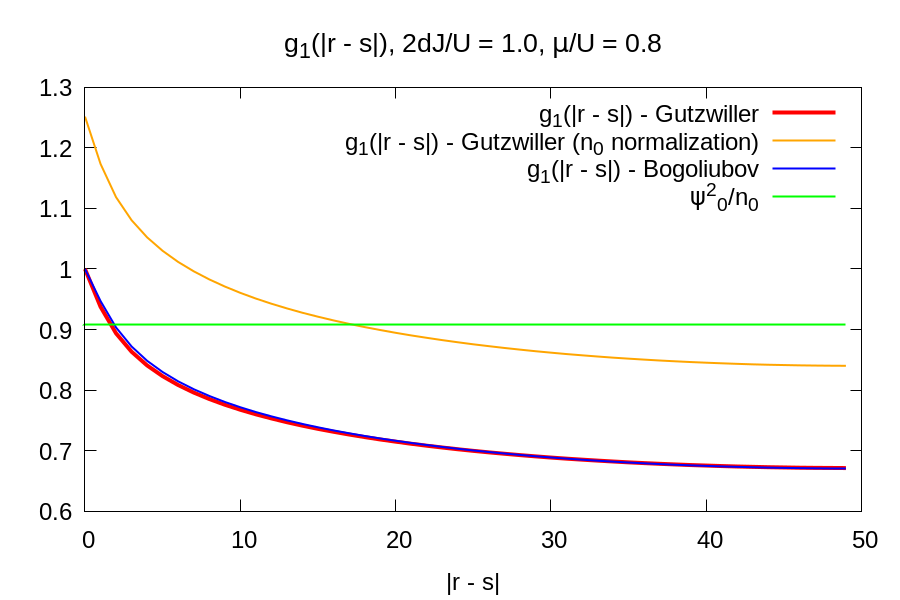}
		\caption{Two-point correlation function $g_1{\left( \mathbf{r}, \mathbf{s} \right)}$ compared with the tentative Bogoliubov prediction for $2 d J/U = 1$ in 1D. The red line indicates the quantity (\ref{4.45}) naturally normalized by $N = \psi^2_0 + D_g$, while the orange one refers to the correlation function normalized by the mean-field density $n_0$.}
		\label{Figure28}
	\end{minipage}
\end{figure}
A first evident result conveyed by Figures \ref{Figure27} and \ref{Figure28} is the expected match between the Gutzwiller prediction in the deep superfluid regime and the corresponding result based on the Bogoliubov approximation, which also turns out to be quite reliable when the hopping energy and the on-site interaction have comparable amplitudes, as quantum correlations are dominated by the low-energy collective excitations. \\
As suggested by the analysis of the quantum depletion data, the order of the gap between the asymptotic value of $g_1{\left( \mathbf{r}, \mathbf{s} \right)}$ and the mean-field condensate fraction $\rho^{MF}_s = \psi^2_0/n_0$ decreases as the amplitude quantum corrections is reduced towards the weakly-interacting limit. Obviously, the asymptotic value of the correlation function normalized by the mean-field density $n_0$ is closer to the $\rho^{MF}_s$, but the normalization rule $g_1{\left( \mathbf{r}, \mathbf{s} \right)} \le 1$ is not satisfied in this case.
\begin{figure}[!htp]
	\begin{minipage}[t]{8.0cm}
		\centering
		\includegraphics[width=9cm]{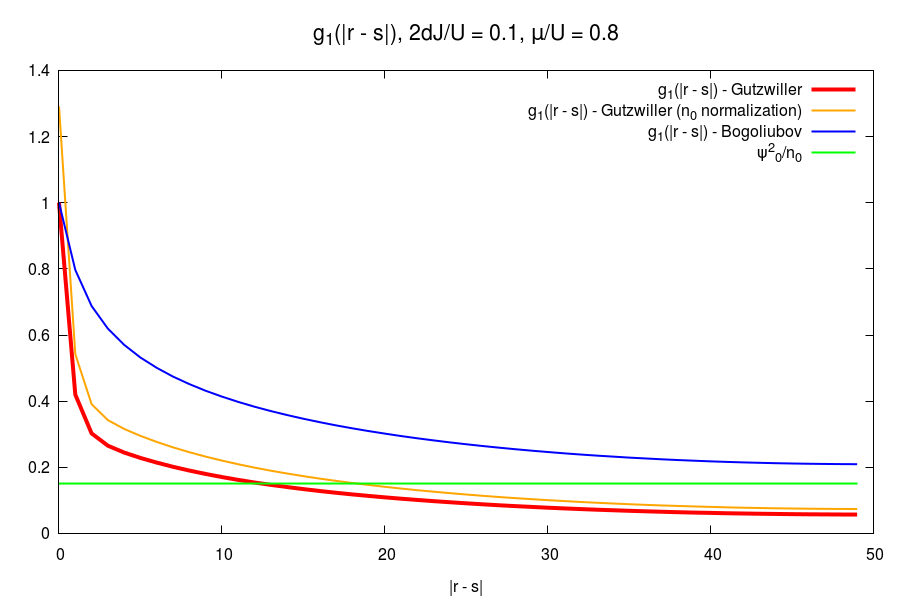}
		\caption{Two-point correlation function $g_1{\left( \mathbf{r}, \mathbf{s} \right)}$ compared with the tentative Bogoliubov prediction in the strongly-interacting superfluid regime for $2 d J/U = 0.1$ in 1D. The red line indicates the quantity (\ref{4.45}) naturally normalized by $N = \psi^2_0 + D_g$, while the orange one refers to the correlation function normalized by the mean-field density $n_0$.}
		\label{Figure29}
	\end{minipage}
	\ \hspace{2mm} \hspace{2mm} \
	\begin{minipage}[t]{8.0cm}
		\centering
		\includegraphics[width=9cm]{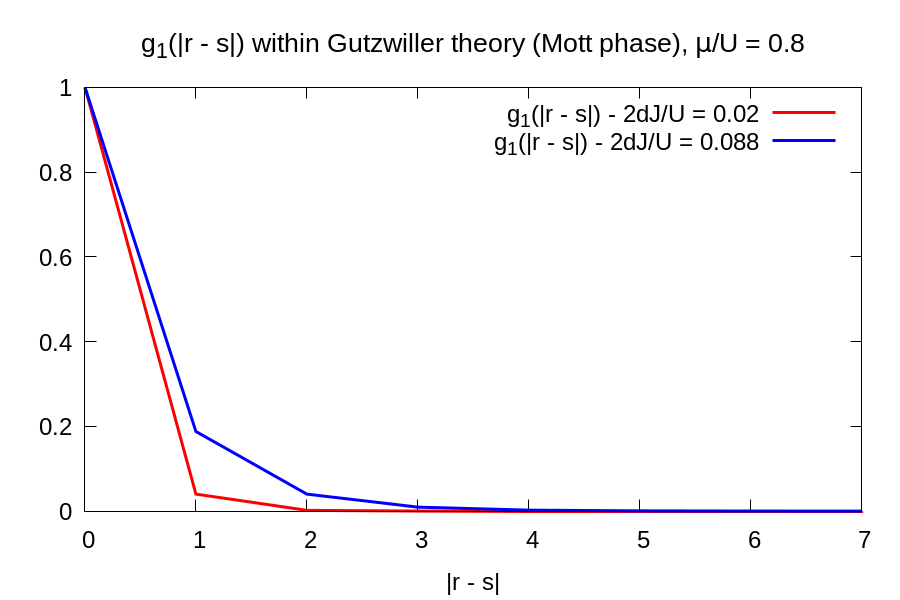}
		\caption{Two-point correlation function $g_1{\left( \mathbf{r}, \mathbf{s} \right)}$ in the Mott phase for $2 d J/U = 0.02$ (red line) and $2 d J/U = 0.088$ (blue line) in 1D.}
		\label{Figure30}
	\end{minipage}
\end{figure} \\
Approaching the strongly-interacting regime (see Figure \ref{Figure29}) the Bogoliubov approximation overestimates spatial correlations, while the normalizations fixed by mean-field theory and quantum depletion tend to return the same asymptotic value. \\
\textsl{Interesting features emerge also in the Mott phase.} \uline{As a consequence of the fact that the amplitudes $U_{\alpha, \mathbf{k}}$ and $V_{\alpha, \mathbf{k}}$ do not vanish in the particle and hole channels, we obtain finite one-body correlation functions for short distances. Noticeably, as in the case of the RPA approximation \cite{menotti_trivedi}, the Gutzwiller quantum theory is able to quantify the actual dimension of the particle-hole couple and then predict the increase of the correlation length in the vicinity of the critical transition} (see Figure \ref{Figure30}).
\begin{figure}[!htp]
	\begin{minipage}[t]{7.4cm}
		\centering
		\includegraphics[width=9cm]{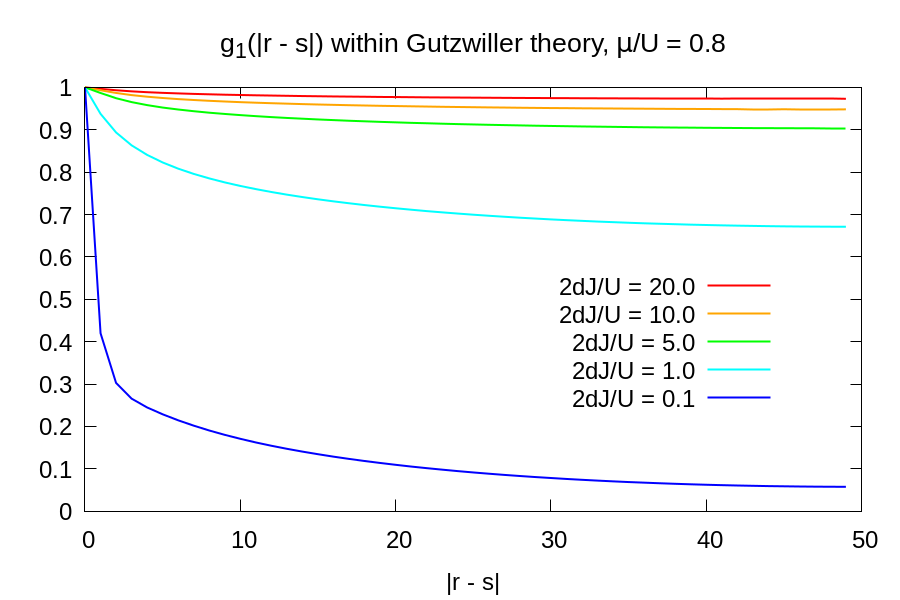}
		\caption{Two-point correlation function $g_1{\left( \mathbf{r}, \mathbf{s} \right)}$ according to the Gutzwiller prediction for different values of the scaled hopping strength $2 d J/U$ in 1D.}
		\label{Figure31}
	\end{minipage}
	\ \hspace{7mm} \hspace{7mm} \
	\begin{minipage}[t]{8.5cm}
		\centering
		\includegraphics[width=9cm]{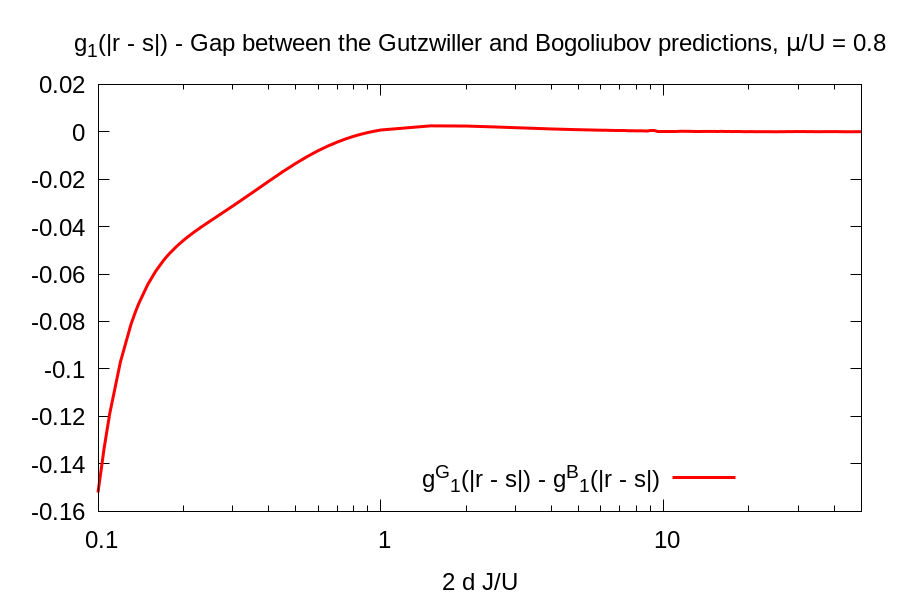}
		\caption{Plot of the difference between the asymptotic value of the two-point correlation function $g^G_1{\left( \left| \mathbf{r} - \mathbf{s} \right| \to \infty \right)}$ within the Gutzwiller theory and the corresponding Bogoliubov-based prediction as a function of $2 d J/U$ in 1D.}
		\label{Figure32}
	\end{minipage}
\end{figure} \\
Regarding the behaviour of the correlation length in the parameter space of the theory, Figure \ref{Figure30} show that in \textsl{the superfluid phase $g_1{\left( \mathbf{r}, \mathbf{s} \right)}$ is a decaying function whose decorrelation length increases with the relative hopping strength $2 d J/U$}.

\vspace{1cm}

\subsubsection{3D case, \texorpdfstring{$8 \times 10^3$}{} sites, T = 0}
For completeness, in the following we report also the numerical results for the one-body correlation function $g_1{\left( \mathbf{r}, \mathbf{s} \right)}$ in the case of a 3D system. Since the isocorrelation curves do not necessarily coincide with spherical surfaces, spatial correlations have been evaluated along an arbitrary lattice direction for simplifying the calculations.

\medskip

\begin{figure}[!htp]
	\begin{minipage}[t]{8.2cm}
		\centering
		\includegraphics[width=9cm]{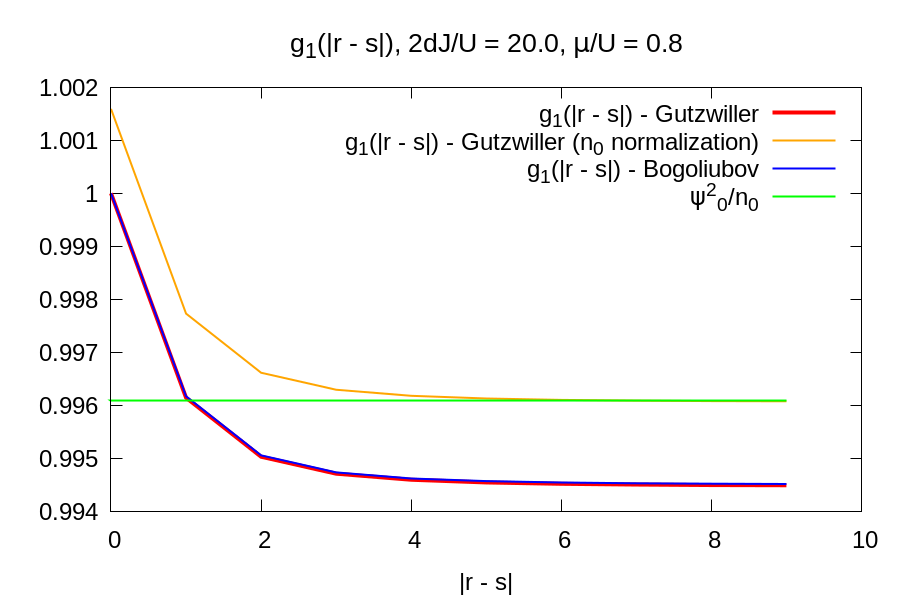}
		\caption{Two-point correlation function $g_1{\left( \mathbf{r}, \mathbf{s} \right)}$ compared with the weakly-interacting Bogoliubov prediction in the deep superfluid regime for $2 d J/U = 20$ in 3D. The red line indicates the quantity (\ref{4.45}) naturally normalized by $N = \psi^2_0 + D_g$, while the orange refers to the correlation function normalized by the mean-field density $n_0$.}
		\label{Figure33}
	\end{minipage}
	\ \hspace{3mm} \hspace{3mm} \
	\begin{minipage}[t]{7.8cm}
		\centering
		\includegraphics[width=9cm]{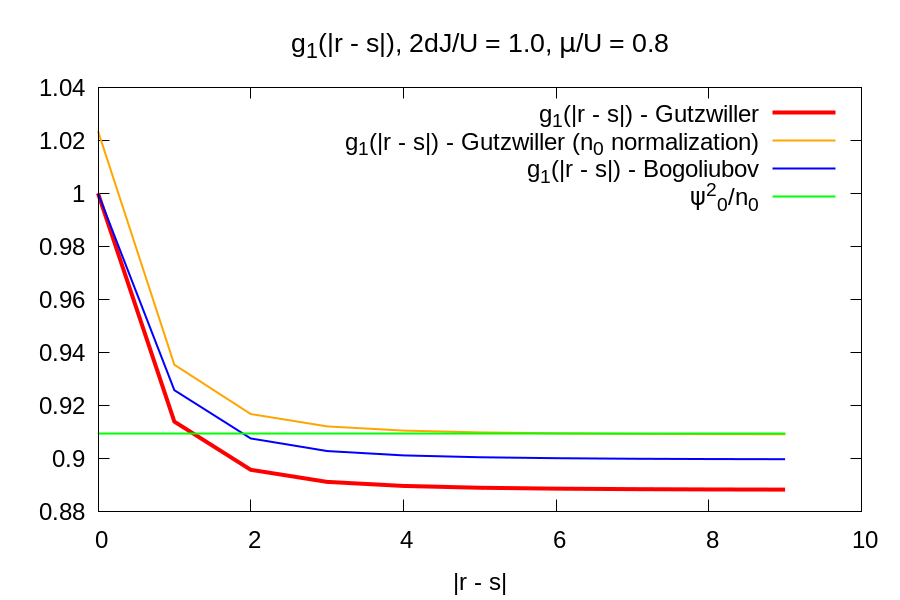}
		\caption{Two-point correlation function $g_1{\left( \mathbf{r}, \mathbf{s} \right)}$ compared with the tentative Bogoliubov prediction for $2 d J/U = 1$ in 3D. The red line indicates the quantity (\ref{4.45}) naturally normalized by $N = \psi^2_0 + D_g$, while the orange refers to the correlation function normalized by the mean-field density $n_0$.}
		\label{Figure34}
	\end{minipage}
\end{figure}

\begin{figure}[!htp]
	\begin{minipage}[t]{8.0cm}
		\centering
		\includegraphics[width=9cm]{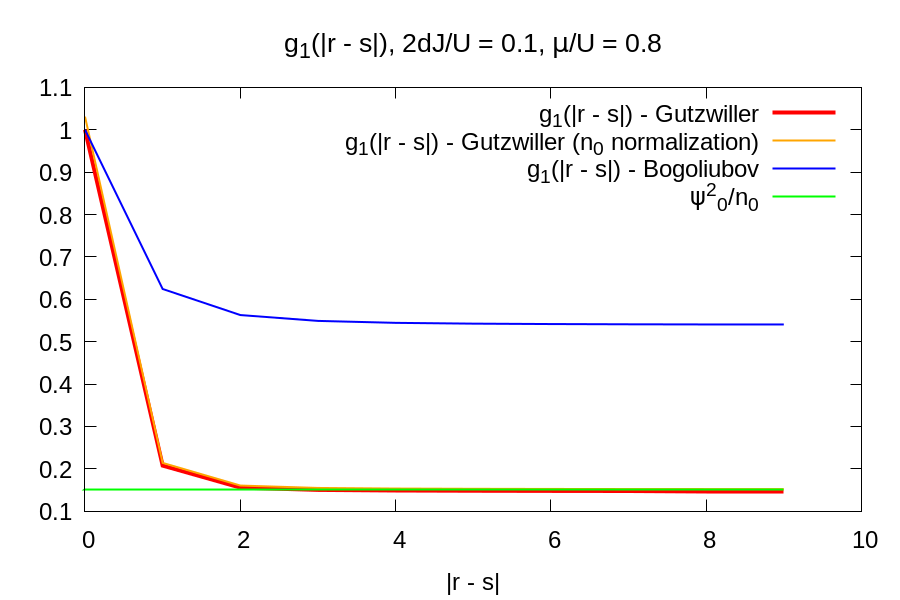}
		\caption{Two-point correlation function $g_1{\left( \mathbf{r}, \mathbf{s} \right)}$ compared with the tentative Bogoliubov prediction in the strongly-interacting superfluid regime for $2 d J/U = 0.1$ in 3D. The red line indicates the quantity (\ref{4.45}) naturally normalized by $N = \psi^2_0 + D_g$, while the orange one refers to the correlation function normalized by the mean-field density $n_0$.}
		\label{Figure35}
	\end{minipage}
	\ \hspace{3mm} \hspace{3mm} \
	\begin{minipage}[t]{8.0cm}
		\centering
		\includegraphics[width=9cm]{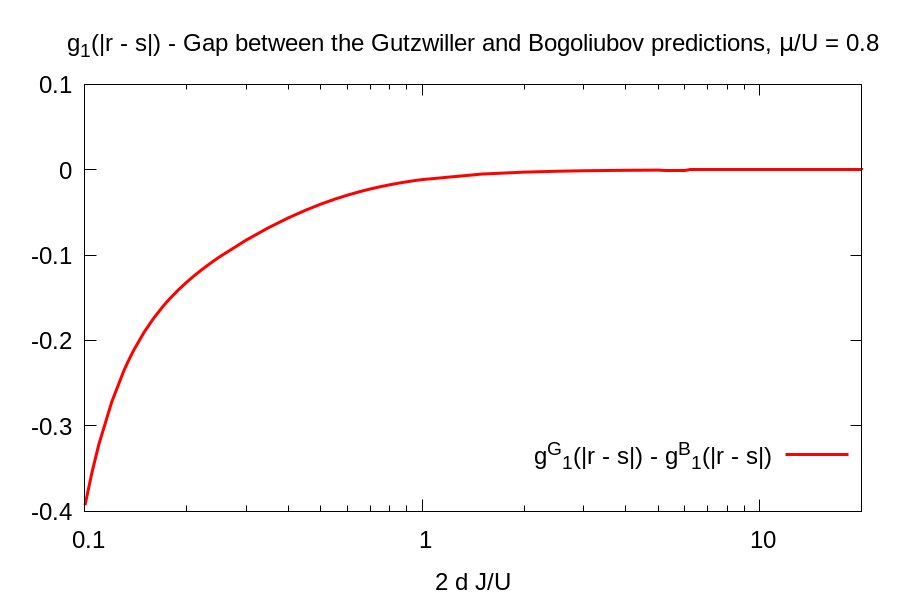}
		\caption{Plot of the difference between the asymptotic value of the two-point correlation function $g^G_1{\left( \left| \mathbf{r} - \mathbf{s} \right| \to \infty \right)}$ within the Gutzwiller theory and the corresponding Bogoliubov-based prediction as a function of $2 d J/U$ in 3D.}
		\label{Figure36}
	\end{minipage}
\end{figure}

Figures \ref{Figure33}-\ref{Figure36} present only slight qualitative variations with respect to the correlation data in the 1D case, except for peculiar aspects related to the reduced amplitude of the quantum corrections. These observations confirm the fact that at this level our Gutzwiller approach, which preserves a strong connection with the underlying mean-field theory, is not able to reproduce the peculiar features of the Bose-Hubbard model for low-dimensional systems. We expect that more refined methods, such as the cluster Gutzwiller method \cite{cluster_gutzwiller} or DMFT studies, could capture the missing fingerprints of properties of strong correlations which are expected to emerge in low dimensions.

\newpage
\subsection{Density-density correlations and pair correlation function}
A second meaningful quantity is provided by the density-density two-point correlation function, whose normalized expression is given by:
\begin{equation}\label{4.47}
g_2{\left( \mathbf{r}, \mathbf{s} \right)} = \frac{\langle \hat{n}{\left( \mathbf{r} \right)} \, \hat{n}{\left( \mathbf{s} \right)} \rangle}{\langle \hat{n}{\left( \mathbf{r} \right)} \langle \hat{n}{\left( \mathbf{s} \right)} \rangle}
\end{equation}
Our quantum theory allows to calculate the density-density correlation function (\ref{4.47}) through two different channels.
\begin{enumerate}
\item The first approach consists in expressing the density operator in terms of the quantized Bose field, so that (\ref{4.47}) has the following explicit expression:
\begin{equation}\label{4.48}
g^{\psi}_2{\left( \mathbf{r}, \mathbf{s} \right)} = \frac{\langle \hat{\psi}^{\dagger}{\left( \mathbf{r} \right)} \, \hat{\psi}{\left( \mathbf{r} \right)} \, \hat{\psi}^{\dagger}{\left( \mathbf{s} \right)} \, \hat{\psi}{\left( \mathbf{s} \right)} \rangle}{\langle \hat{\psi}^{\dagger}{\left( \mathbf{r} \right)} \, \hat{\psi}{\left( \mathbf{r} \right)} \rangle^2}
\end{equation}
which is related to the canonical normal-ordered pair correlation function by the application of a commutation relation, since:
\begin{equation}\label{4.49}
\begin{aligned}
\overline{g}^{\psi}_2{\left( \mathbf{r}, \mathbf{s} \right)} &= \frac{\langle \hat{\psi}^{\dagger}{\left( \mathbf{r} \right)} \, \hat{\psi}^{\dagger}{\left( \mathbf{s} \right)} \, \hat{\psi}{\left( \mathbf{s} \right)} \, \hat{\psi}{\left( \mathbf{r} \right)} \rangle}{\langle \hat{\psi}^{\dagger}{\left( \mathbf{r} \right)} \, \hat{\psi}{\left( \mathbf{r} \right)} \rangle^2} = \\
&= \frac{\langle \hat{\psi}^{\dagger}{\left( \mathbf{r} \right)} \, \hat{\psi}{\left( \mathbf{r} \right)} \, \hat{\psi}^{\dagger}{\left( \mathbf{s} \right)} \, \hat{\psi}{\left( \mathbf{s} \right)} \rangle - \delta{\left( \mathbf{r} - \mathbf{s} \right)} \, \langle \hat{\psi}^{\dagger}{\left( \mathbf{r} \right)} \, \hat{\psi}{\left( \mathbf{r} \right)} \rangle}{\langle \hat{\psi}^{\dagger}{\left( \mathbf{r} \right)} \, \hat{\psi}{\left( \mathbf{r} \right)} \rangle^2} = \\
&= g^{\psi}_2{\left( \mathbf{r}, \mathbf{s} \right)} - \frac{ \delta{\left( \mathbf{r} - \mathbf{s} \right)}}{\langle \hat{\psi}^{\dagger}{\left( \mathbf{r} \right)} \, \hat{\psi}{\left( \mathbf{r} \right)} \rangle}
\end{aligned}
\end{equation}
In general the expectation value of the four-field operator in the first line of (\ref{4.49}) can be evaluated by resorting to Wick's theorem. Actually, considering only Gaussian fluctuations, in a similar way we can expand the pair-correlation function in the form:
\begin{equation}\label{4.50}
\begin{aligned}
&\langle \hat{\psi}^{\dagger}{\left( \mathbf{r} \right)} \, \hat{\psi}^{\dagger}{\left( \mathbf{s} \right)} \, \hat{\psi}{\left( \mathbf{s} \right)} \, \hat{\psi}{\left( \mathbf{r} \right)} \rangle \approx \\
&\approx \bigg\langle \psi^4_0 + \psi^2_0 \, \bigg[ \delta_1 \psi^{\dagger}{\left( \mathbf{r} \right)} \, \delta_1 \psi{\left( \mathbf{r} \right)} + \delta_1 \psi^{\dagger}{\left( \mathbf{s} \right)} \, \delta_1 \psi{\left( \mathbf{s} \right)} + \delta_1 \psi^{\dagger}{\left( \mathbf{r} \right)} \, \delta_1 \psi{\left( \mathbf{s} \right)} + \delta_1 \psi^{\dagger}{\left( \mathbf{s} \right)} \, \delta_1 \psi{\left( \mathbf{r} \right)} + \\
&+ \delta_1 \psi^{\dagger}{\left( \mathbf{r} \right)} \, \delta_1 \psi^{\dagger}{\left( \mathbf{s} \right)} + \delta_1 \psi{\left( \mathbf{r} \right)} \, \delta_1 \psi{\left( \mathbf{s} \right)} \bigg] \bigg\rangle = \\
&= \psi^4_0 + 2 \, \psi^2_0 \left[ D_g + C{\left( \mathbf{r}, \mathbf{s} \right)} + A{\left( \mathbf{r}, \mathbf{s} \right)} \right]
\end{aligned}
\end{equation}
where:
\begin{equation}\label{4.51}
\begin{aligned}
&D_g = \langle \delta_1 \psi^{\dagger}{\left( \mathbf{r} \right)} \, \delta_1 \psi{\left( \mathbf{r} \right)} \rangle \\
&C{\left( \mathbf{r}, \mathbf{s} \right)} = \langle \delta_1 \psi^{\dagger}{\left( \mathbf{r} \right)} \, \delta_1 \psi{\left( \mathbf{s} \right)} \rangle = \langle \delta_1 \psi^{\dagger}{\left( \mathbf{s} \right)} \, \delta_1 \psi{\left( \mathbf{r} \right)} \rangle \\
&A{\left( \mathbf{r}, \mathbf{s} \right)} = \langle \delta_1 \psi^{\dagger}{\left( \mathbf{r} \right)} \, \delta_1 \psi^{\dagger}{\left( \mathbf{s} \right)} \rangle = \langle \delta_1 \psi{\left( \mathbf{r} \right)} \, \delta_1 \psi{\left( \mathbf{s} \right)} \rangle = \\
&= \frac{1}{M} \sum_{\alpha} \sum_{\mathbf{k}} \left( 2 \, \langle \hat{b}^{\dagger}_{\alpha, \mathbf{k}} \,\hat{b}_{\alpha, \mathbf{k}} \rangle + 1 \right)  U_{\alpha, \mathbf{k}} \, V_{\alpha, \mathbf{k}} \, \cos{\left[ \mathbf{k} \cdot \left( \mathbf{r} - \mathbf{s} \right) \right]}
\end{aligned}
\end{equation}
It is worth noticing that the anomalous averages $\langle \delta_1 \psi^{\dagger}{\left( \mathbf{r} \right)} \, \delta_1 \psi^{\dagger}{\left( \mathbf{s} \right)} \rangle$ and $\langle \delta_1 \psi{\left( \mathbf{r} \right)} \, \delta_1 \psi{\left( \mathbf{s} \right)} \rangle$ do not vanish and give a significant contribution to (\ref{4.50}).
\item A second way of calculating (\ref{4.47}) exploits directly the first-order fluctuations of the density operator defined in (\ref{4.17}). \\
As underlined before in Section \ref{quantization_discussion}, it is necessary to pay attention to the calculation of (\ref{4.47}) when the density operators are defined over coincident points.
\begin{itemize}
	\item If $\left| \mathbf{r} - \mathbf{s} \right| = 0$, in evaluating (\ref{4.47}) the double-density operator $\hat{D}{\left( \mathbf{r} \right)} \equiv \hat{n}{\left( \mathbf{r} \right)} \, \hat{n}{\left( \mathbf{r} \right)}$ has to be considered as a whole object to be expressed in the Gutzwiller quantum formalism. Its mean-field expectation value is given by the expression:
	\begin{equation}\label{4.52}
	D_0 = \sum_n n^2 \left| c^0_n \right|^2
	\end{equation}
	so that the zero-separation value of (\ref{4.47}) reads:
	\begin{equation}\label{4.53}
	g^{n}_2{\left( \mathbf{r}, \mathbf{r} \right)} = \frac{D_0}{n^2_0}
	\end{equation}
	where again second-order corrections to the mean-field averages are neglected for the reasons explained in Section \ref{second_order_corrections}.
	\item If $\left| \mathbf{r} - \mathbf{s} \right| \neq 0$ the full second-order expansion of (\ref{4.47}) is given by:
	\begin{equation}\label{4.54}
	g^n_2{\left( \mathbf{r}, \mathbf{s} \right)} = \frac{\langle \hat{n}{\left( \mathbf{r} \right)} \, \hat{n}{\left( \mathbf{s} \right)} \rangle}{\langle \hat{n}{\left( \mathbf{r} \right)} \rangle \langle \hat{n}{\left( \mathbf{s} \right)} \rangle} \approx \frac{n^2_0 + \langle \delta_1 \hat{n}{\left( \mathbf{r} \right)} \, \delta_1 \hat{n}{\left( \mathbf{s} \right)} \rangle}{n^2_0} = 1 + \frac{\langle \delta_1 \hat{n}{\left( \mathbf{r} \right)} \, \delta_1 \hat{n}{\left( \mathbf{s} \right)} \rangle}{n^2_0}
	\end{equation}
	The form of expression (\ref{4.54}) does not change if second-order local density corrections, namely $\langle \delta_2 \hat{n}{\left( \mathbf{r} \right)} \rangle$ (see Section \ref{second_order_corrections}), are included in the expansion. \\
	As a final result, recalling the exact expression (\ref{4.49}), the pair correlation function given by the average value of density operators only is given by:
	\begin{equation}\label{4.55}
	\overline{g}^n_2{\left( \mathbf{r}, \mathbf{s} \right)} \approx g^n_2{\left( \mathbf{r}, \mathbf{s} \right)} - \frac{\delta{\left( \mathbf{r} - \mathbf{s} \right)}}{n_0}
	\end{equation}
\end{itemize}
\textsl{It is important to underline that the calculation of $g^n_2{\left( \mathbf{r}, \mathbf{s} \right)}$ through the Gutzwiller density operator occurs on an equal footing with respect to the evaluation of $g_1{\left( \mathbf{r}, \mathbf{s} \right)}$, since both of these calculations are performed at the second-order level. On the contrary, an accurate calculation of $g^{\psi}_2{\left( \mathbf{r}, \mathbf{s} \right)}$ would require the inclusion of higher-order terms, which turn to be significant in the strongly-interacting regime. Therefore, we expect that $g^n_2{\left( \mathbf{r}, \mathbf{s} \right)}$ and $g^{\psi}_2{\left( \mathbf{r}, \mathbf{s} \right)}$ coincide in the deep superfluid limit, where the quantum fluctuations above the mean-field solution are small.}
\end{enumerate}

\subsubsection{3D case, \texorpdfstring{$8 \times 10^3$}{} sites, T = 0}
In the following we report the 3D numerical results for the canonical pair correlation function $\overline{g}_2{\left( \mathbf{r}, \mathbf{s} \right)}$ obtained through equations (\ref{4.49}) and (\ref{4.55}), as it describes how density varies as a function of the distance from a reference particle.
\begin{figure}[!htp]
	\begin{minipage}[t]{8.2cm}
		\centering
		\includegraphics[width=9cm]{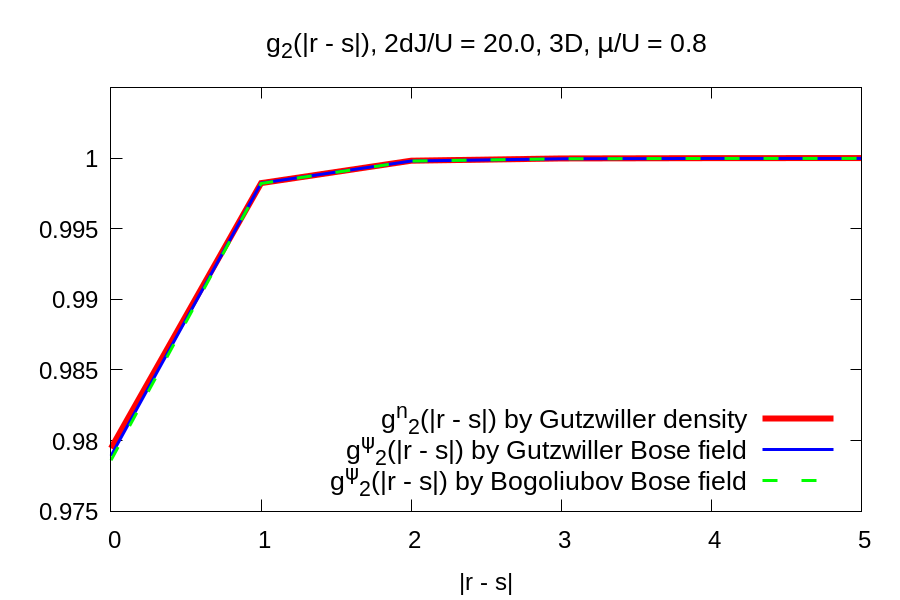}
		\caption{Comparison between the pair correlation functions $g^n_2{\left( \mathbf{r}, \mathbf{s} \right)}$ and $g^{\psi}_2{\left( \mathbf{r}, \mathbf{s} \right)}$ in the deep superfluid regime for $2 d J/U = 20$ in 3D. The blue line refers to $g^{\psi}_2{\left( \mathbf{r}, \mathbf{s} \right)}$ calculated through the Bose field extracted by the Gutzwiller theory, the green dashed line is obtained by the ordinary Bogoliubov expansion of the same object.}
		\label{Figure37}
	\end{minipage}
	\ \hspace{2mm} \hspace{2mm} \
	\begin{minipage}[t]{8cm}
		\centering
		\includegraphics[width=9cm]{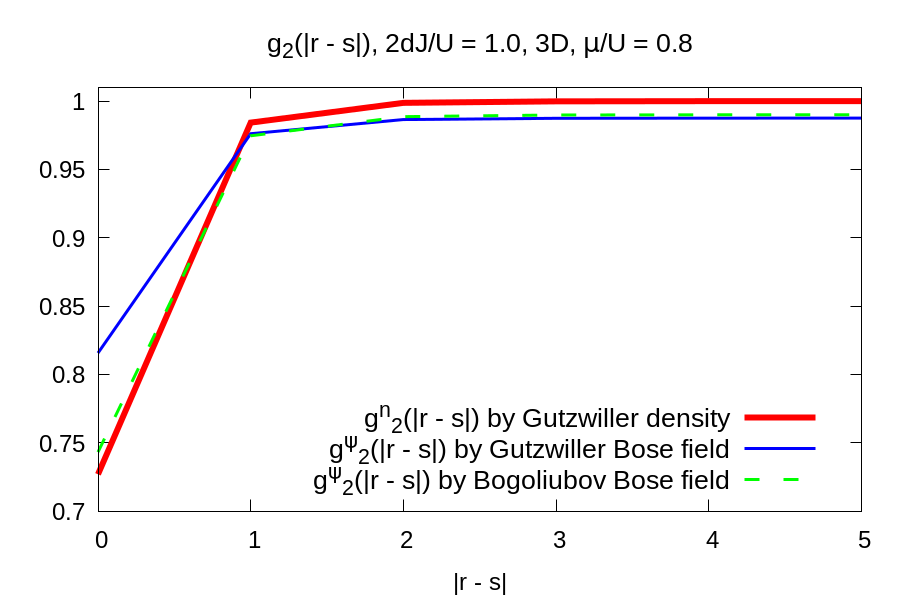}
		\caption{Comparison between the density-density correlation functions $g^n_2{\left( \mathbf{r}, \mathbf{s} \right)}$ and $g^{\psi}_2{\left( \mathbf{r}, \mathbf{s} \right)}$ in the strongly-interacting superfluid regime for $2 d J/U = 1.0$ in 3D. The blue line refers to $g^{\psi}_2{\left( \mathbf{r}, \mathbf{s} \right)}$ calculated through the Bose field extracted by the Gutzwiller theory, the green dashed line is obtained by the ordinary Bogoliubov expansion of the same object.}
		\label{Figure38}
	\end{minipage}
\end{figure} \\
As we have predicted on the basis of simple considerations, density-density correlations given by the expression of $g^{\psi}_2{\left( \mathbf{r}, \mathbf{s} \right)}$ reproduce the behaviour of $g^n_2{\left( \mathbf{r}, \mathbf{s} \right)}$ when quantum corrections represent only small variations with respect to the saddle point solution. This condition loses completely its validity while reaching the critical line separating the superfluid and the Mott phase. Considerably, Figure \ref{Figure38} displays how the discrepancy between the two approaches starts to appear at the approximate boundary between weak and strong interactions. \\
Actually, $g^n_2{\left( \mathbf{r}, \mathbf{s} \right)}$ remains the most reliable prediction for density-density correlations, as its exact expansion includes only second-order terms and its asymptotic value always converges to 1. \\
\uline{Consistently with the increasing of the interaction strength, the probability of removing and recreating a particle in the same position decreases as the critical region is approached from the superfluid side.} The effect of strong interactions can be recognised also in the decrease of the decorrelation length along the same direction.

\chapter{Susceptibilities and response functions}\label{response_functions}
\markboth{\MakeUppercase{6. Response functions}}{\MakeUppercase{6. Response functions}}
A good fraction of the experimental efforts to characterize the physics of cold atoms consists in probing the system under study by means of appropriate perturbations of physical observables. Structure factors and spectral functions are ones of the most direct piece of information that we can gain about the properties of quantum many-body systems. \\
On the other hand, easily accessible quantities could also help shed light on the nature of the excitation modes that populate the many-body energy spectrum. \\
The widely explored Bose-Hubbard mean-field theory developed by Krutitsky and Navez \cite{krutitsky_navez} has posed a number of questions about the role of modes living above the Goldstone excitation, in particular the long-standing debate about the detectability of the Higgs mode in the superfluid phase and the nature of the narrow-banded higher modes. \\
The following investigation has been precisely inspired by such questions with an eye to possible experimental applications which could give an answer to the open problems of the excitation dynamics of the Bose-Hubbard configuration. In particular, drawing on the quantum formalism developed in Chapter \ref{quantum}, we perform a more rigorous treatment of linear response of the superfluid phase with respect to \cite{krutitsky_navez} and focus on the response of the local density or related observables to different physical probes, in order to identify possible good candidates in experimental studies aiming at characterizing the spectral properties of the model discussed in this work.

\section{Density response to a density perturbation}
\markboth{\MakeUppercase{6. Response functions}}{\MakeUppercase{6.1 Density-density response}}
The simplest response function which one can think about is the density susceptibility, which can be defined as a function of momentum and frequency:
\begin{equation}\label{4.56}
\chi_{n}{\left( \mathbf{k}, \omega \right)} = -i \int dt \, e^{-i \omega t} \, e^{-\eta t} \, \theta{\left( t \right)} \sum_{\mathbf{r}} \, e^{-i \mathbf{k} \cdot \mathbf{r}} \, \big\langle \left[ \hat{n}{\left( \mathbf{r}, t \right)}, \hat{n}{\left( \mathbf{0}, 0 \right)} \right] \big\rangle
\end{equation}
where the exponential factor $e^{-\eta \, t}$ acts as an analytical regularization. Since expression (\ref{4.56}) is strictly related to the propagator of the density field, $\chi_{n}{\left( \mathbf{k}, \omega \right)}$ is essentially a measure of the compressibility of the system. \\
As response functions depend only on the commutator of two operators, the second-order expansion of (\ref{4.56}) reads a single term due to the first-order operator $\delta_1 \hat{n}{\left( \mathbf{r}, t \right)}$, so that the complete calculation of (\ref{4.56}) reads:
\begin{equation}\label{4.57}
\begin{aligned}
\chi_{n}{\left( \mathbf{k}, \omega \right)} &= -i \int dt \, e^{-i \omega t} \, e^{-\eta t} \, \theta{\left( t \right)} \sum_{\mathbf{r}} \, e^{-i \mathbf{k} \cdot \mathbf{r}} \, \big\langle \left[ \hat{n}{\left( \mathbf{r}, t \right)}, \hat{n}{\left( 0, 0 \right)} \right] \big\rangle \approx \\
&\approx -i \int dt \, e^{-i \omega t} \, e^{-\eta t} \, \theta{\left( t \right)} \sum_{\mathbf{r}} \, e^{-i \mathbf{k} \cdot \mathbf{r}} \, \big\langle \left[ \delta_1 \hat{n}{\left( \mathbf{r}, t \right)}, \delta_1 \hat{n}{\left( 0, 0 \right)} \right] \big\rangle  \approx \\
&\approx \sum_{\alpha} \bigg[ \sum_{n m} n \, m \, c^0_n \, c^0_m \big( u_{\alpha, \mathbf{k}, n} \, u_{\alpha, \mathbf{k}, m} + v_{\alpha, \mathbf{k}, n} \, v_{\alpha, \mathbf{k}, m} + u_{\alpha, \mathbf{k}, n} \, v_{\alpha, \mathbf{k}, m}  + \\
&+ u_{\alpha, \mathbf{k}, m} \, v_{\alpha, \mathbf{k}, n} \big) \frac{1}{i z - \omega_{\alpha, \mathbf{k}} + i \eta} - \sum_{n m} n \, m \, c^0_n \, c^0_m \big( u_{\alpha, \mathbf{k}, n} \, u_{\alpha, \mathbf{k}, m} + v_{\alpha, \mathbf{k}, n} \, v_{\alpha, \mathbf{k}, m} + \\
&+ u_{\alpha, \mathbf{k}, n} \, v_{\alpha, \mathbf{k}, m} + u_{\alpha, \mathbf{k}, m} \, v_{\alpha, \mathbf{k}, n} \big) \frac{1}{i z + \omega_{\alpha, \mathbf{k}} + i \eta} \bigg] = \\
&= 2 \sum_{\alpha} \frac{\omega_{\alpha, \mathbf{k}} \sum_{n m} n \, m \, c^0_n \, c^0_m \left( u_{\alpha, \mathbf{k}, n} + v_{\alpha, \mathbf{k}, n} \right) \left( u_{\alpha, \mathbf{k}, m} + v_{\alpha, \mathbf{k}, m} \right)}{\left( i z + i \eta \right)^2 - \omega^2_{\alpha, \mathbf{k}}}
\end{aligned}
\end{equation}
hence, considering the limit $\eta \to 0$ and interpreting $z = i \, \omega$ as a generalized complex frequency, we obtain the final result:
\begin{equation}\label{4.58}
\begin{aligned}
\chi_{n}{\left( \mathbf{k}, \omega \right)} = 2 \sum_{\alpha} \frac{N^2_{\alpha, \mathbf{k}} \, \omega_{\alpha, \mathbf{k}}}{\omega^2 - \omega^2_{\alpha, \mathbf{k}}}
\end{aligned}
\end{equation}
where $z = i \, \omega$ is defined to be a generalized complex frequency. The expression (\ref{4.57}) is directly calculated from the following identity:
\begin{equation}\label{4.59}
C{\left( \mathbf{r}, t \right)} \equiv \left[ \delta_1 \hat{n}{\left( \mathbf{r}, t \right)}, \delta_1 \hat{n}{\left( \mathbf{0}, 0 \right)} \right] = \frac{2 i}{M} \sum_{\alpha} \sum_{\mathbf{k}} N^2_{\alpha, \mathbf{k}} \, \sin{\left( \mathbf{k} \cdot \mathbf{r} - \omega_{\alpha, \mathbf{k}} t \right)}
\end{equation}
The response amplitude attributed to the $\alpha$-mode excitation is quantified by:
\begin{equation} \label{4.60}
\chi^n_{\alpha, \mathbf{k}} \equiv N^2_{\alpha, \mathbf{k}}
\end{equation}
so that we can distinguish the contribution of different modes to the compressibility.

\vspace{0.7cm}

\fbox{\begin{minipage}{16.5cm}
\textbf{\textcolor{blue}{Observation} -} Interestingly, the result (\ref{4.58}) coincide with the expression found by Krutitsky and Navez at the mean-field level within the Gutzwiller framework \cite{krutitsky_navez}. In the following, \uline{we will discover that all the response functions calculated within the actual Gaussian quantum theory coincide with the corresponding mean-field expression in the context of the classical linear-response formalism}. \textsl{This result is not surprising, as the second-order expansions of the form (\ref{4.56}) involve only linear-order perturbations in the Gutzwiller operators.}
\end{minipage}}
\begin{figure}[!htp]
	\begin{minipage}[b]{7.7cm}
		\centering
		\includegraphics[width=9cm]{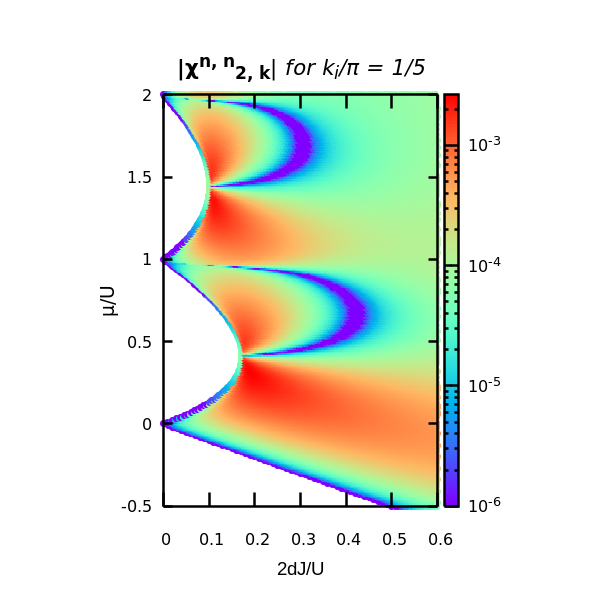}
		\caption{Amplitude of the density-density response function $\left| \chi^{n n}_{2, \mathbf{k}} \right|$ associated with the Higgs mode at momentum $k_x = k_y = k_z = \pi/5$.}
		\label{Figure39}
	\end{minipage}
	\ \hspace{3mm} \hspace{3mm} \
	\begin{minipage}[b]{7.7cm}
		\centering
		\includegraphics[width=9cm]{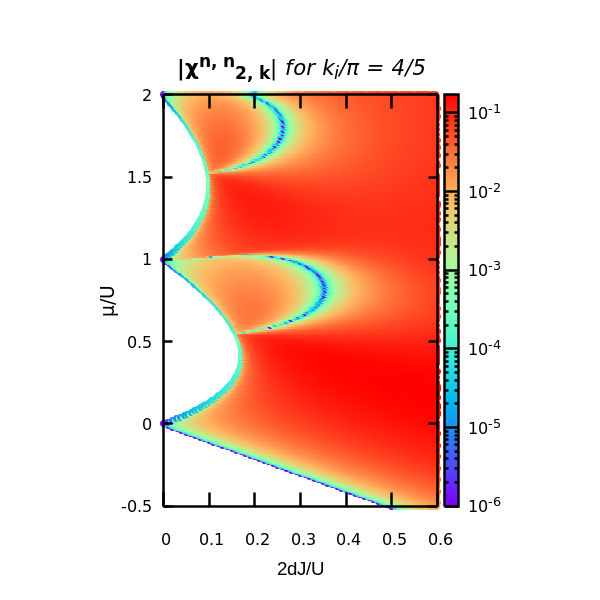}
		\caption{Amplitude of the density-density response function $\left| \chi^{n n}_{2, \mathbf{k}} \right|$ associated with the Higgs mode at momentum $k_x = k_y = k_z = 4 \pi/5$.}
		\label{Figure40}
	\end{minipage}
\end{figure}

\vspace{0.7cm}

According to the results concerning particle-hole symmetry presented in \cite{BEC}, the density-density response due to the Higgs mode is suppressed along arc-shaped lines which connect the boundary of the Mott lobes to the upper hard-core limits of the model and, in particular, depart from the Lorentz-invariant points of the critical line at low momentum (see Figure \ref{Figure39}). \uline{This fact confirms the possibility of tracking the action of the Higgs mode by analysing the response of the system to density perturbations in suitable experiments}, for example through techniques involving Bragg scattering. \\
As pointed out while commenting the behaviour of the particle-hole amplitudes $U_{\alpha, \mathbf{k}}$ and $V_{\alpha, \mathbf{k}}$, in the regime of strong interactions the response of the Higgs and higher modes is comparable to the phononic one near the border of the Brillouin zone (see Figure \ref{Figure40}).
\begin{figure}[!htp]
	\begin{minipage}[b]{7.7cm}
		\centering
		\includegraphics[width=9cm]{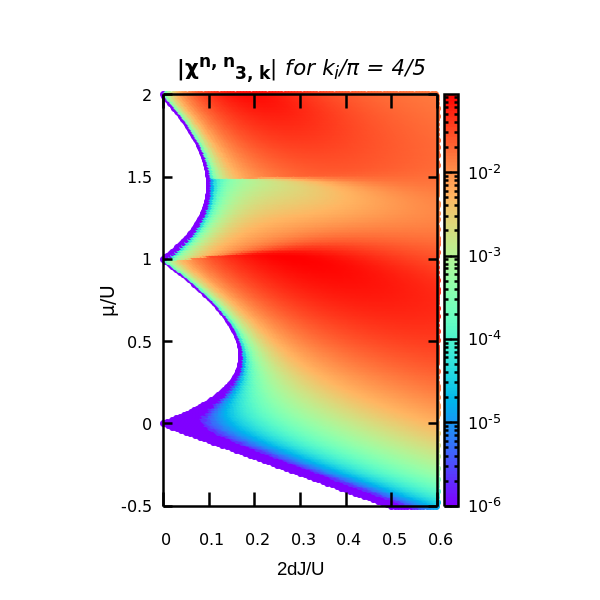}
		\caption{Amplitude of the density-density response function $\left| \chi^{n n}_{3, \mathbf{k}} \right|$ associated with third mode at momentum $k_x = k_y = k_z = 4 \pi/5$.}
		\label{Figure41}
	\end{minipage}
	\ \hspace{3mm} \hspace{3mm} \
	\begin{minipage}[b]{7.7cm}
		\centering
		\includegraphics[width=9cm]{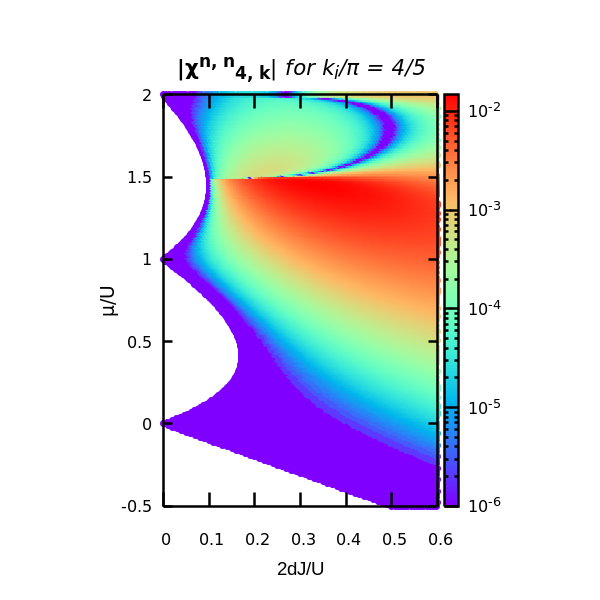}
		\caption{Amplitude of the density-density response function $\left| \chi^{n n}_{4, \mathbf{k}} \right|$ associated with fourth mode at momentum $k_x = k_y = k_z = 4 \pi/5$.}
		\label{Figure42}
	\end{minipage}
\end{figure} \\
Proceeding in this direction, Figures \ref{Figure41} and \ref{Figure42} show that the response of the first two modes living above the Higgs excitation presents non-negligible contributions to $\chi_{n}{\left( \mathbf{k}, \omega \right)}$ at large momentum. In particular, \textsl{the weight of the $\alpha = 3$ mode is enhanced over peculiar regions confined above half-integer filling, while the $\alpha = 4$ mode contribute below half-filling, where the former excitations is suppressed}. \\
More interestingly, \textsl{the contribution of the $\alpha = 4$ mode is suppressed over arc-shaped lines which appear only from the level of the second Mott lobe as expected and strongly resemble the structures exhibited by the Higgs mode} in Figures \ref{Figure39} and \ref{Figure40}.

\paragraph*{Observation}
The most evident difference between the Higgs channel and higher excitations relies in the sharp discontinuity characterizing the passage between their response domains (see Figure \ref{Figure42}). This result, together with the emergence of the structure mentioned above, suggest that \uline{the spectra of these modes cross each other in particle-hole-symmetric points and interchange the nature of the dominant excitation in the density-density channel}. The fact that particle-hole symmetry can be associated with crossing and anti-crossing points in the excitation spectrum has been highlighted in \cite{BEC} and now finds important consequences in measurable quantities such as response functions.
\begin{figure}[!htp]
	\begin{minipage}[b]{7.7cm}
		\centering
		\includegraphics[width=9cm]{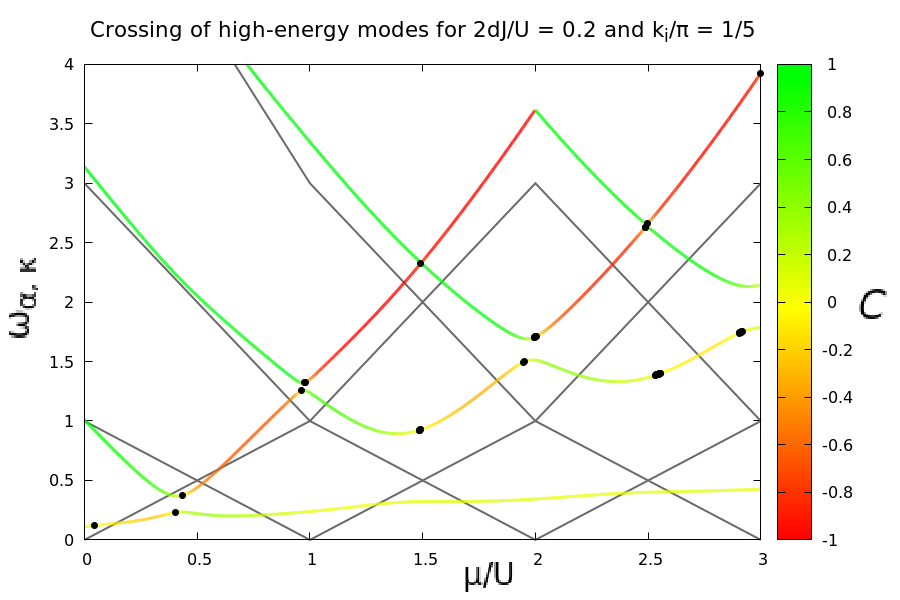}
	\end{minipage}
	\ \hspace{4mm} \hspace{4mm} \
	\begin{minipage}[b]{7.7cm}
		\centering
		\includegraphics[width=9cm]{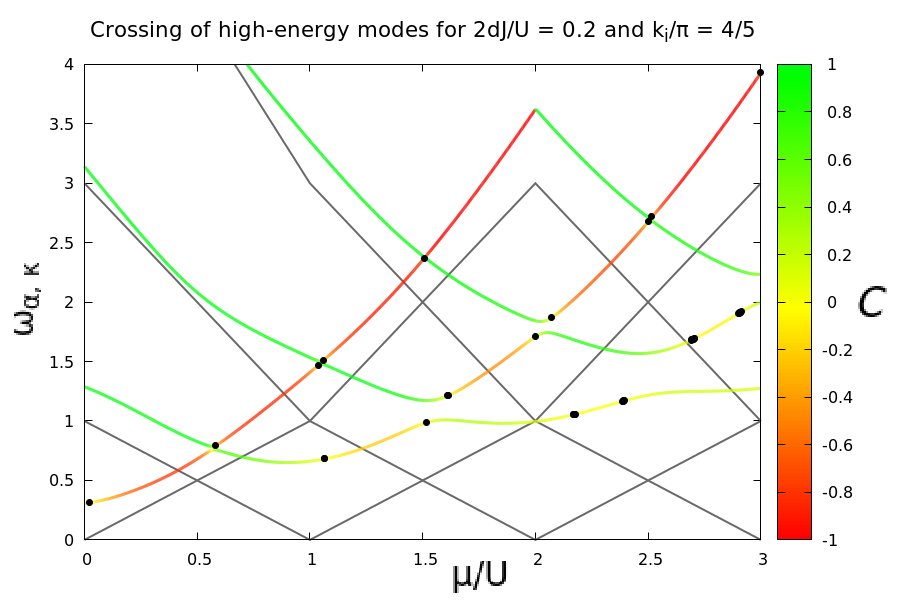}
	\end{minipage}
	\caption{Grey dotted lines $m \, \mu/U$ are the $m$-holes (or $m$-particles) excitation energies in the Mott phase at $J = 0$. The thick lines indicate the excitation energies $\omega_{\alpha, \mathbf{k}}$ of the first four modes at $k_x = k_y = k_z = \pi/5$ (left), $4 \pi/5$ (right) as a function of $\mu/U$ for $2 \, d \, J/U = 0.2$. The points of particle-hole symmetry are highlighted by the black dots, while the particle-hole character of the excitation is given by $\mathcal{C} = \left( \left| U_{\alpha, \mathbf{k}} \right| - \left| V_{\alpha, \mathbf{k}} \right| \right)/\left( \left| U_{\alpha, \mathbf{k}} \right| - \left| V_{\alpha, \mathbf{k}} \right| \right)$.}
	\label{Figure43}
\end{figure}
These observations can be supported by analysing in detail the excitation spectrum near the supposed crossing contours. In Figure \ref{Figure43} we can immediately identify the response discontinuities in Figures \ref{Figure41} and \ref{Figure42} with the black dot at $\mu/U \approx 1.5$ and $\omega_{3, \mathbf{k}} \approx \omega_{4, \mathbf{k}} \approx 2.5$, where only a thin gap separates the two excitation modes. \\
\uline{We can notice that, differently from the Goldstone and Higgs modes which are coherent superpositions of would-be one-particle and one-hole modes in character, the next excitations inherit their pure particle or hole nature from their counterparts in the Mott phase, as they are continuously connected.} According to the expectations, the particle-hole characters tends to hybridize only for weak interactions or large values of $\mu/U$, as well as to anti-crossing points of perfect particle-hole symmetry.

\bigskip

Having identified the nature and the role of higher excitation modes in the compressibility channel, we are now interested in finding suitable physical probes for exciting such modes in a significant manner with respect to the Goldstone excitations.

\markboth{\MakeUppercase{6. Response functions}}{\MakeUppercase{6.1 Density-interaction response}}
\section{Density response to a perturbation of the interaction}\label{response_density_interaction}
\markboth{\MakeUppercase{6. Response functions}}{\MakeUppercase{6.1 Density-interaction response}}
Since we have appreciated the strong particle or hole character of higher excitation modes, interesting results could emerge from the determination of the system response to the perturbation of local observables connected to the density channel. \\
For instance, a similar calculation as above allows to determine the lowest-order response of the density to a perturbation of the on-site interaction:
\begin{equation}\label{4.61}
\begin{aligned}
\chi_{n U}{\left( \mathbf{k}, z \right)} &= -i \int dt \, e^{-i \omega t} \, \theta{\left( t \right)} \sum_{\mathbf{r}} \, e^{-i \mathbf{k} \cdot \mathbf{r}} \, \big\langle \left[ \hat{n}{\left( \mathbf{r}, t \right)}, \hat{n}{\left( \mathbf{0}, 0 \right)} \left\{ \hat{n}{\left( \mathbf{0}, 0 \right)} - 1 \right\} \right] \big\rangle \approx \\
&\approx 2 \sum_{\alpha} \frac{N_{\alpha, \mathbf{k}} \left( D_{\alpha, \mathbf{k}} - N_{\alpha, \mathbf{k}} \right) \omega_{\alpha, \mathbf{k}}}{z^2 - \omega^2_{\alpha, \mathbf{k}}}
\end{aligned}
\end{equation}
where:
\begin{equation}\label{4.62}
D_{\alpha, \mathbf{k}} = \sum_n n^2 \, c^0_n \left( u_{\alpha, \mathbf{k}, n}  + v_{\alpha, \mathbf{k}, n} \right)
\end{equation}
which quantifies the amplitude of the first-order fluctuations of the double-density operator $\hat{n}{\left( \mathbf{r}, t \right)} \, \hat{n}{\left( \mathbf{r}, \mathbf{t} \right)}$ expressed in its Gutzwiller form:
\begin{equation}\label{4.63}
\hat{n}{\left( \mathbf{r}, t \right)} \, \hat{n}{\left( \mathbf{r}, \mathbf{t} \right)} = \sum_n n^2 \, \hat{c}^{\dagger}_n{\left( \mathbf{r}, t \right)} \, \hat{c}_n{\left( \mathbf{r}, t \right)}
\end{equation}
in analogy with the usual operator expansions performed in this work. \\
Then, the amplitude of the response function $\chi_{n U}{\left( \mathbf{k}, z \right)}$ for each mode is provided by:
\begin{equation}\label{4.64}
\chi^{n U}_{\alpha, \mathbf{k}} \equiv N_{\alpha, \mathbf{k}} \left( D_{\alpha, \mathbf{k}} - N_{\alpha, \mathbf{k}} \right)
\end{equation}
\paragraph*{Observation}
It is important to notice that only the quantization rule for the full double-density operator given by (\ref{4.63}) leads to a consistent result. In fact, if the two density operators in (\ref{4.63}) were expanded separately, the second-order expansion of $\chi_{n U}{\left( \mathbf{k}, z \right)}$ would lead to a negative density response for $n_0 \le 1/2$. With reference to the discussion of Section \ref{quantization_discussion}, this result gives additional hints about the open problem of calculating expectation values of Gutzwiller operators defined in coincident points.
\begin{figure}[!htp]
	\begin{minipage}[b]{7.7cm}
		\centering
		\includegraphics[width=9cm]{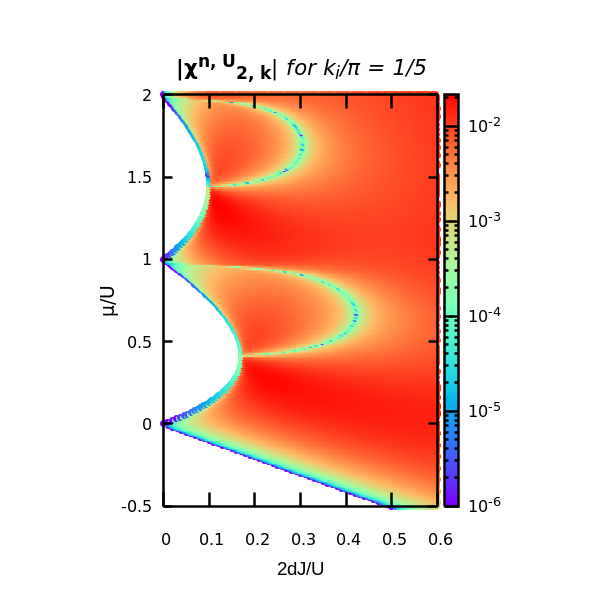}
		\caption{Amplitude of the density-density response function $\left| \chi^{n U}_{2, \mathbf{k}} \right|$ associated with the Higgs mode at momentum $k_x = k_y = k_z = \pi/5$.}
		\label{Figure44}
	\end{minipage}
	\ \hspace{3mm} \hspace{3mm} \
	\begin{minipage}[b]{7.7cm}
		\centering
		\includegraphics[width=9cm]{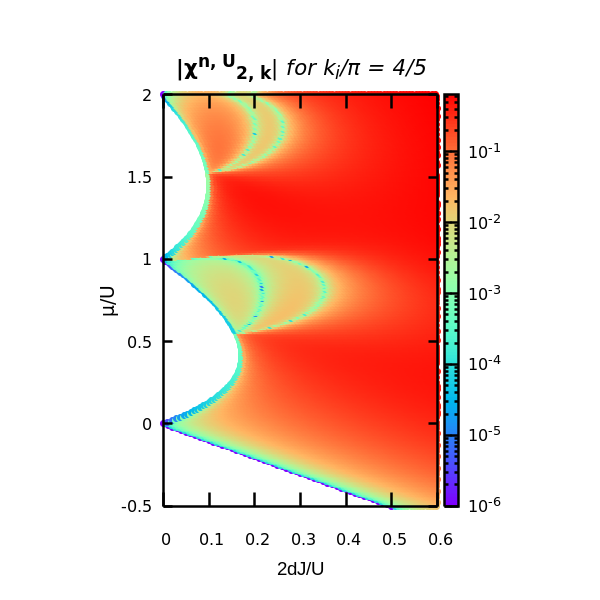}
		\caption{Amplitude of the density-density response function $\left| \chi^{n U}_{2, \mathbf{k}} \right|$ associated with the Higgs mode at momentum $k_x = k_y = k_z = 4 \pi/5$.}
		\label{Figure45}
	\end{minipage}
\end{figure}
\begin{figure}[!htp]
	\begin{minipage}[b]{7.7cm}
		\centering
		\includegraphics[width=9cm]{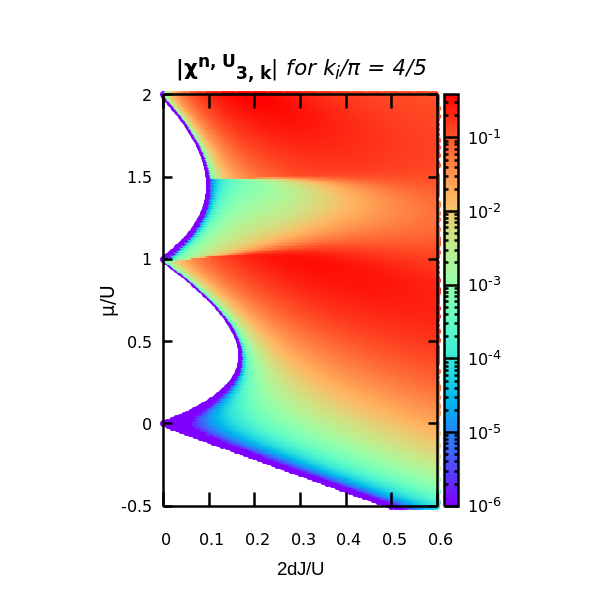}
		\caption{Amplitude of the density-density response function $\left| \chi^{n U}_{3, \mathbf{k}} \right|$ associated with the third mode at momentum $k_x = k_y = k_z = 4 \pi/5$.}
		\label{Figure46}
	\end{minipage}
	\ \hspace{3mm} \hspace{3mm} \
	\begin{minipage}[b]{7.7cm}
		\centering
		\includegraphics[width=9cm]{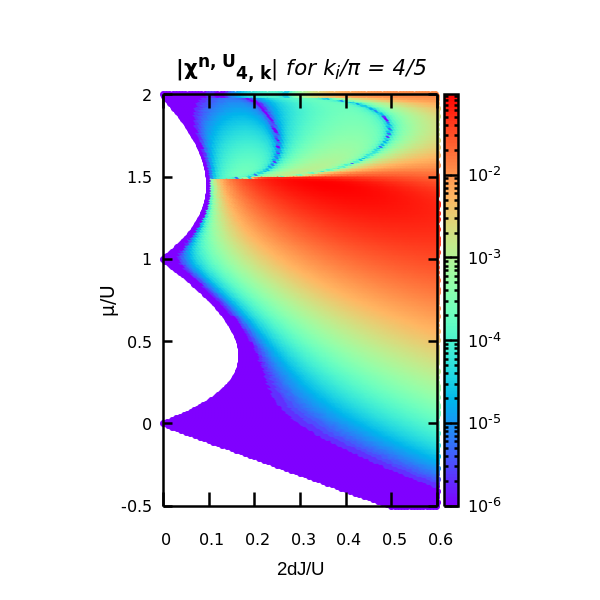}
		\caption{Amplitude of the density-density response function $\left| \chi^{n U}_{4, \mathbf{k}} \right|$ associated with the fourth mode at momentum $k_x = k_y = k_z = 4 \pi/5$.}
		\label{Figure47}
	\end{minipage}
\end{figure}
\newpage
From the point of view of the impact of particle-hole symmetry, the response of the system to a perturbation of the on-site interaction exhibit the same structural features as $\chi_{n}{\left( \mathbf{k}, \omega \right)}$ (see Figures \ref{Figure44} and \ref{Figure45}). On the other hand, as one can deduce from Figures \ref{Figure46} and \ref{Figure47}, in the interaction channel the response of the high-energy particle or hole-dominated modes, if compared to the Goldstone and Higgs mode, is amplified with respect to the density-density channel (see again Figures \ref{Figure40}-\ref{Figure42}). \\
However, \textsl{tuning the on-site interaction is experimentally more difficult than modifying other parameters of the theory, such as the hopping rate between adjacent sites. This is the main reason why our investigation includes also the study of the response of the Bose-Hubbard system to perturbations of the kinetic part of the Hamiltonian.}

\markboth{\MakeUppercase{6. Response functions}}{\MakeUppercase{6.1 Density-particle/hole response}}
\section{Density response to a particle-hole perturbation}\label{ph_perturbation}
\markboth{\MakeUppercase{6. Response functions}}{\MakeUppercase{6.1 Density-particle/hole response}}
Let us consider the following interaction term as a perturbation to the Bose-Hubbard Hamiltonian:
\begin{equation}\label{4.65}
\hat{H}_s{\left( t \right)} = \sum_{\mathbf{r}} V_{\mathbf{k}, \omega} \sin{\left( \mathbf{k} \cdot \mathbf{r} - \omega \, t \right)} \left( \hat{a}_{\mathbf{r}} + \hat{a}^{\dagger}_{\mathbf{r}} \right)
\end{equation}
The Hamiltonian term (\ref{4.65}) can be seen as a perturbation creating particle-hole pairs with equal amplitude and explicitly breaks the $U{\left( 1 \right)}$ symmetry. \\
In our Gutzwiller quantum formalism the second-order expansion of the response function of the density with respect to the perturbation (\ref{4.65}) reads:
\begin{equation}\label{4.66}
\begin{aligned}
\chi_{n s}{\left( \mathbf{k}, z \right)} &= -i \int dt \, e^{-i \omega t} \theta{\left( t \right)} \sum_{\mathbf{r}} \, e^{-i \mathbf{k} \cdot \mathbf{r}} \, \big\langle \left[ \hat{n}{\left( \mathbf{r}, t \right)}, \hat{\psi}{\left( \mathbf{0}, 0 \right)} + \hat{\psi}^{\dagger}{\left( \mathbf{0}, 0 \right)} \right] \big\rangle \approx \\
&\approx 2 \sum_{\alpha} \frac{N_{\alpha, \mathbf{k}} \left( U_{\alpha, \mathbf{k}} + V_{\alpha, \mathbf{k}} \right) \omega_{\alpha, \mathbf{k}}}{z^2 - \omega^2_{\alpha, \mathbf{k}}}
\end{aligned}
\end{equation}
while the response function of the density variance is simply given by:
\begin{equation}\label{4.67}
\begin{aligned}
\chi_{n n s}{\left( \mathbf{k}, z \right)} &= -i \int dt \, e^{-i \omega t} \theta{\left( t \right)} \sum_{\mathbf{r}} \, e^{-i \mathbf{k} \cdot \mathbf{r}} \, \big\langle \left[ \hat{n}{\left( \mathbf{r}, t \right)} \, \hat{n}{\left( \mathbf{r}, t \right)}, \hat{\psi}{\left( 0, 0 \right)} + \hat{\psi}^{\dagger}{\left( 0, 0 \right)} \right] \big\rangle \approx \\
&\approx 2 \sum_{\alpha} \frac{D_{\alpha, \mathbf{k}} \left( U_{\alpha, \mathbf{k}} + V_{\alpha, \mathbf{k}} \right) \omega_{\alpha, \mathbf{k}}}{z^2 - \omega^2_{\alpha, \mathbf{k}}}
\end{aligned}
\end{equation}
The amplitude of the density response to particle or hole creations for each excitation mode is defined to read:
\begin{equation}\label{4.68}
\chi^{n s}_{\alpha, \mathbf{k}} \equiv N_{\alpha, \mathbf{k}} \left( U_{\alpha, \mathbf{k}} + V_{\alpha, \mathbf{k}} \right)
\end{equation}
while the density variance reacts according to the following amplitude:
\begin{equation}\label{4.69}
\chi^{n n s}_{\alpha, \mathbf{k}} \equiv D_{\alpha, \mathbf{k}} \left( U_{\alpha, \mathbf{k}} + V_{\alpha, \mathbf{k}} \right)
\end{equation}
\textsl{It is interesting to notice that the response functions (\ref{4.68}) and (\ref{4.69}) depend on the quantity $U_{\alpha, \mathbf{k}} + V_{\alpha, \mathbf{k}}$, which has been proven to follow an hydrodynamic behaviour in the low-energy limit and present peculiar properties related to single-mode particle-hole symmetry in the strongly-interacting regime \cite{BEC}.}

\subsection{\texorpdfstring{$\chi^{n s}_{\alpha, \mathbf{k}}$}{} - Numerical results}
\vspace{-0.5cm}
\begin{figure}[!htp]
	\begin{minipage}[b]{7.7cm}
		\centering
		\includegraphics[width=9cm]{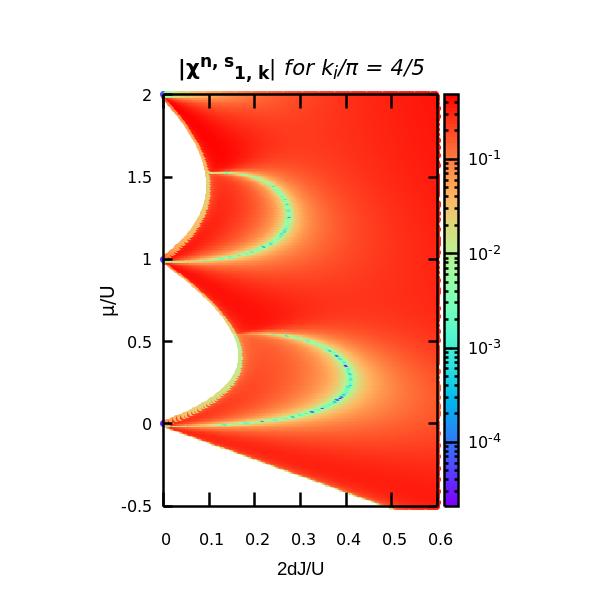}
		\caption{Parametric plot of the response amplitude $\left| \chi^{n s}_{1, \mathbf{k}} \right|$ associated with the Goldstone mode at momentum $k_x = k_y = k_z = 4 \pi/5$.}
		\label{Figure48}
	\end{minipage}
	\ \hspace{3mm} \hspace{3mm} \
	\begin{minipage}[b]{7.7cm}
		\centering
		\includegraphics[width=9cm]{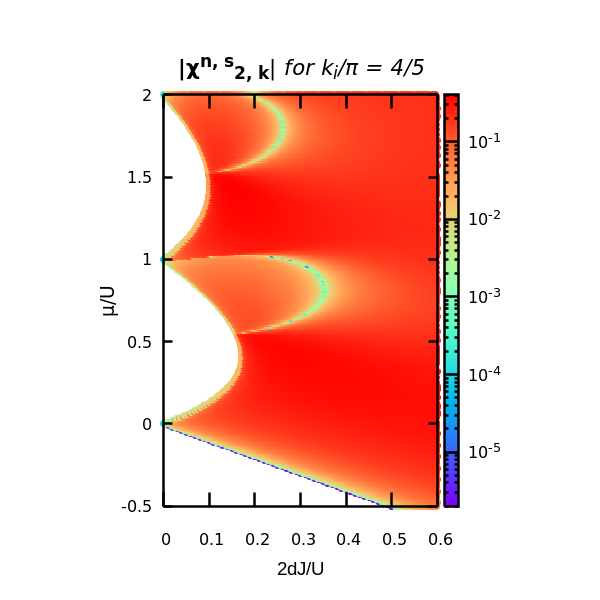}
		\caption{Parametric plot of the response amplitude $\left| \chi^{n s}_{2, \mathbf{k}} \right|$ associated with the Higgs mode at momentum $k_x = k_y = k_z = 4 \pi/5$.}
		\label{Figure49}
	\end{minipage}
\end{figure}
\begin{figure}[!htp]
	\begin{minipage}[t]{7.7cm}
		\centering
		\includegraphics[width=9cm]{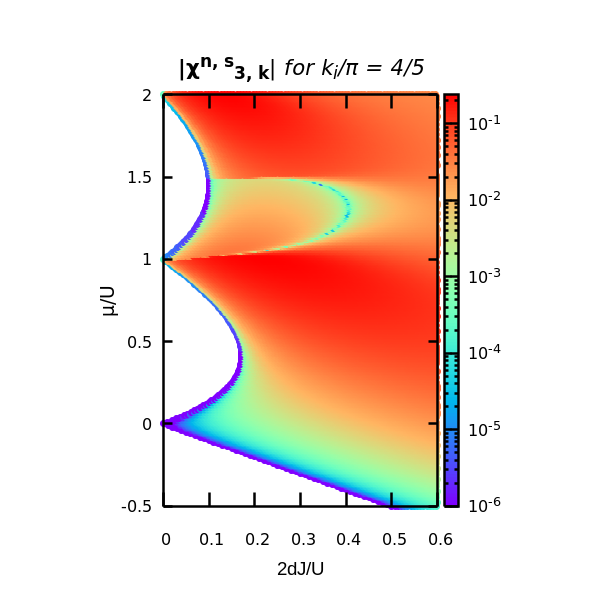}
		\caption{Parametric plot of the response amplitude $\left| \chi^{n s}_{3, \mathbf{k}} \right|$ associated to the third excitation mode at momentum $k_x = k_y = k_z = 4 \pi/5$.}
		\label{Figure50}
	\end{minipage}
	\ \hspace{3mm} \hspace{3mm} \
	\begin{minipage}[t]{7.7cm}
		\centering
		\includegraphics[width=9cm]{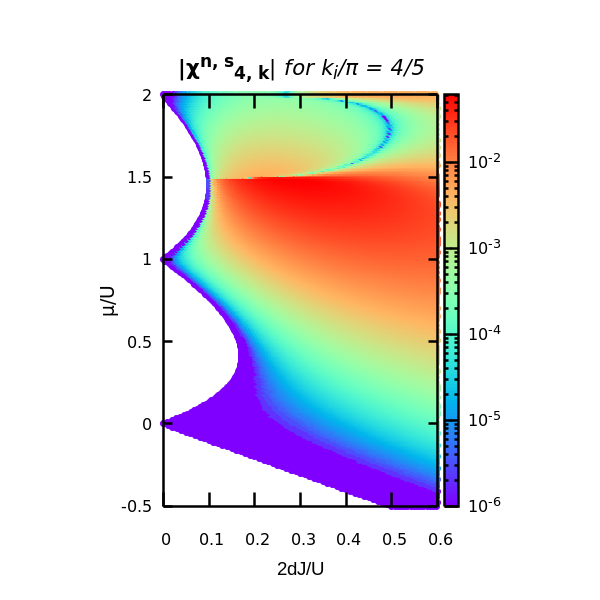}
		\caption{Parametric plot of the response amplitude $\left| \chi^{n s}_{4, \mathbf{k}} \right|$ associated to the fourth excitation mode at momentum $k_x = k_y = k_z = 4 \pi/5$.}
		\label{Figure51}
	\end{minipage}
\end{figure}
Since the response amplitude $\chi^{n s}_{\alpha, \mathbf{k}}$ is given by the product of both $N_{\alpha, \mathbf{k}}$ and $U_{\alpha, \mathbf{k}} + V_{\alpha, \mathbf{k}}$, the Goldstone and the Higgs components show all the properties connected to particle-hole symmetry that we have introduced previously and already observed in the density-density channel. \\
\uline{The same considerations apply to the higher modes (see Figures \ref{Figure50} and \ref{Figure51}), whose sensitivity to a particle-hole perturbation is of the same order of the phononic and Higgs contributions.} Again, high-energy modes have non-negligible sensitivity to external perturbations only in limited regions of the phase diagram of strong interactions. \\
\textsl{In this sense, the possibility of localizing specific regions where the higher excitation modes are insensitive to external perturbations like (\ref{4.65}) should be taken into account as a possible measurement channel for determining characterizing quantum dynamical properties at high-energies.} For example, \uline{this information can be useful for isolating the response associated to the Higgs modes in lattice modulation experiments, where the detection of the amplitude mode still remains a challenging measurement, as we will discuss at the end of the following Section}.

\subsection{\texorpdfstring{$\chi^{n n s}_{\alpha, \mathbf{k}}$}{} - Numerical results}
\begin{figure}[!htp]
	\begin{minipage}[b]{7.7cm}
		\centering
		\includegraphics[width=9cm]{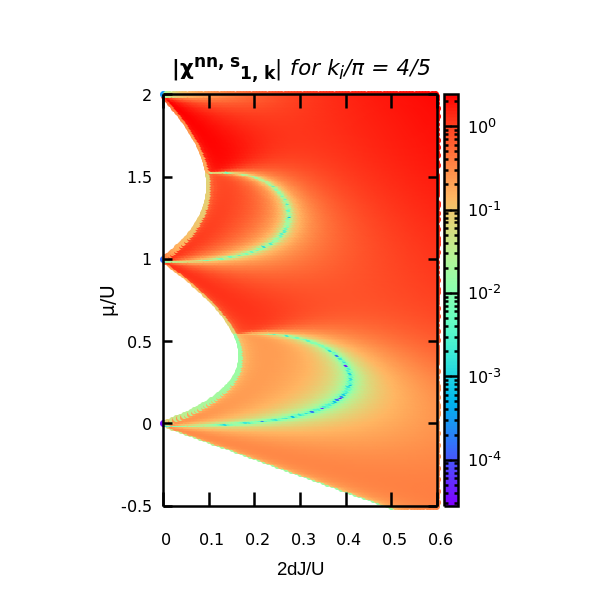}
		\caption{Parametric plot of the response amplitude $\left| \chi^{n n s}_{1, \mathbf{k}} \right|$ associated with the Goldstone mode at momentum $k_x = k_y = k_z = 4 \pi/5$.}
		\label{Figure52}
	\end{minipage}
	\ \hspace{3mm} \hspace{3mm} \
	\begin{minipage}[b]{7.7cm}
		\centering
		\includegraphics[width=9cm]{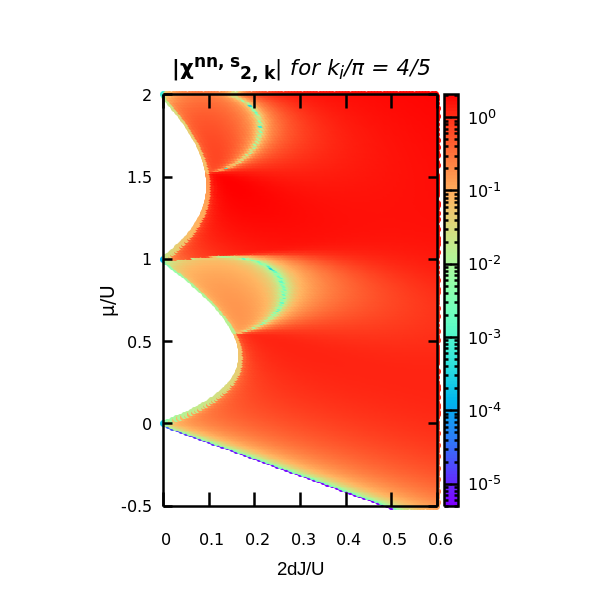}
		\caption{Parametric plot of the response amplitude $\left| \chi^{n n s}_{2, \mathbf{k}} \right|$ associated with the Higgs mode at momentum $k_x = k_y = k_z = 4 \pi/5$.}
		\label{Figure53}
	\end{minipage}
\end{figure}
\begin{figure}[!htp]
	\begin{minipage}[b]{7.7cm}
		\centering
		\includegraphics[width=9cm]{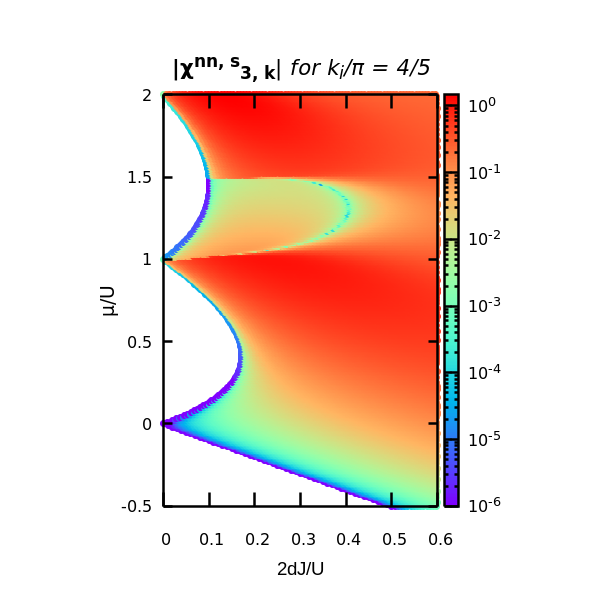}
		\caption{Parametric plot of the response amplitude $\left| \chi^{n n s}_{3, \mathbf{k}} \right|$ associated with the third excitation mode at momentum $k_x = k_y = k_z = 4 \pi/5$.}
		\label{Figure54}
	\end{minipage}
	\ \hspace{3mm} \hspace{3mm} \
	\begin{minipage}[b]{7.7cm}
		\centering
		\includegraphics[width=9cm]{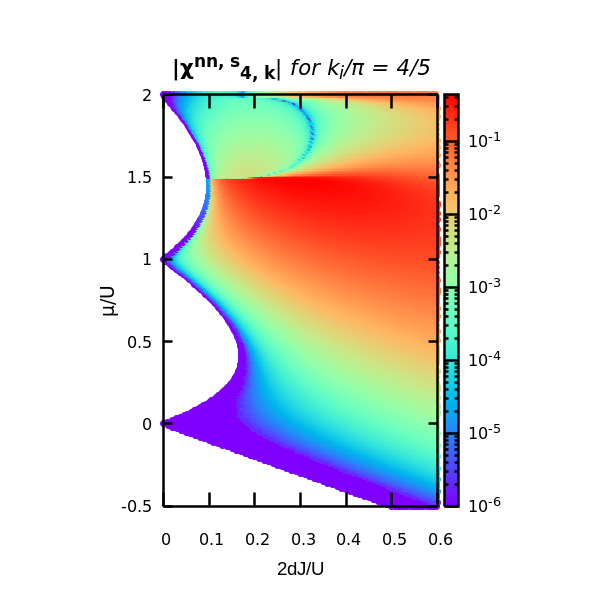}
		\caption{Parametric plot of the response amplitude $\left| \chi^{n n s}_{4, \mathbf{k}} \right|$ associated with the fourth excitation mode at momentum $k_x = k_y = k_z = 4 \pi/5$.}
		\label{Figure55}
	\end{minipage}
\end{figure}
The response of the density variance to a particle-hole creation process does not display significant differences with respect to the previous case. Remarkably, as one can notice in Figures \ref{Figure53} and \ref{Figure54}, the regions of response suppressions of the even-order excitation modes shrink towards criticality. It is also worth noticing that we cannot identify a dominant excitation determining the behaviour of the system, as the contributions of the first four modes have to the same order of magnitude.

\markboth{\MakeUppercase{6. Response functions}}{\MakeUppercase{6.1 Density-hopping response}}
\section{Density response to a modulation of the hopping}\label{hopping_perturbation}
\markboth{\MakeUppercase{6. Response functions}}{\MakeUppercase{6.1 Density-hopping response}}
As a final example, let us consider the following modulation of the hopping term:
\begin{equation}\label{4.70}
\hat{H}_J{\left( t \right)} = \sum_{\langle \mathbf{r}, \mathbf{s} \rangle} \delta J_{\mathbf{k}, \omega} \sin{\left( \mathbf{k} \cdot \mathbf{r} - \omega \, t \right)} \left( \hat{a}^{\dagger}_{\mathbf{r}} \, \hat{a}_\mathbf{s} + \text{h.c.} \right)
\end{equation}
The second-order expansion of the response function of the density with respect to a modulation of the hopping rate in our Gutzwiller quantum formalism strongly recalls the response to the local creation of the particle-hole pairs:
\begin{equation}\label{4.71}
\begin{aligned}
\chi_{n J}{\left( \mathbf{k}, z \right)} &= -i \sum_{s{\left( \mathbf{r} \right)}} \int dt \, e^{-i \omega t} \theta{\left( t \right)} \sum_{\mathbf{x} - \mathbf{r}} \, e^{-i \mathbf{k} \cdot \left( \mathbf{x} - \mathbf{r} \right)} \, \big\langle \left[ \hat{n}{\left( \mathbf{x}, t \right)}, \hat{\psi}^{\dagger}{\left( \mathbf{r}, 0 \right)} \, \hat{\psi}{\left( s, 0 \right)} + \text{h.c.} \right] \big\rangle \approx \\
&\approx \psi_0 \, j{\left( \mathbf{k} \right)} \sum_{\alpha} \frac{N_{\alpha, \mathbf{k}} \left( U_{\alpha, \mathbf{k}} + V_{\alpha, \mathbf{k}} \right) \omega_{\alpha, \mathbf{k}}}{z^2 - \omega^2_{\alpha, \mathbf{k}}}
\end{aligned}
\end{equation}
where the function $s{\left( \mathbf{r} \right)}$ takes into account the nearest-neighbouring sites with respect to a given site $r$ and $j{\left( \mathbf{k} \right)} = 4 \, d - 4 \sum_{a = 1}^{d} \sin^2{\left( k_a/2 \right)}$. \\
The analogous response function of the density variance is simply given by:
\begin{equation}\label{4.72}
\small
\begin{aligned}
\chi_{n n J}{\left( \mathbf{k}, z \right)} &= -i \sum_{s{\left( \mathbf{r} \right)}} \int dt \, e^{-i \omega t} \theta{\left( t \right)} \sum_{\mathbf{x} - \mathbf{r}} \, e^{-i \mathbf{k} \cdot \left( \mathbf{x} - \mathbf{r} \right)} \, \big\langle \left[ \hat{n}{\left( x, t \right)} \hat{n}{\left( x, t \right)}, \hat{\psi}^{\dagger}{\left( r, 0 \right)} \, \hat{\psi}{\left( s, 0 \right)} + \text{h.c.} \right] \big\rangle \approx \\
&\approx \psi_0 \, j{\left( \mathbf{k} \right)} \sum_{\alpha} \frac{D_{\alpha, \mathbf{k}} \left( U_{\alpha, \mathbf{k}} + V_{\alpha, \mathbf{k}} \right) \omega_{\alpha, \mathbf{k}}}{z^2 - \omega^2_{\alpha, \mathbf{k}}}
\end{aligned}
\end{equation}

\fbox{\begin{minipage}{16.5cm}
\textbf{\textcolor{blue}{Observation} -} \textsl{We notice that the response functions of physical observables constructed over the density operator with respect to a perturbation of the hopping or kinetic term (\ref{4.70}) is proportional to the response of the same quantities to a perturbation that locally creates particle-hole pairs (\ref{4.65})}, apart from a factor depending on the excitation momentum. As far as Gaussian fluctuations are considered, \textsl{it follows that experimentally the linear response of the system to a modulation of the hopping strength could be equivalently probed through the particle-hole excitation channel}.
\end{minipage}}

\vspace{0.5cm}

\subsection{Experimentally-oriented considerations}
\begin{wrapfigure}{l}{0.5\textwidth}
	\centering
	\includegraphics[width=0.4\textwidth]{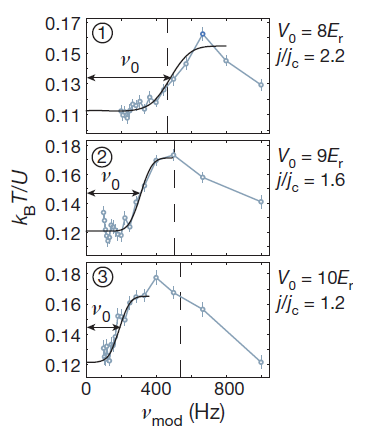}
	\caption{Temperature response (energy absorption) to a 2D lattice modulation (circles and connecting blue line) and fit with an error function (solid black line) for the three different points of the phase diagram, reported on the right. As the coupling $J$ approaches the critical value $J_c$, the change in the Higgs gap values $\nu_0$ to lower frequencies is clearly visible (from panel 1 to panel 3). Vertical dashed lines mark the frequency $U/\hbar$ corresponding to the on-site interaction. Figure taken from \cite{lattice_depth_modulation_2}.}
	\label{endres}
\end{wrapfigure}
In Section \ref{ph_perturbation} we have circumscribed the range of parameters defining strongly-interacting superfluids where the susceptibility of weakly-hybridized particle and hole excitations on top of the Higgs modes is significant for large momenta. Successively, in Section \ref{hopping_perturbation}, we have realized that the results deriving a modulation of the hopping term of the Hamiltonian correspond are related to the application of a simple particle-hole perturbation. \\
Having recognized this connection, we can suppose that this class of information could find important applications in improving the analysis of real data regarding the indirect observation of the Higgs mode. Although the most recent experimental efforts \cite{lattice_depth_modulation_2} has allowed to gain an excellent estimation of the low-energy scalar susceptibility of the order parameter, it has been observed that the resonance peak due to the Higgs excitation is stretched into a threshold line separating the low-energy domain from the energy scale of the on-site interaction (see Figure \ref{endres}). In \cite{lattice_depth_modulation_2} and \cite{pekker_varma} it is argued that there here are two possible explanations for this observation. \\
One reason is related to the non-uniform boson density due to the trapping field which confines the system. The presence of a potential trap modifies the structure of the energy spectrum slightly above the amplitude excitation by introducing normal modes that are supposed to absorb a significant part of the energy injected by the modulation of the lattice depth used for probing the scalar susceptibility. \\
On the other hand, it is not excluded the possibility that part of the anomalous energy absorption is also due to the high-energy particle-hole modes analysed along this Chapter and exhibiting similar energy thresholds as the Higgs excitation. \\
Following the results proposed in the present work, we are confident that the localization of specific parametric region where such excitations are relatively insensitive to external probes as the hopping modulation and a systematic estimation of their contribution to the observable quantities could enhance the analysis of new experimental data. In this sense, our extended Gutzwiller approach represents a suitable method of investigation, which could be further improved for comprising higher-order corrections in a number of experimental contexts (see the Chapter \textit{Conclusions and Perspectives}).

\chapter{Quantization beyond the second order: decay processes}\label{third_order}
\markboth{\MakeUppercase{7. Beyond the second-order}}{\MakeUppercase{7. Beyond the second-order}}
The quantization of the Gutzwiller ansatz has allowed to expand the Hamiltonian operator up to the second-order in the quasi-particle operators in order to study Gaussian fluctuations around the saddle-point solution of the Gutzwiller approach. This method has allowed to determine the dispersion relations of the collective modes characterizing the Bose-Hubbard dynamics and characterize observable quantum corrections depending on the interplay between these excitations. \\
Since the quasi-particle operators defined in (\ref{4.13}) are associated to the quantum fluctuations living above the Gutzwiller saddle-point solution, such operators are also useful in studying higher-order terms of the Lagrangian expansion which are responsible for the interaction between the fundamental excitations described above. Quite generally, this procedure leads to the determination of the band widths characterizing the spectral function of the system, which cannot be provided by a second-order description. In particular, the final scope of this Chapter is the identification of possible decay processes that involve Higgs or amplitude modes dissipating into Goldstone excitations in a sufficiently wide range of the phase diagram of strong interactions. In fact, the amplitude mode is expected to be long-lived only at the classical (weak-coupling) level, while quantum corrections allows for its decay into pairs of phase modes.

\section{Third-order Bogoliubov expansion}
The third-order terms of the quantized Gutzwiller Hamiltonian given by the functional (\ref{4.6}) derive both from the hopping part and the local terms, so that:
\begin{equation}\label{4.88}
\begin{aligned}
&\hat{H}^{\left( 3 \right)} = \hat{H}^{\left( 3 \right)}_1 + \hat{H}^{\left( 3 \right)}_2 + \hat{H}^{\left( 3 \right)}_3 = \\
&= -J \sum_{\langle \mathbf{r}, \mathbf{s} \rangle} \left[ \delta_1 \, \hat{\psi}^{\dagger}{\left( \mathbf{r} \right)} \, \delta_2 \, \hat{\psi}{\left( \mathbf{s} \right)} + \delta_2 \, \hat{\psi}^{\dagger}{\left( \mathbf{r} \right)} \, \delta_1 \, \hat{\psi}{\left( \mathbf{s} \right)} + \delta_1 \, \hat{\psi}^{\dagger}{\left( \mathbf{s} \right)} \, \delta_2 \, \hat{\psi}{\left( \mathbf{r} \right)} + \delta_2 \, \hat{\psi}^{\dagger}{\left( \mathbf{s} \right)} \, \delta_1 \, \hat{\psi}{\left( \mathbf{r} \right)} \right] + \\
&- J \sum_{\langle \mathbf{r}, \mathbf{s} \rangle} \left[ \psi_0{\left( \mathbf{r} \right)} \, \delta_3 \, \hat{\psi}{\left( \mathbf{s} \right)} +  \delta_3 \, \hat{\psi}^{\dagger}{\left( \mathbf{r} \right)} \, \psi_0{\left( \mathbf{s} \right)} + \psi_0{\left( \mathbf{s} \right)} \, \delta_3 \, \hat{\psi}{\left( \mathbf{r} \right)} + \delta_3 \, \hat{\psi}^{\dagger}{\left( \mathbf{s} \right)} \, \psi_0{\left( \mathbf{r} \right)} \right] + \\
&- \frac{1}{2} \sum_{\mathbf{r}} \sum_n H_n \, \delta c^{\dagger}_n{\left( \mathbf{r} \right)} \sum_m \delta c^{\dagger}_m{\left( \mathbf{r} \right)} \, \delta c_m{\left( \mathbf{r} \right)} - \frac{1}{2} \sum_{\mathbf{r}} \sum_m \delta c^{\dagger}_m{\left( \mathbf{r} \right)} \, \delta c_m{\left( \mathbf{r} \right)} \sum_n H_n \, \delta c_n{\left( \mathbf{r} \right)} = \\
\end{aligned}
\end{equation}
$$
\begin{aligned}
&= -J \sum_{\mathbf{r}, s{\left( \mathbf{r} \right)}} \left[ \delta_1 \, \hat{\psi}^{\dagger}{\left( \mathbf{r} \right)} \, \delta_2 \, \hat{\psi}{\left( s \right)} + \delta_2 \, \hat{\psi}^{\dagger}{\left( \mathbf{r} \right)} \, \delta_1 \, \hat{\psi}{\left( s \right)} \right] + \\
&- J \sum_{\mathbf{r}, s{\left( \mathbf{r} \right)}} \left[ \psi_0{\left( \mathbf{r} \right)} \, \delta_3 \, \hat{\psi}{\left( s \right)} +  \delta_3 \, \hat{\psi}^{\dagger}{\left( \mathbf{r} \right)} \, \psi_0{\left( s \right)} \right] + \\
&- \frac{1}{2} \sum_{\mathbf{r}} \sum_n H_n \, c^0_n \, \delta c^{\dagger}_n{\left( \mathbf{r} \right)} \sum_m \delta c^{\dagger}_m{\left( \mathbf{r} \right)} \, \delta c_m{\left( \mathbf{r} \right)} - \frac{1}{2} \sum_{\mathbf{r}} \sum_m \delta c^{\dagger}_m{\left( \mathbf{r} \right)} \, \delta c_m{\left( \mathbf{r} \right)} \sum_n H_n \, c^0_n \, \delta c_n{\left( \mathbf{r} \right)}
\end{aligned}
$$
where the third-order operators $\delta_3 \, \hat{\psi}{\left( \mathbf{r} \right)}$ have been defined in equation (\ref{4.79}) and the function $s{\left( \mathbf{r} \right)}$ counts the nearest-neighbour sites around the site labelled by $\mathbf{r}$. It is important to notice that the third-order corrections to the Bose field operator $\delta_3 \, \hat{\psi}{\left( \mathbf{r} \right)}$ can be now included in the theory consistently with the order of the expansion. \\
The expressions of the terms composing $\hat{H}^{\left( 3 \right)}_2$ simply read:
\begin{equation}\label{4.89}
\begin{aligned}
- J \sum_{\mathbf{r}, s{\left( \mathbf{r} \right)}} \psi_0{\left( \mathbf{r} \right)} \, \delta_3 \, \hat{\psi}{\left( s \right)} &= \frac{J \, \psi_0}{2} \sum_{\mathbf{r}, s{\left( \mathbf{r} \right)}} \sum_{n m} \sqrt{n} \, c^0_n \, \delta \hat{c}^{\dagger}_{n - 1}{\left( s \right)} \, \delta \hat{c}^{\dagger}_m{\left( s \right)} \, \delta \hat{c}_m{\left( s \right)} + \\
&+ \frac{J \, \psi_0}{2} \sum_{\mathbf{r}, s{\left( \mathbf{r} \right)}} \sum_{n m} \sqrt{n} \, c^0_{n - 1} \, \delta \hat{c}^{\dagger}_m{\left( s \right)} \, \delta \hat{c}_m{\left( s \right)} \, \delta \hat{c}_n{\left( s \right)} =
\end{aligned}
\end{equation}
$$
\begin{aligned}
&= \frac{J z \, \psi_0}{2 \sqrt{M}} \sum_{\mathbf{k}, \mathbf{p}} \sum_{n} \sqrt{n} \, c^0_n \, \delta \hat{c}^{\dagger}_{n - 1}{\left( \mathbf{k} \right)} \, \delta \hat{c}^{\dagger}_m{\left( \mathbf{p} \right)} \, \delta \hat{c}_m{\left( \mathbf{k} + \mathbf{p} \right)} + \\
&+ \frac{J z \, \psi_0}{2 \sqrt{M}} \sum_{\mathbf{k}, \mathbf{p}} \sum_{n m} \sqrt{n} \, c^0_{n - 1} \, \delta \hat{c}^{\dagger}_m{\left( \mathbf{p} \right)} \, \delta \hat{c}_m{\left( \mathbf{p} - \mathbf{k} \right)} \, \delta \hat{c}_n{\left( \mathbf{k} \right)}
\end{aligned}
$$
\begin{equation}\label{4.90}
\begin{aligned}
- J \sum_{\mathbf{r}, s{\left( \mathbf{r} \right)}} \delta_3 \, \hat{\psi}^{\dagger}{\left( \mathbf{r} \right)} \, \psi_0{\left( s \right)} &= \frac{J z \, \psi_0}{2} \sum_{\mathbf{r}} \sum_{n m} \sqrt{n} \, c^0_{n - 1} \, \delta \hat{c}^{\dagger}_n{\left( \mathbf{r} \right)} \, \delta \hat{c}^{\dagger}_m{\left( \mathbf{r} \right)} \, \delta \hat{c}_m{\left( \mathbf{r} \right)} + \\
&+ \frac{J z \, \psi_0}{2} \sum_{\mathbf{r}} \sum_{n m} \sqrt{n} \, c^0_n \, \delta \hat{c}^{\dagger}_m{\left( \mathbf{r} \right)} \, \delta \hat{c}_m{\left( \mathbf{r} \right)} \, \delta \hat{c}_{n - 1}{\left( \mathbf{r} \right)} = \\
&= \frac{J z \, \psi_0}{2 \sqrt{M}} \sum_{\mathbf{k}, \mathbf{p}} \sum_{n} \sqrt{n} \, c^0_{n - 1} \, \delta \hat{c}^{\dagger}_n{\left( \mathbf{k} \right)} \, \delta \hat{c}^{\dagger}_m{\left( \mathbf{p} \right)} \, \delta \hat{c}_m{\left( \mathbf{k} + \mathbf{p} \right)} + \\
&+ \frac{J z \, \psi_0}{2 \sqrt{M}} \sum_{\mathbf{k}, \mathbf{p}} \sum_{n m} \sqrt{n} \, c^0_n \, \delta \hat{c}^{\dagger}_m{\left( \mathbf{p} \right)} \, \delta \hat{c}_m{\left( \mathbf{p} - \mathbf{k} \right)} \, \delta \hat{c}_{n - 1}{\left( \mathbf{k} \right)}
\end{aligned}
\end{equation}
A direct substitution of the momenta variables $\tilde{\mathbf{p}} = \mathbf{k} + \mathbf{p}$ reveals that expression (\ref{4.90}) is the hermitian conjugate counterpart of (\ref{4.89}). \\
The same kind of calculations and considerations can be applied to $\hat{H}^{\left( 3 \right)}_1$, whose explicit form is given by the following terms:
\begin{equation}\label{4.91}
\begin{aligned}
-J \sum_{\mathbf{r}, s{\left( \mathbf{r} \right)}} \delta_1 \, \hat{\psi}^{\dagger}{\left( \mathbf{r} \right)} \, \delta_2 \, \hat{\psi}{\left( s \right)} &= -J \sum_{\mathbf{r}, s{\left( \mathbf{r} \right)}} \sum_{n m} \sqrt{n m} \, c^0_{n - 1} \, \delta \hat{c}^{\dagger}_n{\left( \mathbf{r} \right)} \, \delta \hat{c}^{\dagger}_{m - 1}{\left( s \right)} \, \delta \hat{c}_{m}{\left( s \right)} + \\
&-J \sum_{\mathbf{r}, s{\left( \mathbf{r} \right)}} \sum_{n m} \sqrt{n m} \, c^0_n \, \delta \hat{c}_{n - 1}{\left( \mathbf{r} \right)} \, \delta \hat{c}^{\dagger}_{m - 1}{\left( s \right)} \, \delta \hat{c}_{m}{\left( s \right)} + \\
\end{aligned}
\end{equation}
$$
\begin{aligned}
&+ J \psi^2_0 \sum_{\mathbf{r}, s{\left( \mathbf{r} \right)}} \sum_{n m} \sqrt{n} \, c^0_{n - 1} \, \delta \hat{c}^{\dagger}_n{\left( \mathbf{r} \right)} \, \delta \hat{c}^{\dagger}_m{\left( s \right)} \, \delta \hat{c}_{m}{\left( s \right)} + \\
&+ J \psi^2_0 \sum_{\mathbf{r}, s{\left( \mathbf{r} \right)}} \sum_{n m} \sqrt{n} \, c^0_n \, \delta \hat{c}_{n - 1}{\left( \mathbf{r} \right)} \, \delta \hat{c}^{\dagger}_m{\left( s \right)} \, \delta \hat{c}_{m}{\left( s \right)} = \\
&= -\frac{1}{\sqrt{M}} \sum_{\mathbf{k}, \mathbf{p}} J_{\mathbf{k}} \sum_{n m} \sqrt{n m} \, c^0_{n - 1} \, \delta \hat{c}^{\dagger}_n{\left( \mathbf{k} \right)} \, \delta \hat{c}^{\dagger}_{m - 1}{\left( \mathbf{p} \right)} \, \delta \hat{c}_{m}{\left( \mathbf{k} + \mathbf{p} \right)} + \\
&-\frac{1}{\sqrt{M}} \sum_{\mathbf{k}, \mathbf{p}} J_{\mathbf{k}} \sum_{n m} \sqrt{n m} \, c^0_n \, \delta \hat{c}_{n - 1}{\left( \mathbf{k} \right)} \, \delta \hat{c}^{\dagger}_{m - 1}{\left( \mathbf{p} \right)} \, \delta \hat{c}_{m}{\left( \mathbf{p} - \mathbf{k} \right)} + \\
&+ \frac{\psi^2_0}{\sqrt{M}} \sum_{\mathbf{k}, \mathbf{p}} J_{\mathbf{k}} \sum_{n m} \sqrt{n} \, c^0_{n - 1} \, \delta \hat{c}^{\dagger}_n{\left( \mathbf{k} \right)} \, \delta \hat{c}^{\dagger}_m{\left( \mathbf{p} \right)} \, \delta \hat{c}_{m}{\left( \mathbf{k} + \mathbf{p} \right)} + \\
&+ \frac{\psi^2_0}{\sqrt{M}} \sum_{\mathbf{k}, \mathbf{p}} J_{\mathbf{k}} \sum_{n m} \sqrt{n} \, c^0_n \, \delta \hat{c}_{n - 1}{\left( \mathbf{k} \right)} \, \delta \hat{c}^{\dagger}_m{\left( \mathbf{p} \right)} \, \delta \hat{c}_{m}{\left( \mathbf{p} - \mathbf{k} \right)}
\end{aligned}
$$
\begin{equation}\label{4.92}
\begin{aligned}
-J \sum_{\mathbf{r}, s{\left( \mathbf{r} \right)}} \delta_2 \, \hat{\psi}^{\dagger}{\left( \mathbf{r} \right)} \, \delta_1 \, \hat{\psi}{\left( s \right)} &= -J \sum_{\mathbf{r}, s{\left( \mathbf{r} \right)}} \sum_{n m} \sqrt{n m} \, c^0_{n - 1} \, \delta \hat{c}^{\dagger}_{m}{\left( \mathbf{r} \right)} \, \delta \hat{c}_{m - 1}{\left( \mathbf{r} \right)} \, \delta \hat{c}_n{\left( s \right)} + \\
&-J \sum_{\mathbf{r}, s{\left( \mathbf{r} \right)}} \sum_{n m} \sqrt{n m} \, c^0_n \, \delta \hat{c}^{\dagger}_{m}{\left( \mathbf{r} \right)} \, \delta \hat{c}_{m - 1}{\left( \mathbf{r} \right)} \, \delta \hat{c}_{n - 1}^{\dagger}{\left( s \right)} + \\
&+ J \, \psi^2_0 \sum_{\mathbf{r}, s{\left( \mathbf{r} \right)}} \sum_{n m} \sqrt{n} \, c^0_{n - 1} \, \delta \hat{c}^{\dagger}_m{\left( \mathbf{r} \right)} \, \delta \hat{c}_{m}{\left( \mathbf{r} \right)} \, \delta \hat{c}_n{\left( s \right)} + \\
&+ J \, \psi^2_0 \sum_{\mathbf{r}, s{\left( \mathbf{r} \right)}} \sum_{n m} \sqrt{n} \, c^0_n \, \delta \hat{c}^{\dagger}_m{\left( \mathbf{r} \right)} \, \delta \hat{c}_{m}{\left( \mathbf{r} \right)} \, \delta \hat{c}^{\dagger}_{n - 1}{\left( s \right)} = \\
&= -\frac{1}{\sqrt{M}} \sum_{\mathbf{k}, \mathbf{p}} J_{\mathbf{k}} \sum_{n m} \sqrt{n m} \, c^0_{n - 1} \, \delta \hat{c}^{\dagger}_m{\left( \mathbf{p} \right)} \, \delta \hat{c}_{m - 1}{\left( \mathbf{p} - \mathbf{k} \right)} \, \delta \hat{c}_n{\left( \mathbf{k} \right)} + \\
&-\frac{1}{\sqrt{M}} \sum_{\mathbf{k}, \mathbf{p}} J_{\mathbf{k}} \sum_{n m} \sqrt{n m} \, c^0_n \, \delta \hat{c}^{\dagger}_m{\left( \mathbf{p} \right)} \, \delta \hat{c}_{m - 1}{\left( \mathbf{k} + \mathbf{p} \right)} \, \delta \hat{c}^{\dagger}_{n - 1}{\left( \mathbf{k} \right)}+ \\
&+ \frac{\psi^2_0}{\sqrt{M}} \sum_{\mathbf{k}, \mathbf{p}} J_{\mathbf{k}} \sum_{n m} \sqrt{n} \, c^0_{n - 1} \, \delta \hat{c}^{\dagger}_m{\left( \mathbf{p} \right)} \, \delta \hat{c}_{m}{\left( \mathbf{p} - \mathbf{k} \right)} \, \delta \hat{c}_n{\left( \mathbf{k} \right)} + \\
&+ \frac{\psi^2_0}{\sqrt{M}} \sum_{\mathbf{k}, \mathbf{p}} J_{\mathbf{k}} \sum_{n m} \sqrt{n} \, c^0_n \, \delta \hat{c}^{\dagger}_m{\left( \mathbf{p} \right)} \, \delta \hat{c}_{m}{\left( \mathbf{k} + \mathbf{p} \right)} \, \delta \hat{c}^{\dagger}_{n - 1}{\left( \mathbf{k} \right)}
\end{aligned}
\end{equation} \\
where again (\ref{4.92}) is the hermitian conjugate operator of (\ref{4.91}). \\
Finally, the last contribution $\hat{H}^{\left( 3 \right)}_3$ has the form:
\begin{equation}\label{4.93}
\begin{aligned}
\hat{H}^{\left( 3 \right)}_3 &= - \frac{1}{2 \, \sqrt{M}} \sum_{\mathbf{k}, \mathbf{p}} \sum_{n m} H_n \, c^0_n \, \delta c^{\dagger}_n{\left( \mathbf{k} \right)} \, \delta c^{\dagger}_m{\left( \mathbf{p} \right)} \, \delta c_m{\left( \mathbf{k} + \mathbf{p} \right)} + \\
&- \frac{1}{2 \, \sqrt{M}} \sum_{\mathbf{k}, \mathbf{p}} \sum_{n m} H_n \, c^0_n \, \delta c^{\dagger}_m{\left( \mathbf{p} \right)} \, \delta c_m{\left( \mathbf{p} - \mathbf{k} \right)} \, \delta c_n{\left( \mathbf{k} \right)}
\end{aligned}
\end{equation}
After a straightforward calculation we can rewrite the third-order expansion of the Hamiltonian (\ref{4.88}) in a more familiar form:
\newpage
\begin{equation}\label{4.94}
\begin{aligned}
\hat{H}^{\left( 3 \right)} &= -\frac{1}{\sqrt{M}} \sum_{\mathbf{k}, \mathbf{p}} J_{\mathbf{k}} \sum_{\alpha} U_{\alpha, \mathbf{k}} \, \hat{b}^{\dagger}_{\alpha, \mathbf{k}} \sum_m \sqrt{m} \, \delta \hat{c}^{\dagger}_{m - 1}{\left( \mathbf{p} \right)} \, \delta \hat{c}_{m}{\left( \mathbf{k} + \mathbf{p} \right)} + \\
&- \frac{1}{\sqrt{M}} \sum_{\mathbf{k}, \mathbf{p}} J_{\mathbf{k}} \sum_{\alpha} V_{\alpha, \mathbf{k}} \, \hat{b}_{\alpha, -\mathbf{k}} \sum_m \sqrt{m} \, \delta \hat{c}^{\dagger}_{m - 1}{\left( \mathbf{p} \right)} \, \delta \hat{c}_m{\left( \mathbf{k} + \mathbf{p} \right)} + \\
&+ \frac{\psi^2_0}{\sqrt{M}} \sum_{\mathbf{k}, \mathbf{p}} J_{\mathbf{k}} \sum_{\alpha} U_{\alpha, \mathbf{k}} \, \hat{b}^{\dagger}_{\alpha, \mathbf{k}} \sum_m \delta \hat{c}^{\dagger}_m{\left( \mathbf{p} \right)} \, \delta \hat{c}_{m}{\left( \mathbf{k} + \mathbf{p} \right)} + \\
&+ \frac{\psi^2_0}{\sqrt{M}} \sum_{\mathbf{k}, \mathbf{p}} J_{\mathbf{k}} \sum_{\alpha} V_{\alpha, \mathbf{k}} \, \hat{b}_{\alpha, -\mathbf{k}} \sum_m \delta \hat{c}^{\dagger}_m{\left( \mathbf{p} \right)} \, \delta \hat{c}_m{\left( \mathbf{k} + \mathbf{p} \right)} + \\
&+ \frac{J z \, \psi_0}{2 \sqrt{M}} \sum_{\mathbf{k}, \mathbf{p}} \sum_{\alpha} U_{\alpha, \mathbf{k}} \, \hat{b}_{\alpha, -\mathbf{k}} \sum_m \delta \hat{c}^{\dagger}_m{\left( \mathbf{p} \right)} \, \delta \hat{c}_m{\left( \mathbf{k} + \mathbf{p} \right)} + \\
&+ \frac{J z \, \psi_0}{2 \sqrt{M}} \sum_{\mathbf{k}, \mathbf{p}} \sum_{\alpha} V_{\alpha, \mathbf{k}} \, \hat{b}^{\dagger}_{\alpha, \mathbf{k}} \sum_m \delta \hat{c}^{\dagger}_m{\left( \mathbf{p} \right)} \, \delta \hat{c}_m{\left( \mathbf{k} + \mathbf{p} \right)} + \\
&- \frac{1}{2 \sqrt{M}} \sum_{\mathbf{k}, \mathbf{p}} \sum_{\alpha} H_{\alpha, \mathbf{k}} \, \hat{b}^{\dagger}_{\alpha, \mathbf{k}} \sum_m \delta \hat{c}^{\dagger}_m{\left( \mathbf{p} \right)} \, \delta \hat{c}_m{\left( \mathbf{k} + \mathbf{p} \right)} + \\
&- \frac{1}{2 \sqrt{M}} \sum_{\mathbf{k}, \mathbf{p}} \sum_{\alpha} K_{\alpha, \mathbf{k}} \, \hat{b}_{\alpha, -\mathbf{k}} \sum_m \delta \hat{c}^{\dagger}_m{\left( \mathbf{p} \right)} \, \delta \hat{c}_m{\left( \mathbf{k} + \mathbf{p} \right)} + \\
&+ \text{h.c.} = \\
&= -\frac{1}{\sqrt{M}} \sum_{\mathbf{k}, \mathbf{p}} J_{\mathbf{k}} \sum_{\alpha} U_{\alpha, \mathbf{k}} \, \hat{b}^{\dagger}_{\alpha, \mathbf{k}} \sum_m \sqrt{m} \, \delta \hat{c}^{\dagger}_{m - 1}{\left( \mathbf{p} \right)} \, \delta \hat{c}_{m}{\left( \mathbf{k} + \mathbf{p} \right)} + \\
&- \frac{1}{\sqrt{M}} \sum_{\mathbf{k}, \mathbf{p}} J_{\mathbf{k}} \sum_{\alpha} V_{\alpha, \mathbf{k}} \, \hat{b}_{\alpha, -\mathbf{k}} \sum_m \sqrt{m} \, \delta \hat{c}^{\dagger}_{m - 1}{\left( \mathbf{p} \right)} \, \delta \hat{c}_m{\left( \mathbf{k} + \mathbf{p} \right)} + \\
&+ \frac{1}{2 \sqrt{M}} \sum_{\mathbf{k}, \mathbf{p}} \sum_{\alpha} \left( 2 \, J_{\mathbf{k}} \, \psi^2_0 \, U_{\alpha, \mathbf{k}} + J \, z \, \psi_0 \, V_{\alpha, \mathbf{k}} - H_{\alpha, \mathbf{k}} \right) \hat{b}^{\dagger}_{\alpha, \mathbf{k}} \sum_m \delta \hat{c}^{\dagger}_m{\left( \mathbf{k} \right)} \, \delta \hat{c}_m{\left( \mathbf{k} + \mathbf{p} \right)} + \\
&+ \frac{1}{2 \sqrt{M}} \sum_{\mathbf{k}, \mathbf{p}} \sum_{\alpha} \left( J \, z \, \psi_0 \, U_{\alpha, \mathbf{k}} + 2 \, J_{\mathbf{k}} \, \psi^2_0 \, V_{\alpha, \mathbf{k}} - K_{\alpha, \mathbf{k}} \right) \hat{b}_{\alpha, -\mathbf{k}} \sum_m \delta \hat{c}^{\dagger}_m{\left( \mathbf{p} \right)} \, \delta \hat{c}_m{\left( \mathbf{k} + \mathbf{p} \right)} + \\
&+ \text{h.c.}
\end{aligned}
\end{equation}
where we have dropped irrelevant terms deriving from the use of the bosonic commutation rules and we have defined:
\begin{equation}\label{4.95}
\begin{aligned}
&H_{\alpha, \mathbf{k}} = \sum_m H_m \, c^0_n \, u_{\alpha, \mathbf{k}, m} \\
&K_{\alpha, \mathbf{k}} = \sum_m H_m \, c^0_n \, v_{\alpha, \mathbf{k}, m}
\end{aligned}
\end{equation}
The particle-hole amplitudes $U_{\alpha, \mathbf{k}}$ and $V_{\alpha, \mathbf{k}}$, which are symmetric under the exchange $\mathbf{k} \to -\mathbf{k}$, have been introduced while analysing the first-order corrections to the Bose field operator $\delta_1 \psi{\left( \mathbf{r} \right)}$ and properties of the theory related to particle-hole symmetry.

\section{Decay processes}
Let us focus on decay processes associated to the vertices of the theory involving the excitation modes found by diagonalizing the Hamiltonian at the quadratic level. The related terms are given by combinations of two creation operators $\hat{b}^{\dagger}_{\alpha, \mathbf{k}}$ and one annihilation operator $\hat{b}_{\beta, \mathbf{p}}$. \\
Expanding all the contributions of equation (\ref{4.94}) and including also the hermitian conjugate terms, we obtain the following expression:
\begin{equation}\label{4.96}
\begin{aligned}
\hat{H}^{\left( 3 \right)}_D &= -\frac{1}{\sqrt{M}} \sum_{\mathbf{k}, \mathbf{p}} J_{\mathbf{k}} \sum_{\alpha, \beta, \gamma} U_{\alpha, \mathbf{k}} \, \hat{b}^{\dagger}_{\alpha, \mathbf{p}} \sum_m \sqrt{m} \bigg( u_{\beta, \mathbf{p}, m - 1} \, u_{\gamma, \mathbf{k} + \mathbf{p}, m} \, \hat{b}^{\dagger}_{\beta, \mathbf{p}} \, \hat{b}_{\gamma, \mathbf{k} + \mathbf{p}} + \\
&+ v_{\beta, -\mathbf{p}, m - 1} \, v_{\gamma, -\mathbf{k} - \mathbf{p}, m} \, \hat{b}_{\beta, -\mathbf{p}} \, \hat{b}^{\dagger}_{\gamma, -\mathbf{k} - \mathbf{p}} \bigg) + \\
&- \frac{1}{\sqrt{M}} \sum_{\mathbf{k}, \mathbf{p}} J_{\mathbf{k}} \sum_{\alpha} V_{\alpha, \mathbf{k}} \, \hat{b}_{\alpha, -\mathbf{k}} \sum_m \sqrt{m} \, u_{\beta, \mathbf{p}, m - 1} \, v_{\gamma, -\mathbf{k} - \mathbf{p}, m} \, \hat{b}^{\dagger}_{\beta, \mathbf{p}} \, \hat{b}^{\dagger}_{\gamma, -\mathbf{k} - \mathbf{p}} + \\
&+ \frac{1}{2 \sqrt{M}} \sum_{\mathbf{k}, \mathbf{p}} \sum_{\alpha} \left( 2 \, J_{\mathbf{k}} \, \psi^2_0 \, U_{\alpha, \mathbf{k}} + J \, z \, \psi_0 \, V_{\alpha, \mathbf{k}} - H_{\alpha, \mathbf{k}} \right) \hat{b}^{\dagger}_{\alpha, \mathbf{k}} \sum_m \bigg( u_{\beta, \mathbf{p}, m} \, u_{\gamma, \mathbf{k} + \mathbf{p}, m} \, \hat{b}^{\dagger}_{\beta, \mathbf{p}} \, \hat{b}_{\gamma, \mathbf{k} + \mathbf{p}} + \\
&+ v_{\beta, -\mathbf{p}, m} \, v_{\gamma, -\mathbf{k} - \mathbf{p}, m} \, \hat{b}_{\beta, -\mathbf{p}} \, \hat{b}^{\dagger}_{\gamma, -\mathbf{k} - \mathbf{p}} \bigg) + \\
&+ \frac{1}{2 \sqrt{M}} \sum_{\mathbf{k}, \mathbf{p}} \sum_{\alpha} \left( J \, z \, \psi_0 \, U_{\alpha, \mathbf{k}} + 2 \, J_{\mathbf{k}} \, \psi^2_0 \, V_{\alpha, \mathbf{k}} - K_{\alpha, \mathbf{k}} \right) \hat{b}_{\alpha, -\mathbf{k}} \sum_m \, u_{\beta, \mathbf{p}, m} \, v_{\gamma, -\mathbf{k} - \mathbf{p}, m} \, \hat{b}^{\dagger}_{\beta, \mathbf{p}} \, \hat{b}^{\dagger}_{\gamma, -\mathbf{k} - \mathbf{p}} + \\
&- \frac{1}{\sqrt{M}} \sum_{\mathbf{k}, \mathbf{p}} J_{\mathbf{k}} \sum_{\alpha, \beta, \gamma} U_{\alpha, \mathbf{k}} \, \sum_m \sqrt{m} \, u_{\gamma, \mathbf{k} + \mathbf{p}, m} \, v_{\beta, -\mathbf{p}, m - 1} \, \hat{b}^{\dagger}_{\gamma, \mathbf{k} + \mathbf{p}} \, \hat{b}^{\dagger}_{\beta, -\mathbf{p}} \, \hat{b}_{\alpha, \mathbf{k}} + \\
&- \frac{1}{\sqrt{M}} \sum_{\mathbf{k}, \mathbf{p}} J_{\mathbf{k}} \sum_{\alpha} V_{\alpha, \mathbf{k}} \sum_m \sqrt{m} \bigg( u_{\gamma, \mathbf{k} + \mathbf{p}, m} \, u_{\beta, \mathbf{p}, m - 1} \, \hat{b}^{\dagger}_{\gamma, \mathbf{k} + \mathbf{p}} \, \hat{b}_{\beta, \mathbf{p}} + \\
&+ v_{\gamma, -\mathbf{k} - \mathbf{p}, m} \, v_{\beta, -\mathbf{p}, m - 1} \, \hat{b}_{\gamma, -\mathbf{k} - \mathbf{p}} \, \hat{b}^{\dagger}_{\beta, -\mathbf{p}} \bigg) \hat{b}^{\dagger}_{\alpha, -\mathbf{k}} + \\
&+ \frac{1}{2 \sqrt{M}} \sum_{\mathbf{k}, \mathbf{p}} \sum_{\alpha} \left( 2 \, J_{\mathbf{k}} \, \psi^2_0 \, U_{\alpha, \mathbf{k}} + J \, z \, \psi_0 \, V_{\alpha, \mathbf{k}} - H_{\alpha, \mathbf{k}} \right) \sum_m u_{\gamma, \mathbf{k} + \mathbf{p}, m} \, v_{\beta, -\mathbf{p}, m} \, \hat{b}^{\dagger}_{\gamma, \mathbf{k} + \mathbf{p}} \, \hat{b}^{\dagger}_{\beta, -\mathbf{p}} \, \hat{b}_{\alpha, \mathbf{k}} + \\
&+ \frac{1}{2 \sqrt{M}} \sum_{\mathbf{k}, \mathbf{p}} \sum_{\alpha} +\left( J \, z \, \psi_0 \, U_{\alpha, \mathbf{k}} + 2 \, J_{\mathbf{k}} \, \psi^2_0 \, V_{\alpha, \mathbf{k}} - K_{\alpha, \mathbf{k}} \right) \sum_m \bigg( u_{\gamma, \mathbf{k} + \mathbf{p}, m} \, u_{\beta, \mathbf{p}, m} \, \hat{b}^{\dagger}_{\gamma, \mathbf{k} + \mathbf{p}} \, \hat{b}_{\beta, \mathbf{p}} + \\
&+ v_{\gamma, -\mathbf{k} - \mathbf{p}, m} \, v_{\beta, -\mathbf{p}, m} \, \hat{b}_{\gamma, -\mathbf{k} - \mathbf{p}} \, \hat{b}^{\dagger}_{\beta, -\mathbf{p}} \bigg) \hat{b}^{\dagger}_{\alpha, -\mathbf{k}}
\end{aligned}
\end{equation}
Introducing the notation:
\begin{equation}\label{4.97}
\begin{aligned}
&U^{\alpha \beta}_{\mathbf{k}, \mathbf{p}} = \sum_m u_{\alpha, \mathbf{k}, m} \, u_{\beta, \mathbf{p}, m} \\
&V^{\alpha \beta}_{\mathbf{k}, \mathbf{p}} = \sum_m v_{\alpha, \mathbf{k}, m} \, v_{\beta, \mathbf{p}, m} \\
&W^{\alpha \beta}_{\mathbf{k}, \mathbf{p}} = \sum_m u_{\alpha, \mathbf{k}, m} \, v_{\beta, \mathbf{p}, m} \\
\end{aligned}
\end{equation}
$$
\begin{aligned}
&\overline{U}^{\alpha \beta}_{\mathbf{k}, \mathbf{p}} = \sum_m \sqrt{m} \, u_{\alpha, \mathbf{k}, m - 1} \, u_{\beta, \mathbf{p}, m} \\
&\overline{V}^{\alpha \beta}_{\mathbf{k}, \mathbf{p}} = \sum_m \sqrt{m} \, v_{\alpha, \mathbf{k}, m - 1} \, v_{\beta, \mathbf{p}, m} \\
&\overline{W}^{\alpha \beta}_{\mathbf{k}, \mathbf{p}} = \sum_m \sqrt{m} \, u_{\alpha, \mathbf{k}, m - 1} \, v_{\beta, \mathbf{p}, m} \\
&\underline{W}^{\alpha \beta}_{\mathbf{k}, \mathbf{p}} = \sum_m \sqrt{m} \, u_{\alpha, \mathbf{k}, m} \, v_{\beta, \mathbf{p}, m - 1}
\end{aligned}
$$
we obtain:
\begin{equation}\label{4.98}
\begin{aligned}
\hat{H}^{\left( 3 \right)}_D &= -\frac{1}{\sqrt{M}} \sum_{\mathbf{k}, \mathbf{p}} J_{\mathbf{k}} \sum_{\alpha, \beta, \gamma} U_{\alpha, \mathbf{k}} \, \hat{b}^{\dagger}_{\alpha, \mathbf{k}} \bigg( \overline{U}^{\beta \gamma}_{\mathbf{p}, \mathbf{k} + \mathbf{p}} \, \hat{b}^{\dagger}_{\beta, \mathbf{p}} \, \hat{b}_{\gamma, \mathbf{k} + \mathbf{p}} + \overline{V}^{\beta \gamma}_{\mathbf{p}, \mathbf{k} + \mathbf{p}} \, \hat{b}_{\beta, -\mathbf{p}} \, \hat{b}^{\dagger}_{\gamma, -\mathbf{k} - \mathbf{p}} \bigg) + \\
&- \frac{1}{\sqrt{M}} \sum_{\mathbf{k}, \mathbf{p}} J_{\mathbf{k}} \sum_{\alpha, \beta, \gamma} V_{\alpha, \mathbf{k}} \, \overline{W}^{\beta \gamma}_{\mathbf{p}, \mathbf{k} + \mathbf{p}} \, \hat{b}_{\alpha, -\mathbf{k}} \, \hat{b}^{\dagger}_{\beta, \mathbf{p}} \, \hat{b}^{\dagger}_{\gamma, -\mathbf{k} - \mathbf{p}} +
\end{aligned}
\end{equation}
$$
\begin{aligned}
&+ \frac{1}{2 \sqrt{M}} \sum_{\mathbf{k}, \mathbf{p}} \sum_{\alpha, \beta, \gamma} \left( 2 \, J_{\mathbf{k}} \, \psi^2_0 \, U_{\alpha, \mathbf{k}} + J \, z \, \psi_0 \, V_{\alpha, \mathbf{k}} - H_{\alpha, \mathbf{k}} \right) \hat{b}^{\dagger}_{\alpha, \mathbf{k}} \bigg( U^{\beta \gamma}_{\mathbf{p}, \mathbf{k} + \mathbf{p}} \, \hat{b}^{\dagger}_{\beta, \mathbf{p}} \, \hat{b}_{\gamma, \mathbf{k} + \mathbf{p}} + \\
&+ V^{\beta \gamma}_{\mathbf{p}, \mathbf{k} + \mathbf{p}} \, \hat{b}_{\beta, -\mathbf{p}} \, \hat{b}^{\dagger}_{\gamma, -\mathbf{k} - \mathbf{p}} \bigg) + \\
&+ \frac{1}{2 \sqrt{M}} \sum_{\mathbf{k}, \mathbf{p}} \sum_{\alpha, \beta, \gamma} \left( J \, z \, \psi_0 \, U_{\alpha, \mathbf{k}} + 2 \, J_{\mathbf{k}} \, \psi^2_0 \, V_{\alpha, \mathbf{k}} - K_{\alpha, \mathbf{k}} \right) \, W^{\beta \gamma}_{\mathbf{p}, \mathbf{k} + \mathbf{p}} \, \hat{b}_{\alpha, -\mathbf{k}} \, \hat{b}^{\dagger}_{\beta, \mathbf{p}} \, \hat{b}^{\dagger}_{\gamma, -\mathbf{k} - \mathbf{p}} + \\
&- \frac{1}{\sqrt{M}} \sum_{\mathbf{k}, \mathbf{p}} J_{\mathbf{k}} \sum_{\alpha, \beta, \gamma} U_{\alpha, \mathbf{k}} \, \underline{W}^{\gamma, \beta}_{\mathbf{k} + \mathbf{p}, \mathbf{p}} \, \hat{b}^{\dagger}_{\gamma, \mathbf{k} + \mathbf{p}} \, \hat{b}^{\dagger}_{\beta, -\mathbf{p}} \, \hat{b}_{\alpha, \mathbf{k}} + \\
&- \frac{1}{\sqrt{M}} \sum_{\mathbf{k}, \mathbf{p}} J_{\mathbf{k}} \sum_{\alpha} V_{\alpha, \mathbf{k}} \sum_m \bigg( \overline{U}^{\beta, \gamma}_{\mathbf{p}, \mathbf{k} + \mathbf{p}} \, \hat{b}^{\dagger}_{\gamma, \mathbf{k} + \mathbf{p}} \, \hat{b}_{\beta, \mathbf{p}} + \overline{V}^{\beta, \gamma}_{\mathbf{p}, \mathbf{k} + \mathbf{p}} \, \hat{b}_{\gamma, -\mathbf{k} - \mathbf{p}} \, \hat{b}^{\dagger}_{\beta, -\mathbf{p}} \bigg) \hat{b}^{\dagger}_{\alpha, -\mathbf{k}} + \\
&+ \frac{1}{2 \sqrt{M}} \sum_{\mathbf{k}, \mathbf{p}} \sum_{\alpha} \left( 2 \, J_{\mathbf{k}} \, \psi^2_0 \, U_{\alpha, \mathbf{k}} + J \, z \, \psi_0 \, V_{\alpha, \mathbf{k}} - H_{\alpha, \mathbf{k}} \right) W^{\gamma, \beta}_{\mathbf{k} + \mathbf{p}, \mathbf{p}} \, \hat{b}^{\dagger}_{\gamma, \mathbf{k} + \mathbf{p}} \, \hat{b}^{\dagger}_{\beta, -\mathbf{p}} \, \hat{b}_{\alpha, \mathbf{k}} + \\
&+ \frac{1}{2 \sqrt{M}} \sum_{\mathbf{k}, \mathbf{p}} \sum_{\alpha} \left( J \, z \, \psi_0 \, U_{\alpha, \mathbf{k}} + 2 \, J_{\mathbf{k}} \, \psi^2_0 \, V_{\alpha, \mathbf{k}} - K_{\alpha, \mathbf{k}} \right) \bigg( U^{\beta, \gamma}_{\mathbf{p}, \mathbf{k} + \mathbf{p}} \, \hat{b}^{\dagger}_{\gamma, \mathbf{k} + \mathbf{p}} \, \hat{b}_{\beta, \mathbf{p}} + \\
&+ V^{\beta, \gamma}_{\mathbf{p}, \mathbf{k} + \mathbf{p}} \, \hat{b}_{\gamma, -\mathbf{k} - \mathbf{p}} \, \hat{b}^{\dagger}_{\beta, -\mathbf{p}} \bigg) \hat{b}^{\dagger}_{\alpha, -\mathbf{k}}
\end{aligned}
$$
Renaming the momenta and modes indices of expression (\ref{4.98}) appropriately, we can rewrite it in the following compact form:
\begin{equation}\label{4.99}
\begin{aligned}
\hat{H}^{\left( 3 \right)}_D &= \frac{1}{2 \sqrt{M}} \sum_{\alpha, \beta, \gamma} \sum_{\mathbf{k}, \mathbf{p}} \bigg[ \left( J \, z \, \psi_0 \, U_{\alpha, \mathbf{k}} + 2 \, J_{\mathbf{k}} \, \psi^2_0 \, V_{\alpha, \mathbf{k}} - K_{\alpha, \mathbf{k}} \right) W^{\beta \gamma}_{\mathbf{p}, \mathbf{k} - \mathbf{p}} - 2 \, J_{\mathbf{k}} \, V_{\alpha, \mathbf{k}} \, \overline{W}^{\beta \gamma}_{\mathbf{p}, \mathbf{k} - \mathbf{p}} + \\
&- 2 \, J_{\mathbf{k} - \mathbf{p}} \, U_{\gamma, \mathbf{k} - \mathbf{p}} \, \overline{U}^{\beta \alpha}_{\mathbf{p}, \mathbf{k}} - 2 \, J_{\mathbf{p}} \, U_{\beta, \mathbf{p}} \, \overline{V}^{\alpha \gamma}_{\mathbf{k}, \mathbf{k} - \mathbf{p}} + \\
&+ \left( 2 \, J_{\mathbf{k} - \mathbf{p}} \, \psi^2_0 \, U_{\gamma, \mathbf{k} - \mathbf{p}} + J \, z \, \psi_0 \, V_{\gamma, \mathbf{k} - \mathbf{p}} - H_{\gamma, \mathbf{k} - \mathbf{p}} \right) U^{\beta \alpha}_{\mathbf{p}, \mathbf{k}} + \\
&+ \left( 2 \, J_{\mathbf{p}} \, \psi^2_0 \, U_{\beta, \mathbf{p}} + J \, \, z \, \psi_0 \, V_{\beta, \mathbf{p}} - H_{\beta, \mathbf{p}} \right) V^{\alpha \gamma}_{\mathbf{k}, \mathbf{k} - \mathbf{p}} + \\
\end{aligned}
\end{equation}
$$
\begin{aligned}
&- 2 \, J_{\mathbf{k}} \, U_{\alpha, \mathbf{k}} \, \underline{W}^{\gamma \beta}_{\mathbf{k} - \mathbf{p}, \mathbf{p}} - 2 \, J_{\mathbf{p}} \, V_{\beta, \mathbf{p}} \, \overline{U}^{\alpha, \gamma}_{\mathbf{k}, \mathbf{k} - \mathbf{p}} - 2 \, J_{\mathbf{k} - \mathbf{p}} \, V_{\gamma, \mathbf{k} - \mathbf{p}} \, \overline{V}^{\beta, \alpha}_{\mathbf{p}, \mathbf{k}} + \\
&+ \left( 2 \, J_{\mathbf{k}} \, \psi^2_0 \, U_{\alpha, \mathbf{k}} + J \, z \, \psi_0 \, V_{\alpha, \mathbf{k}} - H_{\alpha, \mathbf{k}} \right) W^{\gamma \beta}_{\mathbf{k} - \mathbf{p}, \mathbf{p}} + \\
&+ \left( J \, z \, \psi_0 \, U_{\beta, \mathbf{p}} + 2 \, J_{\mathbf{p}} \, \psi^2_0 \, V_{\beta, \mathbf{p}} - K_{\beta, \mathbf{p}} \right) U^{\alpha \gamma}_{\mathbf{k}, \mathbf{k} - \mathbf{p}} + \\
&+ \left( J \, z \, \psi_0 \, U_{\gamma, \mathbf{k} - \mathbf{p}} + 2 \, J_{\mathbf{k} - \mathbf{p}} \, \psi^2_0 \, V_{\gamma, \mathbf{k} - \mathbf{p}} - K_{\gamma, \mathbf{k} - \mathbf{p}} \right) V^{\beta, \alpha}_{\mathbf{p}, \mathbf{k}} \bigg] \hat{b}^{\dagger}_{\beta, \mathbf{p}} \, \hat{b}^{\dagger}_{\gamma, \mathbf{k} - \mathbf{p}} \, \hat{b}_{\alpha, \mathbf{k}} = \\
&= \frac{1}{2 \sqrt{M}} \sum_{\alpha, \beta, \gamma} \sum_{\mathbf{k}, \mathbf{p}} H^{i f}_{\alpha, \beta, \gamma}{\left( \mathbf{k}, \mathbf{p} \right)} \, \hat{b}^{\dagger}_{\beta, \mathbf{p}} \, \hat{b}^{\dagger}_{\gamma, \mathbf{k} - \mathbf{p}} \, \hat{b}_{\alpha, \mathbf{k}}
\end{aligned}
$$
The central aim of this machinery is the determination of the life time of a given decaying excitation $\alpha$ with respect to the general decay channel $\alpha_{\mathbf{k}} \to \beta_{\mathbf{p}} + \gamma_{\mathbf{k} - \mathbf{p}}$. \\
The zero-order total probability amplitude of this kind of process is provided by the well-known Fermi golden rule:
\begin{equation}\label{4.100}
\begin{aligned}
P_{\alpha, \beta, \gamma}{\left( \mathbf{k} \right)} &= \left( 1 - \frac{1}{2} \, \delta_{\beta \gamma} \right) \frac{1}{4 M} \sum_{\mathbf{p}} \left| H^{i f}_{\alpha, \beta, \gamma}{\left( \mathbf{k}, \mathbf{p} \right)} \right|^2 2 \pi \, \delta{\left( \omega_{\beta, \mathbf{p}} + \omega_{\gamma, \mathbf{k} - \mathbf{p}} - \omega_{\alpha, \mathbf{k}} \right)} =
\end{aligned}
\end{equation}
$$
\begin{aligned}
&= \frac{\pi}{2 M} \left( 1 - \frac{1}{2} \, \delta_{\beta \gamma} \right) \sum_{\mathbf{p}} \left| H^{i f}_{\alpha, \beta, \gamma}{\left( \mathbf{k}, \mathbf{p} \right)} \right|^2 \, \delta{\left( \omega_{\beta, \mathbf{p}} + \omega_{\gamma, \mathbf{k} -\mathbf{p}} - \omega_{\alpha, \mathbf{k}} \right)}
\end{aligned}
$$
where the factor $1/2$ accounts for the momentum double counting when the product excitations belong to the same quasi-particle species and the $\delta$-function in the sum over momenta imposes the conservation of energy. If $\beta = \gamma$ the expression of $H^{i f}_{\alpha, \beta, \gamma}{\left( \mathbf{k}, \mathbf{p} \right)}$ needs also to be symmetrized with respect to the momenta $\mathbf{p}$ and $\mathbf{k} - \mathbf{p}$. \\
The life time of the $\alpha$-excitation will be given by the inverse the quantity (\ref{4.100}) and is expected to be a function of the relevant Hamiltonian parameters of the theory.

\section{Life time of the Higgs excitation}
Since the argument of the sum in (\ref{4.100}) includes a Dirac delta function which imposes the conservation of energy, the discrete momentum integration can be numerically attacked by the use of its Gaussian distribution representation:
\begin{equation}\label{4.101}
\small
P_{\alpha, \beta, \gamma}{\left( \mathbf{k} \right)} = \frac{\pi}{2 M} \left( 1 - \frac{1}{2} \, \delta_{\beta \gamma} \right) \lim_{\varepsilon \to 0} \frac{1}{\sqrt{2 \pi} \varepsilon} \sum_{\mathbf{p}} \left| H^{i f}_{2, 1, 1}{\left( \mathbf{k}, \mathbf{p} \right)} \right|^2 \, \exp{\left[ -\frac{\left( \omega_{\beta, \mathbf{p}} + \omega_{\gamma, \mathbf{k} - \mathbf{p}} - \omega_{\alpha, \mathbf{k}} \right)^2}{2 \, \varepsilon^2} \right]}
\end{equation}
where small parameter $\varepsilon$ has to be chosen carefully: it has to be sufficiently small with respect to the energy scale separating two different modes, but also large enough when compared to the momentum discretization due to the finite size description, in order to obtain an optimized resolution.

\vspace{0.5cm}

The formulas derived above have been applied to the simple but relevant case of the decay of the Higgs excitation into two Goldstone modes. The calculation of respective probability amplitude has been performed over different portions of the phase diagram around the Mott lobe, with the additional aim of identifying the regions where the Higgs excitation is likely to observed experimentally. \\
For simplicity we have considered the decay process of a Higgs excitation with $\mathbf{k} = \mathbf{0}$ and, after symmetrizing the expression of $H^{i f}_{\alpha, \beta, \gamma}{\left( \mathbf{k}, \mathbf{p} \right)}$ with respect to the outgoing momenta, we have performed the calculation of:
\begin{equation}\label{4.102}
\Gamma \equiv P_{2, 1, 1}{\left( \mathbf{0} \right)} = \frac{\pi}{4 M} \lim_{\varepsilon \to 0} \frac{1}{\sqrt{2 \pi} \varepsilon} \sum_{\mathbf{p}} \left| H^{i f}_{2, 1, 1}{\left( \mathbf{0}, \mathbf{p} \right)} \right|^2 \, \exp{\left[ -\frac{\left( \omega_{1, \mathbf{p}} + \omega_{1, -\mathbf{p}} - \omega_{2, \mathbf{0}} \right)^2}{2 \, \varepsilon^2} \right]}
\end{equation}
by choosing the Gaussian width of integration through the following rule of thumb:
\begin{equation}\label{4.103}
\varepsilon \approx \left| \mathbf{v}_{group} \right| \cdot \left| \Delta \mathbf{k} \right| = \left| \frac{\dpar \, \omega_{1, \mathbf{p}}}{\dpar \, \mathbf{p}} \right| \cdot \left| \Delta \mathbf{k} \right|
\end{equation}
where $\mathbf{v}_{group}$ is the group velocity of the outgoing Goldstone excitations and $\left| \Delta \mathbf{k} \right|$ is the momentum discretization interval of the Brillouin zone. \\
In the thermodynamic limit ($M \to \infty$) the vertex rate (\ref{4.101}) turns into a $d$-dimensional integral:
\begin{equation}\label{4.104}
\Gamma = \frac{1}{4 \, \pi^{d - 1}} \lim_{\varepsilon \to 0} \frac{1}{\sqrt{2 \pi} \varepsilon} \int d\mathbf{p} \, \left| H^{i f}_{2, 1, 1}{\left( \mathbf{0}, \mathbf{p} \right)} \right|^2 \, \exp{\left[ -\frac{\left( \omega_{1, \mathbf{p}} + \omega_{1, -\mathbf{p}} - \omega_{2, \mathbf{0}} \right)^2}{2 \, \varepsilon^2} \right]}
\end{equation}
In principle $\Gamma$ is expected to be a function of the Hamiltonian parameters of the model depending on the dimensionality of the system. In our formalism the Bogoliubov coefficients $u_{\alpha, \mathbf{k}, n}$, $v_{\alpha, \mathbf{k}, n}$ and spectra $\omega_{\alpha, \mathbf{k}}$ depend on the excitation momentum only through the reduced quantity:
\begin{equation}\label{4.105}
x = \left[ \frac{1}{d} \sum_{a = 1}^{d} \sin^2{\left( \frac{k_a}{2} \right)} \right]^{1/2} \in \left[ 0, 1 \right]
\end{equation}
for every dimensionality. Therefore, we can transform the momentum integral (\ref{4.104}) into:
\begin{equation}\label{4.106}
\Gamma = \frac{1}{4 \, \pi^{d - 1}} \lim_{\varepsilon \to 0} \frac{1}{\sqrt{2 \pi} \varepsilon} \int dx \, \rho_d{\left( x \right)} \, \left| H^{i f}_{2, 1, 1}{\left( 0, x \right)} \right|^2 \, \exp{\left[ -\frac{\left( 2 \, \omega_{1, x} - \omega_{2, 0} \right)^2}{2 \, \varepsilon^2} \right]}
\end{equation}
where $\rho_d{\left( x \right)}$ is the density of momentum states for a given value of $x$. This quantity can be easily determined from the general equation:
\begin{equation}\label{4.107}
\rho_d{\left( x \right)} = \int d^d p \ \delta{\left[ x - x{\left( p \right)} \right]} = \int d^d p \ \delta{\left\{ x - \left[ \frac{1}{d} \sum_{a = 1}^{d} \sin^2{\left( \frac{p_a}{2} \right)} \right]^{1/2} \right\}}
\end{equation}
Introducing the variable change $p_a \to y_a = \sin{\left( p_a/2\right)}$ for every momentum component and calculating the respective Jacobian determinant, the integral (\ref{4.107}) can be rewritten in the form:
\begin{equation}\label{4.108}
\rho_d{\left( x \right)} = 2^d \int d^d y \ \prod_{a = 1}^{d} \frac{1}{\sqrt{1 - y^2_a}} \ \delta{\left( x - \frac{\left| y \right|}{\sqrt{d}} \right)}
\end{equation}
The $d$-dimensional integral (\ref{4.108}) can be easily solved in 1D and leads to:
\newpage
\begin{equation}\label{4.109}
\rho_1{\left( x \right)} = \frac{4}{\sqrt{1 - x^2}}
\end{equation}
Since we are particularly interested in the Higgs mode decay around the $O{\left( 2 \right)}$-invariant points of the transition, where this process has low-energy character, for any arbitrary dimension we can perform a low-momentum expansion of the integrand in (\ref{4.108}):
\begin{equation}\label{4.110}
\begin{aligned}
\rho_d{\left( x \right)} &= 2^d \int d^d y \ \prod_{a = 1}^{d} \frac{1}{\sqrt{1 - y^2_a}} \ \delta{\left( x - \frac{\left| y \right|}{\sqrt{d}} \right)} \approx 2^d \int d^d y \ \left( 1 + \frac{1}{2} y^2 \right) \ \delta{\left( x - \frac{\left| y \right|}{\sqrt{d}} \right)} = \\
&= 2^d \, S_d \, \sqrt{d} \left( \sqrt{d} \, x \right)^{d - 1} \left( 1 + \frac{d}{2} \, x^2 \right)
\end{aligned}
\end{equation}
where $S_d$ is the hypersurface of the $d$-dimensional sphere. It follows that:
\begin{equation}\label{4.111}
\Gamma = \frac{1}{4 \, \pi^{d - 1}} \lim_{\varepsilon \to 0} \frac{1}{\sqrt{2 \pi} \varepsilon} \int dx \, \rho_d{\left( x \right)} \, \left| H^{i f}_{2, 1, 1}{\left( 0, x \right)} \right|^2 \, \exp{\left[ -\frac{\left( 2 \, \omega_{1, x} - \omega_{2, 0} \right)^2}{2 \, \varepsilon^2} \right]}
\end{equation}
is expected to present different behaviours in the low-energy limit with respect to the dimensionality of the system. In particular, approximating the Higgs mode gap as $\Delta_H \approx 2 \times \left( 2 \, \sqrt{d} \, c^{tip}_s \, x \right)$, the relative damping rate:
\begin{equation}\label{4.112}
\Gamma_r \equiv \frac{\Gamma}{\Delta_H} \sim x^{d - 2} \, \left| H^{i f}_{2, 1, 1}{\left( 0, x \right)} \right|^2
\end{equation}
is expected to diverge at the $O{\left( 2 \right)}$-criticality when the vertex amplitude $\left| H^{i f}_{2, 1, 1}{\left( 0, x \right)} \right|^2$ diverges as
\begin{equation}\label{4.113}
\left| H^{i f}_{2, 1, 1}{\left( 0, x \right)} \right|^2 \approx A_h \, x^{-\alpha}
\end{equation}
with $\alpha > d - 2$, hence for $d < \alpha + 2$.

\section{Numerical calculations and results}
\begin{figure}[!htp]
	\centering
	\includegraphics[width=10cm]{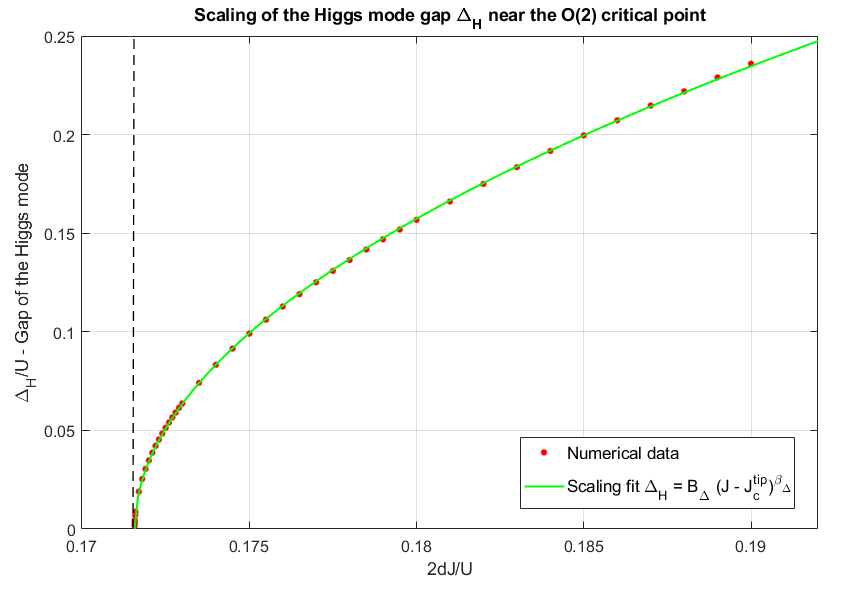}
	\caption{Scaling behaviour of the gap of the Higgs mode $\Delta_H$ in the vicinity of the Lorentz-invariant critical point located at the tip of the first Mott lobe. The black dashed line indicates the critical value of the hopping strength $(2 d J/U)^{tip}_c$.}
	\label{Figure67}
\end{figure}

Adopting the low-energy approximation $\overline{x} \approx \Delta_H/\left( 4 \, \sqrt{d} \, c^{tip}_s \right)$ for the outgoing momentum of the Goldstone excitations, equation (\ref{4.111}) leads to:
\begin{equation}\label{4.114}
\begin{aligned}
\Gamma &= \frac{1}{4 \, \pi^{d - 1}} \rho_d{\left( \frac{\Delta_H}{4 \, \sqrt{d} \, c^{tip}_s} \right)} \left| H^{i f}_{2, 1, 1}{\left( 0, \frac{\Delta_H}{4 \, \sqrt{d} \, c^{tip}_s} \right)} \right|^2 \approx \\
&\approx \frac{d^{1/2}}{2^{d - 1} \, \pi^{d/2 - 1} \, \Gamma{\left( \frac{d}{2} \right)}} \left( \frac{\Delta_H}{c^{tip}_s} \right)^{d - 1} \left[ 1 + \frac{1}{2} \left( \frac{\Delta_H}{4 \, c^{tip}_s} \right)^2 \right] \left| H^{i f}_{2, 1, 1}{\left( 0, \frac{\Delta_H}{4 \, \sqrt{d} \, c^{tip}_s} \right)} \right|^2
\end{aligned}
\end{equation}
where the sound velocity is given by equation (\ref{3.41}):
\begin{equation}\label{4.115}
c^{tip}_s = \frac{U}{\hbar} \sqrt{\left( \frac{J}{U} \right)^{tip}_c} \left[ \overline{n} \left( \overline{n} + 1 \right) \right]^{1/4}
\end{equation}
\newpage
and the energy gap of the Higgs mode $\Delta_H$ can be calculated numerically through the usual diagonalization procedure. For practical purposes we will consider the filling case $\overline{n} = 1$. \\
Let us suppose that the Higgs gap scales as a power law around the critical point:
\begin{equation}\label{4.116}
\Delta_H{\left( \overline{J} \right)} \approx B_{\Delta} \left( \overline{J} - \overline{J}_c \right)^{\beta_{\Delta}}
\end{equation}
where we have introduced the notation $\overline{J} \equiv 2 d J/U$. Taking into account the scaling law (\ref{4.104}) and expression (\ref{4.114}), the relative damping rate (\ref{4.112}) becomes:
\begin{equation}\label{4.117}
\Gamma^d_r{\left( \overline{J} \right)} = \frac{A_h \, B^{d - 2 - \alpha}_{\Delta} \, d^{d/2}}{2^{\left( 3 \, d - 7 \, \alpha - 3 \right)/4} \, \pi^{d/2 - 1} \Gamma{\left( \frac{d}{2} \right)}} \frac{\left( \overline{J} - \overline{J}^{tip}_c \right)^{\left( d - 2 - \alpha \right) \beta_{\Delta}}}{\overline{J}^{\left( d - 1 - \alpha \right)/2}} \left[ 1 + \frac{B^2_{\Delta} \, d}{2^{9/2}} \frac{\left( \overline{J} - \overline{J}^{tip}_c \right)^{2 \, \beta_{\Delta}}}{\overline{J}} \right]
\end{equation}
Since we expect that $\beta_{\Delta} > 0$, for $d \neq 2 + \alpha$ expression (\ref{4.117}) has the form:
\begin{equation}\label{4.118}
\Gamma^d_r{\left( \overline{J} \right)} = \frac{A_h \, B^{d - 2 - \alpha}_{\Delta} \, d^{d/2}}{2^{\left( 3 \, d - 7 \, \alpha - 3 \right)/4} \, \pi^{d/2 - 1} \Gamma{\left( \frac{d}{2} \right)}} \frac{\left( \overline{J} - \overline{J}^{tip}_c \right)^{\left( d - 2 - \alpha \right) \beta_{\Delta}}}{\overline{J}^{\left( d - 1 - \alpha \right)/2}} \quad , \quad d \neq 2 + \alpha \quad , \quad \overline{J} \to \overline{J}^+_c
\end{equation}
while for $d = 2 + \alpha$ the lowest-order expansion of (\ref{4.117}) reads:
\begin{equation}\label{4.119}
\Gamma^{d = 2 + \alpha}_r{\left( \overline{J} \right)} = \frac{A_h \, \left( \alpha + 2 \right)^{\left( \alpha + 2 \right)/2}}{2^{-\alpha + 3/4} \, \pi^{\alpha/2} \Gamma{\left( \frac{\alpha}{2} + 1 \right)}} \frac{1}{\overline{J}^{1/2}} \left[ 1 + \frac{\left( \alpha + 2 \right) B^2_{\Delta}}{2^{9/2}} \frac{\left( \overline{J} - \overline{J}^{tip}_c \right)^{2 \, \beta_{\Delta}}}{\overline{J}} \right] \quad , \quad \overline{J} \to \overline{J}^+_c
\end{equation}
As a first step, we can estimate the scaling behaviour of the gap of the Higgs modes $\Delta_H$ by a simple fitting procedure with respect to equation (\ref{4.116}), whose results are reported in Figure \ref{Figure67}.
\newpage
The fit parameters $B_{\Delta}$ and $\beta_{\Delta}$ have been determined within a 95\% interval of confidence:
\begin{equation}\label{4.120}
B_{\Delta} = 1.81 \pm 0.01 \quad , \quad \beta_{\Delta} = 0.511 \pm 0.002
\end{equation}
In accordance with \cite{altman_auerbach, rokhsar_kotliar} and the 2D experimental data \cite{lattice_depth_modulation_2} the Higgs gap presents a square root scaling with respect the relative hopping strength at the commensurate-density critical point. \\
Secondly, we have determined the relative damping rate $\Gamma_r$ through the application of equation (\ref{4.114}) within a full numerical treatment. Such calculations have been performed at parametric points very close to the tip of the Mott lobe, so that their accuracy is limited only by the numerical instability exhibited by the diagonalization outcomes due to the singularity of the Goldstone mode at $\left| \mathbf{k} \right| \approx 0$. Actually, this limitation restricts also the set of parametric points that we can choose close to the tip of the Mott lobe.
\begin{figure}[!htp]
	\begin{minipage}[t]{7.6cm}
		\centering
		\includegraphics[width=9cm]{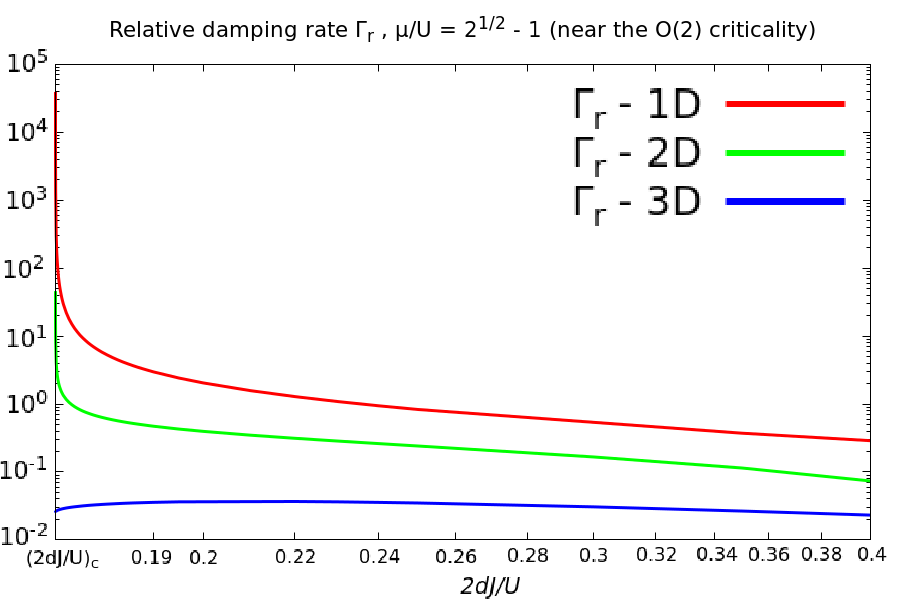}
		\caption{Relative damping rate $\Gamma_r = \Gamma/\Delta_H$ of the Higgs excitation near the $\overline{n} = 1$ tip transition point for $d = 1, 2, 3$ within the Goldstone decay channel.}
		\label{Figure68}
	\end{minipage}
	\ \hspace{3.5mm} \hspace{3.5mm} \
	\begin{minipage}[t]{7.7cm}
		\centering
		\includegraphics[width=9cm]{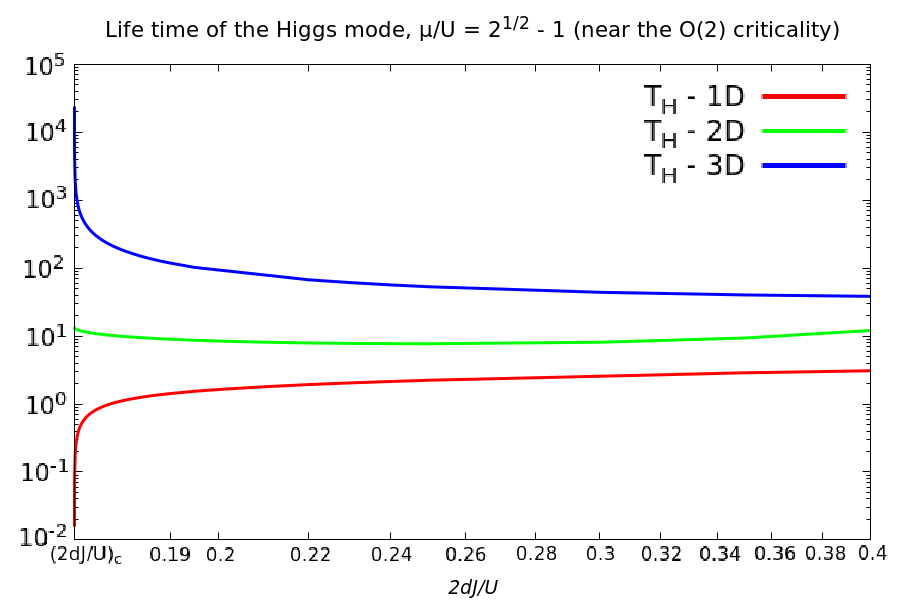}
		\caption{Life time $T_H = \Gamma^{-1}$ of the Higgs mode near the $\overline{n} = 1$ tip transition point for $d = 1, 2, 3$ within the Goldstone decay channel. The time unit is given by $\hbar/U$.}
		\label{Figure69}
	\end{minipage}
\end{figure}
\uline{Figure \ref{Figure68} shows that the critical value of $\Gamma_r$ clearly diverges when $d < 3$, a result that agrees with the (theoretically and experimentally) well-known divergence of the Higgs self-energy given by the Goldstone coupling in low dimensions} \cite{lattice_depth_modulation_2, sachdev_higgs, zwerger_higgs, podolsky_higgs}, \uline{while for $d = 3$ it reaches a finite value or tends slowly to 0 at the transition. These evidences partially agrees with the relevant results obtained by Altman and Auerbach in \cite{altman_auerbach} and allows to fix the approximate scaling exponent of the decay amplitude (\ref{4.112}) as $\alpha \approx 1$.} \\
Nevertheless, given the limited accuracy of our calculations and the slow damping of $\Gamma_r$ near the 3D critical point, we expect the numerical data to provide a value of $\alpha$ slightly smaller than 1.

\medskip

This supposition can be verified by performing a guided fit of the curves in Figure \ref{Figure68} with respect to equations (\ref{4.118}) and (\ref{4.119}). Using the estimation (\ref{4.111}) of the scaling exponent $\beta_{\Delta}$, such a procedure yields:
\begin{equation}\label{4.121}
\alpha_{1D} = 0.94 \pm 0.02 \quad , \quad \alpha_{2D} = 0.96 \pm 0.01 \quad , \quad \alpha_{3D} = 0.923 \pm 0.008
\end{equation}
where the subscripts refer to predictions for different dimensions. As expected, the present calculations based on equation (\ref{4.117}) do not provide the exact value of $\alpha$, but only an approximate estimation.

\medskip

In conclusion, \textsl{our numerical approach is only able to mimic the expected behaviour of the relative damping rate $\Gamma_r$ because of the limited accuracy of the low-energy limit calculations} in the special case $d = 3$.
\textsl{On the other hand, taking into account that our numerical data have been derived from a series of analytical approximations, from the low-energy limit to the expression of the sound velocity, these results represent another confirmation of the surprising versatility of the Gutzwiller theory tested in this work.}

\vspace{0.7cm}

\fbox{\begin{minipage}{16.5cm}
\textbf{\textcolor{blue}{Observation} -} The numerical outcomes presented in Figures \ref{Figure68} and \ref{Figure69} give considerable suggestions about the observability of the Higgs mode depending on the dimensionality of the lattice. For instance, we can claim that, when approaching the critical region, the Higgs mode is unlikely to be observed in low-dimensional systems.
\end{minipage}}

\vspace{0.7cm}

For completeness, we have performed also a more general calculation of $\Gamma_r$ and the Higgs life time $T_H$ over the strongly-interacting region of the phase diagram in the 1D case, which does not need any approximation because of the exact knowledge of $\rho_d{\left( x \right)}$.
\vspace{0.5cm}
\begin{figure}[!htp]
	\begin{minipage}[t]{8.2cm}
		\centering
		\includegraphics[width=9cm]{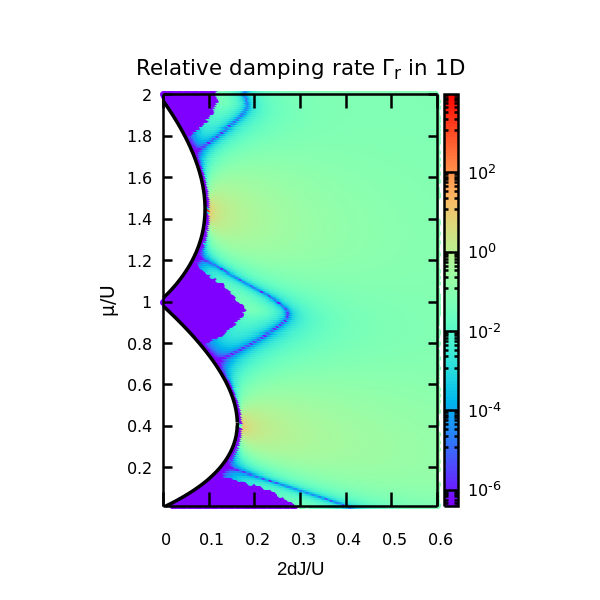}
		\caption{Relative damping rate $\Gamma_r$ of the Higgs excitation for $d = 1$ in the strongly-interacting regime within the Goldstone decay channel. In the Mott phase and in the hard-core limits (wedge-shaped violet regions) of the phase diagram the damping rate is identically zero.}
		\label{Figure70}
	\end{minipage}
	\ \hspace{3.5mm} \hspace{3.5mm} \
	\begin{minipage}[t]{8.2cm}
		\centering
		\includegraphics[width=9cm]{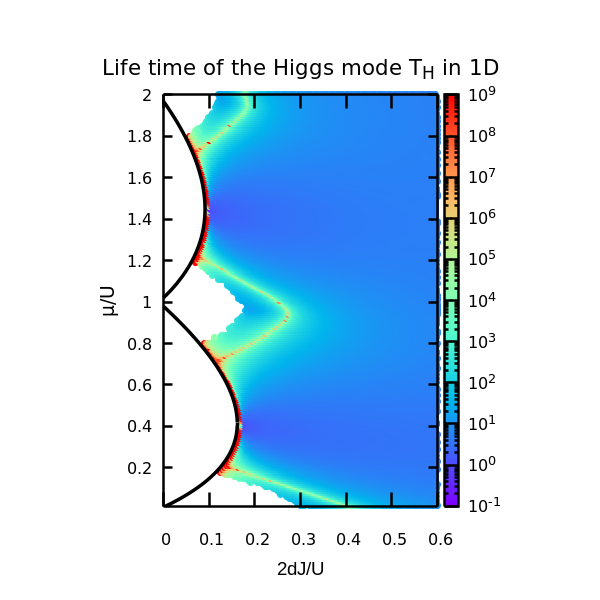}
		\caption{Life time $T_H = 1/\Gamma$ of the Higgs excitation for $d = 1$ in the strongly-interacting regime within the Goldstone decay channel. The white region comprising the Mott lobes and the hard-core limits of the superfluid phase corresponds to $T_H \to \infty$.}
		\label{Figure71}
	\end{minipage}
\end{figure}
\\
Figures \ref{Figure70} and \ref{Figure71} give the opportunity to make important observations about the behaviour of the Higgs mode damping away from the critical region.
\begin{itemize}
	\item Recalling the previous discussion, in Figure \ref{Figure70} we can immediately recognize the divergence of $\Gamma_r$ near the tip of the Mott lobes. Actually, new structures emerge: the decay of the Higgs excitation turns to be highly suppressed along wedge-shaped lines connecting adjacent Mott lobes.
	\item The suppression lines enclose the hard-core limits of the phase diagram, where the decay rate is identically zero. This happens because of purely kinematic reasons, as the Higgs gap is always larger than the dispersion band of the Goldstone mode in the hard-core limit. On the other hand, the decay channel associated to the wedge-shaped lines opens at high momenta (see Figure \ref{Figure72}) and is simply overdamped.
	\item As a long-lived collective excitation, Figures \ref{Figure70} and \ref{Figure71} seem to indicate that the Higgs mode can be observed along the usual wedge-shaped lines and in the vicinity of the Mott lobe, except for the well-known Lorentz-invariant points. This result could give some hints about the role of strong interactions in the fate of the Higgs mode away from criticality.
\end{itemize}
\textbf{Remark -} All the calculations presented in this section have been performed only in the strongly-interacting regime, where the second-order corrections studied in Section \ref{second_order_corrections} are proved to be small compared to their mean-field values. On the other hand, our promising results have to be confirmed or improved by further calculations based on a self-consistent quantum theory of the excitations as argued previously.
\begin{figure}[!htp]
	\centering
	\includegraphics[width=0.45\textwidth]{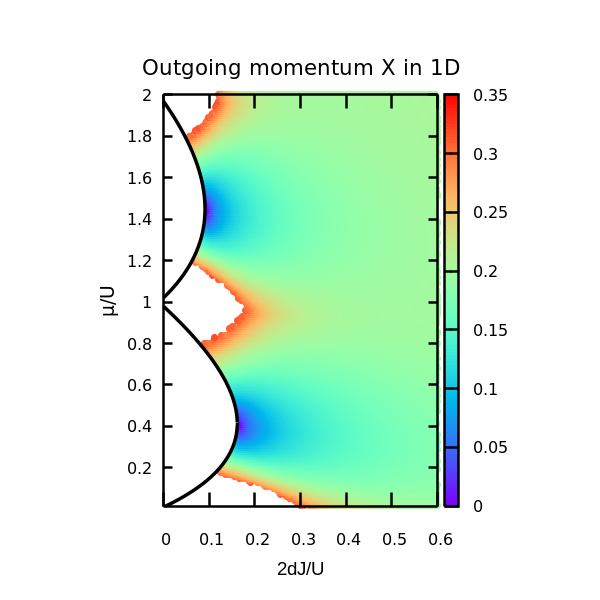}
	\caption{Reduced momentum $x$ of the counterpropagating Goldstone modes produced by the decay of a Higgs excitation at rest in the strongly-interacting regime for $d = 1$. The white colour indicates regions where the instability of the Higgs mode is kinematically forbidden.}
	\label{Figure72}
\end{figure} \\
As regards the effective detectability of the Higgs mode from an operational, it is important to underline that the overdamping of amplitude collective excitations strictly depends on the kind of susceptibility adopted for probing the system dynamics. As demonstrated by Podolsky et al. in \cite{podolsky_higgs}, a characteristic infrared singularity appears if the physical probe couples longitudinally to the order parameter \cite{sachdev, sachdev_higgs, zwerger_higgs}, as in the case of neutron scattering, while in the scalar susceptibility this divergence is avoided and substituted by a soft scaling at low frequencies \cite{lattice_depth_modulation_2, podolsky_higgs}, as expected for lattice modulation techniques. In the latter case, the Higgs mode produces a pronounced response peak in correspondence of its energy gap or mass depicted in Figure \ref{Figure67}, determined by the parameters of the theory.

\addcontentsline{toc}{chapter}{Conclusions and perspectives}
\chapter*{Conclusions and perspectives}
\markboth{\MakeUppercase{Conclusions and perspectives}}{\MakeUppercase{Conclusions and perspectives}}
This thesis work has focused on finding a simple and appropriate method for investigating quantum fluctuations above the mean-field Gutzwiller solution of the Bose-Hubbard model over the whole range of the physical parameters determining its phase diagram. Exploiting the versatility of the well-known Gutzwiller approximation, the present research deviates from other and more traditional approaches as it is intended to offer a global picture of the Bose-Hubbard physics instead of a description concerning a delimited region of parameters.

\markboth{\MakeUppercase{Conclusions and perspectives}}{\MakeUppercase{Main results achievements}}
\section*{Main results and achievements}
\markboth{\MakeUppercase{Conclusions and perspectives}}{\MakeUppercase{Main results and achievements}}
Taking inspiration from the results of Chapter \ref{MF}, where we presented the application of the Gutzwiller mean-field approach to the study of the ground state solutions and the many-body excitation dynamics, in Chapter \ref{quantum} we have identified an approximate mapping between the Bose-Hubbard problem and an effective action of the Gutzwiller parameters $c_n{\left( \mathbf{r} \right)}$. Considering small perturbations around the saddle-point solutions and assuming the parameters $c_n{\left( \mathbf{r} \right)}$ to be complex scalar fields, we expanded the Gutzwiller action up to second order and performed a canonical quantization of the fluctuations through a Bogoliubov diagonalization. The resulting quantum theory turns to be gapless and offers a complete description of the collective excitation spectrum beyond the Goldstone mode. \\
Following the analogy with the Bogoliubov method and promoting the Gutzwiller density and order parameters to operators, the second part of Chapter \ref{quantum} have dealt with the study of linear operator fluctuations and in particular with the particle and hole amplitudes $U_{\alpha, \mathbf{k}}$ and $V_{\alpha, \mathbf{k}}$ emerging in the fluctuation part of the Bose field operator. Analysing the momentum dependence these amplitudes for different points of the phase diagram, we have proved the perfect compatibility of the Gutzwiller theory with the standard Bogoliubov approach to weakly-interacting condensates. \\
Moreover, in accordance with Popov \cite{popov}, we have demonstrated that the peculiar low-momentum behaviour of the amplitudes $U_{\alpha, \mathbf{k}}$ and $V_{\alpha, \mathbf{k}}$ hints at the possibility of describing the low-energy physics of strongly-interacting condensates by using an effective hydrodynamic action with renormalized parameters as they are provided by the Gutzwiller approximation.

\medskip

As a natural prosecution and in analogy with the Bogoliubov treatment of weak interactions, Chapter \ref{quantum_corrections} has been dedicated to the calculation and analysis of relevant observables within our Gutzwiller theory. \\
In Section \ref{depletion} the calculation of the quantum depletion of the condensate over the whole superfluid part of the phase diagram has allowed to identify the interplay between the Goldstone and Higgs modes in determining the quantum fluctuations due to pure interaction effects. In particular, we have discovered that the Higgs excitation saturates the quantum depletion when approaching the MI-SF transition, while the Goldstone mode becomes has a dominant role in the deep superfluid phase, where have seen again a perfect agreement with the standard Bogoliubov theory. As a remarkable result, we have shown that the flexibility of the Gutzwiller approach leads to non-trivial and \textit{a priori} unexpected predictions in the regime of strong interactions, eventually giving a more precise answer to the long-standing question about the fate of the Higgs away from the critical point \cite{pekker_varma}. \\
Similar results have been obtained in Section \ref{superfluid_fraction} on the basis of analysis of the superfluid density, calculated from the current response of the system to a global phase twist and generalizing again the corresponding Bogoliubov calculation. We have shown how the Gutzwiller quantum theory improves the ordinary Bogoliubov predictions by predicting a significant reduction of the superfluid fraction in presence of strong interactions as a result of the band population of the first two collective modes. Although the Gutzwiller prediction shows interesting analytical properties at the critical transition, we have discovered that our quantum theory is not able to account for the vanishing of the superfluid density at the transition, compatibly with the insulating character of the Mott phase. In the discussion of Section \ref{superfluid_fraction_discussion} we have argued that a non-vanishing conductivity in the Mott phase is due to the incorrect estimation of current correlations within non-self-consistent approaches as our Gutzwiller theory and, by similarity, the standard Bogoliubov approach, where the condensate fraction is not consistently renormalized with respect to the addition of depletion effects. \\
Adopting this point of view, the requirement of a more refined treatment of quantum fluctuations has been further investigated in Section \ref{second_order_corrections} through the calculation of second-order quantum corrections to the average density and order parameter; in contrast to the expectations, these higher-order fluctuations exhibit pathological behaviours in the deep superfluid regime, where indeed the Gutzwiller saddle-point equations are know to match with discrete Gross-Pitaevskii description.

\medskip

The study of quadratic corrections to the mean-field theory is concluded by Chapter \ref{response_functions}, which focuses on a rigorous treatment of density susceptibilities to different probes within linear response theory. As main result of this study, we have recognized the strong particle-hole character of the excitations living above the Higgs mode and their connection with the corresponding bands in the Mott phase. Moreover, we have identified the hopping modulation as a possible experimental probe for detecting these high-energy modes, whose significant response amplitudes in the density channel are limited to specific regions of the phase diagram of strong interactions. In conclusion, we have discussed how this information can be exploited in order to enrich the interpretation of experimental measures searching for the dynamical features of the Higgs excitation.

\medskip

The capability of the quantum Gutzwiller theory in giving powerful insights into the physics of strong interactions has been further tested in Chapter \ref{third_order}, which has been finally devoted to the analysis of non-linear quantum effects. For this purpose, performing a third-order expansion of the Gutzwiller action in terms of quasi-particle operators, we have derived an approximate estimation of the damping time of the Higgs mode around the Lorentz-invariant points of the MI-SF criticality. Studying the analytical behaviour of the Higgs damping rate at the transition for different dimensionalities, we have found considerable overlaps between the Gutzwiller predictions and the exact results obtained by applying continuous field theory \cite{altman_auerbach}. Finally, we have argued that the successful application of our formalism could be extended out of the critical region for answering the long-discussed issue of the fate of high-energy excitation modes when approaching the weakly-interacting limit.

\newpage

\markboth{\MakeUppercase{Conclusions and perspectives}}{\MakeUppercase{Experimental implications}}
\section*{Experimental implications}
\markboth{\MakeUppercase{Conclusions and perspectives}}{\MakeUppercase{Experimental implications}}
Besides being a relevant theoretical result, the new information about the modal composition of the quantum depletion and the superfluid fraction that we have obtained in this thesis work could be also taken into consideration as reference data for experimental application aiming at giving a more detailed picture of strongly-interacting superfluids. The strong dependence of quantum corrections on the amplitude mode that we have observed near the critical region suggests the possibility of tracking the impact of massive modes on the zero-point physics of the Bose-Hubbard model simply by looking at the behaviour of the non-condensed and superfluid fractions of the system for a given range of parameters, in accordance with the fact that ultracold bosonic systems in optical lattices offer unique possibilities of parameter tunability even in reduced dimensions. \\
This argument acquires particular importance within the analysis of the behaviour of the superfluid fraction. In Section \ref{superfluid_fraction} we have observed that $f_s$ is always larger than the condensate density at equilibrium; on the other hand, this results does not prevent the occurence of the opposite situation when the system is driven out of equilibrium. Therefore, we can suppose that looking at the evolution of the superfluid density, that is probing current-current correlations according to \cite{scalapino_white_zhang_1, scalapino_white_zhang_2}, after the application of phase twist perturbations or after a quench of the Hamiltonian parameters would give direct insights into the exact relations between the gapped modes, above all the Higgs excitation, and the dynamical properties of strongly-correlated superfluids, even by exploring physical states where not all the condensate particles contribute to superfluidity.

\medskip

The eventuality of measuring non-trivial features as the Higgs mass threshold through the study of conductive properties have been proposed for the $\text{O}{\left( 2 \right)}$ quantum critical point by Podolsky et al. in \cite{podolsky_higgs}. As far as the effective observability of the Higgs mode only is concerned, our attempt to comprehend decay processes within our Gutzwiller treatment hints at the possibility of providing interesting results on the fate of the Higgs-Goldstone coupling away from the critical region, as well as at the boundary between strong and weak interactions. Since a solid prediction of the disappearing of the Higgs response is still lacking, it follows that the outcomes of a self-consistent reformulation of the Gutzwiller approach would contribute new theoretical predictions to be compared with and as required by the experimental evidences extracted at the linear-response level \cite{lattice_depth_modulation_2}. \\
More generally, the new approach to quantum fluctuations proposed in this work could be applied also to the study of crucial quantum effects characterizing real experimental situations by including in the Gutzwiller Hamiltonian density (\ref{4.6}) the local terms given by the trapping potentials. This analytical and numerical treatment is expected to satisfy the necessity of theoretical predictions that account for the experimental observations regarding the Higgs mode detection \cite{lattice_depth_modulation_2}. In particular, taking inspiration from the results of this thesis, we are confident that a similar study of the excitation properties of trapped superfluid systems could allow to characterize the additional in-trap vibrational modes which populate the spectrum above the Higgs excitation in confined systems and, together with the analysis of the response of pure particle-hole excitations presented in Chapter \ref{response_functions}, give an explanation of the spurious effects affecting the experimental data.
\begin{figure*}[!htp]
	\centering
	\includegraphics[width=0.5\textwidth]{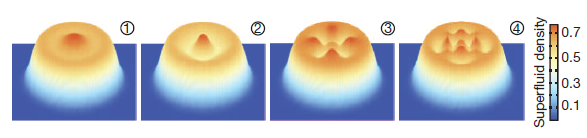}
	\caption*{\textbf{Figure 1 -} In-trap superfluid density distribution for the first four amplitude or vibrational modes characterizing the excitation spectrum of a confined system. This plot has been obtained by calculations based on the Gutzwiller mean-field approximation. Figure taken from \cite{lattice_depth_modulation_2}.}
\end{figure*}

\medskip

Leaving the realm of ultracold atoms and real systems, the capability of our Gutzwiller approach to get a global picture of the quantum effects given by strong correlations can be further exploited within a second research direction which has become very popular in recent years because of its expected predictive power and its interdisciplinary character. \\
Among the other studies involving the field of quantum simulation, the theoretical and practical development of a new generation of arrays of superconducting qubits, performed by the Google research group of P. Roushan with the collaboration of the UCSB (University of California, Santa Barbara), has opened new perspectives for the construction of virtual systems reproducing exactly the behaviour of physical models which cannot be easily studied through the most advanced experimental techniques. Implementing a specific method for resolving the energy levels of the interacting photons contributed by the unit circuits and mapping these levels into single-site occupation states, the qubit system can be used to study the physics of bosonic optical lattices with significant advantages \cite{qubits_1, qubits_2}. In particular, the high-level tunability of the photon hopping parameter $J$ and qubit frequency, corresponding to the chemical potential $\mu$, with single-site resolution on nanosecond time scales allows to explore a wide range of the Bose-Hubbard physics and has allowed to obtain outstanding results in simulating many-body localization and disorder effects \cite{qubits_3}. \\
As far as the application of the Gutzwiller theory is concerned, we suppose that our method could be a good candidate for providing theoretical predictions in parallel with the large class of physical phenomena accessible by these optimized devices, from quantum quenches to the simulation of complex experimental situations.
\begin{figure*}[!htp]
	\centering
	\includegraphics[width=0.4\textwidth]{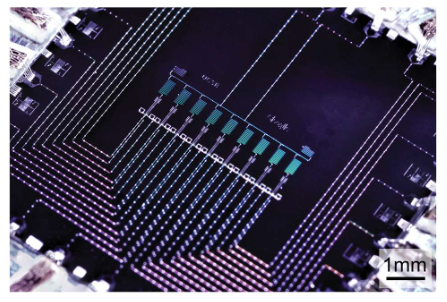}
	\caption*{\textbf{Figure 2 -} Optical micrograph of the nine-qubit array used by P. Roushan and his collaborators in order to study disorder effects as the transition from ergodic to energy-localized phases. Here we can observe the optical couplers joining nearest-neighbour qubits and used for tuning the photon hopping $J$ through microwave pulses.}
\end{figure*}

\markboth{\MakeUppercase{Conclusions and perspectives}}{\MakeUppercase{Outlook}}
\section*{Outlook}
\markboth{\MakeUppercase{Conclusions and perspectives}}{\MakeUppercase{Outlook}}
Although a quantum theory of the Bose-Hubbard excitations based on the Gutzwiller approximation has proved to be a simple and flexible method for studying significant phenomena due to pure interaction effects beyond the ordinary Bogoliubov theory of weakly-interacting condensates, the evaluation of more elaborate quantities as the superfluid density have revealed that our approach and the standard Bogoliubov approximation suffer from similar structural drawbacks in providing a range of higher-order effects away from the weakly-interacting limit. \\
This observation, accompanied by the emergence of unexpected outcomes from the calculation of second order corrections to the average density and order parameter, seems to indicate that, similarly to the Hartree-Fock-Popov theory \cite{proukakis_morgan_choi_burnett}-\cite{morgan}, a self-consistent approach to quantization should be taken into account for future developments. In analogy with the findings of Trivedi and Menotti \cite{menotti_trivedi}, we are confident that a careful treatment of the addition of quantum effects around the Gutzwiller mean-field theory will solve also the problem of density normalization due to the addition of quantum depletion to the condensate fraction and highlighted in Sections \ref{depletion}, \ref{quantization_discussion}. \\
From this perspective, the idea of finding a more sophisticate approach to the calculation of quantum fluctuations, namely including the depletion feedback on the Gutzwiller ground state solution, is not only intended to reach a possible improvement of the actual predictions (for example, in the shape of the MI-SF critical line), but also to check and give additional value to the good results obtained by linear-order fluctuations in this work.

\medskip

Following this research direction, we expect that future studies devoted to the formulation of a consistent quantum theory of the many-body excitations will allow to explore at a deeper level the effective hydrodynamic description of strongly-interacting superfluids introduced in Chapter \ref{quantum}, for example in characterizing and identifying the relevant low-energy vertices of the theory through the knowledge of the particle-hole amplitudes only as simple information, as it occurs in the case of Beliaev decay in continuous superfluid systems \cite{pitaevskii_stringari}. \\
As a second step, a more complete description of relevant quantum effects would consists in extending the Bose-Hubbard phase diagram to finite temperatures, studying how the insulating and superfluid behaviours survive the introduction of thermal excitations and understanding how the physical observables are modified by thermal fluctuations. \\
On the other hand, a whole new interesting regime is expected to arise when decreasing the dimensionality of the optical lattice. In this perspective, since both the mean-field Gutzwiller approximation and the quantum theory used in this work are not reliable in systems with a small coordination number, an improvement of our theory could be given by a quantum reformulation of the cluster Gutzwiller approach \cite{cluster_gutzwiller}, which is able to account for short-range quantum correlations in the vicinity of the MI-SF transition and correct the mean-field picture of the quantum criticality also in two-dimensional lattices. \\
Finally, we cannot exclude that new interesting issues and suggestions could emerge by a comparison between the present results, derived from a perturbative approach, and the application of non-perturbative treatments of local interactions, in particular the DMFT (dynamical mean-field theory) approach, which succeeds in describing the Mott-metal transition in fermionic systems and the behaviour of real materials. Proceeding along this direction, we speculate also the application of our Gutzwiller framework to the case of Fermi-Hubbard systems, where actually the implementation of the canonical quantization procedure could present more mathematical challenges with respect to the context of this work.

\medskip

In spite of its simplicity and perturbative character, the Gutzwiller quantum theory of collective excitations developed in this thesis, born as a tentative new method to address the current problem of non-trivial quantum effects in strong correlations, contributes new elements to the knowledge of the Bose-Hubbard physics and, in general, shines a light on important questions about the physics of strong correlations in Bose systems, as well as on possible methodological problems and research lines that could address these issues.

\appendix
\chapter{Mean-field calculation \texorpdfstring{\\}{} of the Landau-Ginzburg free energy}\label{A}
\markboth{\MakeUppercase{A. Mean-field free energy}}{\MakeUppercase{A. Mean-field free energy}}
Let us reconsider the mean-field expression of the Bose-Hubbard ground state energy derived in Chapter \ref{Introduction}:
\begin{equation}\label{M.1}
\frac{E_0}{M} = \frac{\langle \Psi_{MF} | \hat{H} | \Psi_{MF} \rangle}{M} = \frac{\langle \Psi_{MF} | \hat{H}_{MF} | \Psi_{MF} \rangle}{M} - J \, z \, \langle \hat{b}^{\dagger} \rangle \langle \hat{b} \rangle + \psi^{*} \, \langle \hat{b} \rangle + \psi \, \langle \hat{b}^{\dagger} \rangle
\end{equation}
where $| \Psi_{MF} \rangle$ is the mean-field ground state wave function and:
\begin{equation}\label{M.2}
\hat{H}_{MF} =\sum_{i} \left[ -\psi^{*} \, \hat{b}_i - \psi \, \hat{b}^{\dagger}_i + \frac{U}{2} \, \hat{n}_i \left( \hat{n}_i - 1 \right) - \mu \hat{n}_i \right]
\end{equation}
is the mean-field Hamiltonian of the Bose-Hubbard model. \\
The expansion of (\ref{M.1}) up to the second order in $\psi$ can be performed via perturbation theory applied to the calculation of the expectation values on the RHS of (\ref{M.1}). \\
Let us rewrite the mean-field Hamiltonian (\ref{M.2}) as the sum of an unperturbed part $\hat{H}_0$ corresponding to the pure Mott phase and a perturbation $\hat{H}_1$ given by the decoupled hopping term:
\begin{equation}\label{M.3}
\hat{H}_0 = \sum_{i} \left[ \frac{U}{2} \, \hat{n}_i \left( \hat{n}_i - 1 \right) - \mu \hat{n}_i \right] \quad , \quad \hat{H}_1 = -\sum_{i} \left( \psi^{*} \, \hat{b}_i + \psi \, \hat{b}^{\dagger}_i \right)
\end{equation}
According to perturbation theory, the first-order expansion of the mean-field ground state $| \Psi_{MF} \rangle$ with respect to the order parameter $\psi$ characterizing the perturbation $\hat{H}_1$ is given by:
\begin{equation}\label{M.4}
| \Psi_{MF}{\left( \psi \right)} \rangle = | \Psi_0 \rangle + \sum_{k \neq \Psi_0} \frac{\langle k | \hat{H}_1 | \Psi_0 \rangle}{E^{\left( 0 \right)}_0 - E_k} | k \rangle + O{\left( \psi^2 \right)}
\end{equation}
where $| \Psi_0 \rangle$ is the pure Mott state, $E^{\left( 0 \right)}_0 \equiv \langle \Psi_0 | \hat{H}_{MF} | \Psi_0 \rangle$ and $| k \rangle$ labels the Mott excited states given by the addition of a particle or a hole to the system. Therefore:
\begin{equation}\label{M.5}
| \Psi_{MF}{\left( \psi \right)} \rangle = | \Psi_0 \rangle + \psi^* \sum_{k_-} \frac{\sqrt{n_0{\left( \mu/U \right)}}}{\mu - U \left[ n_0{\left( \mu/U \right)} - 1 \right]} | k_- \rangle + \psi \sum_{k_+} \frac{\sqrt{n_0{\left( \mu/U \right)} + 1}}{\left[ U \, n_0{\left( \mu/U \right)} - \mu \right]} | k_+ \rangle + O{\left( \psi^2 \right)}
\end{equation}
where $| k_+ \rangle$ and $| k_- \rangle$ label particle and hole-doped states respectively. \\
Equation (\ref{M.5}) allows now to calculate the expectation value $J \, z \, \langle \hat{b}^{\dagger} \rangle \langle \hat{b} \rangle$ contributing to the RHS of (\ref{M.1}). In particular:
\begin{equation}\label{M.6}
\langle \Psi_{MF}{\left( \psi \right)} | \hat{b} | \Psi_{MF}{\left( \psi \right)} \rangle = \psi \left\{ \frac{n_0{\left( \mu/U \right)} + 1}{\left[ U \, n_0{\left( \mu/U \right)} - \mu \right]} + \frac{n_0{\left( \mu/U \right)}}{\mu - U \left[ n_0{\left( \mu/U \right)} - 1 \right]} \right\} + O{\left( \psi^2 \right)}
\end{equation}
\begin{equation}\label{M.7}
\langle \Psi_{MF}{\left( \psi \right)} | \hat{b} | \Psi_{MF}{\left( \psi \right)} \rangle = \psi^* \left\{ \frac{n_0{\left( \mu/U \right)} + 1}{\left[ U \, n_0{\left( \mu/U \right)} - \mu \right]} + \frac{n_0{\left( \mu/U \right)}}{\mu - U \left[ n_0{\left( \mu/U \right)} - 1 \right]} \right\} + O{\left( \psi^2 \right)}
\end{equation}
hence:
\begin{equation}\label{M.8}
\begin{aligned}
J \, z \, \langle \hat{b}^{\dagger} \rangle \langle \hat{b} \rangle &= J z \left| \psi \right|^2 \left\{ \frac{n_0{\left( \mu/U \right)} + 1}{\left[ U \, n_0{\left( \mu/U \right)} - \mu \right]} + \frac{n_0{\left( \mu/U \right)}}{\mu - U \left[ n_0{\left( \mu/U \right)} - 1 \right]} \right\}^2 + O{\left( \psi^3 \right)} \equiv \\
&\equiv J z \left| \psi \right|^2 \chi^2_0{\left( \mu/U \right)} + O{\left( \psi^3 \right)}
\end{aligned}
\end{equation}
where $\chi_0{\left( \mu/U \right)}$ is the same function introduced in equation (\ref{1.13}). \\
The first contribution to the ground state energy (\ref{M.1}), that is the expectation value of the mean-field Hamiltonian (\ref{M.2}), can be easily evaluated by using again second-order perturbation theory:
\begin{equation}\label{M.9}
\begin{aligned}
&\frac{\langle \Psi_{MF} | \hat{H}_{MF} | \Psi_{MF} \rangle}{M} = \frac{\langle \Psi_0 | \hat{H}_{MF} | \Psi_0 \rangle}{M} + \sum_{k \neq \Psi_0} \frac{\left| \langle k | \hat{H}_1 | \Psi_0 \rangle \right|^2}{E^{\left( 0 \right)}_0 - E_k} + O{\left( \psi^3 \right)} = \\
&= \frac{E^{\left( 0 \right)}_0}{M} - \left| \psi \right|^2 \left\{ \frac{n_0{\left( \mu/U \right)} + 1}{\left[ U \, n_0{\left( \mu/U \right)} - \mu \right]} + \frac{n_0{\left( \mu/U \right)}}{\mu - U \left[ n_0{\left( \mu/U \right)} - 1 \right]} \right\} + O{\left( \psi^3 \right)} = \\
&= \frac{E^{\left( 0 \right)}_0}{M} - \left| \psi \right|^2 \chi_0{\left( \mu/U \right)} + O{\left( \psi^3 \right)}
\end{aligned}
\end{equation}
As a final result, inserting expressions (\ref{M.6})-(\ref{M.9}) into the RHS of equation (\ref{M.1}), one obtains:
\begin{equation}\label{M.10}
\begin{aligned}
\frac{E_0}{M} &= \frac{\langle \Psi_{MF} | \hat{H}_{MF} | \Psi_{MF} \rangle}{M} - J \, z \, \langle \hat{b}^{\dagger} \rangle \langle \hat{b} \rangle + \psi^{*} \, \langle \hat{b} \rangle + \psi \, \langle \hat{b}^{\dagger} \rangle = \\
&= \frac{E^{\left( 0 \right)}_0}{M} + \left| \psi \right|^2 \chi_0{\left( \mu/U \right)} \left[ 1 - J \, z \, \chi_0{\left( \mu/U \right)} \right] + O{\left( \psi^3 \right)}
\end{aligned}
\end{equation}
which coincides with the well-known Landau-Ginzburg free energy (\ref{1.11}).

\chapter{Mean-field excitations: other results}\label{B}
\markboth{\MakeUppercase{A. Mean-field excitations}}{\MakeUppercase{A. Mean-field excitations}}
\section{Energy gaps}
Let us rewrite equation (\ref{3.22}) in the following operatorial form:
\begin{equation}\label{A.1}
\hbar \omega_{\mathbf{k}}
\begin{pmatrix}
| u_{\mathbf{k}} \rangle \\
| v_{\mathbf{k}} \rangle
\end{pmatrix}
=
\begin{pmatrix}
\hat{A}_{\mathbf{k}} & \hat{B}_{\mathbf{k}} \\
-\hat{B}_{\mathbf{k}} & -\hat{A}_{\mathbf{k}}
\end{pmatrix}
\begin{pmatrix}
| u_{\mathbf{k}} \rangle \\
| v_{\mathbf{k}} \rangle
\end{pmatrix}
\end{equation}
The operators:
\begin{equation}\label{A.2}
\hat{A}_{\mathbf{k}} \equiv -J_{\mathbf{0}} \, \psi_0 \left( \hat{a}^{\dagger} + \hat{a} \right) + \frac{U}{2} \, \hat{n} \left( \hat{n} - 1 \right) - \mu \, \hat{n} - \hbar \omega_0 - J_{\mathbf{k}} \left( \hat{a} | s_0 \rangle \langle s_0 | \hat{a}^{\dagger} + \hat{a}^{\dagger} | s_0 \rangle \langle s_0 | \hat{a} \right)
\end{equation}
\begin{equation}\label{A.3}
\hat{B}_{\mathbf{k}} \equiv -J_{\mathbf{k}} \left( \hat{a} | s_0 \rangle \langle s_0 | \hat{a} + \hat{a}^{\dagger} | s_0 \rangle \langle s_0 | \hat{a}^{\dagger} \right)
\end{equation}
where $| s_0 \rangle$ is the Gutzwiller ground state wave function, act on the ket states defined by:
\begin{equation}\label{A.4}
| u_{\mathbf{k}} \rangle = \sum_n u_{\mathbf{k}, n} | n \rangle \quad , \quad | v_{\mathbf{k}} \rangle = \sum_n v_{\mathbf{k}, n} | n \rangle
\end{equation}
The zeroth-order exact solution of (\ref{A.1}) in the small parameter $U/J$ is given by:
\begin{equation}\label{A.5}
| u^0_{\mathbf{k}, \lambda} \rangle = \hat{D}{\left( \psi_0 \right)} | \lambda \rangle \quad , \quad | u^0_{\mathbf{k}, \lambda} \rangle = 0
\end{equation}
where $\lambda$ is the mode index, $| \lambda \rangle$ is the corresponding occupation state and $\hat{D}{\left( \alpha \right)} = \exp{\left( \alpha \, \hat{a}^{\dagger} - \alpha^{*} \, \hat{a} \right)}$ is the displacement operator, since it satisfies the following composition property:
\begin{equation}\label{A.6}
\hat{D}{\left( \alpha \right)} \, \hat{a} \, \hat{D}{\left( \alpha \right)} = \hat{a} + \alpha
\end{equation}
Finally, first-order perturbation theory in the small parameter $U/J$ and the relation:
\begin{equation}\label{A.7}
\langle u^0_{\mathbf{k}, \lambda} | \hat{B}_{\mathbf{k}} | u^0_{\mathbf{k}, \lambda} \rangle
\end{equation}
valid for $\lambda \ge 2$ allow to establish that:
\begin{equation}\label{A.8}
\hbar \omega^{\left( 1 \right)}_{\mathbf{k}, \lambda} = \langle u^0_{\mathbf{k}, \lambda} | \hat{A}_{\mathbf{k}} | u^0_{\mathbf{k}, \lambda} \rangle
\end{equation}
The explicit calculation of the matrix element (\ref{A.8}) leads to the result (\ref{3.43}) for the gaps in the excitation spectrum.

\section{Derivation of the sound velocity}
In order to work out the sound velocity $c^0_s$, we consider the limit of small $\left| \mathbf{k} \right|$ and look for the lowest-energy solution of equation (\ref{3.22}) as an expansion with respect to $\mathbf{k}$:
\begin{equation}\label{A.9}
\begin{aligned}
&\vec{u}_{\mathbf{k}} = \vec{u}^{\left( 0 \right)}_{\mathbf{k}} + \vec{u}^{\left( 1 \right)}_{\mathbf{k}} + \vec{u}^{\left( 2 \right)}_{\mathbf{k}} + \dots \\
&\vec{v}_{\mathbf{k}} = \vec{v}^{\left( 0 \right)}_{\mathbf{k}} + \vec{v}^{\left( 1 \right)}_{\mathbf{k}} + \vec{v}^{\left( 2 \right)}_{\mathbf{k}} + \dots \\
&\omega_{\mathbf{k}} = \omega^{\left( 0 \right)}_{\mathbf{k}} + \omega^{\left( 1 \right)}_{\mathbf{k}} + \omega^{\left( 2 \right)}_{\mathbf{k}} + \dots
\end{aligned}
\end{equation}
The zeroth-order solution satisfies the equation of motion:
\begin{equation}\label{A.10}
\hbar \omega^{\left( 0 \right)}_{\mathbf{k}}
\begin{pmatrix}
\vec{u}^{\left( 0 \right)}_{\mathbf{k}} \\
\vec{v}^{\left( 0 \right)}_{\mathbf{k}}
\end{pmatrix}
=
\begin{pmatrix}
A_{\mathbf{0}} & B_{\mathbf{0}} \\
-B_{\mathbf{0}} & -A_{\mathbf{0}}
\end{pmatrix}
\begin{pmatrix}
\vec{u}^{\left( 0 \right)}_{\mathbf{k}} \\
\vec{v}^{\left( 0 \right)}_{\mathbf{k}}
\end{pmatrix}
\end{equation}
and is non-trivial only in the SF phase with the form:
\begin{equation}\label{A.11}
u^{\left( 0 \right)}_{\mathbf{k}, n} = \left( n + \hbar \frac{\dpar \omega_0}{\dpar \mu} \right) c^0_n \quad , \quad v^{\left( 0 \right)}_{\mathbf{k}, n} = -u^{\left( 0 \right)}_{\mathbf{k}, n} \quad , \quad \omega^{\left( 0 \right)}_{\mathbf{k}} = 0
\end{equation}
The quantities of the first order are governed by the equation:
\begin{equation}\label{A.12}
\hbar \omega^{\left( 1 \right)}_{\mathbf{k}}
\begin{pmatrix}
\vec{u}^{\left( 0 \right)}_{\mathbf{k}} \\
\vec{v}^{\left( 0 \right)}_{\mathbf{k}}
\end{pmatrix}
=
\begin{pmatrix}
A_{\mathbf{0}} & B_{\mathbf{0}} \\
-B_{\mathbf{0}} & -A_{\mathbf{0}}
\end{pmatrix}
\begin{pmatrix}
\vec{u}^{\left( 1 \right)}_{\mathbf{k}} \\
\vec{v}^{\left( 1 \right)}_{\mathbf{k}}
\end{pmatrix}
\end{equation}
Taking into account the identity:
\begin{equation}\label{A.13}
\sum_m \left( A^{n m}_0 + B^{n m}_0 \right) \frac{\dpar c^0_m}{\dpar \mu} = \left( n + \hbar \frac{\dpar \omega_0}{\dpar \mu} \right) c^0_n
\end{equation}
the first-order solution can be written as:
\begin{equation}\label{A.14}
u^{\left( 1 \right)}_{\mathbf{k}, n} = v^{\left( 1 \right)}_{\mathbf{k}, n} = \hbar \omega^{\left( 1 \right)}_{\mathbf{k}} \frac{\dpar c^0_n}{\dpar \mu}
\end{equation}
We substitute all these results in the equation for the quantities of the second order in $\left| \mathbf{k} \right|$:
\begin{equation}\label{A.15}
\hbar \omega^{\left( 1 \right)}_{\mathbf{k}}
\begin{pmatrix}
\vec{u}^{\left( 1 \right)}_{\mathbf{k}} \\
\vec{v}^{\left( 1 \right)}_{\mathbf{k}}
\end{pmatrix}
+ \hbar \omega^{\left( 2 \right)}_{\mathbf{k}}
\begin{pmatrix}
\vec{u}^{\left( 0 \right)}_{\mathbf{k}} \\
\vec{v}^{\left( 0 \right)}_{\mathbf{k}}
\end{pmatrix}
=
\begin{pmatrix}
A_{\mathbf{0}} & B_{\mathbf{0}} \\
-B_{\mathbf{0}} & -A_{\mathbf{0}}
\end{pmatrix}
\begin{pmatrix}
\vec{u}^{\left( 2 \right)}_{\mathbf{k}} \\
\vec{v}^{\left( 2 \right)}_{\mathbf{k}}
\end{pmatrix}
+
\begin{pmatrix}
A^{\left( 2 \right)}_{\mathbf{k}} & B^{\left( 2 \right)}_{\mathbf{k}} \\
-B^{\left( 2 \right)}_{\mathbf{k}} & -A^{\left( 2 \right)}_{\mathbf{k}}
\end{pmatrix}
\begin{pmatrix}
\vec{u}^{\left( 0 \right)}_{\mathbf{k}} \\
\vec{v}^{\left( 0 \right)}_{\mathbf{k}}
\end{pmatrix}
\end{equation}
Multiplying equation (\ref{A.15}) by the conjugate vector $\left( \vec{u}^{\left( 0 \right)}_{\mathbf{k}}, -\vec{v}^{\left( 0 \right)}_{\mathbf{k}} \right)$ from the left side and taking into account that:
\begin{equation}\label{A.16}
\vec{u}^{\left( 0 \right)}_{\mathbf{k}} \cdot \vec{u}^{\left( 0 \right)}_{\mathbf{k}} - \vec{v}^{\left( 0 \right)}_{\mathbf{k}} \cdot \vec{v}^{\left( 0 \right)}_{\mathbf{k}} = 0
\end{equation}
\begin{equation}\label{A.17}
\sum_n \left( u^{\left( 0 \right)}_{\mathbf{k}, n} - v^{\left( 0 \right)}_{\mathbf{k}, n} \right) \frac{\dpar c^0_n}{\dpar \mu} = \frac{\dpar \langle \hat{n} \rangle}{\dpar \mu} \equiv \kappa
\end{equation}
where $\kappa$ is the compressibility, we arrive at equation (\ref{3.44}) for the sound velocity.

\section{Bogoliubov dispersion relation}
We look for the solution of equation (\ref{3.22}) for arbitrary momentum $\mathbf{k}$ in the form:
\begin{equation}\label{A.18}
\vec{u}_{\mathbf{k}} = \vec{u}^{\left( 0 \right)}_{\mathbf{k}} \, a_{\mathbf{k}} + \vec{u}^{\left( 1 \right)}_{\mathbf{k}} \, b_{\mathbf{k}} \quad , \quad \vec{v}_{\mathbf{k}} = \vec{v}^{\left( 0 \right)}_{\mathbf{k}} \, a_{\mathbf{k}} + \vec{v}^{\left( 1 \right)}_{\mathbf{k}} \, b_{\mathbf{k}}
\end{equation}
where $\vec{u}^{\left( 0, 1 \right)}$ and $\vec{v}^{\left( 0, 1 \right)}$ are given by equations (\ref{A.11}) and (\ref{A.14}) with $\omega^{\left( 1 \right)}_{\mathbf{k}}$ being replaced by $\omega_{\mathbf{k}}$. Plugging (\ref{A.18}) into equation (\ref{3.22}) and multiplying the resulting equations by vectors $\vec{u}^{\left( 0 \right)}_{\mathbf{k}}$ and $\vec{u}^{\left( 1 \right)}_{\mathbf{k}}$, we obtain linear homogeneous equations for the amplitudes $a_{\mathbf{k}}$ and $b_{\mathbf{k}}$:
\begin{equation}\label{A.19}
\hbar \omega_{\mathbf{k}} \left( \vec{u}^{\left( 0 \right)}_{\mathbf{k}} \cdot \vec{u}^{\left( 1 \right)}_{\mathbf{k}} \right) a_{\mathbf{k}} = b_{\mathbf{k}} \sum_{n m} u^{\left( 1 \right)}_{\mathbf{k}, n} \left( A^{n m}_{\mathbf{k}} + B^{n m}_{\mathbf{k}} \right) u^{\left( 1 \right)}_{\mathbf{k}, m}
\end{equation}
\begin{equation}\label{A.20}
\hbar \omega_{\mathbf{k}} \left( \vec{u}^{\left( 0 \right)}_{\mathbf{k}} \cdot \vec{u}^{\left( 1 \right)}_{\mathbf{k}} \right) b_{\mathbf{k}} = a_{\mathbf{k}} \sum_{n m} u^{\left( 0 \right)}_{\mathbf{k}, n} \left( A^{n m}_{\mathbf{k}} + B^{n m}_{\mathbf{k}} \right) u^{\left( 0 \right)}_{\mathbf{k}, m}
\end{equation}
Finally, the relations:
\begin{equation}\label{A.21}
\vec{u}^{\left( 0 \right)}_{\mathbf{k}} \cdot \vec{u}^{\left( 1 \right)}_{\mathbf{k}} = \frac{\hbar \omega_{\mathbf{k}}}{2} \kappa
\end{equation}
\begin{equation}\label{A.22}
\sum_{n m} u^{\left( 1 \right)}_{\mathbf{k}, n} \left( A^{n m}_{\mathbf{k}} + B^{n m}_{\mathbf{k}} \right) u^{\left( 1 \right)}_{\mathbf{k}, m} = \left( \hbar \omega_{\mathbf{k}} \right)^2 \left[ \frac{\kappa}{2} + \left( \frac{\dpar \psi_0}{\dpar \mu} \right)^2 \varepsilon_{\mathbf{k}} \right]
\end{equation}
\begin{equation}\label{A.23}
\sum_{n m} u^{\left( 0 \right)}_{\mathbf{k}, n} \left( A^{n m}_{\mathbf{k}} + B^{n m}_{\mathbf{k}} \right) u^{\left( 0 \right)}_{\mathbf{k}, m} = \psi^2_0 \, \varepsilon_{\mathbf{k}}
\end{equation}
lead us to the result:
\begin{equation}\label{A.24}
\hbar \omega_{\mathbf{k}} = \sqrt{\frac{2 \psi^2_0}{\kappa} \varepsilon_{\mathbf{k}} + \left( \frac{\dpar \psi^2_0}{\dpar \langle \hat{n} \rangle} \right)^2 \varepsilon^2_{\mathbf{k}}}
\end{equation}
In the limit of small $U/J$ one has $\langle \hat{n} \approx \psi^2_0 \rangle$, so that we recover the well-know Bogoliubov dispersion relation providing the sound velocity (\ref{3.44}). For arbitrary values of $U/J$ equation (\ref{A.24}) better describes the lowest excitation branch of the SF phase than the standard Bogoliubov theory, but gives higher values of energy compared to the exact numerical data.

\chapter{Diagonalization of the operator \texorpdfstring{$\hat{\mathcal{L}}_{\mathbf{k}}$}{}}\label{C}
Let us consider the general form of the pseudo-Hermitian operator $\hat{\mathcal{L}}_{\mathbf{k}}$ in equation (\ref{4.10}):
\begin{equation}\label{B.1}
\mathcal{L}_k =
\begin{pmatrix}
A_k & B_k \\
-A^{*}_k & -B^{*}_k
\end{pmatrix}
\end{equation}
As $\mathcal{L}$ is not a Hermitian operator, its eigenbasis is not orthogonal because of the minus sign in the second line of (\ref{B.1}) due to Bose statistics. Since the knowledge of is \textsl{a priori} not sufficient to write $\mathcal{L}$ in diagonal form, one has to follow the following theorem.

\vspace{0.5cm}

\paragraph{Theorem}
\textsl{Let $M$ be a diagonalizable but not necessarily Hermitian operator. Then the diagonal form of $M$ can be written as:}
\begin{equation}\label{B.2}
M = \sum_k m_k | \psi^R_k \rangle \langle \psi^L_k |
\end{equation}
\textsl{where $| \psi^R_k \rangle$ is the right eigenvector of $M$ with eigenvalue $m_k$:}
\begin{equation}\label{B.3}
M | \psi^R_k \rangle = m_k | \psi^R_k \rangle
\end{equation}
\textsl{and $\langle \psi^L_k |$ is the left eigenvector of $M$ with the same eigenvalue $m_k$:}
\begin{equation}\label{B.4}
\langle \psi^L_k | M = m_k \langle \psi^L_k |
\end{equation}
\textsl{or equivalently:}
\begin{equation}\label{B.5}
M^{\dagger} | \psi^L_k \rangle = m^{*}_k | \psi^L_k \rangle
\end{equation}
\textsl{The normalization of the left and right eigenvectors is such that:}
\begin{equation}\label{B.6}
\langle \psi^L_k | \psi^R_p \rangle = \delta_{k, p}
\end{equation}
\textsl{$| \psi^L_k \rangle$ is then called the adjoint vector of $| \psi^R_k \rangle$.}

\vspace{0.5cm}

Let us apply this theorem to the case of $\mathcal{L}_k$. Defining the right eigenvectors:
\begin{equation}\label{B.7}
\langle \psi^R_k | M =
\begin{pmatrix}
| u_k \rangle \\
| v_k \rangle
\end{pmatrix}
\end{equation}
with eigenvalues $\varepsilon_k$ and exploiting the spectral property (\ref{B.4}), we obtain the left eigenvectors up to a normalization factor:
\begin{equation}\label{B.8}
\langle \psi^L_k | M = \mathcal{N}_k
\begin{pmatrix}
| u_k \rangle \\
-| v_k \rangle
\end{pmatrix}
\end{equation}
with the same real eigenvalue $\varepsilon_k$. The normalization condition (\ref{B.6}) imposes:
\begin{equation}\label{B.9}
\langle \psi^L_k | \psi^R_p \rangle = \mathcal{N}^{*}_k \left[ \langle u_k | u_k \rangle - \langle v_k | v_k \rangle \right] = 1
\end{equation}
It is then natural to normalize the right eigenvectors in such a way that the quantity between  brackets in equation (\ref{B.9}) is $\pm 1$, leading to $\mathcal{N}_k = \pm 1$. \\
Therefore, we group the eigenvectors of $\mathcal{L}_k$ in three families:
\begin{itemize}
	\item the $+$ family, such that $\langle u_k | u_k \rangle - \langle v_k | v_k \rangle = 1$;
	\item the $-$ family, such that $\langle u_k | u_k \rangle - \langle v_k | v_k \rangle = -1$;
	\item the $0$-family, such that $\langle u_k | u_k \rangle - \langle v_k | v_k \rangle = 0$.
\end{itemize}
From the spectral property (\ref{B.3}) we see that there is a duality between the $+$ family and the $-$ family. Conventionally we refer to the eigenvectors of the $+$ family as $\left( u_k, v_k \right)$ with eigenvalue $\varepsilon_k$ and the eigenvectors of the $-$ family are expressed as $\left( v^{*}_k, u^{*}_k \right)$ with eigenvalue $-\varepsilon_k$. Generally, the $0$-family is composed by only two member of the form $\left( \phi_0, 0 \right)$ and $\left( 0, \phi^{*}_0 \right)$. One can check that these two vectors are the only eigenvectors of $\mathcal{L}_k$ with zero eigenvalue. In the case of the Gutzwiller equations of motion (\ref{3.22}) the $0$-family vectors are precisely given by the ground-state distribution $c^0_n$ in the form $\left( \underline{c}_0, 0 \right)$ and $\left( 0, \underline{c}^{*}_0 \right)$. \\
As this special type of vectors are also eigenmodes of $\mathcal{L}^{\dagger}_k$ with zero eigenvalue, they are actually their own conjugate vectors. From equation (\ref{B.6}) it follows then the important property:
\begin{equation}\label{B.10}
\langle \phi_0 | u_k \rangle = \langle \phi^{*}_k | v_k \rangle = 0
\end{equation}
for all the eigenmodes $k$ in the $+$ family. \\
This final result explains the choice (\ref{4.3}) of perturbing the Gutzwiller ground state parameters $c^0_n$ along orthogonal directions when studying the corresponding quantum fluctuations.

\chapter{Drude weight in the Mott phase}\label{D}
Let us consider the ground state wave function of the Mott insulator at $J = 0$:
\begin{equation}\label{C.1}
| \Psi_0 \rangle = \prod_{\mathbf{r}} | n_0{\left( \mu/U \right)} \rangle_{\mathbf{r}}
\end{equation}
where $n_0{\left( \mu/U \right)}$ is the integer-filling function specified in equation (\ref{1.8}). The corresponding Hamiltonian reads:
\begin{equation}\label{C.2}
\hat{H}_0 \equiv \sum_{\mathbf{r}} \left( \frac{U}{2} \hat{a}^{\dagger}_{\mathbf{r}} \, \hat{a}^{\dagger}_{\mathbf{r}} \, \hat{a}_{\mathbf{r}} \, \hat{a}_{\mathbf{r}} - \mu \, \hat{a}^{\dagger}_{\mathbf{r}} \, \hat{a}_{\mathbf{r}} \right)
\end{equation}
The ground-state properties of the Mott lobes can be investigated perturbatively when the relative tunnelling rate $J/U$ of the hopping term:
\begin{equation}\label{C.3}
\hat{H}_1 \equiv \sum_{\langle \mathbf{r}, \mathbf{s} \rangle} \left( \hat{a}^{\dagger}_{\mathbf{r}} \, \hat{a}_{\mathbf{s}} + \hat{a}^{\dagger}_{\mathbf{s}} \, \hat{a}_{\mathbf{r}} \right)
\end{equation}
is small compared to the critical threshold $\left( J/U \right)_c$ at a given chemical potential $\mu/U$. Then the perturbative solution of the Bose-Hubbard model can be obtained in powers of $J/U$ and goes under the name of strong-coupling expansion. \\
Since the Mott ground state is not degenerate except for integer values of the chemical potential $\mu/U$, we can employ non-degenerate perturbation theory. Given a generic Hamiltonian correction $\hat{V}$ and the corresponding expansion parameter $\lambda$ measuring the strength of the perturbation, the normalized second-order correction of the ground state wave function reads:
\begin{equation}\label{C.4}
\begin{aligned}
| \Psi{\left( \lambda \right)} \rangle &\approx | \Psi_0 \rangle + \lambda \sum_{k \neq \Psi_0} \frac{\langle k | \hat{V} | \Psi_0 \rangle}{E_0 - E_k} | k \rangle + \lambda^2 \sum_{k \neq \Psi_0} \sum_{p \neq \Psi_0} \frac{\langle k | \hat{V} | p \rangle \langle p | \hat{V} | \Psi_0 \rangle}{\left( E_0 - E_k \right) \left( E_0 - E_p \right)} | k \rangle + \\
&- \lambda^2 \sum_{k \neq \Psi_0} \frac{\langle \Psi_0 | \hat{V} | \Psi_0 \rangle \langle k | \hat{V} | \Psi_0 \rangle}{\left( E_0 - E_k \right)^2} | k \rangle - \frac{\lambda^2}{2} | \Psi_0 \rangle \sum_{k \neq \Psi_0} \frac{\langle \Psi_0 | \hat{V} | k \rangle \langle k | \hat{V} | \Psi_0 \rangle}{\left( E_0 - E_k \right)^2}
\end{aligned}
\end{equation}
where $k$ and $p$ label the excited states of the unperturbed Hamiltonian, $E_0$ is the ground state energy of the unperturbed state and $E_k$ is the energy of the $k$-th excited state. \\
In the case of the pure Mott state, the perturbation $\hat{H}_1$ can couple only the ground state to excited states with one particle-hole creation involving two neighbouring sites. It follows that the third and fourth term on the RHS of (\ref{C.4}) vanish identically, hence:
\begin{equation}\label{C.5}
| \Psi{\left( J/U \right)} \rangle = | \Psi_0 \rangle - J \sum_{k \neq \Psi_0} \frac{\langle k | \hat{H}_1 | \Psi_0 \rangle}{E_0 - E_k} | k \rangle - \frac{J^2}{2} | \Psi_0 \rangle \sum_{k \neq \Psi_0} \frac{\left| \langle k | \hat{H}_1 | \Psi_0 \rangle \right|^2}{\left( E_0 - E_k \right)^2} + \text{O}{\left( J^3 \right)}
\end{equation}
where now $k$ labels only states characterized by a particle-hole perturbation at neighbouring sites. A straightforward calculation leads to:
\begin{equation}\label{C.6}
\begin{aligned}
&E_k = E{\left[ n_0{\left( \mu/U \right)} + 1 \right]} + E{\left[ n_0{\left( \mu/U \right)} - 1 \right]} - 2 \, E{\left[ n_0{\left( \mu/U \right)} \right]} = U \\
&\hspace{2cm} \langle k | \hat{H}_1 | \Psi_0 \rangle = \sqrt{n_0{\left( \mu/U \right)} \left[ n_0{\left( \mu/U \right)} + 1 \right]}
\end{aligned}
\end{equation}
where $n_0$ is the Mott integer filling, so that we obtain:
\begin{equation}\label{C.7}
| \Psi{\left( J/U \right)} \rangle = | \Psi_0 \rangle + \left( \frac{J}{U} \right) \sqrt{n_0 \left( n_0 + 1 \right)} \sum_{k \neq \Psi_0} | k \rangle - \frac{1}{2} \left( \frac{J}{U} \right)^2 z \, n_0 \left( n_0 + 1 \right) \, | \Psi_0 \rangle + \text{O}{\left[ \left( J/U \right)^3 \right]}
\end{equation}
where we have replaced $n_0 \equiv n_0{\left( \mu/U \right)}$ for simplicity. \\
The average value of the kinetic energy operator $\hat{T}$ calculated over the perturbative ground state (\ref{C.7}) is provided by:
\begin{equation}\label{C.8}
\langle \Psi{\left( J/U \right)} | \hat{T} | \Psi{\left( J/U \right)} \rangle = -J \langle \Psi{\left( J/U \right)} | \hat{H}_1 | \Psi{\left( J/U \right)} \rangle \approx -2 \, M \, J \left( \frac{J}{U} \right) z \, n_0 \left( n_0 + 1 \right)
\end{equation}
where we have retained only the first-order term because second-order terms do not appear from the expansion of $| \Psi{\left( J/U \right)} \rangle$. Here $M$ is the lattice volume, while $z$ is the coordination number. It follows that the kinetic contribution to the Drude weight is given by:
\begin{equation}\label{C.9}
f^{\left( 1 \right)}_s = -\frac{1}{2 \, M \, J} = \left( \frac{J}{U} \right) z \, n_0 \left( n_0 + 1 \right) + \text{O}{\left[ \left( J/U \right)^3 \right]}
\end{equation}
On the other hand, the conductivity fraction due to current correlations reads:
\begin{equation}\label{C.10}
f^{\left( 2 \right)}_s = \frac{1}{M \, J} \sum_{\nu} \frac{\left| \langle \Psi_{\nu}{\left( J/U \right)} | \hat{J} | \Psi{\left( J/U \right)} \rangle \right|^2}{E_{\nu}{\left( J/U \right)} - E{\left( J/U \right)}}
\end{equation}
where $| \Psi_{\nu}{\left( J/U \right)} \rangle$ indicates the $\nu$-th perturbed excited state, $E_{\nu}{\left( J/U \right)}$ is the corresponding energy and $\hat{J}$ is the current operator given by equation (\ref{2.7}). Since $\hat{J} \propto J$ and the argument of the sum in (\ref{C.10}) is of the second order in the hopping strength, it is sufficient to consider the zeroth-order expansions of the perturbed eigenstates and energies:
\begin{equation}\label{C.11}
\begin{aligned}
&\hspace{1.2cm} | \Psi{\left( J/U \right)} \rangle = | \Psi_0 \rangle + \text{O}{\left( J/U \right)} \\
&\hspace{1.2cm} | \Psi_{\nu}{\left( J/U \right)} \rangle = | \nu \rangle + \text{O}{\left( J/U \right)} \\
&E_{\nu}{\left( J/U \right)} - E{\left( J/U \right)} = U + \text{O}{\left[ \left( J/U \right)^2 \right]}
\end{aligned}
\end{equation}
therefore:
\begin{equation}\label{C.12}
f^{\left( 2 \right)}_s = \frac{M \, J^2 \, z \, n_0 \left( n_0 + 1 \right)}{M \, J \, U} + \text{O}{\left[ \left( J/U \right)^3 \right]} = \left( \frac{J}{U} \right) z \, n_0 \left( n_0 + 1 \right) + \text{O}{\left[ \left( J/U \right)^3 \right]}
\end{equation}
We can immediately see that $f^{\left( 1 \right)}_s = f^{\left( 2 \right)}_s$ up to the third order in the relative hopping rate, so that the Drude weight $D_c = f^{\left( 1 \right)}_s - f^{\left( 2 \right)}_s$ vanishes at the same order of expansion. Since in the Mott phase exact calculations yield $D_c = 0$, we can conclude that perturbation theory provides the same result at any order.

\medskip

In conclusion, a simple perturbative calculations has allowed to verify that the conductivity of the Mott insulator vanishes identically because of the counterbalance between the kinetic and current contributions to the Drude weight of the lattice. It is important to observe that this result is crucially due to the fact that the perturbative calculation in the Mott phase is able to distinguish the dimension of the particle-hole pair locally, while the Bogoliubov and Gutzwiller approaches lose this information through global translational invariance. \\
The perturbative dimension of the particle-hole couple can be deduced from the calculation of the two-point correlation function $g_1{\left( \left| \mathbf{r}_1 - \mathbf{r}_2 \right| \right)}$ up to the third order in $J/U$:
\begin{equation}\label{C.13}
\small
\begin{aligned}
&g_1{\left( \left| \mathbf{r}_1 - \mathbf{r}_2 \right| \right)} = \frac{\langle \hat{a}^{\dagger}_{\mathbf{r}_1} \, \hat{a}_{\mathbf{r}_2} \rangle}{\langle \hat{a}^{\dagger}_{\mathbf{r}_1} \, \hat{a}_{\mathbf{r}_1} \rangle} = \frac{\langle \Psi{\left( J/U \right)} | \hat{a}^{\dagger}_{\mathbf{r}_1} \, \hat{a}_{\mathbf{r}_2} | \Psi{\left( J/U \right)} \rangle}{\langle \Psi{\left( J/U \right)} | \hat{a}^{\dagger}_{\mathbf{r}_1} \, \hat{a}_{\mathbf{r}_1} | \Psi{\left( J/U \right)} \rangle} \approx \\
&\approx
\begin{cases}
2 \left( J/U \right) \left( n_0 + 1 \right) + \left( J/U \right)^3 \left( n_0 + 1 \right) \left[ 16 \, z - 34 + \left( 76 \, z - 157 \right) n_0 \left( n_0 + 1 \right) \right]/3 \quad &\left| \mathbf{r}_1 - \mathbf{r}_2 \right| = 1 \\
3 \left( J/U \right)^2 \left( n_0 + 1 \right) \left( 2 \, n_0 + 1 \right) \quad &\left| \mathbf{r}_1 - \mathbf{r}_2 \right| = 2 \\
4 \left( J/U \right)^3 \left( n_0 + 1 \right) \left( 5 \, n^2_0 + 5 \, n_0 + 1 \right) \quad &\left| \mathbf{r}_1 - \mathbf{r}_2 \right| = 3
\end{cases}
\end{aligned}
\end{equation}
where we have used the exact result $\langle a^{\dagger}_{\mathbf{r}} \, a_{\mathbf{r}} \rangle = n_0$. The introduction of higher-order contributions allows to identify how the perturbation theory accounts for the enlargement of the particle-hole couple dimension beyond adjacent lattice sites as the hopping rate is increased starting from the deep Mott phase, despite a vanishing Drude weight. Then, we can imagine that the two-point correlation range increases with $J/U$ until reaching a macroscopic magnitude at the superfluid transition, where long-range order occurs and $D_c > 0$. This is exactly what we have observed when calculating $g_1{\left( \left| \mathbf{r}_1 - \mathbf{r}_2 \right| \right)}$ within the Gutzwiller formalism.

\chapter{Drude weight and superfluid fraction}\label{E}
\markboth{\MakeUppercase{C. Drude weight and superfluid fraction}}{\MakeUppercase{C. Drude weight and superfluid fraction}}
\section{Current response}
Let us imagine to perturb the Bose-Hubbard system by the application of a spatially and temporally varying phase $\theta_x{\left( \mathbf{r}, t \right)}$ along the $x$-direction in the form of a unitary transformation. In accordance with equation (\ref{2.4}) the hopping part of the Hamiltonian is modified by the usual Peierls phase factors given by the application of a gauge potential:
\begin{equation}\label{D.1}
\hat{H}_{\theta} = -J \sum_{\langle \mathbf{r}, \mathbf{s} \rangle_x} \left[ e^{i \, \theta{\left( \mathbf{r} \right)}} \hat{\psi}^{\dagger}{\left( \mathbf{s} \right)} \, \hat{\psi}{\left( \mathbf{r} \right)} + e^{-i \, \theta{\left( \mathbf{r} \right)}} \hat{\psi}^{\dagger}{\left( \mathbf{r} \right)} \, \hat{\psi}{\left( \mathbf{s} \right)} \right] + \frac{U}{2} \sum_{\mathbf{r}} \hat{n}{\left( \mathbf{r} \right)} \left[ \hat{n}{\left( \mathbf{r} \right)} - 1 \right] - \mu \sum_{\mathbf{r}} \hat{n}{\left( \mathbf{r} \right)}
\end{equation}
The expansion of equation (\ref{D.1}) to order $\theta^2{\left( \mathbf{r} \right)}$ gives:
\begin{equation}\label{D.2}
\hat{H}_{\theta} \simeq \hat{H}_0 - \sum_{\mathbf{r}} \left[ \theta{\left( \mathbf{r} \right)} \, \hat{J}^p_x{\left( \mathbf{r} \right)} - \frac{\theta^2{\left( \mathbf{r} \right)}}{2} \hat{K}_x{\left( \mathbf{r} \right)} \right]
\end{equation}
where $\hat{H}_0$ is the original Bose-Hubbard Hamiltonian, while the operators:
\begin{equation}\label{D.3}
\hat{K}_x{\left( \mathbf{r} \right)} = -J \left[ \hat{\psi}^{\dagger}{\left( \mathbf{r} + \mathbf{e}_x \right)} \, \hat{\psi}{\left( \mathbf{r} \right)} + \hat{\psi}^{\dagger}{\left( \mathbf{r} \right)} \, \hat{\psi}{\left( \mathbf{r} + \mathbf{e}_x \right)} \right]
\end{equation}
\begin{equation}\label{D.4}
\hat{J}^p_x{\left( \mathbf{r} \right)} = i \, J \left[ \hat{\psi}^{\dagger}{\left( \mathbf{r} + \mathbf{e}_x \right)} \, \hat{\psi}{\left( \mathbf{r} \right)} - \hat{\psi}^{\dagger}{\left( \mathbf{r} \right)} \, \hat{\psi}{\left( \mathbf{r} + \mathbf{e}_x \right)} \right]
\end{equation}
are the local kinetic and current operator respectively associated to the lattice links oriented along the $x$-direction. The whole current operator consists of the two terms provided by the functional derivative of the perturbed Hamiltonian with respect to the gauge field given by the phase variation:
\begin{equation}\label{D.5}
\hat{J}_x{\left( \mathbf{r} \right)} = -\frac{\delta \hat{H}_{\theta}}{\delta \theta{\left( \mathbf{r} \right)}} = \hat{J}^p_x{\left( \mathbf{r} \right)} - \theta{\left( \mathbf{r} \right)} \, \hat{K}_x{\left( \mathbf{r} \right)}
\end{equation}
The linear current response produced by the phase shift $\theta{\left( \mathbf{r}, t \right)}$ is given by the expression
\begin{equation}\label{D.6}
\langle \hat{J}_x{\left( \mathbf{q}, \omega \right)} \rangle = \left[ \Lambda_J{\left( \mathbf{q}, \omega \right)} - \langle \hat{K}_x{\left( \mathbf{r} \right)} \rangle \right] \tilde{\theta}{\left( \mathbf{q}, \omega \right)}
\end{equation}
which can be derived by the use of linear response theory. Here $\langle \hat{K}_x{\left( \mathbf{r} \right)} \rangle$ is the average kinetic energy per site divided by the number of linear lattice dimensions $z = 2$, $\tilde{\theta}{\left( \mathbf{q}, \omega \right)}$ is the Fourier transform of the phase shift potential and $\Lambda_J{\left( \mathbf{q}, \omega \right)}$ is the current-current response function:
\begin{equation}\label{D.7}
\Lambda_J{\left( \mathbf{q}, \omega \right)} = -i \int_{0}^{\infty} d\tau \, e^{-\eta \, \tau }e^{-i \, \omega \, \tau} \int d\mathbf{r} \, e^{-i \, \mathbf{q} \cdot \mathbf{r}} \big\langle \left[ \hat{J}^p_x{\left( \mathbf{r}, \tau \right)}, \hat{J}^p_x{\left( \mathbf{0}, 0 \right)} \right] \big\rangle
\end{equation}
Adopting the Gutzwiller description, the expansion (\ref{4.21}) allows again to obtain:
\begin{equation}\label{D.8}
\begin{aligned}
\langle \hat{K}_x{\left( \mathbf{r} \right)} \rangle &= -J \left\{ \left| \psi_0 \right|^2 + \frac{1}{M} \sum_{\alpha > 0} \sum_{\mathbf{k}} \left[ \left( \left| U_{\alpha, \mathbf{k}} \right|^2 + \left| V_{\alpha, \mathbf{k}} \right|^2 \right) \langle \hat{b}^{\dagger}_{\alpha, \mathbf{k}} \,\hat{b}_{\alpha, \mathbf{k}} \rangle + \left| V_{\alpha, \mathbf{k}} \right|^2 \right] \cos{\left( k_x \right)} \right\} \longrightarrow \\
&\xrightarrow[T = 0]{} -J \left[ \left| \psi_0 \right|^2 + \frac{1}{M} \sum_{\alpha > 0} \sum_{\mathbf{k}} \left| V_{\alpha, \mathbf{k}} \right|^2 \cos{\left( k_x \right)} \right]
\end{aligned}
\end{equation}
while the second-order expansion of (\ref{C.7}) reads:
\begin{equation}\label{D.9}
\Lambda_J{\left( \mathbf{q}, \omega \right)} = 4 \, J^2 \, \left| \psi_0 \right|^2 \left[ 1 - \cos{\left( q_x \right)} \right] \sum_{\alpha > 0} \frac{\left( \left| U_{\alpha, \mathbf{q}} \right|^2 + \left| V_{\alpha, \mathbf{q}} \right|^2 \right) \omega_{\alpha, \mathbf{q}}}{\omega^2 - \omega^2_{\alpha, \mathbf{q}}}
\end{equation}
The following step consists in analysing the behaviour of the average current response (\ref{D.6}) and the current-current susceptibility (\ref{D.7}) in the Mott and superfluid phase depending on the order in which the bulk and static limits are approached.

\section{Mott phase}
As far as Gaussian fluctuations are included in the theory, we notice that the current susceptibility vanishes identically:
\begin{equation}\label{D.10}
\Lambda^{\text{MI}}_J{\left( \mathbf{q}, \omega \right)} = 0
\end{equation}
because of the condition $\psi_0 = 0$. The Drude weight is defined as the coefficient of the delta function contribution to the real part of the conductivity $\sigma_{x x}{\left( \mathbf{q}, \omega \right)}$ in the uniform case $\mathbf{q} = \mathbf{0}$. As briefly mentioned in Chapter \ref{Bogoliubov}, $\sigma_{x x}{\left( \mathbf{q} = \mathbf{0}, \omega \right)}$ is formally given by the current response to a an acceleration potential applied to the lattice:
\begin{equation}\label{D.11}
\tilde{a}_x{\left( \mathbf{q} = \mathbf{0}, \omega \right)} = i \, \omega \, \tilde{\theta}\left( \mathbf{q} = \mathbf{0}, \omega \right)
\end{equation}
so that:
\begin{equation}\label{D.12}
\sigma_{x x}{\left( \mathbf{q}, \omega \right)} = \frac{\langle \hat{K}_x{\left( \mathbf{r} \right)} \rangle}{i \left( \omega + i \delta \right)}
\end{equation}
by analytic continuation. Considering now the static limit $\omega \to 0$, it follow that the Drude weight is given by:
\begin{equation}\label{D.13}
D_{\text{MI}} = -\pi \langle \hat{K}_x{\left( \mathbf{r} \right)} \rangle
\end{equation}
which is proportional to the same quantity calculated in Section \ref{superfluid_fraction} when we searched for the superfluid fraction of the system. \\
It is important to notice that, besides taking into account the value of $\psi_0$, $\Lambda^{\text{MI}}_J{\left( \mathbf{q}, \omega \right)}$ would vanish also when considering the limits $\mathbf{q} \to \mathbf{0}$ and $\omega \to 0$, no matter their order. It follows that the kinetic contribution (\ref{D.13}) is not counterbalanced by the current term and we obtain the incorrect result $D_{MI} > 0$, which is incompatible with the fact that the Mott spectrum is gapped. This outcome is a clearcut contradiction of the results found in the case of the half-filled repulsive Hubbard model, which displays a spin-density-wave gap $\Delta_{SDW}$ and where the kinetic and current terms cancel each other perfectly at the mean-field level \cite{scalapino_white_zhang_1, scalapino_white_zhang_2}. \\
Having verified that this unexpected results is not due to finite size effects, we can only conclude that our second-order quantum theory is not able to predict the conductive properties of the Mott phase. \\
As highlighted in Chapter \ref{quantum_corrections}, this problem could be fixed by considering higher order contributions to the Gutzwiller fluctuations in a self-consistent treatment.

\section{Superfluid phase}
\subsection{Drude weight}
Let us consider the zero-momentum value of (\ref{D.9}) in the superfluid phase. \\
Observing that:
\begin{equation}\label{D.14}
1 - \cos{\left( q_x \right)} \sim \frac{q^2_x}{2} \quad , \quad \left| U_{1, \mathbf{q}} \right|^2 + \left| V_{1, \mathbf{q}} \right|^2 \sim \frac{C_{UV}}{\left| \mathbf{q} \right|} \quad , \quad \omega_{1, \mathbf{q}} \sim c_s \left| \mathbf{q} \right| \quad \text{as} \quad \mathbf{q} \to \mathbf{0}
\end{equation}
In the uniform limit we always obtain that $\Lambda_J{\left( \mathbf{q} \to \mathbf{0}, \omega \right)} = 0$ independently of the momentum direction along which the limit is performed. It follows that:
\begin{equation}\label{D.15}
D_{\text{SF}} = -\pi \langle \hat{K}_x{\left( \mathbf{r} \right)} \rangle
\end{equation}
As expected, the Drude weight does not vanish in the superfluid phase and is again provided by the average value of the hopping energy per site within the Gutzwiller framework.

\subsection{Superfluid fraction}
The superfluid fraction is usually defined as the stiffness of the system to the application of a static phase twist over its entire length, for example by the imposition of twisted boundary conditions. \\
The static limit of (\ref{D.9}) reads:
\begin{equation}\label{D.16}
\Lambda_J{\left( \mathbf{q}, \omega = 0 \right)} = 4 \, J^2 \, \left| \psi_0 \right|^2 \left[ \cos{\left( q_x \right)} - 1 \right] \sum_{\alpha > 0} \frac{\left( \left| U_{\alpha, \mathbf{q}} \right|^2 + \left| V_{\alpha, \mathbf{q}} \right|^2 \right)}{\omega_{\alpha, \mathbf{q}}}
\end{equation}
Once again, the current response vanishes when the uniform limit is approached transversally with respect to the direction of phase variation:
\begin{equation}\label{D.17}
\Lambda_J{\left( q_x = 0, q_{i \neq x} \to 0, \omega = 0 \right)} = 0
\end{equation}
while a different result emerges in the longitudinal case. Recalling the asymptotic behaviours (\ref{D.14}) and the low-energy expansion of the same quantities for higher excitation modes, we obtain that only the Goldstone mode produces a finite contribution to the current response. In particular:
\begin{equation}\label{D.18}
\Lambda_J{\left( q_x \to 0, q_{i \neq x} = 0, \omega = 0 \right)} = -\frac{2 \, J^2 \, \left| \psi_0 \right|^2 C_{UV}}{c_s} = -\sqrt{2 \, J^3 \, \kappa} \, \psi_0 \, C_{UV}
\end{equation}
where we have introduced the general expression of the sound velocity (\ref{3.44}) with $\hbar = 1$ and $\kappa = \frac{\dpar \langle \hat{n} \rangle}{\dpar \mu}$ is the compressibility. \\

It is worth exploring the limiting values of (\ref{D.18}) in the Bose-Hubbard phase diagram. \\
\begin{itemize}
	\item In the weakly-interacting limit:
	\begin{equation}\label{D.19}
	c^B_s = \sqrt{2 \, J \, U} \psi_0 \quad , \quad C_{UV} = \sqrt{\frac{U}{2 \, J}} \psi_0
	\end{equation}
	so that:
	\begin{equation}\label{D.20}
	\Lambda^B_J{\left( q_x \to 0, q_{i \neq x} = 0, \omega = 0 \right)} = -J \left| \psi_0 \right|^2
	\end{equation}
	which is exactly equal to minus the mean kinetic energy in the non-interacting limit and cancels the contribution of $\langle \hat{K}_x{\left( \mathbf{r} \right)} \rangle$.
	\item In correspondence of the tip of the Mott lobes:
	\begin{equation}\label{D.21}
	c^{tip}_s = U \sqrt{\left( \frac{J}{U} \right)^{tip}_c} \left[ \overline{n} \left( \overline{n} + 1 \right) \right]^{1/4}
	\end{equation}
	hence:
	\begin{equation}\label{C.22}
	\Lambda^{tip}_J{\left( q_x \to 0, q_{i \neq x} = 0, \omega = 0 \right)} = -\frac{2 \, J \, \sqrt{\left( J/U \right)^{tip}_c} \, \left| \psi_0 \right|^2 C^{tip}_{UV}}{\left[ \overline{n} \left( \overline{n} + 1 \right) \right]^{1/4}}
	\end{equation}
\end{itemize}
It is important to notice that $\Lambda_J{\left( q_x \to 0, q_{i \neq x} = 0, \omega = 0 \right)} \to 0$ for every point of the critical line between the Mott and the superfluid phase. This occurs because $c_s$ vanishes elsewhere proportionally to $\psi_0$, except for the $O{\left( 2 \right)}$-invariant points of the Mott lobes, hence $\Lambda^{tip}_J{\left( q_x \to 0, q_{i \neq x} = 0, \omega = 0 \right)} \sim -\left| \psi_0 \right|^2$.

\medskip

In the following we report the 3D numerical results for $\left| -\langle \hat{K}_x{\left( \mathbf{r} \right)} \rangle \right|$ and $\left| \Lambda_J{\left( q_x \to 0, q_{i \neq x} = 0, \omega = 0 \right)} \right|$ normalized by the usual density norm at zero temperature:
\begin{equation}\label{D.23}
N = \psi^2_0 + \frac{1}{M} \sum_{\mathbf{k}} \sum_{\alpha > 0} \left| V_{\alpha, \mathbf{k}} \right|^2
\end{equation}
as a function of $2 d J/U$ for three different values of the chemical potential. Obviously, the normalization of $\left| \Lambda_J{\left( q_x \to 0, q_{i \neq x} = 0, \omega = 0 \right)} \right|$ takes into account only the condensate depletion along the $x$-direction.
\begin{figure}[!htp]
	\begin{minipage}[b]{7.7cm}
		\centering
		\includegraphics[width=8.8cm]{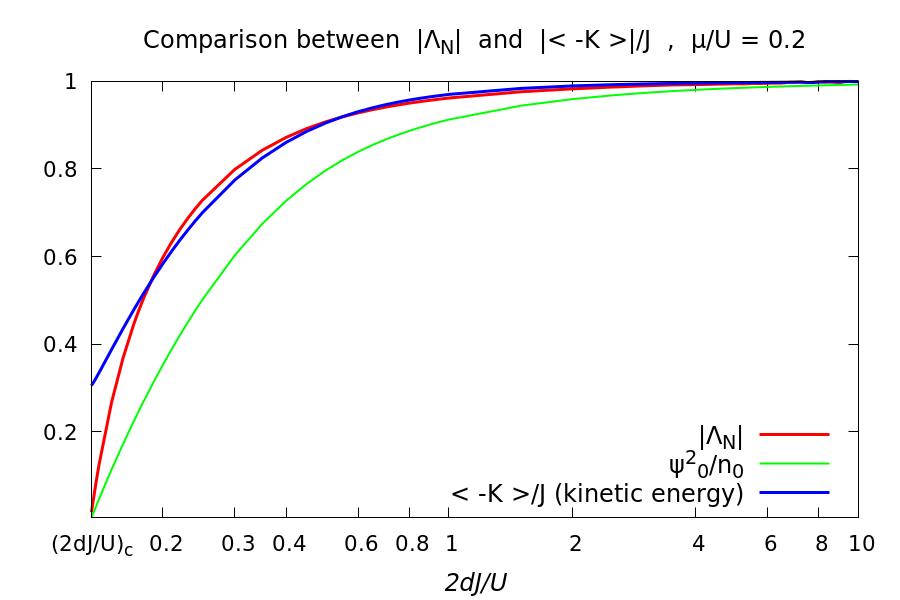}
		\caption{Longitudinal uniform limit of the static current response function $\Lambda_J{\left( q_x \to 0, q_{i \neq x}, \omega = 0 \right)}$ normalized by $N$ in the superfluid phase as a function of $2 d J/U$ for $\mu/U = 0.2$.}
		\label{FigureD1}
	\end{minipage}
	\ \hspace{2mm} \hspace{2mm} \
	\begin{minipage}[b]{8.5cm}
		\centering
		\includegraphics[width=8.8cm]{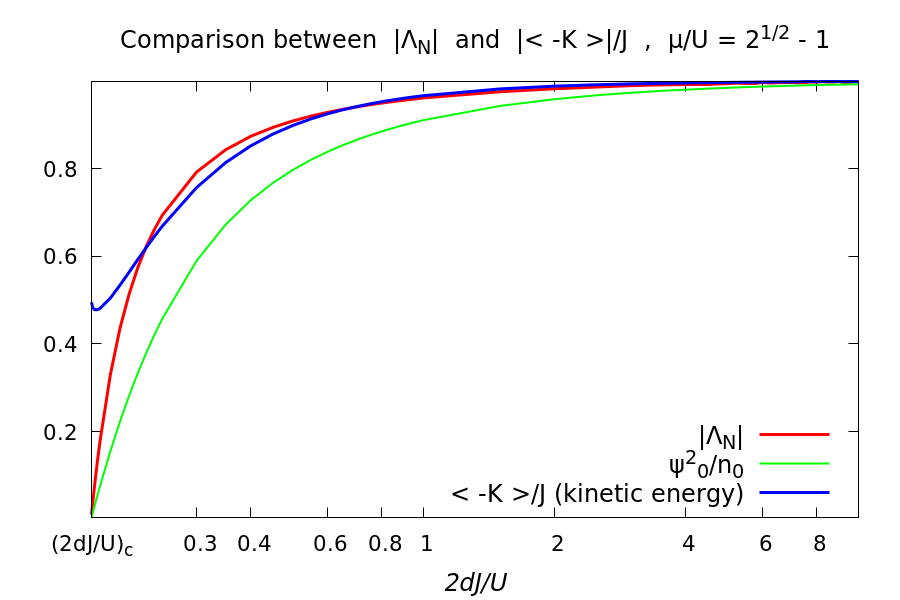}
		\caption{Longitudinal uniform limit of the static current response function $\Lambda_J{\left( q_x \to 0, q_{i \neq x}, \omega = 0 \right)}$ normalized by $N$ in the superfluid phase as a function of $2 d J/U$ for $\mu/U = \sqrt{2} - 1$ (across the tip of the first Mott lobe).}
		\label{FigureD2}
	\end{minipage}
\end{figure}
\begin{figure}[!htp]
	\centering
	\includegraphics[width=8.8cm]{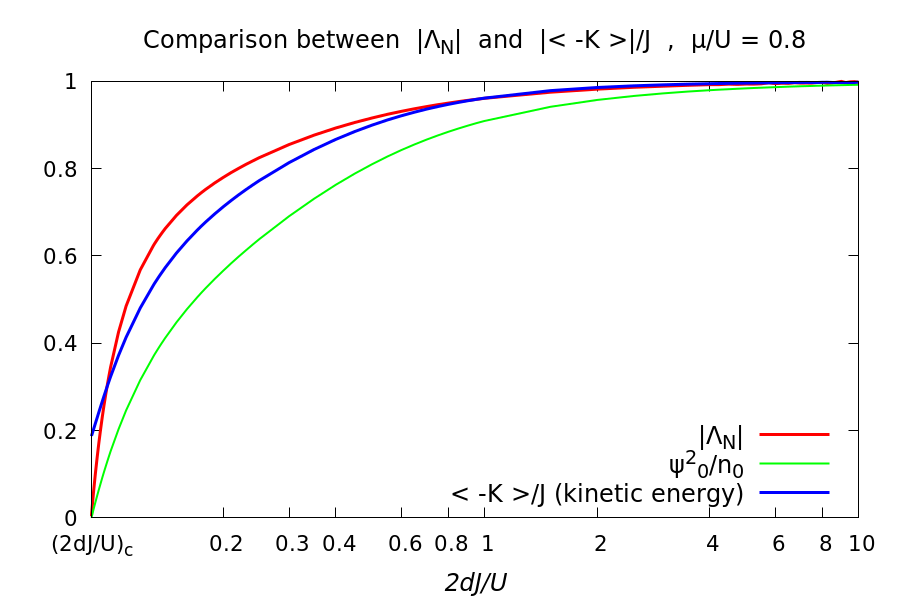}
	\caption{Longitudinal uniform limit of the static current response function $\Lambda_J{\left( q_x \to 0, q_{i \neq x}, \omega = 0 \right)}$ normalized by $N$ in the superfluid phase as a function of $2 d J/U$ for $\mu/U = 0.8$.}
	\label{FigureD3}
\end{figure}
Remarkably, the quantity $\Lambda_N \equiv \Lambda_J{\left( q_x \to 0, q_{i \neq x} = 0, \omega = 0 \right)}/(J \, N)$ turns to be always larger than the mean-field condensate fraction as in the case of the quantity $f_s$ calculated in Section \ref{superfluid_fraction}. Moreover, we observe that $\Lambda_N \equiv \Lambda_J{\left( q_x \to 0, q_{i \neq x}  0, \omega = 0 \right)}$ and $\langle -\hat{K}_x{\left( \mathbf{r} \right)} \rangle$ cancel each other in the regime of weak interactions $2 \, d \, J/U \gtrsim 1$, so that the longitudinal current response (\ref{D.6}) vanishes identically, in analogy with the preservation of gauge invariance along the longitudinal direction for the Meissner effect \cite{scalapino_white_zhang_1, scalapino_white_zhang_2}.

\section{Conclusive observations}
The calculations performed above have allowed to identify the distinction between the Drude weight and the superfluid fraction in the two phases of the Bose-Hubbard model into the value of the current response $\Lambda_J{\left( \mathbf{q}, \omega \right)}$. \\
In particular, we have found that in the superfluid phase the Drude weight can be calculated through the average value of the hopping energy, while an incorrect result is obtained in the Mott regime, presumably because higher-order corrections due to the lack of self-consistency are neglected within our approach. It follows that the second-order estimations of both the Drude weight and the superfluid density are not able to distinguish between the two phases within our Gutzwiller description, as we have anticipated in Section \ref{superfluid_fraction_discussion}. \\
Interestingly, Figures \ref{FigureD1}-\ref{FigureD3} show that the longitudinal limit $\Lambda_J{\left( q_x \to 0, q_{i \neq x} = 0, \omega = 0 \right)}$ perfectly cancels the kinetic contribution due to $\langle \hat{K}_x{\left( \mathbf{r} \right)} \rangle$ in the Bogoliubov limit, so that the current response turns to be suppressed along the direction of the phase twist. On the other hand, we obtain a non-vanishing value for the superfluid fraction $f_s$ when the thermodynamic limit is approached transversally. \\
It is also important to notice that $\left| \Lambda_J{\left( q_x \to 0, q_{i \neq x} = 0, \omega = 0 \right)} \right|$ and $\left| \langle -\hat{K}_x{\left( \mathbf{r} \right)} \rangle \right|$ exhibit a significant discrepancy exactly in the strongly-interacting regime, where the Gutzwiller corrections have a larger amplitude with respect to the mean-field density. \\
This fact confirms what have been already observed while commenting the results on the quantum depletion and the Drude weight in the Mott phase: a number of evidences suggest that Gaussian fluctuations around the Gutzwiller mean-field theory account for the Bogoliubov predictions and go in the right direction in predicting the properties of the strongly-interacting regime, but are not adequate to capture the vanishing of the superfluid fraction at the transition and the absence of conductivity in the Mott phase. \\
However, we can see how the role of higher-order corrections emerges when more complex quantities are calculated. In this sense, determining the superfluid fraction provides a first measure of the reliability of our quantum theory near the critical transition.

\newpage
\pagestyle{plain}

\end{document}